\documentclass[graybox,envcountchap,sectrefs]{svmono}

\usepackage{type1cm}
\usepackage{mathtools}
\usepackage{hyperref}
\usepackage{amsmath}
\usepackage{inconsolata}
\usepackage{natbib}
\usepackage{makeidx}         
\usepackage{graphicx}        
\usepackage{multicol}        
\usepackage[bottom]{footmisc}
\usepackage{physics}
\usepackage[nameinlink]{cleveref}
\usepackage{booktabs}
\usepackage{newtxtext}       %
\usepackage[varvw]{newtxmath}       
\usepackage[most]{tcolorbox}
\usepackage{mathrsfs}
\usepackage{varwidth}
\definecolor{colorboxcolor}{HTML}{006896}
\newtcolorbox[auto counter, number within=chapter]{ex}[2][]{
enhanced, breakable, left=0.1cm, right=0.1cm, attach boxed title to top left={yshift=-0.5mm}, colframe=colorboxcolor, colback=colorboxcolor!2,colbacktitle=colorboxcolor,
varwidth boxed title,coltitle=white,fonttitle=\bfseries, 
boxed title style={sharp corners, rounded corners=north},
rightrule=0pt, sharp corners, rounded corners=northwest, outer arc=0pt, arc = 0.3cm, leftrule=4pt, title={Example \thetcbcounter~({#2})}, before lower={\textcolor{colorboxcolor}{\textbf{Solution \thetcbcounter}}~}, #1}
\usepackage{float}
\usepackage{enumerate}
\usepackage{multirow}
\usepackage{listings}
\definecolor{codegreen}{rgb}{0,0.6,0}
\definecolor{codegray}{rgb}{0.5,0.5,0.5}
\definecolor{codepurple}{rgb}{0.58,0,0.82}
\definecolor{backcolour}{rgb}{0.96,0.96,0.93}
\definecolor{bluecol}{rgb}{0.11,0.67,0.84}
\lstdefinestyle{mystyle}{
    backgroundcolor=\color{backcolour},   
    commentstyle=\color{codegreen},
    keywordstyle=\color{bluecol},
    numberstyle=\tiny\color{codegray},
    stringstyle=\color{codepurple},
    basicstyle=\ttfamily\footnotesize,
    breakatwhitespace=false,         
    breaklines=true,                 
    captionpos=b,                    
    keepspaces=true,                 
    numbers=right,                    
    numbersep=1pt,                  
    showspaces=false,                
    showstringspaces=false,
    showtabs=false,                  
    tabsize=2
}
\makeindex             

\definecolor{ForestGreen}{RGB}{18, 159, 87}
\definecolor{RoyalBlue}{RGB}{0, 104, 150}
\definecolor{Cerulean}{RGB}{0, 173, 231}
\hypersetup{
    colorlinks=true,            
    linkcolor=RoyalBlue,        
    citecolor=ForestGreen,      
    urlcolor=Cerulean,          
    bookmarks=true,             
    bookmarksnumbered=true,     
    pdfstartview={FitH},        
    pdfpagemode={UseOutlines},  
    breaklinks=true             
}

\begin{document}

\author{Seyyed Ali Emami, Paolo Castaldi, Mohammad Narimani, Mahdi Shahrajabian, Mostafa Ezabadi, Alireza Ettelaei, Hosein Saadatmanesh, and Zahra Shahbazi}
\title{Neural Network-based Control Systems with Application to Flight Control }
\subtitle{From Classical Neural Control to Reinforcement Learning}
\maketitle

\frontmatter

\tableofcontents

\mainmatter
%
%
%
\chapter{Introduction}\label{intro} 
\section{Intelligent control systems}
Although the words \emph{Intelligence} and \emph{Autonomy} have been vastly employed interchangeably, they have an essential conceptual difference \citep{Clough.2002}. Different definitions have been given for both concepts in the literature \citep{Gottfredson.1997,Long.2009}. However, in a general view, intelligence may be defined as ``a very general mental capability that involves the ability to reason, plan, solve problems, think abstractly, comprehend complex ideas, learn quickly and learn from experience'' \citep{Long.2009}. On the other hand, ``the ability to generate one's own purposes without any instruction from outside'' can be interpreted as the autonomy of a system \citep{Clough.2002}.

Nevertheless, concerning these two general definitions, in some cases, distinguishing the high Level of Autonomy (LoA) from the low level of intelligence is not trivial at all.
   As systems become increasingly automated, they exhibit behaviors that resemble intelligence. This suggests a continuum between high levels of automation and simple, low-level intelligent systems.
Within the control theory framework, generally, the final goal is to develop an autonomous system (rather than an intelligent system), which can adequately fulfill a set of predefined missions. However, like the common literature, one may interpret the high level of autonomy of some Unmanned Aerial Vehicles (UAVs) as intelligence. \citep{emami.2022}. 
Accordingly, despite the conceptual difference between these two words, in the current study (like existing literature,) we will use the words Intelligence and Autonomy interchangeably \citep{emami.2022}.

Various criteria have been provided in the literature to specify a system's LoA \citep{Eze.2012}, especially a UAV. Despite the lack of a unique categorization \citep{Vagia.2016}, a beneficial division has been given in \citet{Clough.2002,Protti.2007} for UAVs. Considering it, the highest LoA for a single UAV (level $4$) indicates the self-accomplishment of an assigned tactical plan, where it is capable of on-board trajectory replanning, event-driven self-resource management, and compensating for most disturbances and faults in various flight conditions. This, in turn, requires different learning-based mechanisms in the whole control system, including the entire range from low-level control, such as attitude control, to high-level control, such as path planning. Such a scheme will be known as an Intelligent Control System (ICS) in this book.

Today's academic landscape boasts a plethora of advanced Control Mechanisms (CMs) stemming from diverse methodologies like neural networks, fuzzy systems  \citep{Sarabakha.2018b}, decision trees  \citep{Scheper.2016,Croon.2016}, reinforcement learning \citep{Hwangbo.2017}, different data-driven approaches \citep{Cheng.2011b,Ju.2013}, evolutionary algorithms \citep{Floreano.2005,Silva.2016}, etc.
Given such vast and varied CM explorations in scholarly literature, it is imperative to provide a comprehensive overview, tracing the journey of CMs from conceptual blueprints to their tangible implementations in today's world. Leading the way, the authors of \citet{Santoso.2018} offer insights into the multifaceted technical and operational aspects of CMs, exploring avenues like fuzzy logic, neural networks, evolutionary algorithms, collective intelligence tactics, and hybrid optimization methods.

However, an in-depth theoretical exposition was noticeably absent given the topic's vast scope. To bridge this void, our current examination pivots to Neural Network (NN)-based control systems, a predominant theme in contemporary research.
NNs have distinctly showcased their prowess in controller design processes and dynamic model identification. Their intrinsic ability (based on the universal approximation property) to estimate various non-linearities in dynamic frameworks sets them apart. Unlike traditional fuzzy control approaches, which heavily rely on expert knowledge and experiential data to construct their fuzzy rule base \citep{Lee.1990}, NNs intuitively understand system or controller dynamics' intricacies, even without prior knowledge. Their adaptability is commendable, harmoniously integrating with diverse adaptive learning methods.
NNs are applicable in both \emph{direct} and \emph{indirect} control structures. 
In direct control, NNs are directly used to construct the control command, while in indirect control, they are used to estimate unknown functions within the system equations, e.g., model uncertainties, which are then used to design the control command (See section \ref{dirvsindir} for more details).
Furthermore, thanks to their inherent parallel processing capabilities, neural architectures are especially suited for real-time implementations \citep{Campa.2002,Kacprzyk.2010b}.

The genesis of ICS in flying vehicles can be traced back to the 1990s when the world of flight control began embracing NNs as adaptive tools, responding dynamically to unexpected faults and unique flying conditions \citep{Tomayko.2003}.
Though the fusion of NNs with flight systems began in the early '90s, real-world applications employing dynamic NNs in practical flight controls only gained traction in 2001, stymied previously by technological and procedural barriers \citep{Steinberg.2005b}.
A monumental undertaking in this arena was the collaboration between NASA and Boeing, which spanned ten years, from 1999 to 2009, birthing the trailblazing Intelligent Flight Control System (IFCS) program  \citep{Schumann.2010,Kacprzyk.2010b,Kaminski.2017}. This endeavor was bifurcated into two pivotal phases. The first phase, focusing on indirect adaptive flight control, saw its inaugural flights using a modified F-15B model in 1999. These initial runs employed pre-configured NNs to estimate the aircraft's stability and control derivatives. In a significant evolution, the mutable cell structure of NNs was incorporated into the control scheme, granting online modifications to these estimated derivatives. Flights harnessing this online identification mode took to the skies in 2003
\citep{Williams.2004b}. Although closed-loop control simulations showed promise, genuine adaptive control was notably absent, as flight controls refrained from making full use of the identified models in actual flight \citep{Hageman.2003}.
In a shift of focus, the project's second phase was dedicated to crafting a direct adaptive control strategy, intertwining a dynamic inversion element with an online NN \citep{Smith.2010}. This stage's test flights began in 2006 and extended into 2008. These trials drew comparisons between performances with and without the dynamic NN augmentation, especially during structural challenges and control surface faults. Evaluation criteria ranged from pilot rating feedback to performance measurements. As underscored in \citet{Bosworth.2007,Burken.2009}, NN-based improvements notably overall pitch performance amidst structural damages. Conversely, control surface faults produced varied results from slight enhancement in pitch rate to occasional pilot-induced roll oscillations. This latter stage also saw refinements such as alternating NN inputs to mitigate issues like excessive correlation and high gain in the base design, introducing a weight decrement factor (during NN weight updates) to avoid over-fitting, and adding singular dead zones in the adaptation laws for more fluid operations. The 2009 findings, mirroring these refinements, exhibited significant advancements over the project's outset \citep{Burken.2009}. This ambitious journey concluded with the F-15B's retirement in January 2009 \citep{Kaminski.2017}.

The revelations from the IFCS initiative shed light on the potential concerns and challenges of implementing adaptive control mechanisms, particularly in real-world applications like human-crewed aircraft. Among these challenges were the complex intricacies of control design and unpredictable control behaviors in unexpected flight and fault conditions \citep{Hanson.2012,Hanson.2014}. In response to these findings, NASA embarked on a new journey in 2009 with the Integrated Resilient Aircraft Control (IRAC) project. A primary objective was to design adaptive control strategies that were both intuitive and effective while ensuring that these adaptive flight controls underwent thorough validation and verification for critical safety checks. A comprehensive discussion of this project is beyond the scope of this book. Nevertheless, inspired by the context at hand, our study seeks to deeply explore both direct and indirect NN-driven adaptive control methods, charting their evolution towards increased efficiency and reliability. Intriguingly, while direct and indirect adaptive control methodologies stem from different vantage points, they converge in their mathematical structures and dynamic NN update principles, echoing the spirit of NASA's IFCS initiative. Hence, this book organizes ICSs into two primary categories: model-based and model-free NN control strategies. Moreover, our investigation extends to various model-free control mechanisms grounded in alternative machine learning paradigms, including Reinforcement Learning (RL). In particular, we will investigate using the introduced structures in designing IFCS for applications such as satellites, quadrotors, and fixed-wing aircraft.

\section{Direct versus indirect adaptive control}\label{dirvsindir}
The scholarly landscape unveils many control approaches wherein NNs are utilized to address online optimization hurdles in the control domain \citep{Dalamagkidis.2011,Li.2016} or to imitate conventional controller actions  \citep{Efe.2011b}. Our discussion will center on control methodologies where NNs are directly integrated into the design phase, serving as intelligent mediators that embed an added layer of intelligence into the feedback mechanism. This integration can manifest in several ways: Some research initiatives employ NNs to estimate model uncertainties, subsequently shaping the control command. This approach is termed indirect neural control.
Training NNs can be executed in online mode using established learning techniques such as Feedback Error Learning (FEL) or be preset offline to provide a pre-trained dynamic system model. 
Within the realm of direct neural control, NNs are responsible for directly shaping the control command. 

For instance, considering model-based control strategies where system dynamics are expressed as $\dot{x}=f(x)+g(x)u$, where $f(x)$ and $g(x)$ signify unknown nonlinear functions of the system states and control input, the control command $u$ can be designed in two primary manners. The first approach employs NNs to estimate the unknown functions f and g, which are subsequently integrated into the designed control command. Alternatively, the second approach may directly determine the control command by evaluating $g^{-1}(\dot{x}_d-f)$) via an NN (where $x_d$ denotes the desired trajectory).
In scholarly discussions, the first approach is typically labeled as indirect adaptive control, while the second is recognized as a direct adaptive control technique \citep{Xu.2015,Xu.2018}. Nevertheless, as will be elaborated in Chapter 2, both these model-based control methods bear remarkable similarities in mathematical formulation and in the updating mechanisms of NNs. 

On the other hand, delineating model-independent adaptive control strategies brings about its own complexities, given the diverse direct and indirect adaptive control approaches detailed in prior research. A common model-free indirect ICS approach entails employing NNs to identify the complete system dynamics, creating an adaptive control strategy grounded in this newly elucidated model, which was identified online. Though derived from various design philosophies, direct model-free control predominantly revolves around two main paradigms in this book: adaptive optimal and RL-based control methodologies.

While the direct adaptive control is widely recognized for its simplicity and computational efficiency \citep{KarlJohanAstrom.2008}, its widespread application across various domains has also raised eyebrows, especially when applied to critical missions. In particular, aggressive learning, a common tactic in direct adaptive control, accelerates the reduction of dynamic inversion error by increasing the learning rate \citep{Gu.2019}.
However, high-gain control due to aggressive learning in direct adaptive control is a problematic issue which can cause actuator saturation, excite unmodeled dynamics, and other well-known problems of high learning rates \cite{Ioannou.2014}.
Moreover, system dynamics can witness significant alterations, especially in the case of failure. The absence of reliable and solid insights into the current system dynamics can result in incompetent control input, particularly when the control system merges a \emph{nominal} controller with an adaptive NN-driven control command \citep{Nguyen.2008}.

\section{Model-based versus Model-free control}
Now, we present a more detailed categorization of NN-based control systems, laying the groundwork for subsequent chapters. 

Over three decades, NNs have increasingly influenced the control engineering community, particularly the aviation sector in its flight control mechanisms \citep{Calise.1996}.
Broadly, these control systems can be bifurcated into two main types: those rooted in a predefined model (which is called model-based) and those that function independently of any pre-established model (which is called model-free).

\subsection{Model-based approach}
Over recent decades, the model-based neural control approach has blossomed, drawing upon a nominal model of the system during the control design phase. A standout learning approach in this domain is FEL, which has been integrated into numerous advanced intelligent control frameworks. This approach is an unsupervised learning approach (distinguished by the training of the NN to recognize specific patterns without relying on explicit output examples \citep{Hagan.2016}) that uses various system data such as the tracking error, the prediction error of the model, the output of a baseline controller, or a combination of them to develop a Lyapunov-based control command in such a way that guarantee the tracking error and the estimation error of the NNs to remain bounded.
Various interpretations of FEL-driven ICSs have surfaced from academic and research quarters. Within this schema, the NN might be utilized to estimate model uncertainty and external disturbance or direct its energies toward generating the control command itself. The former typically aligns with an indirect control strategy, whereas the latter is more indicative of a direct adaptive control approach. Various network structures, such as Radial Basis Function networks, multi-layer perceptrons, High-Order NNs (HONNs) \citep{Xu.2015c}, Extreme Learning Machine (ELM) \citep{Xu.2014}, and Elman NNs, have been integrated within FEL-based methodologies. Moreover, when the system dynamic is non-affine, or the dimension of system states and control inputs is not equalizing, it needs to use a particular design method like a pseudo-control strategy or backstepping scheme. 
A comprehensive dissection of these topics with their multifaceted interpretations in the flight control system and a brief discussion on different aspects of the stability analysis in such a control framework is earmarked for Chapter 2.

However, it is worth noting that FEL-based control systems are not free from obstacles. One intrinsic drawback is the NN's duty to estimate all model uncertainty under a single term, which might reduce training performance and impede optimal training, especially when signals Persistent Exciting (PE) isn't sufficient. Furthermore, many FEL-base controls employ a foundational controller derived from the system's nominal model with the NN acting as a complementary enhancer. Such an arrangement can sometimes yield amplified control commands, particularly in situations with intense structural failure, actuator faults, and dynamic changes. Therefore, we need additional supplement structures. For example, when the control direction of the system is unknown, Nussbaum functions can be used. It is also helpful to use common ideas in Fault-Tolerant Control (FTC) structures to face possible system failure. Moreover, in many cases, it is crucial to consider constraints on control commands, or system states/outputs. Therefore, using methods such as the modified tracking error or Lyapunov Barrier functions can be a way to consider system constraints. On the other hand, one of the primary assumptions in Chapter 2 was the measurability of all the system states, but in many applications, this assumption is not valid. So, in such a situation, a wise idea is to design the state estimators such as High Gain Observer (HGO) for use in output feedback control structures. Of course, in these mentioned schemes, at least the unobservable dynamics of the system must be stable. The next limiting assumption was that all the terms that the NN should estimate only depend on the states of the system, but in many cases, including disturbances imposed on the system that can depend on time, this assumption is not correct. Therefore, by using the robust adaptive approach through the combination of NN and disturbance observers, more general cases can also be considered. Another limitation is processing power. This is especially important in real-time implementation. The idea of using the minimal learning parameter method, which estimates an upper limit for the norm of uncertainties, will reduce the cost of calculations and actually create a trade-off between adaptive and robust aspects. These considerations and other design subtleties contribute to the refinements introduced to the foundational FEL structure, which will be elaborated upon in Chapter 3.

\subsection{Model-free approach}
Model-based methods, despite their advantages, also have important challenges. The first significant problem is that the unmodeled dynamics of the system are not necessarily time-invariant (for example, actuator dynamics). Furthermore, the variation in  $g(x)$ ($\dot{x}=f(x)+g(x)u$) due to different causes, such as exhaustion, is also the second significant challenge. These problems can overshadow the optimal response and even cause the system to diverge in some cases.
In contrast, model-free strategies function without the benefit of initial insights into system dynamics during the control development phase. As a basic concept, a single NN would often design the entire control structure in this realm \citep{Rinaldi.2013,Artale.2016}, generally reliant on error back-propagation for training the NN. While such methods occasionally displayed satisfactory responses, despite significant external disturbances \citep{Juang.2006}, a solid mathematical basis for closed-loop system stability often remained out of reach \citep{Kurnaz.2010}. Moreover, the inherent tendencies of these training techniques to settle into local optima raised concerns about their suitability for critical missions.

\subsubsection{Neural network-based system identification}
Academic literature abounds with examples of indirect, model-free ICS designs. These typically feature a distinct identification stage during control formulation. Various types of neural structures, from nonlinear autoregressive with exogenous inputs network (NARX) \citep{Rinaldi.2013}, and recurrent Elman networks to Convolutional NNs, Wavelet NNs, ELMs \citep{Emami.2018b}, and even Fuzzy NNs  \citep{Kayacan.2016b} are employed in this phase to identify different unmodeled dynamics. Online modifications can be made to these identified models, adapting to changes in system dynamics. The subsequent control design is then informed by the insights drawn from the NN. Determining closed-loop stability in this tiered control design might pose challenges. However, the results often outperform FEL-based strategies, particularly in the wake of drastic dynamic changes. This variant of indirect IFCSs receives comprehensive coverage in Chapter 4.

\subsubsection{Neuroadaptive optimal control and Direct adaptive control using Reinforcement learning}
The contemporary landscape has seen a surge in model-free ICSs, gleaning wisdom from Approximate Dynamic Programming (ADP) and RL concept. Although these concepts hail from diverse academic origins, ranging from control theory to machine learning and data theory, their foundational operational philosophies are remarkably congruent. A predominant motif within many of these techniques is the actor-critic framework, wherein NNs frequently serve dual roles in both actor and critic function estimation. Here, the critic embodies the anticipated cost function (or value function or cost to go), while the actor determines the system's control command. In other words, since Hamiltonian-Jacobi-Bellman’s (HJB) optimality equations (similar to the Hamilton-Jacobi-Isaacs equations) do not have an analytical solution, it is necessary to approximate a solution for these equations through ADP methods such as policy iteration and policy improvement.
 In an RL context, the control design transforms into a Markov Decision Process (MDP), with the value function representing a cumulative, discounted system (agent) reward based on current policies. The primary objective centers around policy optimization to maximize this value function, which may be done by different ADP-based (such as Q-learning, SARSA, and Monte Carlo) or policy gradient methods. Advancements in computational prowess have accelerated the rise of RL-based ICSs. New algorithms were introduced that are very suitable for control applications. These algorithms, which have continuous action and observation space, use deep neural networks in their structures. In the RL literature, these algorithms are called Deep RL. The first algorithms in this field are Deep Deterministic Policy Gradient (DDPG) and Trust Region Policy Optimization (TRPO), which were introduced around 2014. Although these algorithms have similarities, they have significant differences, especially in the way of updating the policy. The Soft Actor-Critic (SAC) algorithm is an improved version of the DDPG algorithm, which generally has a higher learning efficiency, a lower estimation bias, and a more suitable convergence rate. Also, the Proximal Policy Optimization (PPO) algorithm, which has a more convenient and straightforward policy update process, is considered as an improved version of the TRPO algorithm.
Concluding this discussion, it's crucial to note that RL agents are typically trained in a simulation environment and seamlessly transferred to real-world applications. A thorough examination of this design paradigm awaits in Chapters 5 and 6.

\begin{figure}[!h]
  \begin{center}
    \includegraphics[width=\textwidth]{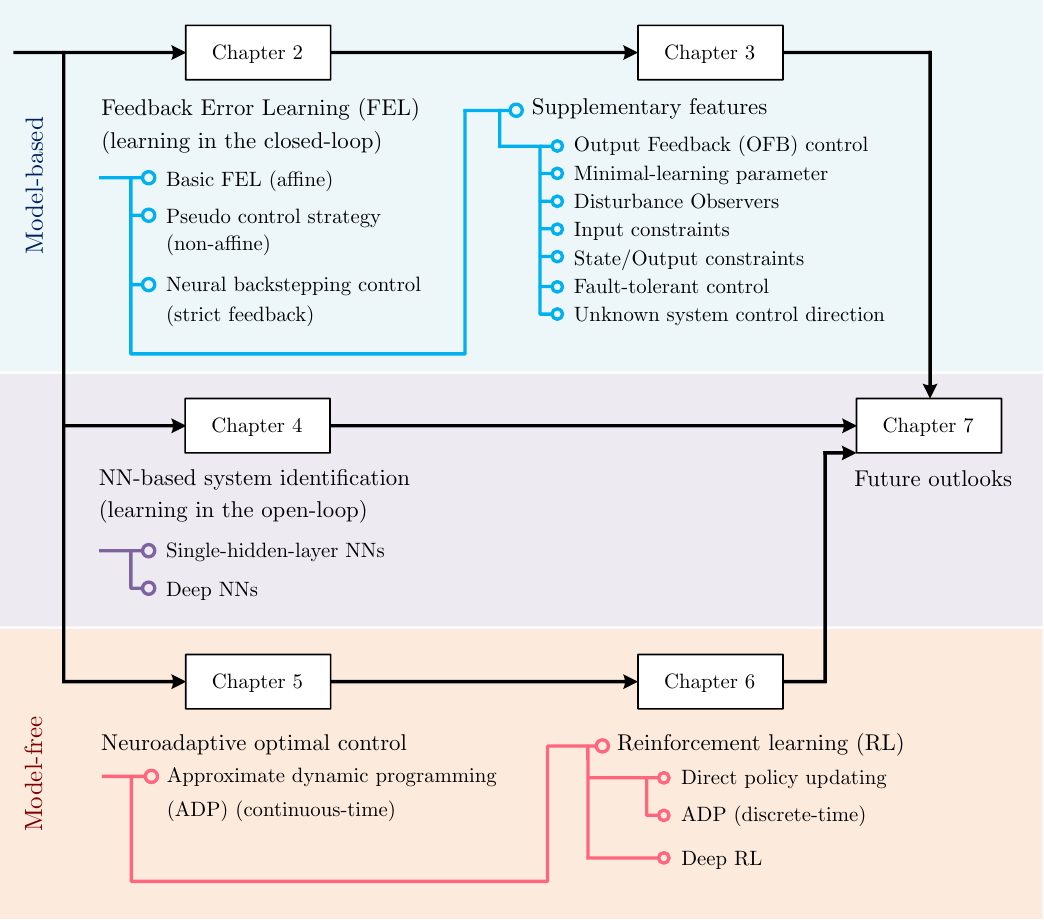}
  \end{center}
  \caption{Roadmap and structure of the book.} 
  \label{fig:contents}
\end{figure}

To round off, Fig. \ref{fig:contents} gives a brief glimpse at the most important topics discussed in this book, and Chapter 7 presents a synthesis of the discussions, paired with foresight for upcoming NN-based ICS innovations. It underscores the ongoing pursuit of an ICS that harmonizes reliability, effectiveness, and computational frugality, which can be utilized in serious tasks.


\eject

\begin{partbacktext}
\part{Model-based Control Design}

\end{partbacktext}
%
%
%
\chapter{Foundations of Model-based Intelligent Control}
\label{Sec:Model_based_IFCS} 

In this chapter, we present the foundation of model-based neural network-driven control systems. In this procedure, our main focus is on the use of neural networks as a compensator for uncertainties and disturbances, and this goal is achieved through proper updating of neural network weights. 
To this end, in the following subsection, we will first give a brief introduction to neural networks, their structures, and training methods.

On the other hand, the development of a theoretical framework to guarantee the stability of the closed-loop system including dynamic neural networks is crucial to render a reliable control method. 
In a general view, the control structure can include a conventional controller in the inner loop, which will be responsible for stabilizing the nominal system dynamics. In addition, a neural network can help the main controller as a compensator for the effects of unmodeled dynamics and disturbances. In the process of designing such a structure, we employ the idea of Lyapunov stability. 
For this purpose, in the proposed Lyapunov functions, there will be terms for the tracking error and the estimation error of NN weights so that at least the Bounded-Input Bounded-Output stability (BIBO) can be guaranteed in the presence of model uncertainties and external disturbances.

In this regard, in the first step, we introduce the fundamental theoretical basis for the most commonly used approach to incorporate NNs within the adaptive control design process, i.e. the Feedback Error Learning method, while the application of such an algorithm in flight control systems would be addressed in various example.
According to the type of system dynamics, the design process may undergo changes and different methods may be used. In section \ref{Sec:FEL}, we deal with affine systems that the control input dimension and the state dimension are equal to each other. 
To this end, first the basic process of model-based neural network-driven controller design will be described in detail, and then some existing challenges (such as the drift phenomenon) and how to deal with them will be expressed. Composite and concurrent learning methods will be introduced as effective methods to improve the learning performance of NNs in the lack of persistent exciting signals. 
Besides, the introduced design will be extended to higher-order dynamic systems by incorporating filtered tracking errors into the design.

In section \ref{Sec:PseudoControl}, we discuss about non-affine systems. Most of the intelligent controller design approaches for this class of systems are direct methods. The pseudocontrol strategy that uses the approximate inverse model of the system (it can be a very simple model of the system that has only modeled the input and output connection with the correct sign, which is called the control assignment structure-property) is one of the widely used methods in this field.

Section \ref{Sec:Backstepping} talks about another group of affine systems, which are called strict feedback forms. In this category of systems, each state variable is considered as the input of the previous state. Hypersonic flight vehicles are generally modeled like this. Neural backstepping is one of the most used control strategies for this category of systems. The main challenge of the basic neural backstepping method is the derivation of the state variables at each stage of the control input calculation, which is called the explosion of terms. However, this problem can be solved with simple modifications. The first idea that will be introduced is the dynamic surface control method. In this method, derivation will be done through the first-order filter. Appropriate selection of the filter time constant is very important in obtaining the desired result. Difficulty in stability analysis is the main challenge of this method, so to solve this problem, the idea of command filter backstepping can be used. This idea is based on the introduction of a new error, which is called a compensated error. In addition to this, there are many other methods for approximating the derivative, including the First-Order Sliding Mode Differentiator (FOSMD) method. In this method, the main idea is to be able to write state equations whose time response is the derivative of the desired signal (desired state variable). The main advantage of this approach compared to the previous ones is that if we consider the derivative filter as a type of linear derivative, this type of derivative method can be considered as a non-linear derivative that if the signal is Lipschitz, the error of the derivative tends to zero. In the final part of this section, we will provide direct neural backstepping methods. Direct methods free us from a series of complications, but generally, these types of solutions do not easily pass the safety standards. In addition, the learning rate is more important than before and its inappropriate choice can lead to state divergence.

In section \ref{Sec:Stability}, we will delve into the discussion on the definitions and different aspects of stability. Stability can be categorized in terms of extent and intensity. In the concept of extend, the stability of the system (equilibrium point) is investigated from local to global aspect. Nevertheless, in the concept of intensity, error limitations and how it behaves towards the zero value are evaluated.

Finally, in the last section, which is called  concluding remarks, we will discuss and review the most important content and points presented during this chapter.

\section{Neural networks overview} \label{Sec:NN_Overview}
Artificial NNs have become a powerful tool for modeling complex nonlinear systems and have found widespread use in control theory and aerospace applications.
As a class of machine learning algorithms, neural networks do not need explicit programming but rather rely on training from data, unlike traditional computing methods.

Understanding the basic terminology used in neural networks is essential for grasping their functioning and applications. Here, we introduce some fundamental terms:
\begin{itemize}
\item \textbf{Layers}: NNs are organized into distinct layers of neurons (See Fig. \ref{fig:rbfnn}). Layers are grouped into \emph{input layers} that receive data, \emph{hidden layers} that extract features, and \emph{output layers} that generate predictions.
\item \textbf{Neurons}: The building blocks of a network. Each neuron sums its input signals weighted by the connection strengths and applies an \emph{activation function} to generate an output signal.
\item \textbf{Activation functions}: Nonlinear functions such as sigmoid, tanh, or ReLU (Rectified Linear Unit) are applied to transform a neuron's summed input into an output signal. They introduce nonlinearity enabling complex modeling.
\item \textbf{Weights}: The strength of connections between neurons. Weights are adjustable parameters that represent the knowledge learned by the network. 

\end{itemize}
These core elements enable neural networks to learn complex relationships between high-dimensional inputs and outputs for problems in prediction, classification, and control.

NNs can be categorized based on their structure into single-layer vs. multi-layer, and feedforward vs. feedback networks. Each architecture has its modeling capabilities and is suited to different control tasks.

This section provides a comprehensive overview of neural network fundamentals, architectures, and training methods in each category, with a focus on their application in control theory.

\subsection{Neural network taxonomy}
Neural networks can be broadly classified into two categories: feedforward and recurrent. Feedforward networks allow signals to travel in only one direction from input to output. A \emph{recurrent networks} contains feedback connections, allowing signals to propagate in both directions. This section delves into the mathematical models and common architectures employed for both classes.

\subsubsection{Feedforward NNs}
One of the simplest and widely used variants of NNs.
They consist of multiple layers, with information flowing in a unidirectional manner from the input layer through one or more hidden layers to the output layer. 
The network may or may not have hidden node layers, making their architecture shallow or deep.

\begin{figure}[t]
  \begin{center}
    \includegraphics[width=0.7\textwidth]{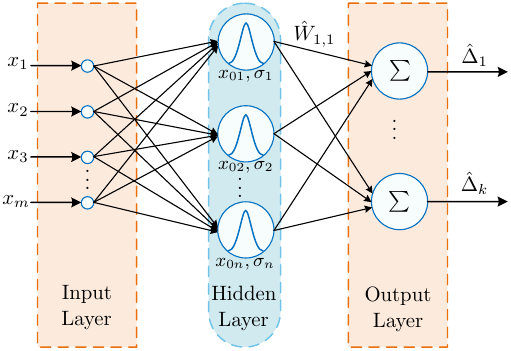}
  \end{center}
  \caption{Typical structure of an RBF neural network.} 
  \label{fig:rbfnn}
\end{figure}

Here, we discuss three popular types of feedforward NNs applicable in closed-loop control.
\paragraph{\textbf{Radial Basis Function Neural Networks (RBFNN)}}
RBFNNs are networks often used in control applications to estimate nonlinearities such as model uncertainties and external disturbances.
As depicted in Fig. \ref{fig:rbfnn}, the estimation $\Delta(x)$ can be described as follows
\begin{equation}
\hat{\Delta}(x)=\hat{W}^T \mu (x),
\end{equation}

\noindent where $x \in \mathbb{R}^{m}$ is the input vector, $\hat{W} \in \mathbb{R}^{n \times k}$ is the adjustable weight matrix, $\mu(X):\mathbb{R}^{m} \rightarrow \mathbb{R}^{n}$ is the vector of basis functions, $n$ is the number of neurons in the network, and $k$ is the number of outputs, which is the length of the vector $\Delta(x)$.

RBFNNs utilize radial basis functions as activation functions. 
The activation function is typically a Gaussian function, given by
\begin{equation}\label{Gaus_Org}
\mu_i(x) = \exp \left( -\frac{\left(x - x_{0i}\right)^T\left(x - x_{0i}\right)}{2\sigma_i^2} \right),
\end{equation}
where $x_{0i}\in \mathbb{R}^{m}$ is the vector of center of the $i$th basis function, and $\sigma$ is the width of the Gaussian function.


\paragraph{\textbf{Multilayer Perceptron (MLP)}}
MLP is a versatile feedforward NN architecture consisting of multiple layers of neurons with nonlinear activation functions, allowing the network to learn complex relationships. The \emph{universal approximation} theorem states that MLPs with just a single hidden layer can approximate any continuous function, given enough hidden units. 
However, they require proper tuning of \emph{hyperparameters} like depth, width, and learning rates.
In the context of machine learning, a hyperparameter is distinct from a parameter in that they are set before the training process begins, whereas parameters are learned by the network during training. Hyperparameters govern the overall architecture and learning behavior of the network, while parameters represent the specific weights and biases of the connections between neurons. Carefully selecting hyperparameters is crucial for achieving optimal performance and generalization to new data.

Some commonly used activation functions are:
\begin{itemize}
\item Sigmoid: 
$f(x) = \dfrac{1}{1+e^{-x}}$,
\item Hyperbolic tangent (tanh),
\item Rectified linear unit (ReLU): 
$f(x) = \max(0, x)$.
\end{itemize}

The sigmoid and tanh functions are nonlinear activation functions that allow the network to learn complex patterns in data.
However, they can lead to a problem called vanishing gradients, where the error signal becomes very small as it propagates through the network, making it difficult for the weights to update effectively.
ReLUs avoid this problem and enable faster training but can get stuck at zero gradients, meaning that the weights are unable to learn further. Variations like leaky ReLUs ($f(x) = \max(0.01x, x)$) help alleviate the latter issue.

\emph{Backpropagation} is an algorithm that allows the network to learn the optimal weights and biases by minimizing a \emph{loss function}.
The loss function represents the error between the network outputs and true targets that we aim to minimize.
The goal of backpropagation is to propagate the error signal backward from the output layer through the network to update the weights in a way that reduces the overall error.
To this end, \emph{gradient descent} is used to update the weights along the negative gradient direction to reduce the loss.

During training, an MLP learns the optimal weights and biases using backpropagation to minimize a loss function.
The key steps include:
\begin{enumerate}
\item Forward pass: Compute the output by passing input features through the network layer-by-layer.
\item Backward pass:
The error gradient is calculated at the output layer and propagated backward through the network using the chain rule of derivatives. This process involves calculating the error gradient for each intermediate layer and using it to update the weights of the corresponding connections.
\item Update weights:
The weights are updated using gradient descent, which involves moving the weights in a direction that reduces the error. The magnitude of the update is determined by the learning rate, which is a hyperparameter that controls the speed of learning.
\end{enumerate}
This process is repeated multiple times over different training data examples, known as \emph{epochs}, until the loss function converges and the network learns to produce accurate outputs for a wide range of inputs.



\paragraph{\textbf{Wavelet Neural Network (WNN)}}
WNNs, introduced in the 1990s by \citeauthor {Zhang.1992}, merge \emph{wavelet theory} with NNs, allowing them to efficiently deal with time-frequency analysis.
WNNs are regarded as a generalization of RBFNNs \citep{Alexandridis.2013}, as they combine the universal approximation capabilities of neural networks with the properties of wavelet decomposition \citep{Emami.2020b}.

The basic structure of a WNN consists of an input layer, a single hidden layer employing wavelets as activation functions, and an output layer.
The hidden layers contain wavelet nodes or \emph{wavelons}, performing wavelet transforms on the inputs.

Wavelet transform is a magnifying glass for signals. It can zoom in to reveal high-frequency components or zoom out to reveal low-frequency components.
Unlike the Fourier transform, which uses a fixed set of sinusoidal basis functions, the wavelet transform uses a set of wavelet basis functions that are \emph{localized} in both time and frequency. This localization allows wavelet transform to extract local spectral and temporal information simultaneously, making it suitable for analyzing non-stationary and transient signals.
The transform operates by convolving the original signal to the scaled and translated versions of the wavelet basis functions, whereby a large transform value is observed if the scale and position of the wavelet match the signal.

\begin{figure}[ht]
\centering
\includegraphics[width=0.85\textwidth]{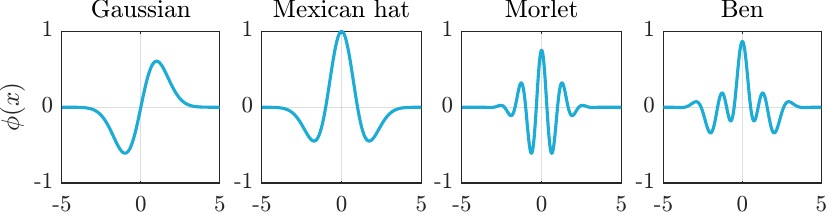}
\caption{Graphical representation of common mother wavelet functions}
\label{fig:wnnfamily}
\end{figure}
Common choices for the wavelet activation function include Gaussian wavelet, Mexican hat and Morlet wavelets, as depicted in Fig. \ref{fig:wnnfamily}. The Mexican Hat wavelet is the second derivative of a Gaussian function and is defined by the following equation
\begin{equation}
\phi(x_i) = (1 - x_i^2)\exp\left(-\frac{x_i^2}{2}\right).
\end{equation}
Each wavelon is obtained by translating and dilating the mother wavelet
\begin{equation}
\Phi_{ij}(x_i) = \phi_{ij}\left(\frac{x_i - c_{ij}}{b_{ij}}\right),
\end{equation}
where $b_{ij}$ and $c_{ij}$ are the dilation and translation factors respectively.

The output of the hidden layer is obtained by taking the product of wavelets from each input dimension
\begin{equation}
\mu_j(x) = \prod_{i=1}^{m}\Phi_{ij}(x_i).
\end{equation}
Finally, the output layer combines the hidden layer outputs linearly to produce the network output
\begin{equation}
y(x) = \sum_{i=1}^{m} \omega_i x_i + \sum_{j=1}^{n} w_j\mu_j(x) + w_{n+1},
\end{equation}
where another set of weights $\omega_i$ is introduced to directly connect the input layer to the output layer. Accordingly, the weighting vector is as follows
\begin{equation}
\hat{W} = \begin{bmatrix}
\omega_i, &w_j, &w_{n+1}
\end{bmatrix}.
\end{equation}

\subsubsection{Recurrent NNs}
Recurrent Neural Networks (RNNs) are a class of artificial neural networks that incorporate feedback connections, introducing dynamics and memory to the networks. Unlike feedforward neural networks, RNNs can process sequences of inputs by maintaining an internal state that takes into account previous inputs and computations. This makes RNNs well-suited for processing time series data and sequence modeling tasks.

\begin{figure}[!h]
  \begin{center}
    \includegraphics[width=0.7\textwidth]{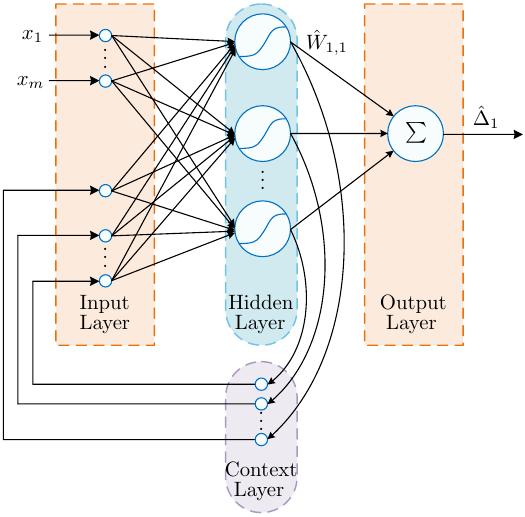}
  \end{center}
  \caption{Elman network architecture.} 
  \label{fig:elman}
\end{figure}

\paragraph{\textbf{Elman Neural Networks (ENN)}}
The Elman network, introduced by Jeffrey \citeauthor{Elman.1990} in 1990, has a simple architecture that includes context units to store the previous hidden state of the network. As shown in Fig. \ref{fig:elman}, the context layer holds a copy of the previous hidden layer output. At each time step, the input is fed into the hidden layer, which computes the activation values that get passed to the output layer to produce the network output. The hidden layer output is also copied to the context layer via a unit delay connection. In the next time step, the contents of the context layer are fed back into the network as additional inputs along with external inputs.
The context units provide the network with dynamic temporal behavior and the ability to retain information from the past. The previous hidden states stored in the context layer allow the network to maintain useful temporal information for generating outputs that depend on prior context. The Elman network is trained via BackPropagation Through Time (BPTT), which unfolds the network through time and runs backpropagation on the unfolded network \citep{Werbos.1990}. BPTT computes the error gradient across multiple time steps to update the weights appropriately. The Elman network's simple architecture makes it easy to implement and train while providing useful sequence learning capabilities. However, it lacks more sophisticated gating mechanisms for controlling information propagation through time.

\begin{figure}[!h]
  \begin{center}
    \includegraphics[width=0.75\textwidth]{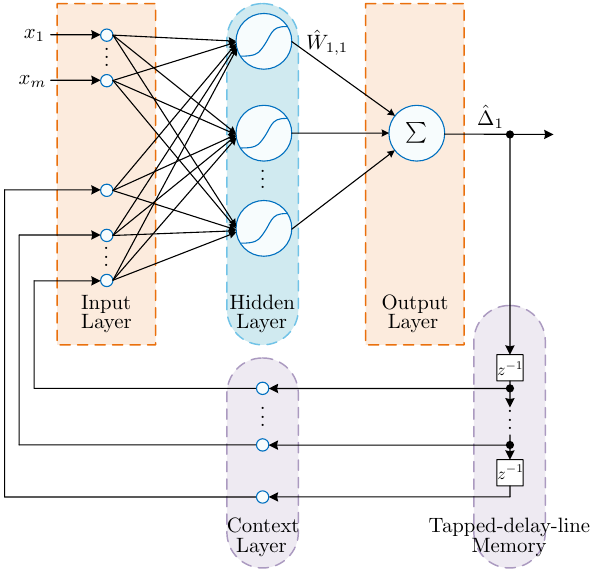}
  \end{center}
  \caption{NARX (Jordan) network architecture.} 
  \label{fig:narx}
\end{figure}

\paragraph{\textbf{NARX (Jordan)}}
Nonlinear AutoRegressive with eXogenous inputs (NARX) and \citeauthor{Jordan.1997} networks have architectures that feed the network output back to the input. As depicted in Fig. \ref{fig:narx}, the NARX model takes current and past input values along with past output values as input to produce the current output.
The tapped delay lines on the input (not shown in the figure) and output store previous values to provide temporal context.

These networks leverage output feedback and temporal memory to generate outputs that depend on past computations. The feedback path allows the networks to learn to make predictions based on prior outputs. NARX/Jordan networks are also trained with BPTT. An advantage over the Elman network is the ability to optimize weight connections from outputs to both hidden and input layers. However, they still lack sophisticated mechanisms for storing information over long time lags.

\begin{figure}[!h]
  \begin{center}
    \includegraphics[width=\textwidth]{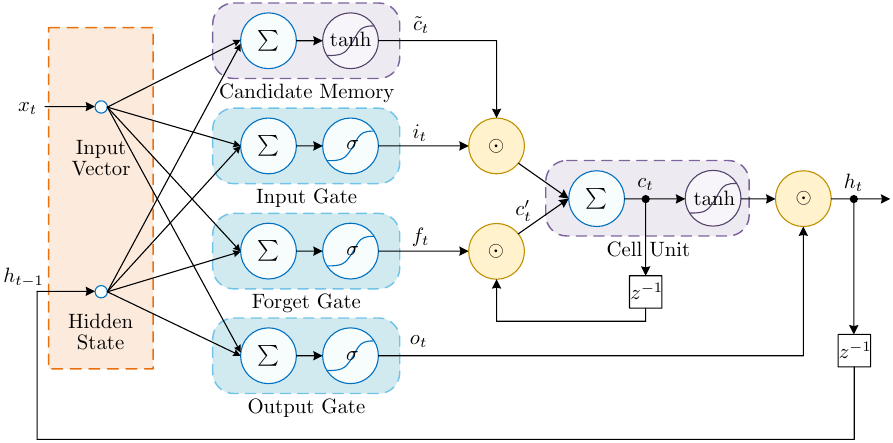}
  \end{center}
  \caption{\small LSTM unit architecture.} 
  \label{fig:lstm}
\end{figure}

\paragraph{\textbf{Long Short-Term Memory (LSTM)}}
LSTM networks were introduced by \citeauthor{Hochreiter.1997} in 1997 to specifically address the problem of learning long-term dependencies. LSTMs contain special units called memory cells that include \emph{gates} to control information flow. As illustrated in Fig. \ref{fig:lstm}, each LSTM unit consists of:
\begin{itemize}
\item A \emph{forget gate} that determines what information to discard from the unit
\item An \emph{input gate} that decides what new information to store in the unit
\item An \emph{output gate} that controls the output activation of the unit
\item A self-connected memory cell unit to store state information
\end{itemize}

These gates adaptively reset or update the memory cell and allow LSTM networks to model long-term sequences in data. The gating mechanisms regulate information flow and mitigate the vanishing gradient problem. Numerous LSTM variants have been developed, but all use gated memory cells for learning temporal dependencies.

As depicted in Fig. \ref{fig:lstm}, the underlying mathematics behind an LSTM network is as follows
\begin{align}
& \text{Input gate:}\!& i_t &= \sigma(W_i x_t + U_i h_{t-1}  + b_i), \\
& \text{Forget gate:}\!& f_t &= \sigma(W_f x_t + U_f h_{t-1} + b_f), \\
& \text{Output gate:}\!& o_t &= \sigma(W_o x_t + U_o h_{t-1} + b_o), \\
& \text{Cell unit update:}\!& \tilde{c}_t &= \tanh(W_c x_t + U_c h_{t-1} + b_c), \\
&\!& c_t &= f_t \odot c_{t-1} + i_t \odot \tilde{c}_t, \\
& \text{Hidden state:}\!& h_t &= o_t \odot \tanh c_t,
\end{align}
where
\begin{itemize}
\item $x_t$ is the input at time step $t$,
\item $h_{t-1}$ is the hidden state output from the previous time step,
\item $W, U, b$ are learnable weight matrices and biases,
\item $\sigma$ is the sigmoid activation function,
\item $\tanh$ is the hyperbolic tangent activation,
\item and, $\odot$ denotes element-wise multiplication.
\end{itemize}

\subsection{Single-layer vs multi-Layer NNs}
The universal approximation theorem shows that a feedforward network with just a single hidden layer containing a finite number of neurons can approximate any continuous function, given appropriate weights. This property enables the modeling of complex uncertainties and dynamics with NNs. However, if the number of hidden nodes is increased excessively, it can result in the overfitting problem \citep{Srivastava.2014}. Overall, deep networks with multiple layers provide greater representation power over shallow single-hidden-layer models. However, they can be more difficult to train due to issues like vanishing gradients.

\subsubsection{Single-hidden-layer NNs}
As we mentioned before, the simplest NN structure consists of an input layer, a single hidden layer, and an output layer.

Despite their simplicity, single hidden layer NNs with enough nodes can approximate any continuous function. They have been applied to model aircraft dynamics \citep{Emami.2020}, design flight controllers \citep{Xu.2021b}, and perform system identification \citep{Talebi.2010b}. The Extreme Learning Machine (ELM) is a popular variant that randomly initializes the hidden layer and only trains the output weights \citep{Emami.2018b}, which simplifies the training process.

To sum up, single-layer NNs offer efficient function fitting but may require more nodes than deep networks. Their simple structure facilitates analysis for control design.

\subsubsection{Deep NNs}
Instead of increasing the number of hidden nodes in a single hidden layer, deep NNs utilize more hidden layers that can efficiently represent highly nonlinear relationships and data patterns.
Common deep architectures include MLPs, Convolutional Neural Networks (CNNs), and RNNs like LSTM.

\paragraph{\textbf{CNN}}
CNNs are a significant type of deep NNs that have a cascade connection structure.
Each CNN cell consists of a convolution layer, a sub-sampling layer (pooling layer), and a fully connected layer as the last layer.

The convolutional layers utilize filters to extract characteristics within the input data. The pooling layers then reduce the size of the convolutional layer's output, decreasing data complexity by reducing the dimensions of the convolved features and improving computational efficiency. This cycle can be repeated multiple times, with additional convolutional and pooling layers to learn increasingly complex representations. Finally, the fully connected layers assess the previous layer's output to make the ultimate classification decision based on the learned features.

CNNs are a powerful tool for the identification of nonlinear systems. They have several features that make them suitable for this task. 

Firstly, the convolution operation in CNNs is similar to the input-output relation of linear time-invariant systems. This allows CNNs to capture the underlying patterns and relationships in the data. 

Secondly, CNNs employ sparse connectivity and shared weights. This means that each neuron in the network is only connected to a small number of neurons in the previous layer, reducing the number of parameters in the network. Additionally, the shared weights allow the network to generalize well to new data, reducing the risk of overfitting.

Another important feature of CNNs is multi-level pooling. Pooling layers downsample the input by summarizing local features, resulting in a more robust identification scheme against measurement noises. This helps to improve the accuracy and reliability of the system identification process.

In summary, a CNN is a type of NN that is commonly used for image processing and computer vision tasks. It consists of multiple layers, able to learn hierarchical representations of the input data, from low-level features such as edges and corners to high-level features such as object parts and whole objects.
While their application in flight control systems is still limited, they have been successfully used in high-level control schemes such as localization and path planning.

\subsection{Training approaches}
Having looked at different structures for NNs, we can discuss two distinct approaches for training them for identification and control purposes.

Consider an RBFNN that needs to be trained. The following methods can be used:
\begin{enumerate}[(a)]
\item \textbf{Learning in the open-Loop structure}
\begin{itemize}
    \item \emph{Hybrid learning method}: Uses k-means clustering (unsupervised learning) to determine the RBF centers and widths in the hidden layer. The output layer weights are tuned through regression (supervised learning).
    \item \emph{ELM}: An ELM can be used to randomly initialize the RBFNN centers and widths and then train the weights of the network in one shot.
\end{itemize}
\item \textbf{Learning in the closed-Loop structure} \\
The RBFNN can be trained online in the closed-loop using the FEL method based on the Lyapunov stability theory. 
This approach introduces a parallel training strategy in relation to supervised or unsupervised learning scenarios.
\begin{itemize}
    \item \emph{ELM}: Similar to the open-loop structure, an ELM can be used in a closed-loop structure, with random initialization of hidden layer parameters. Subsequently, FEL provides with an adaptive update rule to train the network weights.
    \item \emph{Fully-tuned NN}: All parameters including output weights and RBF centers and widths are tuned adaptively based on feedback error signals.
\end{itemize}
\end{enumerate}
Open-loop methods identify neural network parameters using only input-output data, without feedback from the system. Closed-loop techniques, on the other hand, rely on feedback from the real system to train the network. FEL-based tuning utilizes this online feedback to ensure stability during adaptation.
Selecting the right training approach depends on the problem, objectives (identification/control), constraints, availability of data, and performance requirements.

\section{Feedback Error Learning} \label{Sec:FEL}
\begin{figure}[!h]
  \begin{center}
    \includegraphics[width=0.9\textwidth]{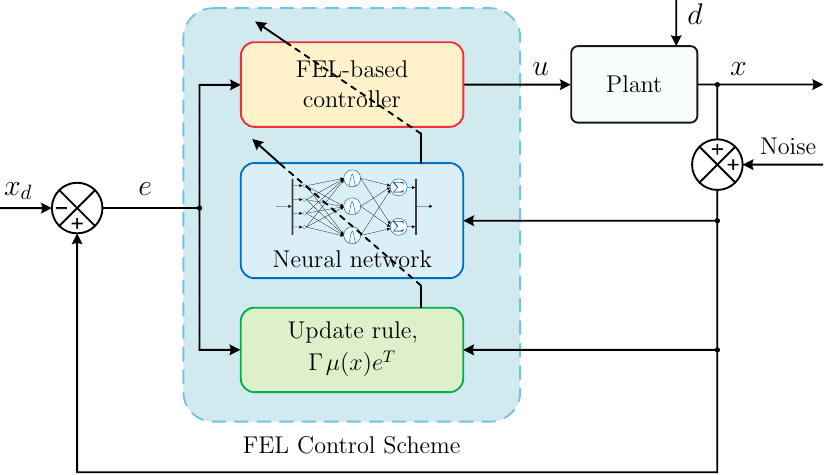}
  \end{center}
  \caption{Schematic diagram of FEL control law.}
  \label{fig:FELschem}
\end{figure}

Here, we will introduce the fundamental theoretical basis for the most commonly used approach to incorporate NNs within the adaptive control design process, i.e. the FEL method, while the application of such an algorithm in flight control systems will be addressed in the following subsections.
FEL can effectively integrate the control design procedure and the online updating law for the parameters of the NN, which is utilized to compensate for model uncertainties and disturbances.
Accordingly, in a general view, the control block can consist of a conventional controller in the inner loop to stabilize the system dynamics, and the neural controller acts as an aid to the controller to compensate for model uncertainties.
This arrangement is depicted in Fig. \ref{fig:FELschem}. As shown in the figure, the weights of the neural network are adaptively adjusted using an update rule, which will be further elaborated upon. These updated weights then modify the control law implemented by the conventional controller.
Thus, employing a composite Lyapunov function including both the tracking error and the estimation error of NN parameters, all signals in the closed-loop system remain bounded, even in the presence of model uncertainties and external disturbances.
To be more precise, consider the dynamic model of an aircraft (in the affine form) as follows
\begin{equation}\label{Model_1}
\dot{x}=F(x)+B(x) u+ \Delta(x),
\end{equation}
where $x,u \in \mathbb{R}^n$ 
and $\Delta(x)$ represents a lumped uncertainty that encapsulates all state-dependent model uncertainties and external disturbances. It's worth noting that the portion of the uncertainty which is a explicit function of time is not included here and will be addressed in Chapter 3.
Defining the desired trajectory as $x_d$, the tracking error is obtained as $e=x-x_d$. Now, the control command can be defined as
\begin{equation}\label{Ctrl_0}
u=B^{-1}\left(-F(x)-\Delta+\dot{x}_d-k_1 e\right),
\end{equation}
where $k_1$ is a positive-definite matrix. However, the vector $\Delta$ is unknown. Thus, it is approximated by an NN (such as RBFNN or MLP) as $\hat{\Delta}=\hat{W}^T \mu (x)$, where $\mu$ represents the vector of basis functions (corresponding to hidden layers of the NN) and $W$ indicates the matrix of unknown weights that should be identified.
Such a formulation can be used to represent different feedforward and recurrent NNs. Here, we will use such a general formulation, and for brevity, do not address different possible network structures and their advantages and disadvantages in flight control systems (for more details, see \citep{Gu.2020}).
Accordingly, due to the universal approximations property of NNs, we have
\begin{equation}\label{NN_1}
\Delta = {W^*}^T \mu (x) + \varepsilon,
\end{equation}
where $W^*$ denotes the (unknown) optimal weight and $\varepsilon$ indicates the \emph{bounded} approximation error ($\|\varepsilon\|\leq \varepsilon_M$).
The control command can now be constructed as follows
\begin{equation}\label{Ctrl_1}
u=B^{-1}\left(-F(x)-\hat{\Delta}(x)+\dot{x}_d-k_1 e\right).
\end{equation}
Substituting the above control command into Eq. (\ref{Model_1}) gives
\begin{align}
\begin{split}\label{e_dot}
\dot{e} & = \dot{x} - \dot{x}_d \\
 & = F(x)+B(x) u+ \Delta - \dot{x}_d \\
 & = \tilde{\Delta}+ k_1 e,
\end{split}
\end{align}
where, $\tilde{\Delta}=\hat{\Delta} - \Delta$.
Now, consider a Lyapunov function as
\begin{equation}\label{Lyap_0}
  V=\frac{1}{2}e^T e + \frac{1}{2} \tr \left(\tilde{W}^T \Gamma^{-1} \tilde{W}\right),
\end{equation}
where $\tilde{W}=\hat{W} - W^*$ and $\Gamma$ is a positive definite matrix. Next, we have
\begin{align}
\begin{split}\label{Lyap_1}
   \dot{V} & = e^T \dot{e}+ \tr \left(\tilde{W}^T \Gamma^{-1} \dot{\tilde{W}}\right)  \\
     & = - e^T \left(\tilde{\Delta}+ k_1 e\right)+ \tr \left(\tilde{W}^T \Gamma^{-1} \dot{\tilde{W}}\right).
\end{split}
\end{align}
Accordingly, defining
\begin{equation}\label{Learn_1}
 \dot{\hat{W}}=\Gamma \mu e^T,
\end{equation}
and considering $\dot{\tilde{W}}\approx  \dot{\hat{W}}$ (as a consequence of assuming a \emph{constant} optimal weight $W^*$, while such an assumption is reasonable even in the case of a time-dependent uncertain term $\Delta={W^*}^T(t) \mu (x) + \varepsilon$ with $\dot{W}^* \ll \dot{\hat{W}}$), we have
\begin{align}
\begin{split}\label{Lyap_2}
\dot{V} & = - e^T \left(\tilde{\Delta}+ k_1 e\right)+ \tr \left(\tilde{W}^T \Gamma^{-1} \dot{\tilde{W}}\right) \\
 & = - e^T \left(\tilde{\Delta} + k_1 e\right) + \tr \left(\tilde{W}^T \mu e^T \right).
\end{split}
\end{align}
Using the property $\tr \left( b{{a}^{T}} \right)={{a}^{T}}b$, where ${{b}_{n\times 1}}=\tilde{W}^T \mu $ and ${{a}_{n\times 1}}=e$ we get
\begin{align}
\begin{split}\label{Lyap_3_1}
\dot{V} & = - e^T \left(\tilde{\Delta} + k_1 e\right) + e^T \tilde{W}^T \mu \\
 & = - k_1 e^T e + e^T \varepsilon = - e^T (k_1 e - \varepsilon).
\end{split}
\end{align}
This Equation leads to $\dot{V}<0$ for $\|k_1 e\| > \|\varepsilon\|$ thereby guaranteeing the bounded tracking error. 

\subsection{Persistent Excitation (PE) condition}
If we neglect the estimation error $\varepsilon$, we can derive that $\dot{V}=-k_1 e^T e$, which results in asymptotic stability of the tracking error when $e$ approaches zero.
However, it raises the question of whether $\tilde{W}$ also converges to zero, as the Lyapunov function's time derivative remains negative unless $e=0$.
Therefore, the system is likely to achieve zero tracking error, which will keep the Lyapunov function constant. However, the estimation error may still exist.
This can also lead to problems such as parameter drift where $\hat{W}$ drifts indefinitely (while the tracking error remains bounded) in the presence of the estimation error $\varepsilon$.
 
To address the issue, we define an augmented vector composed of tracking and uncertainty estimation errors as
\begin{equation}
 z=\begin{bmatrix}
e\\ \tilde{W}
\end{bmatrix}.
\end{equation}

Combining Eqs. (\ref{e_dot}) and (\ref{Learn_1}) gives
\begin{equation}
\dot{z}=\begin{bmatrix}
-k_1 & -\mu^T(x) \\ 
\Gamma\mu(x) & 0 
\end{bmatrix}z + \begin{bmatrix} \varepsilon\\ 0 \end{bmatrix}.
\end{equation}
Since the minimum estimation error $\varepsilon$ is bounded, the boundedness of $z$ can be concluded if $$A=\begin{bmatrix}
-k_1 & -\mu^T(x) \\ 
\Gamma\mu(x) & 0 
\end{bmatrix}$$ is negative definite. This can be guaranteed if the regressor vector $\mu(x)$ is persistently exciting, i.e. (\citet{PE1987,Narendra.2005})
\begin{align}
\int_{t}^{t+T} \mu\mu^T d\tau \geq \alpha I, \quad \forall t \geq 0,
\end{align}
for some $\alpha > 0$ and $T > 0$, where $I$ is the identity matrix. This condition means that the regressor vector $\mu(x)$ should be sufficiently rich so that it persistently excites all directions in $\mathbb{R}^m$ to ensure identifiability of the weights.

Practical realization of this condition is not straightforward. In practice, perturbations are introduced to the system during the initial moments to deliberately move it away from equilibrium. This process helps the convergence of the weights to their correct values, enabling subsequent control execution.

The determination of optimal design parameters such as $k_1$ and $\Gamma$ is not an easy task as it is typically done by trial and error. It is also possible to define an optimization problem in terms of these parameters and solve it using well-known optimization methods (such as evolutionary algorithms) to determine their optimal values according to predefined criteria \citep{MohdBasri.2015}. 

\subsection{Uncertain control gain function}
In cases where the matrix $B$ is also uncertain, we have
\begin{equation}\label{Model_2}
\dot{x}=F(x)+\left(B_n(x)+\Delta B\left(x\right)\right) u+ \Delta,
\end{equation}
where $B_n$ represents the nominal part and $\Delta B = B - B_n$ denotes the uncertainty in the control gain matrix. Thus, it is possible to define
\begin{equation}
  \bar{\Delta}:=\Delta B u+ \Delta,
\end{equation}
and estimate it using an NN as $\hat{\Delta}(x,u)=\hat{W}^T \mu (x,u)$. Consequently, the control command can be calculated as follows
\begin{equation}\label{Ctrl_2}
u=B_n^{-1}\left(-F(x)-\hat{\Delta}(x,u)+\dot{x}_d-k_1 e\right).
\end{equation}

As seen, the control command results in an implicit equation as $u=h(\cdot,u)$. The existence and uniqueness of a solution for $u$ can be guaranteed by employing a contraction mapping assumption \citep{Calise.2001b}.
Sufficient conditions for satisfying this assumption are given in \citet{NakwanKim.2003}. Notably, this assumption implicitly requires the sign of the control gain function to be known \citep{Chowdhary.2014}.

A contraction mapping is a function that takes a set as input and produces a set as output such that the distance between any two outputs is strictly less than the distance between the corresponding inputs. This can be stated using the following condition:
\begin{equation}
\left|\frac{\partial h(x,u)}{\partial u}\right|<1,
\end{equation}
which simplifies to
\begin{equation}
\left| \Delta B \; B^{-1} \right|=\left| \left( B-B_n \right)B^{-1} \right|=\left| B_nB^{-1}-I \right|<1.
\end{equation}

This can be ensured by satisfying:
\begin{equation}
\begin{cases}
\text{sign}(B_n) = \text{sign}(B) \\
|B_n| > |\frac{B}{2}| > 0.
\end{cases}
\end{equation}

Therefore, a unique fixed point solution for $h(x,u)$ exists if the model accurately captures the direction of control effectiveness and loosely approximates the magnitude of the matrix $B$.

Note that, it is also possible to update the weights of the hidden layer to provide more effective learning. 
This can be performed using a similar Lyapunov stability analysis by taking advantage of the Taylor expansion of the hidden layer output ($\mu(x)$) \citep{Lee.2001}. However, due to the more complicated formulation and excessive computational burden, in this book, typically, we will  only update the output weights, $\hat{W}$, and the other parts of the NN remain unchanged. For an instance of updating all parameters of the NN using FEL mechanism, see Example \ref{fullytunedNN}.

Besides, one can replace the tracking error $e$ in (\ref{Lyap_0}) and (\ref{Learn_1}) by a filtered tracking error as $s=e+\lambda \int e \enspace dt$ (with $\lambda$ denotes a positive constant or a positive definite matrix) to compensate for the steady-state tracking error \citep{Li.2016d}.
Further, the introduced FEL neural control scheme can be applied to a second-order system, i.e. $\ddot{x}=F(x)+B(x) u+ \Delta$ by substituting $e$ by a filtered tracking error as $s=\dot{e}+\lambda e$ \citep{Zeghlache.2018,Song.2019}.

\begin{ex}[label=Example:quadFEL]{FEL-based control of quadrotor UAV} 
Fig. \ref{fig:quad} shows the schematics of a 6-Degrees of Freedom (6-DOF) quadrotor UAV, depicting its body-fixed coordinates, as well as the inertial coordinate frame.
\begin{figure}[H]
  \begin{center}
    \includegraphics[width=0.85\textwidth]{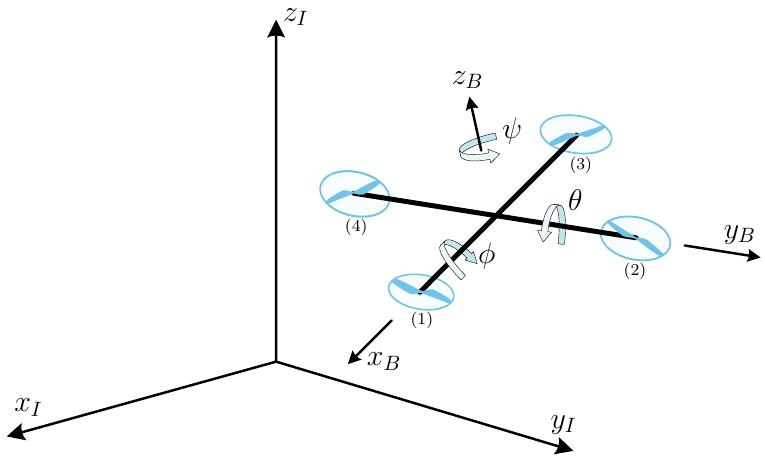}
  \end{center}
  \caption{Schematic diagram of a quadrotor.} 
  \label{fig:quad}
\end{figure}

Derive the dynamic equations for the rotational part and the altitude ($z_I$, or $z$ for brevity) as
\begin{equation*}
  \ddot{x} = \bar{f} + \bar{B}u + \Delta(x,u),
\end{equation*}
where $x=\begin{bmatrix} z,& \phi,& \theta,& \psi \end{bmatrix}^T$. Excluding $x_I$ and $y_I$ from the dynamic model in the inner control loop of a quadrotor is a typical approach in the literature, where $x_I$ and $y_I$ are indirectly controlled by an outer control loop.

\begin{enumerate}[(a)]
\item 
Linearize the governing equations and employ it as the nominal model of the system in the control design process.
\item Design an adaptive RBF neural network to compensate for uncertain dynamics and the nonlinear terms in the model based on the Lyapunov stability. 
Concerning the hidden nodes (neurons of the hidden layer), choose the center and width of each neuron as a (constant) random number, where $x_{0i}\in \left [-1, 1\right ]$ and $\sigma_i \in \left [1.2, 4.2\right ]$. Also, the input of the NN can be defined as $X=\left [x^T,\dot{x}^T,u^T \right]^T$.
\item Design an FEL-based control law and analyze the closed-loop stability.
\item Evaluate the closed-loop performance using the reference command
\begin{equation}\label{FEL_trajectory_1}
{{x}_{d}}\left( t \right)=\left[ \begin{matrix}
   {{z}_{d}}\left( t \right)  \\
   {{\phi }_{d}}\left( t \right)  \\
   {{\theta }_{d}}\left( t \right)  \\
   {{\psi }_{d}}\left( t \right)  \\
\end{matrix} \right]=\left[ \begin{matrix}
   0.5e^{-t}\sin\frac{2\pi t}{5} \\
   0.2\sin \frac{\pi t}{10} + \frac{\pi}{4} \\
   0.2\left( 1-\cos \frac{\pi t}{15} + \frac{\pi}{2} \right)\\
   0  \\
\end{matrix} \right].
\end{equation}
Consider the following uncertainties in your simulation
\begin{itemize}
    \item Wind disturbance,
    \item Ground effects,
    \item Actuator dynamics,
    \item Unmodeled dynamics.
\end{itemize}
Report the Mean Absolute Tracking Error (MAE) for closed-loop simulation, where MAE is defined as
\begin{equation}\label{eq:MAE}
\text{MAE}=\frac{1}{T}\mathop{\int }_{0}^{T}\left| x\left( t \right)-{{x}_{d}}\left( t \right) \right|dt.
\end{equation}
Compare the value of the estimated uncertain dynamics ($ \hat{\Delta}=\hat{W}^T \mu\left( X \right) $) with real uncertain term 
\begin{equation}\label{Exc21:Delta}
\Delta = \bar{d} + \left(f\left(x\right)-\bar{f}\right) + \left(B\left(x\right)-\bar{B}\right)u + B\left(x\right)(u_c-u),
\end{equation}
where $u$ and $u_c$ denote the real and the commanded thrust forces, respectively, and the overbar represents the linearized system.

Report Mean Absolute Estimation Error (MAEE) as
\begin{equation}\label{eq:MAEE}
\text{MAEE} = \frac{1}{T}\mathop{\int }_{0}^{T}\left\| \hat{\Delta} - \Delta \right\|dt,
\end{equation}
where $\left\| \cdot \right\|$ is the Euclidean norm.

\item Perform a parameter study on MAE and MAEE for the number of neurons, learning rate $\Gamma$ and gain $k_1$.
\item Study the effect of normalizing the NN inputs on the estimation and tracking accuracy.
\end{enumerate}
\tcblower
The governing equations of the UAV is derived in Appendix \ref{sec:A1}, including ground effects, wind disturbances and actuator dynamics. Regarding to the unmodeled dynamics, the linearized dynamic equations will be used, along with the assumption of an ideal actuator, which differs from the actual model. Here, we will proceed with the design of the control law.
\begin{enumerate}[(a)]
\item \textbf {Linearized model:}
In order to enhance the complexity of the control task, our objective is to present the controller with a linearized model of the UAV to determine its capability in addressing nonlinearity in the form of unmodeled dynamics, as well as the external disturbances.

The nonlinear dynamic model can be linearized around hovering conditions to obtain a linear model for control design and analysis. With small angle approximations ($\sin x \approx x$ and $\cos x \approx 1$ for small $x$), the linearized governing equations are obtained as follows
\begin{equation}\label{eq:exc21:FBlinear}
  \ddot{x} = \bar{f} + \bar{B}u + \Delta(x,u),
\end{equation}
where
\begin{equation}
\begin{split}
\bar{f} &=\begin{bmatrix}
-g\\
0\\
0\\
0
\end{bmatrix} \\
\bar{B} &=\underbrace{\begin{bmatrix}
\frac{1}{m} & 0 & 0 & 0\\
0 & \frac{1}{I_{x} } & 0 & 0\\
0 & 0 & \frac{1}{I_{y} } & 0\\
0 & 0 & 0 & \frac{1}{I_{z} }
\end{bmatrix}}_{B_1}
\underbrace{\begin{bmatrix}
1 & 1 & 1 & 1 \\
0 & l & 0 & -l \\
-l & 0 & l & 0 \\
c & -c & c & -c
\end{bmatrix}}_{B_2},
\end{split}
\end{equation}
and $\Delta(x,u)$ is the total uncertainty to be approximated using a neural network.

\item
\textbf {Neural network structure:}
RBF neural networks are a class of networks that are often used to estimate uncertainties. 
To this end, we utilize them to take account for the unmodeled dynamics, along with the disturbances.
As depicted in Fig. \ref{fig:rbfnn}, the uncertain dynamics $\Delta(x,u)$ can be described as follows
\begin{equation}
\hat{\Delta}(x,u)=\hat{W}^T \mu (X),
\end{equation}
where $X \in \mathbb{R}^{m}$ is the input vector, $\hat{W} \in \mathbb{R}^{n \times 4}$ is the adjustable weight matrix, $\mu(X):\mathbb{R}^{m} \rightarrow \mathbb{R}^{n}$ is the vector of basis functions, and $n$ is the number of neurons in the network.
The basis functions in RBFNNs are typically Gaussian function that is presented in Eq. (\ref{Gaus_Org}). As mentioned, $x_{0i}\in \mathbb{R}^{m}$ is the vector of center of the $i$th basis function, and $\sigma$ is the width of the Gaussian function. Additionally, since $x_{0i}$ is chosen in the interval of unity, the input vector is normalized before being fed into the network.

\textbf {Input normalization:}
To improve the training and approximation capabilities of the RBFNN, the inputs can be normalized to fall within the specified range of $[-1, 1]$ which we assigned the centers of the basis functions.

Having defined the input vector as $X$,
we first determine the minimum and maximum values for each input dimension based on the expected operating conditions
\begin{equation}
u_i = \max{x_i}, \quad l_i = \min{x_i},
\end{equation}
where $i = 1,2,\dotsc,m$.

Each element of the normalized input $\bar{X}$ is then calculated as:
\begin{equation}
\bar{x}_i = \frac{2(x_i - l_i)}{u_i - l_i}-1.
\end{equation}

This normalization maps each input dimension to the range $[-1, 1]$. The normalized input $\bar{X}$ is then used as the input to the RBFNN instead of the original $X$ as $\hat{\Delta}(x,u) = \hat{W}^T\mu(\bar{X})$.

\item
\textbf {Controller design:}
Regarding the inner loop, if we define the desired trajectory as $x_d$, we can calculate the tracking error as $e=x-x_d$. To generate the filtered error signal, we can define $s=\dot{e}+\lambda e$ with $\lambda >0$. Therefore, the derivative of $s$ can be expressed as
\begin{equation}\label{Ex21_sd}
 \dot{s} = \ddot{e}+\lambda \dot{e}
      = \bar{f} + \bar{B}u + \Delta(x,u) - \ddot{x}_d + \lambda \dot{e}.
\end{equation}

Subsequently, using the previously mentioned NN to estimate $\Delta(x,u)$ as $\hat{\Delta}(x,u)=\hat{W}^T \mu (X)$, it is obtained that
\begin{equation}
\Delta(x,u) = {W^*}^T \mu (X) + \varepsilon,
\end{equation}
which gives $\tilde{\Delta}(x,u) = {\tilde{W}}^T \mu (X) - \varepsilon$.

As a result, we can construct the following control command
\begin{equation}\label{Ex21_u}
u=\bar{B}^{-1}\left(-\bar{f}-\hat{\Delta}(X)+\ddot{x}_d-\lambda \dot{e}-k_1 s\right).
\end{equation}

Now, we proceed to derive the update rule for the NN parameters.

\textbf {Stability analysis:}
Consider the following Lyapunov function $V$ as
\begin{equation}\label{LYA_01}
V=\frac{1}{2}s^T s + \frac{1}{2} \text{tr} \left(\tilde{W}^T \Gamma^{-1} \tilde{W}\right),
\end{equation}
where $\tilde{W}=\hat{W} - W^*$ and $\Gamma$ is a positive definite matrix. Next, by combining Eqs. \ref{Ex21_sd} and \ref{Ex21_u}, the time derivative of $V$ is obtained as
\begin{align}
\begin{split}
\dot{V} &= s^T \dot{s}+ \text{tr} \left(\tilde{W}^T \Gamma^{-1} \dot{\tilde{W}}\right) \\
&= - s^T \left(\tilde{\Delta}+ k_1 s\right)+ \text{tr} \left(\tilde{W}^T \Gamma^{-1} \dot{\hat{W}}\right),
\end{split}
\end{align}
in which we assumed $\dot{\tilde{W}}\approx  \dot{\hat{W}}$. Considering FEL update rule $\dot{\hat{W}}=\Gamma \mu s^T$, results in
\begin{align}
\begin{split}\label{LYA_02}
\dot{V} &= - s^T \left(\tilde{\Delta}+ k_1 s\right)+ \text{tr} \left(\tilde{W}^T \Gamma^{-1} \dot{\tilde{W}}\right) \\
&= - s^T \left({\tilde{W}}^T \mu - \varepsilon + k_1 s\right) + \text{tr} \left(\tilde{W}^T \mu s^T \right) \\
&= - s^T \left({\tilde{W}}^T \mu - \varepsilon + k_1 s\right) + s^T \tilde{W}^T \mu \\
&= - k_1 s^T s + s^T \varepsilon = - s^T (k_1 s - \varepsilon).
\end{split}
\end{align}
This results in $\dot{V}<0$ for $\left\|k_1 s\right\|>\left\|\varepsilon\right\|$, guaranteeing bounded tracking error.

\item
\textbf {Simulation results:}
For the case with 15 neurons and $k_1=20$, $\lambda=10$, $\Gamma=30$, the time history of attitude and altitude is depicted in Fig. \ref{fig:exc21:stat}. Additionally, dead-zone length is selected as $L=1$ and the time constant of the first-order actuator is $\tau=0.01$.
\begin{figure}[H]
  \begin{center}
    \includegraphics[width=0.9\textwidth]{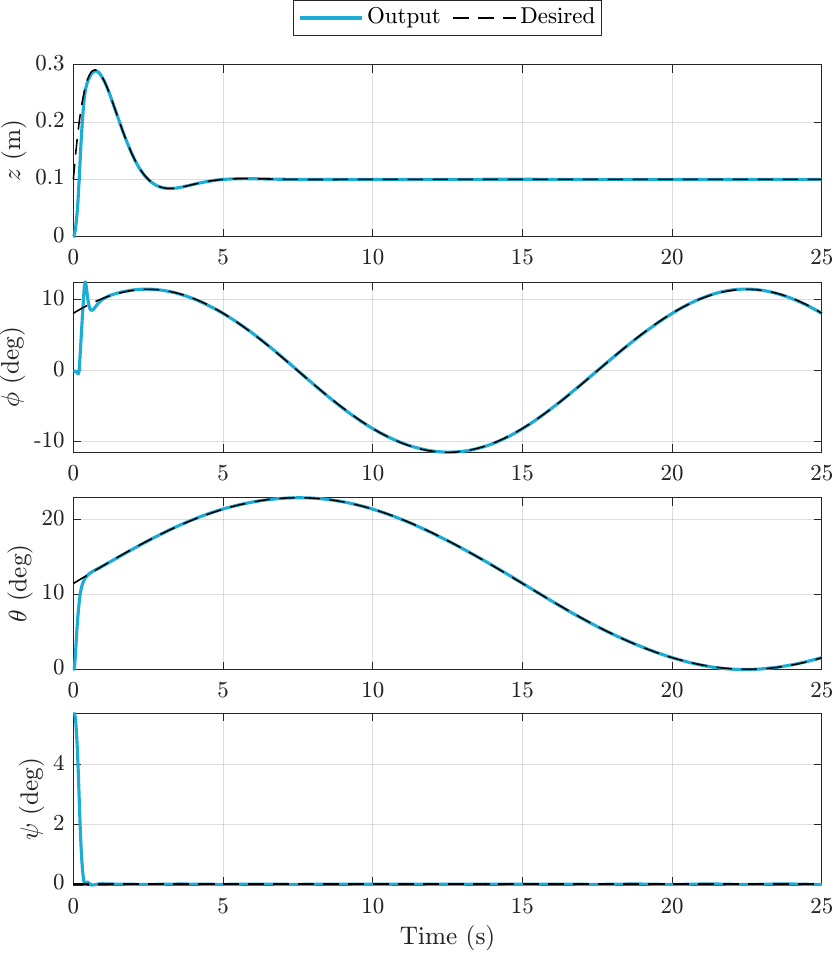}
  \end{center}
  \caption{Time history of attitude and altitude in basic FEL control of quadrotor.} 
  \label{fig:exc21:stat}
\end{figure}
In general, what stands out is that both attitude and altitude responses follow their desired trajectories (dashed lines) while having initial deviations from their target. It should be noted that all states have zero initial conditions, except for $\psi = 0.1$ rad to give an initial deviation from its target value of zero.

Regarding the control effort, it starts with a high demand at the beginning, but due to the presence of first-order filter, the applied control is attenuated. Moreover, due to the presence of dead-zone, there is always a shortage of thrust force, depending on the value of $L$, as shown in Fig. \ref{fig:exc21:eff}.
\begin{figure}[H]
  \begin{center}
    \includegraphics[width=0.9\textwidth]{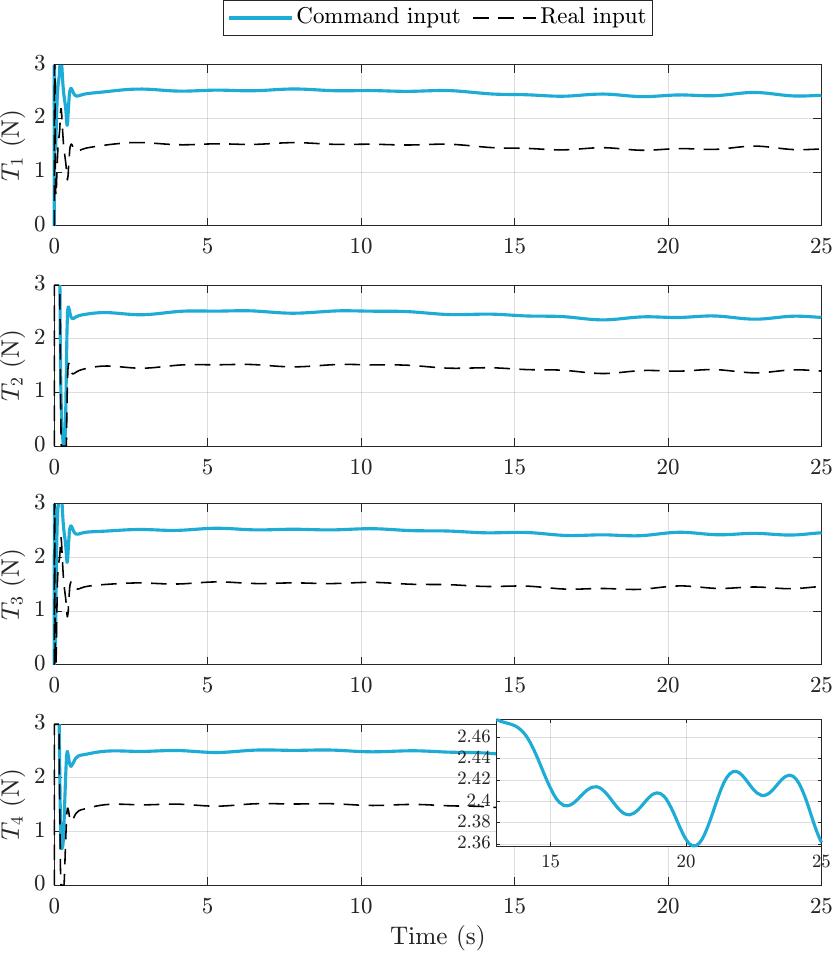}
  \end{center}
  \caption{Time history of control inputs in basic FEL control of quadrotor.} 
  \label{fig:exc21:eff}
\end{figure}

Moving on to the uncertainty estimation, the comparison of real uncertainty from Eq. (\ref{Exc21:Delta}) and the NN estimation is shown in Fig. \ref{fig:exc21:del}, showing an acceptable estimating performance, which can be fine-tuned by using different parameters.
\begin{figure}[H]
  \begin{center}
    \includegraphics[width=0.9\textwidth]{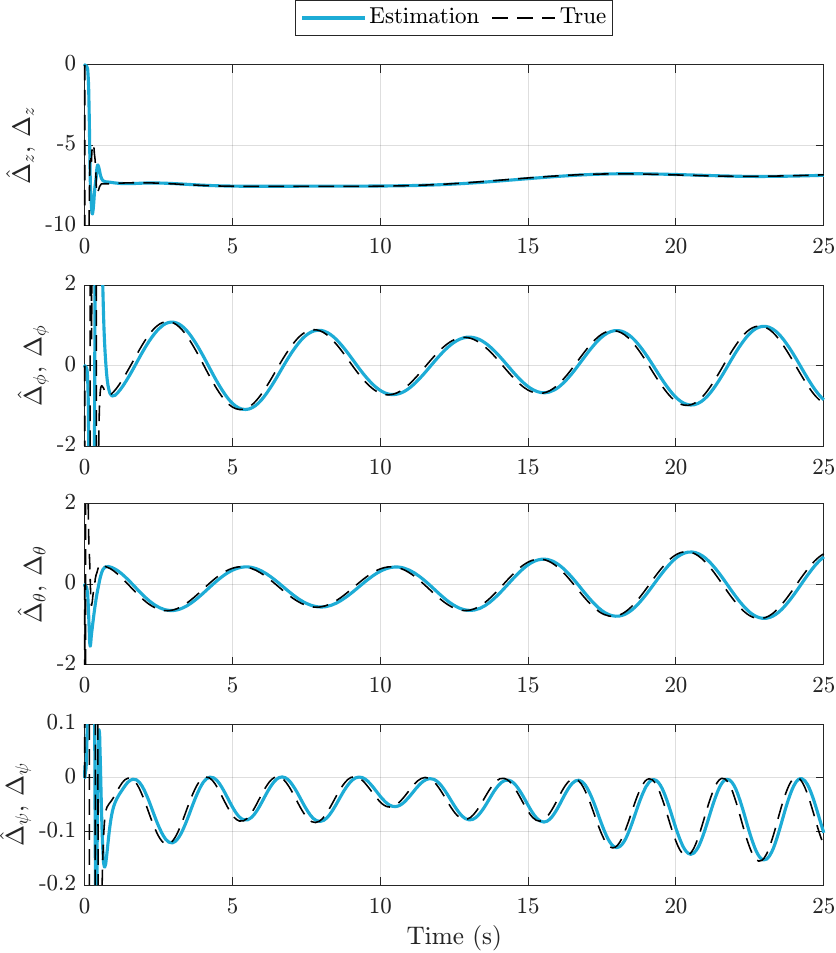}
  \end{center}
  \caption{Uncertainty estimation in basic FEL control of quadrotor.} 
  \label{fig:exc21:del}
\end{figure}

Additionally, Fig. \ref{fig:exc21:weigh} shows the evolution of training weights over time, consisting of oscillating weights, as well as another set of weights which seem to have converged to a constant value at the end of the simulation time.
\begin{figure}[H]
  \begin{center}
    \includegraphics[width=0.9\textwidth]{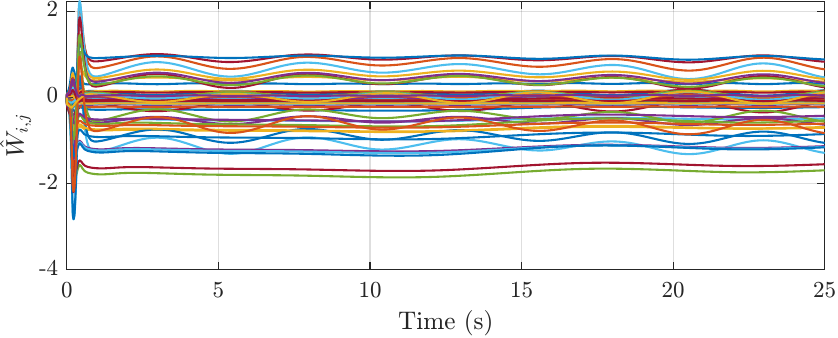}
  \end{center}
  \caption{Evolution of training weights in basic FEL control of quadrotor.} 
  \label{fig:exc21:weigh}
\end{figure}



\item
\textbf {Sensitivity analysis:}
To investigate the effect of employing different parameters on the control performance, which in our case is directly related to the accuracy of the uncertainty estimation, denoted by MAEE, we perform a parameter study. As shown in Fig. \ref{fig:exc21:param}, for a fixed number of neurons, higher learning rates and lower gains of $k_1$ up to some point are advantageous. Additionally, it is observed that by lowering the number of neurons, higher learning rates result in identical MAEE.
\begin{figure}[H]
  \begin{center}
    \includegraphics[width=0.9\textwidth]{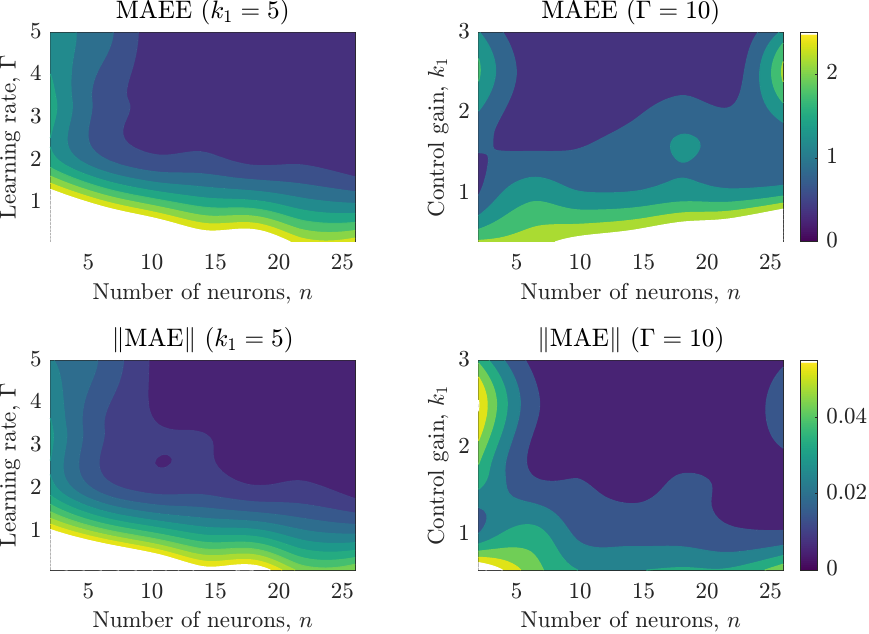}
  \end{center}
  \caption{Sensitivity analysis for number of neurons, learning rate and the gain $k_1$ on MAEE and the norm of MAE in basic FEL control of quadrotor.} 
  \label{fig:exc21:param}
\end{figure}
The effect of these parameters on the norm of MAE is also depicted in the figure, showing a similar trend as MAEE. This is expected because when uncertainty estimation is more precise, the tracking performance improves.

\item
\textbf {Effect of input normalization:}
To investigate the impact of normalizing the inputs to the RBFNN, two cases are compared:
\begin{itemize}
    \item No input normalization
    \item Normalized inputs in the range $[-1, 1]$
\end{itemize}
The tracking performance and uncertainty estimation are evaluated for both cases. The results are summarized in Table \ref{tab:exc01_MAE_MAEE_norm}.
It can be observed that normalizing the inputs leads to reduced tracking error across all four states. The uncertainty estimation accuracy is also improved as evidenced by the lower MAEE value with input normalization.
It should be noted that the precise values of these errors may vary slightly in each simulation run due to the randomness of the RBF activation's parameters.
\begin{table}[H]
    \caption{Effect of RBFNN input normalization on FEL control performance.}
    \label{tab:exc01_MAE_MAEE_norm}
    \centering
    \begin{tabular}{lccccc}
    \hline\noalign{\smallskip}
    & MAE($z$) & MAE($\phi$) & MAE($\theta$) & MAE($\psi$) & MAEE \\
    \noalign{\smallskip}
    \hline
    \noalign{\smallskip}
    No normalization & $2.27\times 10^{-3}$ & $3.52\times 10^{-3}$ & $2.50\times 10^{-3}$ & $1.01\times 10^{-3}$ & $8.72\times 10^{-1}$ \\
    Normalized inputs & $1.39\times 10^{-3}$ & $2.70\times 10^{-3}$ & $1.56\times 10^{-3}$ & $8.52\times 10^{-4}$ & $7.84\times 10^{-1}$ \\
    \noalign{\smallskip}\hline\noalign{\smallskip}
    \end{tabular}
\end{table}

This performance improvement can be attributed to several factors. Normalization allows the inputs to better stimulate the RBF activations. It rescales the inputs so that they have a similar scale and distribution. This makes it easier for the network to learn the relationships between the inputs and outputs.
Moreover, proper normalization enhances the learning process.

The uncertainty estimation error is compared for both cases in Fig. \ref{fig:exc21:delta_Norm}. With normalization, $\tilde{\Delta}$ converges faster to zero while the other case leads to a considerable amount of error.
\begin{figure}[H]
  \begin{center}
    \includegraphics[width=0.9\textwidth]{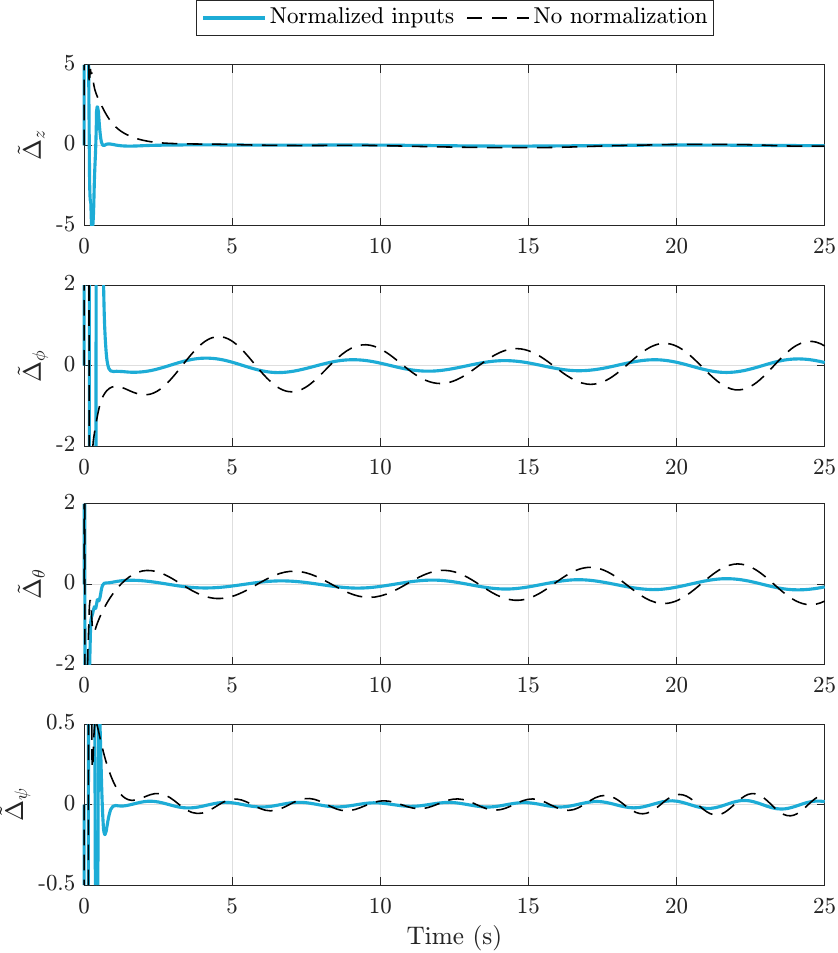}
  \end{center}
  \caption{Analyzing the impact of input normalization on the uncertainty estimation error.} 
  \label{fig:exc21:delta_Norm}
\end{figure}

Figure \ref{fig:exc21:states_Norm} shows the tracking behavior over the first 5 seconds. The normalized input case shows faster convergence to the desired trajectory. Feeding the inputs to the network without normalization prevents the network from being stimulated, making it ineffective. Moreover, it takes longer to reach an acceptable estimation, explaining why there is no overshoot in the response of the roll angle $\phi$.
\begin{figure}[H]
  \begin{center}
    \includegraphics[width=0.9\textwidth]{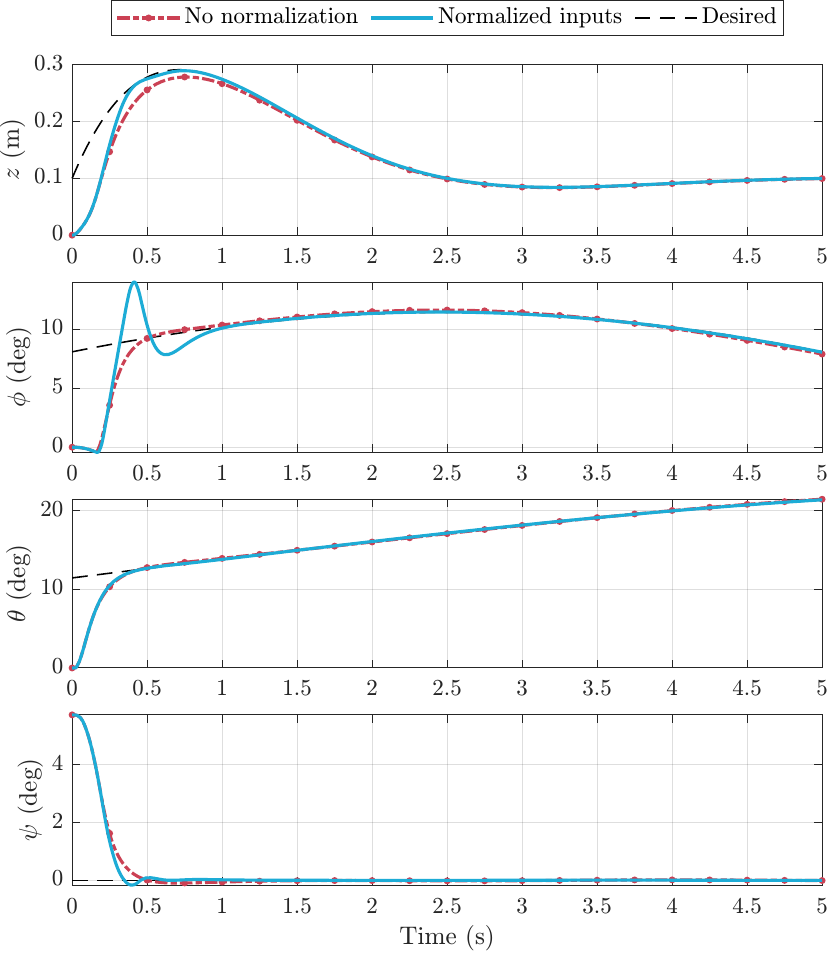}
  \end{center}
  \caption{Analyzing the impact of input normalization on tracking performance in the initial moments.} 
  \label{fig:exc21:states_Norm}
\end{figure}

\end{enumerate}
\end{ex}

\begin{ex}[label=Example:GTM_FEL]{FEL-based control of fixed-wing aircraft}
The NASA Generic Transport Model (GTM) is employed as a fixed-wing aircraft model to investigate the effectiveness of various NN-based control methods presented in this book. The detailed nonlinear model of this aircraft was developed by NASA Langley Research Center for simulation experiments.
The GTM simulation has been implemented in Simulink and is available for download from \citet{GTMGit}. Furthermore, the implementation is described in \citet{Hueschen.2011}. Readers seeking details of the mathematical model of the conventional fixed-wing aircraft can find it in Appendix \ref{app:GTM1}. Additionally for more details about the GTM refer to Appendix \ref{app:GTM2}.

Consider the following linear model of the rotational dynamics of GTM
\begin{align} \label{Eq:ARDE_ex}
\begin{split}
  \underbrace{\begin{bmatrix} \dot{p} \\  \dot{q}\\ \dot{r}  \end{bmatrix} }_{\dot{x}}
    &= \underbrace{\begin{bmatrix} 
    0.0052u-0.5779v-0.0501w-5.159p-0.1373q+1.952r \\[0.8ex]
    0.0183u+0.005v-0.2764w-0.0051p-2.93q+0.0483r \\[0.8ex]
    0.0004u+0.1823v-0.0021w-0.4267p-0.048q-1.191r
    \end{bmatrix}  }_{F}
    +\\
    &
    \underbrace{\begin{bmatrix}
    -0.0127 & -0.7974 & 0.2039 \\
    -0.6891 & 0.0805 & 0.0029 \\
    -0.0007 & -0.0473 & -0.3986 
    \end{bmatrix} 
    }_{B}
    \underbrace{\begin{bmatrix} \delta_e \\  \delta_a \\ \delta_r \end{bmatrix}  }_{u} + \, \Delta(X).
\end{split}
\end{align}

\begin{enumerate}[(a)]
    \item Design a nonlinear controller using the FEL approach to compensate for uncertain dynamics and the nonlinear effects in the GTM model, which are disregarded in the given nominal model presented in Eq. (\ref{Eq:ARDE_ex}).
    \item Evaluate the closed-loop performance using the reference command
\begin{equation}
        x_d(t) = \begin{bmatrix}
        p_d(t) \\
        q_d(t) \\
        r_d(t)
        \end{bmatrix}
        = \begin{bmatrix}
        \begin{cases}
        0 & \text{if } t \leq 40 \, s \\
        0.03 \sin \left (\frac{\pi}{5} t \right ) & \text{if } t > 40 \, s
        \end{cases} \\
        \begin{cases}
        0.02 \sin \left (\frac{\pi}{10} t \right ) & \text{if } t \leq 40 \, s \\
        0 & \text{if } t > 40 \, s
        \end{cases} \\
        \begin{cases}
        0 & \text{if } t \leq 40 \, s \\
        0.01 \sin \left (\frac{\pi}{20} t \right ) & \text{if } t > 40 \, s
        \end{cases} \\
        \end{bmatrix}.
        \end{equation}
        Concerning the hidden nodes (neurons within the hidden layer), consider $20$ hidden neurons and select the center and width of each neuron as a (constant) random number, where $x_{0_{i}} \in [-1, 1]$ and $\sigma_{i} \in [1.1, 2.1]$. Also, the input of the NN can be defined as $X= \begin{bmatrix}u & v & w & p & q & r & \delta_e & \delta_a & \delta_r \end{bmatrix}^T$. Notice that the input vector $X$ must be normalized to the range of $[-1, 1]$, as mentioned in Example \ref{Example:quadFEL}. Furthermore, include wind with a speed of $10$ knots and a direction of $270$ degrees as an external disturbance in the GTM simulation. Additionally, report the MAE for closed-loop simulation.
    \item Plot the change of training weights over time and check the convergence of them to constant values.
    
    \item Report the MAEE and compare the value of the estimated uncertain dynamics ($ \hat{\Delta} (X) =\hat{W}^T \mu\left( X \right) $) with real uncertain term 
        \begin{equation}\label{Exc22:Delta}
        \Delta = \frac{x(k) - x(k-1)}{\Delta t} - F - Bu_c,
        \end{equation}
        where $\frac{x(k) - x(k-1)}{\Delta t}$ represents an estimation of the actual state derivative $\dot{x}$, and $u_c$ denotes the commanded control surface deflections.
        \item Evaluate the effect of design parameters such as the number of hidden nodes of the NN, the learning rate ($\Gamma$) and control gain ($k_1$), on the tracking performance (MAE) as well as estimation performance (MAEE). Furthermore, To simplify the comparison, utilize the Euclidean norm when analyzing the MAE values.
\end{enumerate}
\tcblower
\begin{enumerate}[(a)]
    \item \textbf {Controller design:}
    By Defining the tracking error as $e = x - x_d$, where $x_d$ represents the desired rotational state $(p_d, q_d, r_d)$ and $x$ denotes the actual state $(p, q, r)$, the derivative of $e$ can be expressed as follows
        \begin{equation}
        \label{Ex22_ed}
         \dot{e} = F + Bu + \Delta - \dot{x}_d.
        \end{equation}
        
        Subsequently, we can use a NN to estimate $\Delta$ as $\hat{\Delta}=\hat{W}^T \mu $.
        Additionally the control command can be considered as Eq. (\ref{Ctrl_2}).\\
        \textbf {Stability analysis:}
        The process of stability analysis and the derivation of update rules for neural network weights aligns with the approach presented in Eqs. (\ref{Lyap_0})-(\ref{Lyap_3_1}). Hence, the update rule for neural network weights can be regarded as Eq. (\ref{Learn_1}).
    \item \textbf {Simulation results:}
        For the case with $k_1=18$ and $\Gamma=25$ the time history of angular rates is depicted in Fig. \ref{fig:Ex_2.2_pqr}. In general, it is observed that the angular rates closely follow their desired trajectories in the presence of uncertainties and external disturbances.
        \begin{figure}[H]
            \begin{center}
            \includegraphics[width=0.9\textwidth]{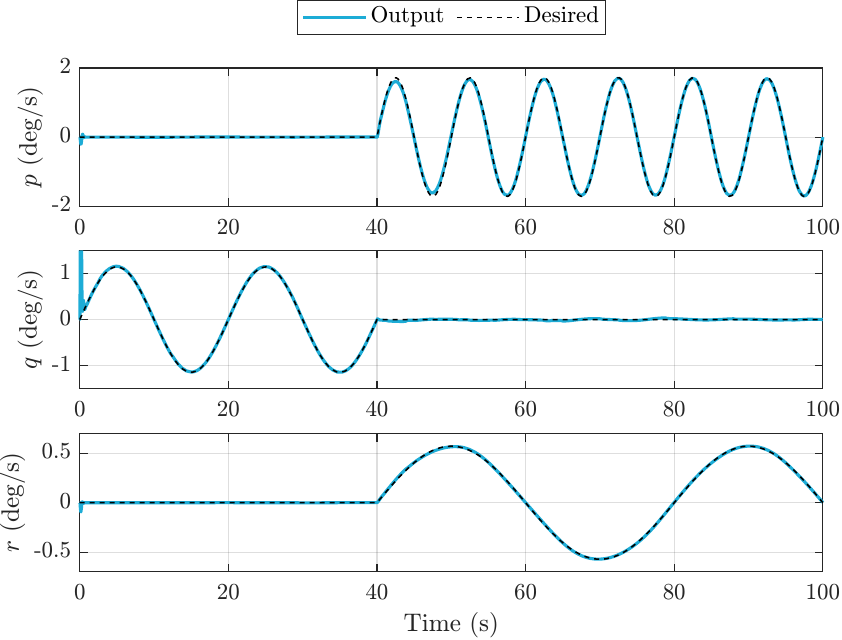}
            \end{center}
            \caption{Time history of angular rates in basic FEL control of GTM.} 
            \label{fig:Ex_2.2_pqr}
        \end{figure}
         Additionally, the corresponding control input (control surface deflections) is shown in Fig. \ref{fig:Ex_2.2_u}. The initial oscillation observed in the time history of elevator deflection can be attributed to a high learning rate. This characteristic is also noticeable in the time history of the pitch rate $(q)$ at the beginning of the simulation.
        \begin{figure}[H]
            \begin{center}
            \includegraphics[width=0.9\textwidth]{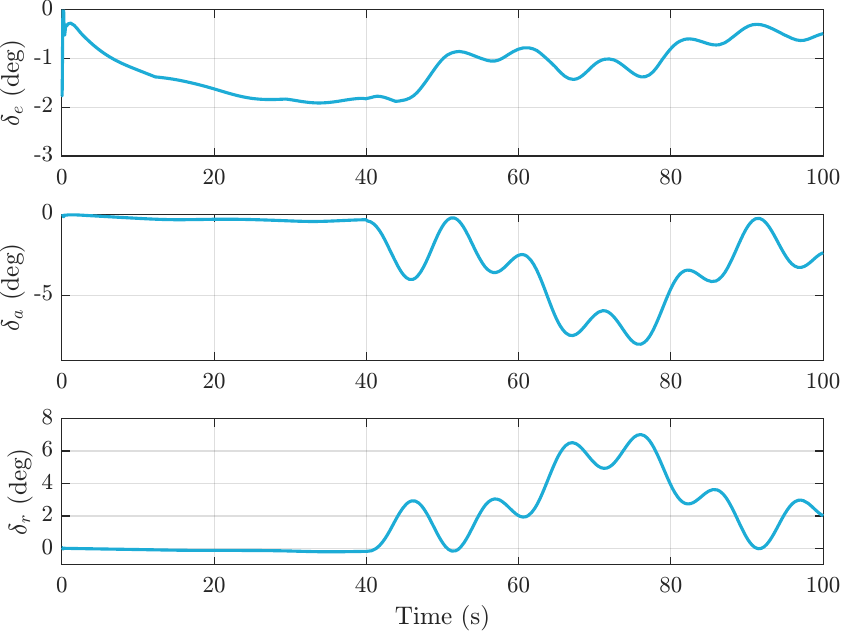}
            \end{center}
            \caption{Time history of control inputs in basic FEL control of GTM.} 
            \label{fig:Ex_2.2_u}
        \end{figure}
        Furthermore, the MAE of tracking for this simulation is given in table \ref{tab:EX_2.2_MAE}.
        \begin{table}[H]
            \caption{MAE of angular rates tracking in basic FEL control of GTM.}
            \label{tab:EX_2.2_MAE}
            \centering
            \begin{tabular}{ccc}
            \hline\noalign{\smallskip}
            MAE ($p$) & MAE ($q$) & MAE ($r$) \\
            \noalign{\smallskip}\hline\noalign{\smallskip}
            $3.2\times 10^{-4}$ & $2.78\times 10^{-4}$  & $5.4\times 10^{-5}$ \\
            \noalign{\smallskip}\hline\noalign{\smallskip}
            \end{tabular}
        \end{table}
        \item The evolution of training weights over time is shown in Fig. \ref{fig:Ex_2.2_W} which seem to have converged to a constant value at the end of the simulation time.
        \begin{figure}[H]
            \begin{center}
            \includegraphics[width=0.9\textwidth]{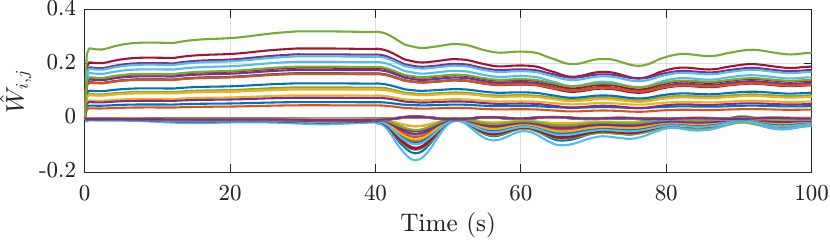}
            \end{center}
            \caption{Evolution of training weights in basic FEL control of GTM.}
            \label{fig:Ex_2.2_W}
        \end{figure}
        \item Regarding uncertainty estimation, the MAEE obtained in this simulation is $1.797\times 10^{-2}$. Furthermore, the comparison of real uncertainty and the NN estimation is shown in Fig \ref{fig:Ex_2.2_Delta}, showing an acceptable estimating performance.
        \begin{figure}[H]
            \begin{center}
            \includegraphics[width=0.9\textwidth]{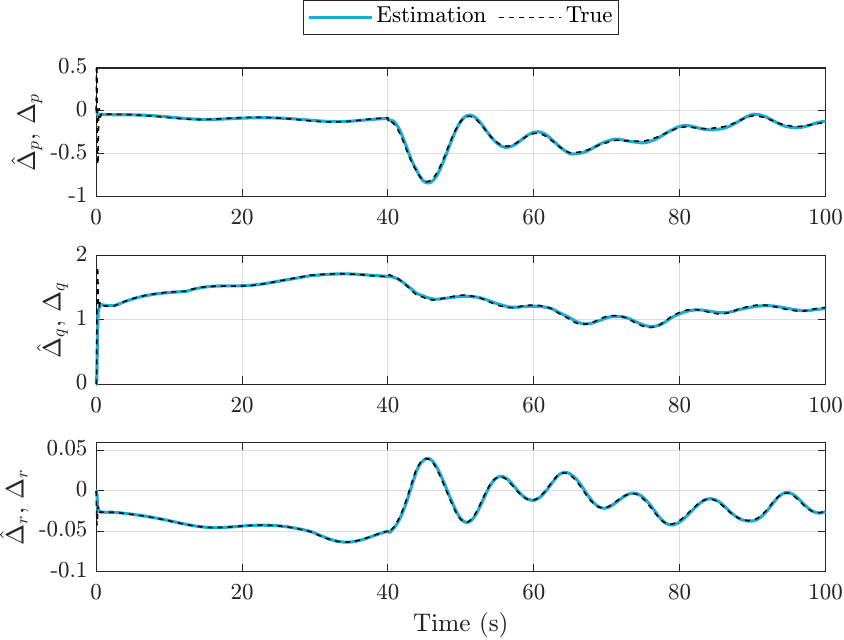}
            \end{center}
            \caption{Uncertainty estimation in basic FEL control of GTM.}
            \label{fig:Ex_2.2_Delta}
        \end{figure}
        \item \textbf{Sensitivity analysis:} A parameter study is conducted to investigate the effects of various parameters on control performance, specifically focusing on the accuracy of uncertainty estimation (measured by MAEE) and the tracking error (measured by MAE). As shown in Fig. \ref{fig:Ex_2.2_SA}, lower learning rates and a lower number of neurons have a clear impact on increasing both the MAEE and MAE (Euclidean norm of MAE). Additionally, a higher number of neurons combined with a higher learning rate can lead to unintended effects on MAE and MAEE. Furthermore, it is worth noting that the high values of MAE and MAEE in cases with high learning rates and a large number of neurons are due to numerical issues. However, this effect can be mitigated by using smaller sample times, which helps to alleviate the impact of numerical issues and oscillations. In the case of control gain, using low control gains leads to an increase in both MAE and MAEE, particularly when a very low or high number of neurons is utilized.
        \begin{figure}[H]
            \begin{center}
            \includegraphics[width=0.9\textwidth]{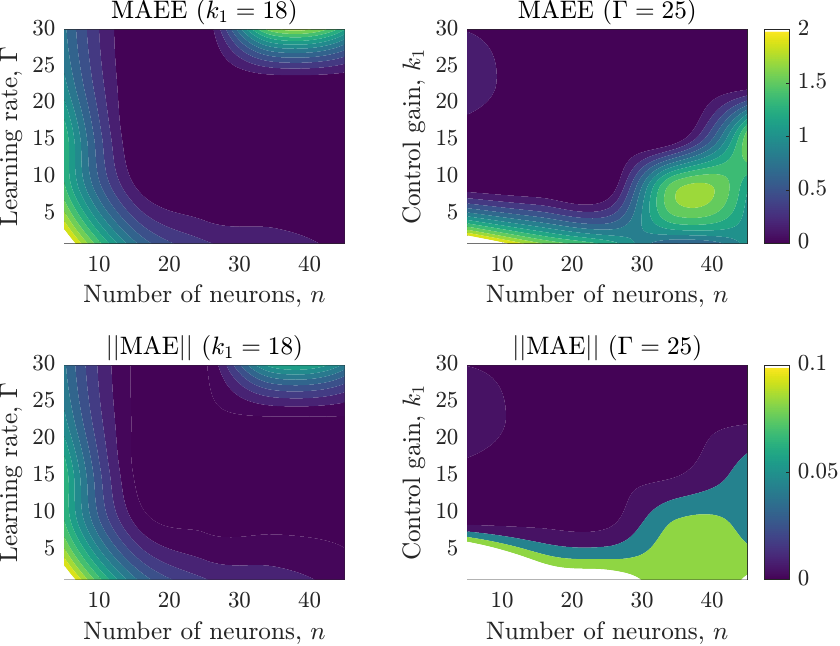}
            \end{center}
            \caption{Sensitivity analysis of MAEE and MAE for the number of neurons, learning rate, and the gain $k_1$ in basic FEL control of GTM.}
            \label{fig:Ex_2.2_SA}
        \end{figure}
\end{enumerate}
\end{ex}

\begin{ex}[label=sat_fel]{FEL-based control of satellite}
Design an FEL controller using an RBF neural network to control the attitude of a cubic satellite with uncertainties.
Using small angles approximations and Euler angles, show that the satellite's attitude equations of motion can be expressed as
\begin{equation}\label{sat_eom}
    \ddot{x} (t) = F (x) + B (x) u + \Delta (x, \dot{x}),
\end{equation}
where $ x = \mqty[\phi & \theta & \psi]^T $ denotes the Euler angles and $ \Delta (x, \dot{x}) $ represents unmodeled dynamics as well as external disturbances. To see the procedure of deriving Eq. \eqref{sat_eom}, refer to Appendix \ref{app:sat}.
\begin{enumerate}[(a)]
\item Design the controller using an RBFNN with 15 neurons. Assign the centers and widths for each neuron randomly in the specified region in Example \ref{Example:quadFEL}.
\item Determine how well the designed control system can follow the following reference input
\begin{equation}\label{sat_desired}
    x_d (t) = 10^\circ \mqty[1 - \cos\left( \dfrac{\pi t}{10} \right), & 1 - \cos \left( \dfrac{\pi t}{5} \right), & 0]^T.
\end{equation}
\item Plot the changes in training weights over time.
\item Report MAE as well as MAEE.
\item Examine how design parameters, including the number of neurons, learning rate, and control gain, impact both the Euclidean norm of the MAE and the MAEE.
\end{enumerate}
\tcblower
In this example, a FEL-based controller is designed for attitude control of a cubic satellite in the presence of uncertainties.
\begin{enumerate}[(a)]
\item \textbf{Control design and stability analysis:} In the following, a neural network is used to estimate the uncertainty $ \Delta (x, \dot{x}) $ as
\begin{equation}
\Delta (x, \dot{x}) = W^{*T}\mu + \varepsilon,
\end{equation}
where $ W^* $ denotes the unknown optimal weight, $ \mu $ represents the vector of basis functions, and $ \varepsilon $ indicates the bounded approximation error ($ \norm{\varepsilon} \leq \varepsilon_{max} $). Since $ W^* $ is not known, it is approximated by $ \hat{W} $, and $ \hat{\Delta} $ is defined as
\begin{equation}
\hat{\Delta} (x, \dot{x}) = \hat{W}^T\mu.
\end{equation}
To design a controller for Eq. (\ref{system}), at first, we define a filtered tracking error as follows
\begin{equation}
s = \dot{e} + \lambda e,
\end{equation}
where $ e = x - x_d $. Now, similar to Eqs. (\ref{LYA_01})-(\ref{LYA_02}), it can be proved for $\left\|k_1 s\right\|>\left\|\varepsilon\right\|$ that $\dot{V}<0$ and guaranteed bounded tracking error. Therefore, the control input can be considered as follows
\begin{equation}
u = B^{-1} \left(-F (\dot{x}) - \hat{\Delta} (x, \dot{x}) - \lambda \dot{e} -k s + \ddot{x}_d (t) \right).
\end{equation}

\item \textbf{Simulation results:} Considering $ k = 7 $, $ \lambda = 5 $, and $  \Gamma = 30 $, Fig. \ref{fig:sat_fel_states} shows time history of the states.
\begin{figure}[H]
\centering
\includegraphics[width = 0.9\textwidth]{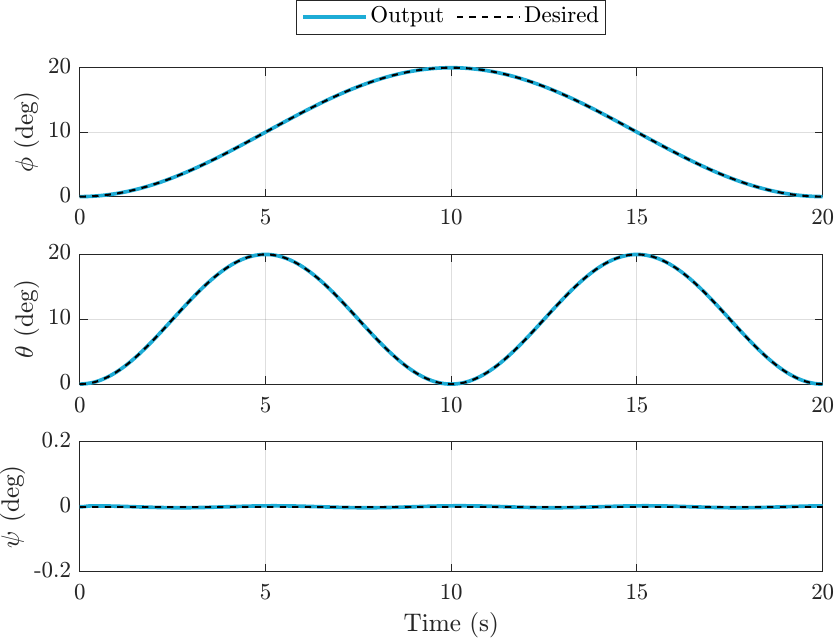}
\caption{Time history of attitude in basic FEL control of satellite.}
\label{fig:sat_fel_states}
\end{figure}

As seen in Fig. \ref{fig:sat_fel_states}, Euler angles have been able to accurately track the reference inputs in the presence of model uncertainties and external disturbances. Additionally, the corresponding control inputs of this example is shown hereunder.
\begin{figure}[H]
\centering
\includegraphics[width=0.9\textwidth]{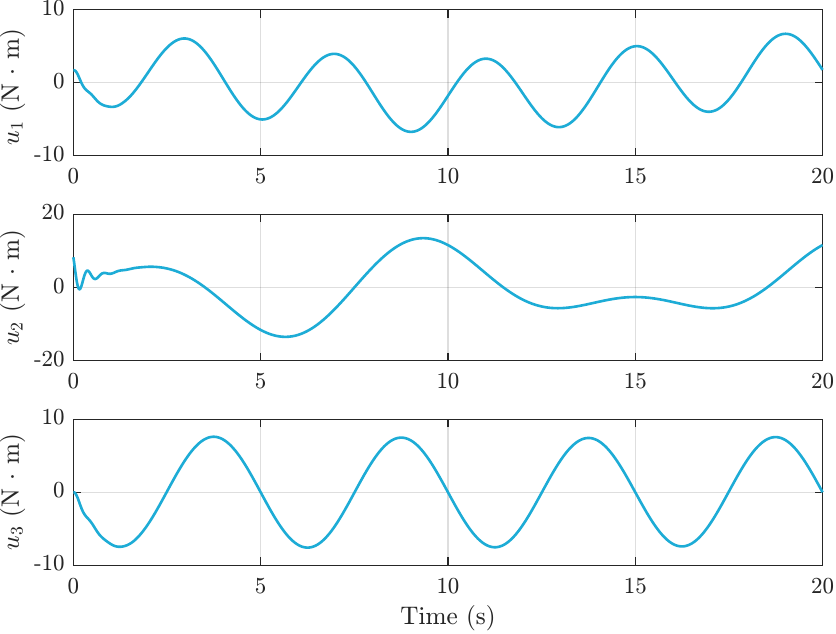}
\caption{Time history of control inputs in basic FEL control of satellite.}
\label{fig:control_sat_fel}
\end{figure}
It is worth mentioning that the oscillatory behavior of control inputs is a consequence of the periodic nature of external disturbances imposed on the satellite's dynamics.
\item The variation of the neural network's weights is shown in Fig \ref{fig:sat_fel_est}:
\begin{figure}[H]
\centering
\includegraphics[width=0.9\textwidth]{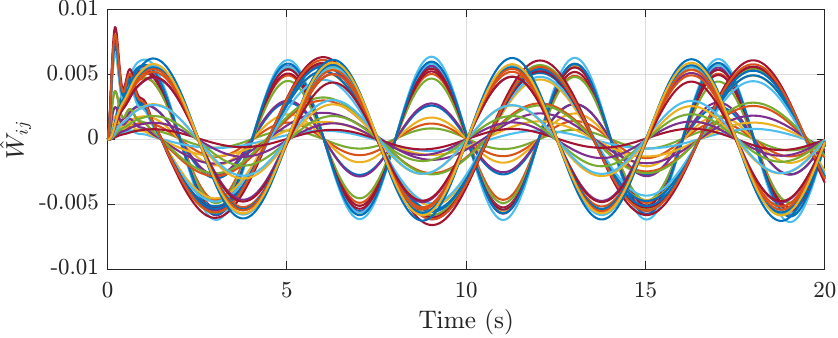}
\caption{Evolution of training weights in basic FEL control of satellite.}
\end{figure}
Since the disturbances acting on the satellite are periodic, then the NN weights enjoy periodic behavior as well.
\item The performance of the estimation of total uncertainties is shown in Fig. \ref{fig:sat_fel_est}.
\begin{figure}[H]
\centering
\includegraphics[width=0.9\textwidth]{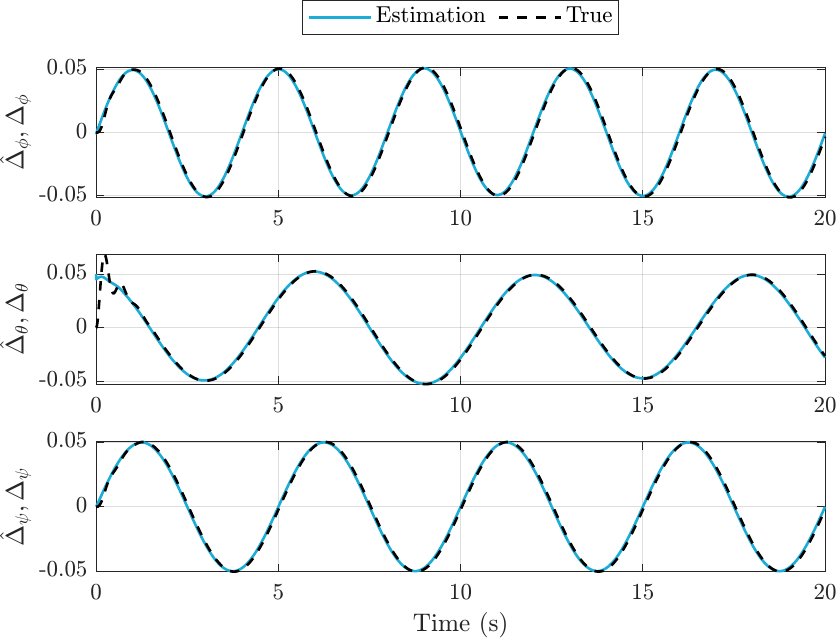}
\caption{Uncertainty estimation in basic FEL control of satellite.}
\label{fig:sat_fel_est}
\end{figure}
As illustrated in Fig. \ref{fig:sat_fel_est}, with the appropriate selection of design parameters, the estimation of uncertainties has been carried out with high precision. It should be noted that the initial oscillation in the estimation of $ \theta $ in the beginning is caused by the high value of learning rate $ \Gamma $. Also, the performance of tracking (MAE) and estimation (MAEE) is reported in Table \ref{tab:sat_fel_param}.
\begin{table}[H]
    \centering
    \caption{Results of MAE and MAEE in basic FEL control of satellite.}
    \label{tab:sat_fel_param}
    \begin{tabular}{c c c c}
    \hline\noalign{\smallskip}
    MAE ($\phi$) & MAE ($\theta$) & MAE ($\psi$) & MAEE \\ 
    \noalign{\smallskip}\hline\noalign{\smallskip}
    $ 3.738 \times 10^{-5} $ & $ 2.748 \times 10^{-5} $ & $ 3.031 \times 10^{-5} $ & $ 2.6 \times 10^{-3} $ \\
    \noalign{\smallskip}\hline\noalign{\smallskip}
    \end{tabular}
\end{table}
\item To investigate the impact of design parameters on MAE and MAEE, a sensitivity analysis has been done, and the results are presented in Fig. \ref{fig:sat_fel_sa}. Based on the results obtained from the sensitivity analysis, it can be observed that increasing the number of neurons and the learning rate both lead to a reduction in both MAE and MAEE values. Meanwhile, it is notable that very large learning rates would result in undesirable fluctuations in the system response. Another point that can be inferred from this sensitivity analysis is that, in a general view, increasing the control gain (when the number of neurons in the neural network is relatively small), leads to increasing the MAE, but decreasing the MAEE. This is due to the fact that increasing the control gain causes the rapid reduction of the tracking error, thereby decreasing the richness of the input signal for training the neural network.  
\begin{figure}[H]
    \centering
    \includegraphics[width=0.9\textwidth]{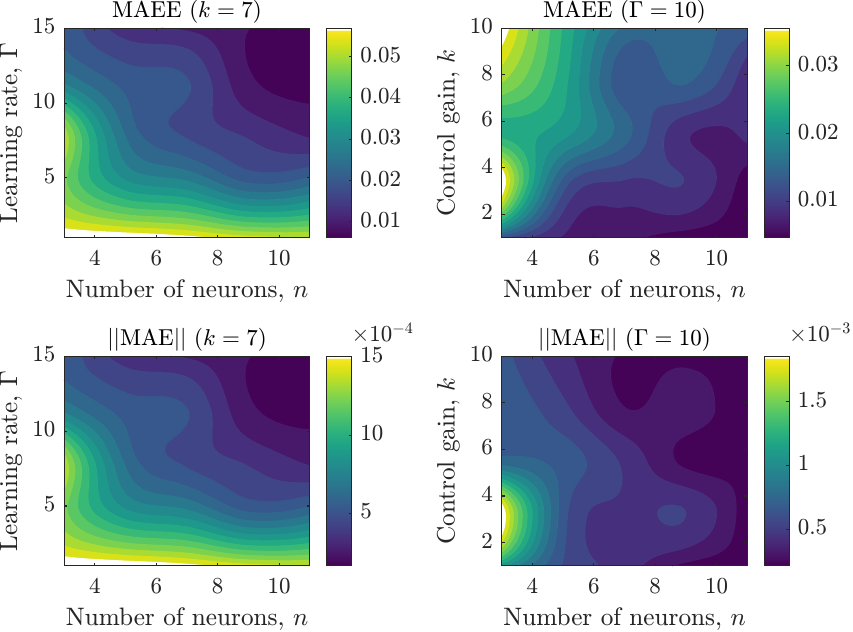}
    \caption{Sensitivity analysis of MAEE and norm of MAE for the number of neurons, learning rate, and the control gain in basic FEL control of satellite.}
    \label{fig:sat_fel_sa}
\end{figure}
\end{enumerate}
\end{ex}

\begin{ex}[label=fullytunedNN]{Fully-tuned NN in FEL framework}
Design an RBF neural network-based control system with adaptive weights, centers, and widths for the system presented in Example \ref{sat_fel}.
	\begin{enumerate}[(a)]
		\item Find a proper update rule for weights, centers and widths of the RBFNN.
		\item Evaluate the tracking performance of the designed control system against the reference input given in Eq. \eqref{sat_desired}.
		\item Plot the changes in NN's parameters, such as training weights, norms of centers, and widths, over time.
		\item Plot estimation of uncertainties and report MAE as well as MAEE for both methods.
		\item Investigate how learning rates affect both Euclidean norm of MAE as well as MAEE.
	\end{enumerate}
	\tcblower
	\begin{enumerate}[(a)]
		\item Similar to Example \ref{sat_fel}, a control signal could be designed as
		\begin{equation}\label{u_fully}
			u = B^{-1} \left(-F (\dot{x}) - \hat{\Delta} (x, \dot{x}) - \lambda \dot{e} -k s + \ddot{x}_d (t) \right).
		\end{equation}
		$ L $ is defined as a vector of RBF NN centers and widths
		\begin{equation}
			L = \mqty[ x_{0_1}^T & x_{0_2}^T & \cdots & x_{0_n}^T & \sigma_1 & \sigma_2 & \cdots & \sigma_{n}]^T,
		\end{equation}
		where $ x_{0_i} $, $ \sigma_i $, and $ n $ are vector of centers, vector of widths, and number on neurons of the RBFNN, respectively.
		We define $ L^* $ as the true value of $ L $ and $ \hat{L} $ as its estimated value. Now, the activation functions of NN are
		\begin{equation}
			\mu = \mu \left( x, L^* \right).
		\end{equation}
		First-order Taylor approximation of $ \mu $ may be written as
		\begin{equation}\label{mu}
			\mu \left( x, L^* \right) = \mu \left( x, \hat{L} \right) + \eval{\pdv{\mu}{L}}_{\hat{L}} \left( L^* - \hat{L} \right).
		\end{equation}
		where $ \pdv*{\mu}{L} $ is defined as
		\begin{equation}
		    \pdv{\mu}{L} = \begin{bmatrix}
    			\left( \dv{\mu_1}{x_{0_1}} \right)^T & \left( \dv{\mu_1}{x_{0_2}} \right)^T & \cdots & \left( \dv{\mu_1}{x_{0_n}} \right)^T &  \dv{\mu_1}{\sigma_1} &  \dv{\mu_1}{\sigma_2} & \cdots &  \dv{\mu_1}{\sigma_{n}} \\
    			\left( \dv{\mu_2}{x_{0_1}} \right)^T &  \left( \dv{\mu_2}{x_{0_2}} \right)^T & \cdots &  \left( \dv{\mu_2}{x_{0_n}} \right)^T &  \dv{\mu_2}{\sigma_1} &  \dv{\mu_2}{\sigma_2} & \cdots &  \dv{\mu_2}{\sigma_{n}} \\
    			\vdots & \vdots & \ddots & \vdots & \vdots & \vdots & \ddots & \vdots \\
    			\left( \dv{\mu_{n}}{x_{0_1}} \right)^T &  \left( \dv{\mu_{n}}{x_{0_2}} \right)^T & \cdots &  \left( \dv{\mu_{n}}{x_{0_n}} \right)^T &  \dv{\mu_{n}}{\sigma_1} &  \dv{\mu_{n}}{\sigma_2} & \cdots &  \dv{\mu_{n}}{\sigma_{n}}.
		    \end{bmatrix}.
		\end{equation}
		In case of RBF NNs, partial derivatives with respect to centers and width are
		\begin{subequations}
		    \begin{equation}
		        \dv{\mu_i}{x_{0_j}} = 
		        \begin{dcases}
		            \begin{matrix*}[l]
		                \frac{x - x_{0_i}}{\sigma_i^2} \mu_i & , & i = j \\
		                0 & , & i \neq j
		            \end{matrix*}
		        \end{dcases},
		    \end{equation}
		    \begin{equation}
		        \dv{\mu_i}{\sigma_j} =
		        \begin{dcases}
		            \begin{matrix*}[l]
		                \frac{\left( x - x_{0_i} \right)^T \left( x - x_{0_i} \right)}{\sigma_i^3} \mu_i & , & i = j \\
		                0 & , & i \neq j
		            \end{matrix*}
		        \end{dcases}
		        .
		    \end{equation}
		\end{subequations}
		A Lyapunov function for this example is defined as
		\begin{equation}
			V = \frac{1}{2} s^T s + \dfrac{1}{2} \mathrm{tr} \left( \tilde{W}^T \Gamma_1^{-1} \tilde{W} \right) + \frac{1}{2} \tilde{L}^T \Gamma_2^{-1} \tilde{L},
		\end{equation}
		where $ \tilde{L} = \hat{L} - L^* $. Taking time derivative, we have
		\begin{equation}\label{Vdot1}
			\begin{split}
				\dot{V} & = s^T \dot{s} + \mathrm{tr} \left( \tilde{W} \Gamma_1^{-1} \dot{\hat{W}} \right) + \tilde{L}^T \Gamma_2^{-1} \dot{\hat{L}} \\
				& = s^T \left( f + B u + \Delta + \lambda \dot{e} - \ddot{x}_d \right) + \mathrm{tr} \left( \tilde{W} \Gamma_1^{-1} \dot{\hat{W}} \right) + \tilde{L}^T \Gamma_2^{-1} \dot{\hat{L}}.
			\end{split}
		\end{equation}
		Substituting Eq. \eqref{u_fully} into Eq. \eqref{Vdot1}, we have
		\begin{equation}\label{e48}
			\dot{V} = s^T \left( -\hat{\Delta} + \Delta - k s \right) + \mathrm{tr} \left( \tilde{W} \Gamma_1^{-1} \dot{\hat{W}} \right) + \tilde{L}^T \Gamma_2^{-1} \dot{\hat{L}}.
		\end{equation}
		Since $ \hat{\Delta} = \hat{W}^T \mu \left( x, \hat{L} \right) $ and $ \Delta = W^{*T} \mu \left( x, L^* \right) $, then Eq. \eqref{e48} could be rewritten
		\begin{equation}
			\dot{V} = s^T \left( - \hat{W}^T \mu \left( x, \hat{L} \right) + W^{*T} \mu \left( x, L^* \right) - k s \right) + \mathrm{tr} \left( \tilde{W} \Gamma_1^{-1} \dot{\hat{W}} \right) + \tilde{L}^T \Gamma_2^{-1} \dot{\hat{L}}.
		\end{equation}
		If $ \tilde{\mu} = \mu \left( x, \hat{L} \right) - \mu \left( x, L^* \right) $, then
		\begin{equation}
			\tilde{\mu} = \mu \left( x, \hat{L} \right) - \mu \left( x, L^* \right) \xrightarrow{\text{Eq. \eqref{mu}}} \tilde{\mu} = \mu' \tilde{L},
		\end{equation}
		where $ \mu' = \eval{\pdv*{\mu}{L}}_{\hat{L}} $. Now, for $ \Delta - \hat{\Delta} $, we have
		\begin{equation}\label{diff}
			\begin{split}
				\Delta - \hat{\Delta} & = W^{*T} \mu \left( x, L^* \right) + \varepsilon - \hat{W}^T \mu \left( x, \hat{L} \right) \\
				& = W^{*T} \mu \left( x, L^* \right) - \tilde{W}^T \mu \left( x, \hat{L} \right) - W^{*T} \mu \left( x, \hat{L} \right) + \varepsilon \\
				& = W^{*T} \left[ \mu \left( x, L^* \right) - \mu \left( x, \hat{L} \right) \right] - \tilde{W}^T \mu \left( x, \hat{L} \right) + \varepsilon \\
				& = - W^{*T} \mu' \tilde{L} - \tilde{W}^T \mu \left( x, \hat{L} \right) + \varepsilon \\
				& = - \hat{W}^T \mu' \tilde{L} + \tilde{W}^T \mu' \tilde{L} - \tilde{W}^T \mu \left( x, \hat{L} \right) + \varepsilon \\
				& = - \hat{W}^T \mu' \tilde{L} + \tilde{W}^T \mu' \hat{L} - \tilde{W}^T \mu \left( x, \hat{L} \right) + w, \\
			\end{split}
		\end{equation}
		where $ w = \varepsilon - \tilde{W}^T \mu \left( x, \hat{L} \right) L^* $ and it could be shown that this parameter is bounded \citep{Lee.2001}. Substituting Eq. \eqref{diff} into Eq. \eqref{e48}, we have
		\begin{equation}
			\begin{split}
			    \dot{V} & = s^T \left[ -\hat{W}^T \mu' \tilde{L} + \tilde{W}^T \mu' \hat{L} - \tilde{W}^T \mu \left( x, \hat{L} \right) + w \right] \\
			    & \quad+ \mathrm{tr} \left( \tilde{W} \Gamma_1^{-1} \dot{\hat{W}} \right) + \tilde{L}^T \Gamma_2^{-1} \dot{\hat{L}}.
			\end{split}
		\end{equation}
		If the update rule for $ \hat{W} $ and $ \hat{L} $ is defined as
		\begin{subequations}
			\begin{align}
				\dot{\hat{W}} &= \Gamma_1 \left[ \mu (x, \hat{L}) - \mu' \hat{L} \right] s^T,\\
				\dot{\hat{L}} &= \Gamma_2 \mu'^T \hat{W} s,
			\end{align}
		\end{subequations}
		and using Cauchy–Schwarz as well as Young inequalities for $ s^T w $, we have
		\begin{equation}
			\dot{V} \leq -K s^T s + C ,
		\end{equation}
		where $ K = k - \dfrac{1}{2} $ and $ C = \dfrac{1}{2} w^T w $, which results bounded tracking error.
		\item Considering $ k = 7 $, $ \lambda = 5 $, $ \Gamma_1 = 30 $, and $ \Gamma_2 = 120 $, Fig. \ref{fig:sat_fully_states} shows time history of Euler angles. As can be seen from this figure, Euler angles are able to tracking their reference inputs with acceptable precision.
		\begin{figure}[H]
            \centering
            \includegraphics[width = 0.9\textwidth]{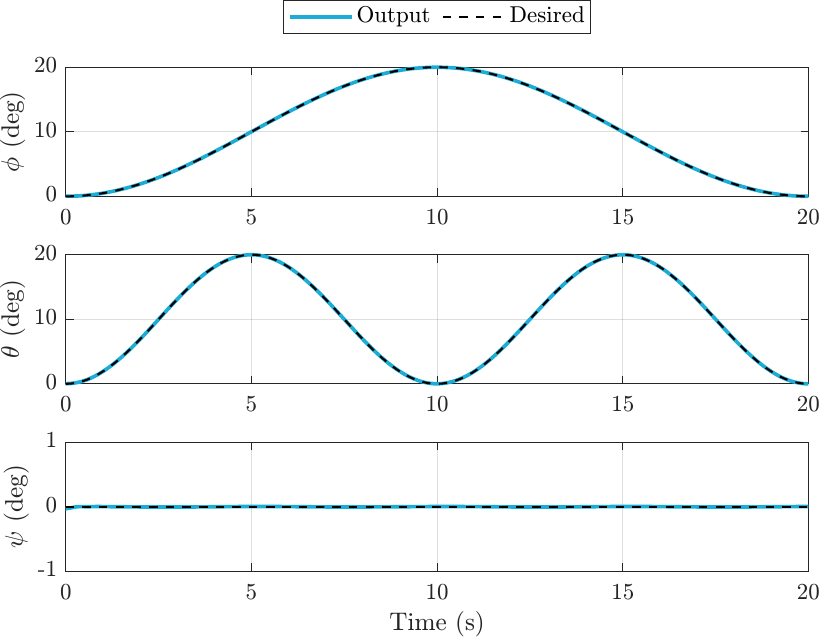}
            \caption{Time history of attitude in fully-tuned NN control of satellite.}
            \label{fig:sat_fully_states}
        \end{figure}
        Additionally, the corresponding control inputs of this method is shown in Fig. \ref{fig:sat_fully_control}.
        \begin{figure}[H]
            \centering
            \includegraphics[width=0.9\textwidth]{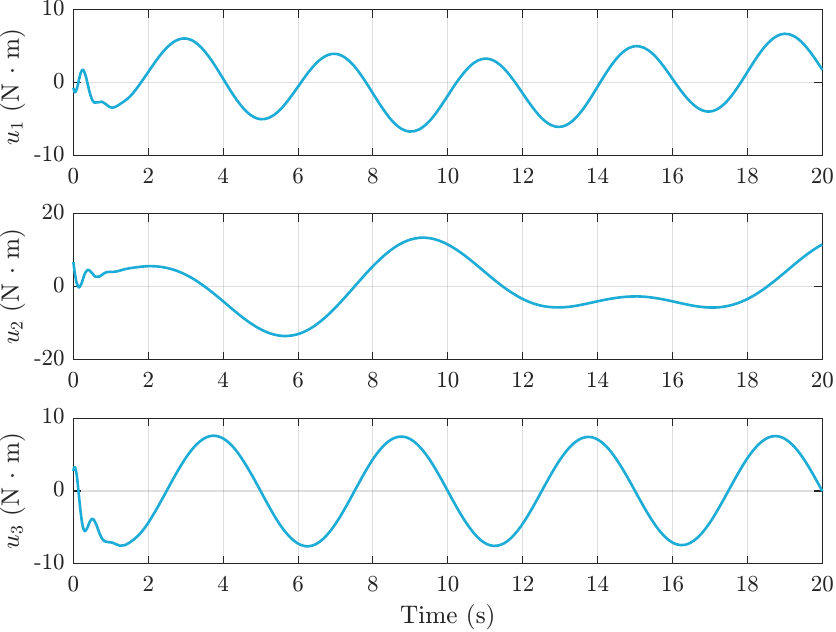}
            \caption{Time history of control inputs in fully-tuned NN control of satellite.}
            \label{fig:sat_fully_control}
        \end{figure}
        It should be noted that te initial oscillation in $ u_2 $ caused by high values of learning rates.
        \item The variation of the neural network's parameters, i.e. weights, centers, and widths, is shown in Fig. \ref{fig:sat_fully_W}.
        \begin{figure}[H]
            \centering
            \includegraphics[width=0.9\textwidth]{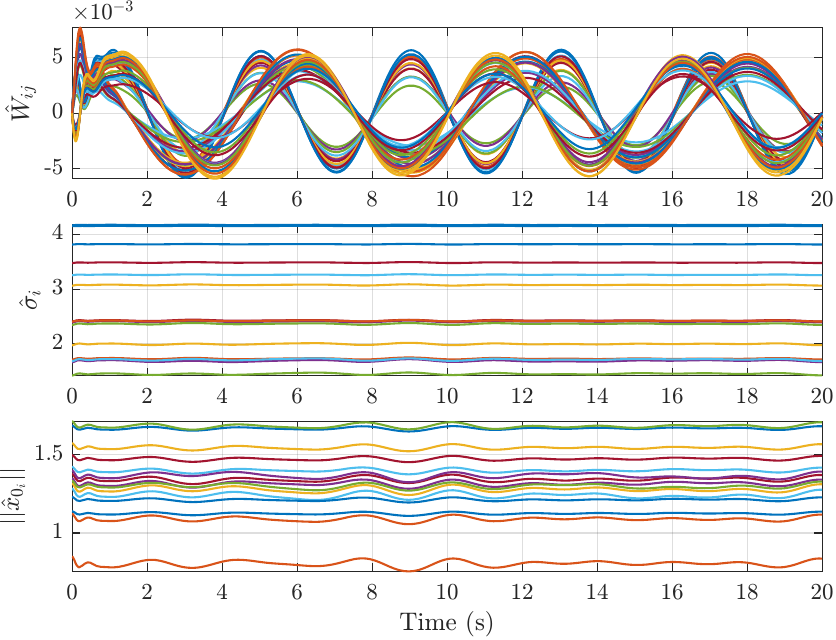}
            \caption{Evolution of training weights, norm of centers, and widths in fully-tuned NN control of satellite.}
            \label{fig:sat_fully_W}
        \end{figure}
        \item The performance of the estimation of total uncertainties is shown in Fig. \ref{fig:sat_fully_est}.
        Similar to the previous example, and illustrated in Fig. \ref{fig:sat_fel_est}, with the appropriate selection of design parameters, the estimation of uncertainties has been carried out with high precision. It should be mentioned that the initial oscillation in the estimation of $ \theta $ in the beginning is caused by the high value of learning rate $ \Gamma $.
        \begin{figure}[H]
            \centering
            \includegraphics[width=0.9\textwidth]{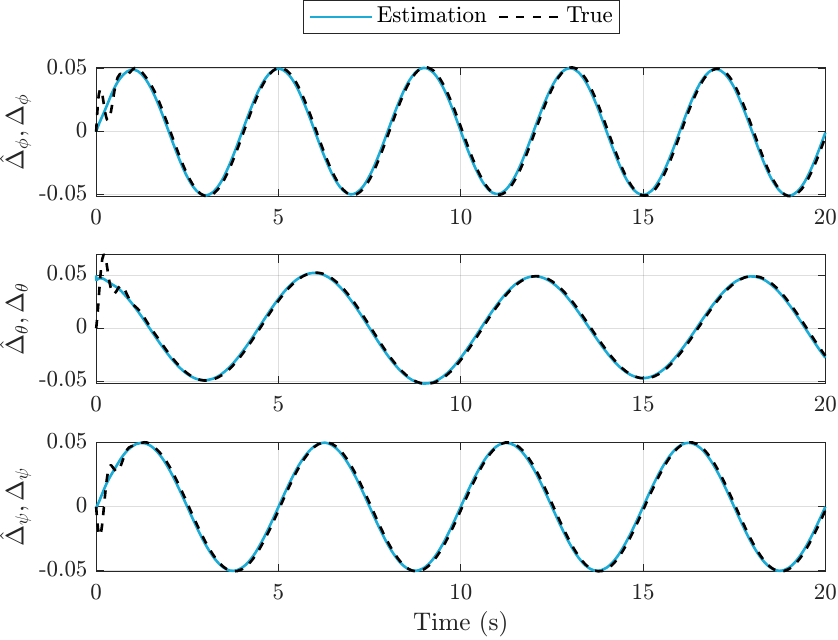}
            \caption{Uncertainty estimation in basic FEL control of satellite.}
            \label{fig:sat_fully_est}
        \end{figure}
        Finally, the performance of tracking (MAE) and estimation (MAEE) is reported in Table \ref{tab:sat_fully_param}.
        \begin{table}[H]
            \centering
            \caption{Results of MAE and MAEE in basic FEL control of satellite.}
            \label{tab:sat_fully_param}
            \begin{tabular}{c c c c}
            \hline\noalign{\smallskip}
            MAE ($\phi$) & MAE ($\theta$) & MAE ($\psi$) & MAEE \\ 
            \noalign{\smallskip}\hline\noalign{\smallskip}
            $ 2.240 \times 10^{-5} $ & $ 1.647 \times 10^{-5} $ & $ 1.818 \times 10^{-5} $ & $ 3.9 \times 10^{-3} $ \\
            \noalign{\smallskip}\hline\noalign{\smallskip}
            \end{tabular}
        \end{table}
        Table \ref{tab:sat_fully_param} shows that updating the centers and widths of the neural network improves the tracking and estimation precision of the FEL control system, compared to Table \ref{tab:sat_fel_param}.
        Specifically, adaptive NN centers and widths reduce the MAE for all three parameters ($\phi$, $\theta$, and $\psi$) by approximately 40\% and the MAEE by 35\%. This suggests that updating the NN centers and widths allows the NN to better learn the dynamics of the system and compensate for uncertainties.
        \item In Fig. \ref{fig:sat_fully_sa}, variation of MAEE as well as Euclidean norm of MAE is shown. As can be seen, increasing both learning rates leads to decrease in both the MAEE and the norm of MAE. However, it is worth mentioning that the effect of training the output weights on the estimation (as well as the tracking) performance is much greater than that of the hidden layer parameters. In other words, the performance improvement achieved by training the hidden layer parameters in almost negligible against that of the output weights. As a result, in most research, only the output weights of the neural network are trained during the operation.     
        \begin{figure}[H]
            \centering
            \includegraphics[width=0.9\textwidth]{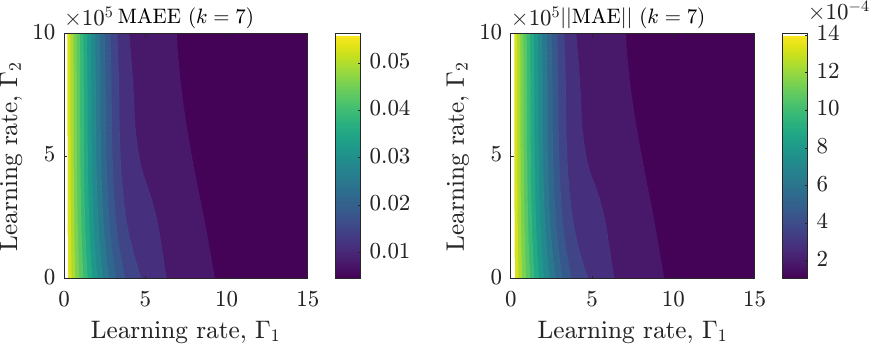}
            \caption{Sensitivity analysis of MAEE and norm of MAE for learning rates in fully-tuned control of satellite.}
            \label{fig:sat_fully_sa}
        \end{figure}
	\end{enumerate}
\end{ex}

\subsection{Discrete-time FEL}
On the other hand, as the designed controller should be programmed on a digital processor in real applications, the development of a control system in the discrete-time domain makes more sense. Using a discrete-time controller, the dependence of the closed-loop \emph{performance} on the sampling rate can also be eliminated. This would be more beneficial in the case of NN-based control systems in which the differential equations for updating the NN weights change to difference equations. Furthermore, in the case of discrete-time controllers, the NN weights' updating rate that guarantees the convergence of the training rule can be computed analytically \citep{DongHoShin.2006}. To illustrate the fundamental structure of a discrete-time FEL scheme, consider the equivalent discrete-time model of (\ref{Model_1}) as follows
\begin{equation}
  x\left(k+1\right)=F_d\left(x\left(k\right)\right)+B_d\left(x\left(k\right)\right)u\left(k\right)+\Delta_d\left(k\right).
\end{equation}
Defining
\begin{gather}\label{Eq:DiscreteFELu}
u(k)=B_d(k)^{-1}\left(-F_d(k)-{\hat{W}}^T \mu(k) +c e(k) +x_d(k+1) \right), \\
\Delta_d(k)={W^*}^T \mu(x(k))+\varepsilon,\\
e(k)=x(k)-x_d(k),
\end{gather}
where $0<c<1$ leads to the following equation
\begin{equation}\label{Eq:15}
  e(k+1)=c e(k) - \tilde{W}^T \mu(k) + \varepsilon.
\end{equation}
By multiplying both side of (\ref{Eq:15}) by $e^T(k+1)$, we have
\begin{align}
\begin{split}
  e^T(k+1)\tilde{W}^T \mu(k)= c e^T(k+1)e(k)
  - e^T(k+1)e(k+1) + e^T(k+1)\varepsilon.
\end{split}
\end{align}
Using Cauchy–Schwarz and Young's inequalities, it is obtained that 
\begin{align}
\begin{split}
  e^T(k+1)\tilde{W}^T \mu(k) \leq  \|e(k+1)\|^2 \Big( -1 + \rho_1 + \rho_2 \Big)
  + \frac{c^2}{4\rho_1} \|e(k)\|^2  + \frac{1}{4\rho_2}\varepsilon_M^2,
\end{split}
\end{align}
with $\rho_1,\rho_2 > 0$.
Thus, if a Lyapunov function is defined as (\ref{Lyap_0}) (without the coefficient $1/2$), using the following updating rule,
\begin{equation}\label{Eq:15_1}
  \hat{W}(k+1)=\hat{W}(k)+\Gamma \mu(k) e^T(k+1),
\end{equation}
the first difference of $V(k)$ is obtained as follows:
\begin{align}
\begin{split}
\Delta V (k) & = V(k+1) - V(k)= \|e(k+1)\|^2- \|e(k)\|^2 \\
& \quad + tr \left(\tilde{W}^T (k+1) \Gamma^{-1} \tilde{W}^T (k+1) -\tilde{W}^T (k) \Gamma^{-1} \tilde{W}^T (k)\right)\\
& \leq \|e(k+1)\|^2 \Big( 1 + 2(-1 + \rho_1 + \rho_2) \Big) + \|e(k)\|^2 \Big( \frac{c^2}{4\rho_1} -1\Big)\\
& \quad + \frac{1}{4\rho_2}\varepsilon_M^2 + \|\mu\|_{\Gamma}^2\|e(k+1)\|^2.
\end{split}
\end{align}
Accordingly, we have
\begin{equation}
 \Delta V (k) \leq -k_1 \|e(k+1)\|^2 -k_2 \|e(k)\|^2 + c_1,
\end{equation}
where,
\begin{gather}
  k_1= 1 -2 \rho_1 -2 \rho_2 - \|\mu\|_{\Gamma}^2,\\
  k_2= 1- \frac{c^2}{4\rho_1}, \\
  c_1= \frac{1}{4\rho_2}\varepsilon_M^2.
\end{gather}
Thus, assuming the boundedness of $\mu(x)$, it is possible to determine $\rho_1$, $\rho_2$, $c$, and $\Gamma$ such that $k_1,k_2 > 0$ thereby guaranteeing  $\Delta V (k)<0$ for $ \|e(k+1)\|^2 > c_1/k_1$.
As seen, although the updating rule (\ref{Eq:15_1}) is similar to that of continuous-time systems, i.e. (\ref{Learn_1}), the stability analysis of the discrete-time FEL neural control is quite different and more complex compared to that of continuous-time systems.
Consequently, in the following, we mainly focus on the continuous-time formulation of control systems, while their discrete-time equivalent would be obtained in a similar manner as discussed above.

\begin{ex}[label=Example:quadFEL_discrete]{Discrete-time FEL-based control of quadrotor UAV}
Repeat Example \ref{Example:quadFEL} in the discrete-time domain for the case of body-rate control of a quadrotor, by first 
\begin{enumerate}[(a)]
\item converting the continuous model to a discrete-time system. Then, 
\item assess the closed-loop performance based on the following target value
\begin{equation}
{v_d}\left( t \right)=\left[ \begin{matrix}
  {v_{z_d}}\left( t \right)  \\
  {p_d}\left( t \right)  \\
  {q_d}\left( t \right)  \\
  {r_d}\left( t \right)  \\
\end{matrix} \right]=\left[ \begin{matrix}
  0.1e^{-t}\sin\left(\frac{2\pi t}{5} \right) + 0.1 \\
  0.05\sin\left( \frac{\pi t}{10} + \frac{\pi}{4}\right) \\
  0.05e^{-0.1t}\left( 1-\cos\left( \frac{3\pi t}{15} + \frac{\pi}{2}\right) \right)\\
  0  \\
\end{matrix} \right].
\end{equation}
Measure the discrepancy between the estimated uncertain dynamics, $\hat{\Delta}_d=\hat{W}^T \mu\left( X \right)$, and the real uncertainty given by
\begin{equation}\label{Exc24:Delta}
\Delta_d = \left[\bar{d} + \left(f\left(x\right)-\bar{f}\right) + \left(B\left(x\right)-\bar{B}\right)u + B\left(x\right)(u_c-u)\right]T_0,
\end{equation}
which in fact is the product of Eq. (\ref{Exc21:Delta}) and the sampling time.
\item Analyze the effect of sampling time $T_0$ on the control performance of the discrete FEL scheme.
\end{enumerate}
\tcblower
The linearized dynamics of the quadrotor UAV, given by Eq. (\ref{eq:exc21:FBlinear}) can be written in the following state-space form
\begin{equation}
  \dot{v} = Av + \bar{B}u,
\end{equation}
where, for the control of angular rates and vertical velocity, the state vector is considered as $v=\begin{bmatrix} v_z,& p,& q,& r \end{bmatrix}^T$. Additionally, the system matrix $A$ is considered to be zero since $\bar{f}$ in Eq. \ref{eq:exc21:FBlinear} only accounts for the constant term $g$ in the vertical dynamics, and can be omitted for a linear, state-space model.

\begin{enumerate}[(a)]
\item \textbf{Discretized model:}
The linear model can be discretized using the forward Euler differentiation method, or more accurately, by assuming a piecewise constant control with the Zero-order Hold (ZoH) method. To this end, it can be simply obtained using the $\texttt{c2d()}$ command in MATLAB. For the case of zero $A$ matrix, the discrete version of it is simply the identity matrix. Moreover, regarding $\bar{B}_d$ we have:
\begin{lstlisting}[language=Octave]
B1 = diag([1/m, 1/I_xx, 1/I_yy, 1/I_zz]);
B2 = [1 1 1 1; 0 l 0 -l; l 0 -l 0; c -c c -c];
Bbar = B1*B2;
sys = ss(zeros(4),Bbar,eye(4),zeros(4));
sys_d = c2d(sys,T_0);
Bd = sys_d.B
\end{lstlisting}
where, $B_1$ and $B_2$ are defined in Eq. (\ref{eq:Apdx:quad:fB}). The above script gives the discretized version of $\bar{B}$ matrix, providing a suitable value for the sampling period $T_0$.

\textbf {Controller design:}
Since the model is of first-order, the control command given by Eq. (\ref{Eq:DiscreteFELu}) is applicable.
Moreover, the update rule is given in Eq. (\ref{Eq:15_1}), and the stability analysis is also similar to the approach followed from there, which will be omitted here.

\item \textbf {Simulation results:}
Fig. \ref{fig:exc24:stat} shows the time history of angular and vertical velocities, for the discrete controller with 15 hidden neurons, $c=0.7$, $\Gamma=0.01$ and $T_0=0.01$, showing an acceptable tracking performance followed by an initial oscillatory response. It is notable that the vertical velocity starts in the opposite direction at the initial time and following the undershoot, there is also an overshoot and the response tends to settle down after nearly half a second, i.e., 50 steps. 
\begin{figure}[H]
  \begin{center}
    \includegraphics[width=0.9\textwidth]{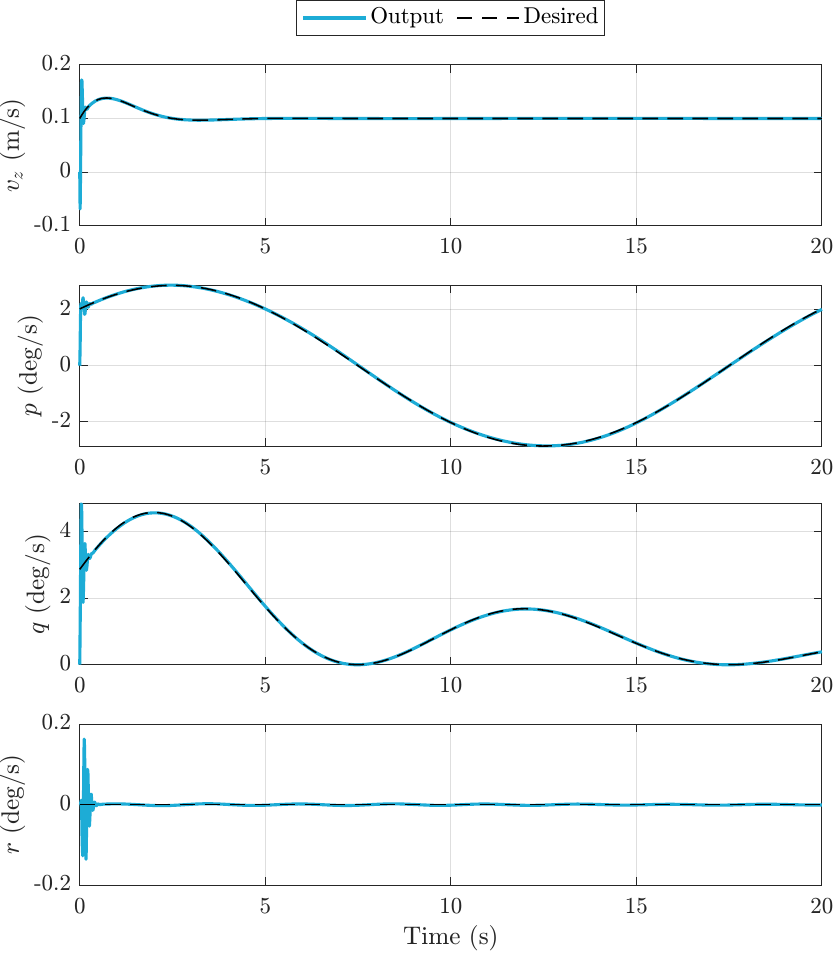}
  \end{center}
  \caption{Time history of angular rates and vertical velocity in discrete FEL control of quadrotor.} 
  \label{fig:exc24:stat}
\end{figure}

Moving on to the control history, the initial oscillation is lowered by the first-order filter corresponding to the actuator dynamics. In addition, due to the presence of dead-zone nonlinearity, the existing bias in the applied control seems to be identified by the NN estimator to compensate for the shortage in the required thrust force, as seen in Fig. \ref{fig:exc24:eff}.
\begin{figure}[H]
  \begin{center}
    \includegraphics[width=0.9\textwidth]{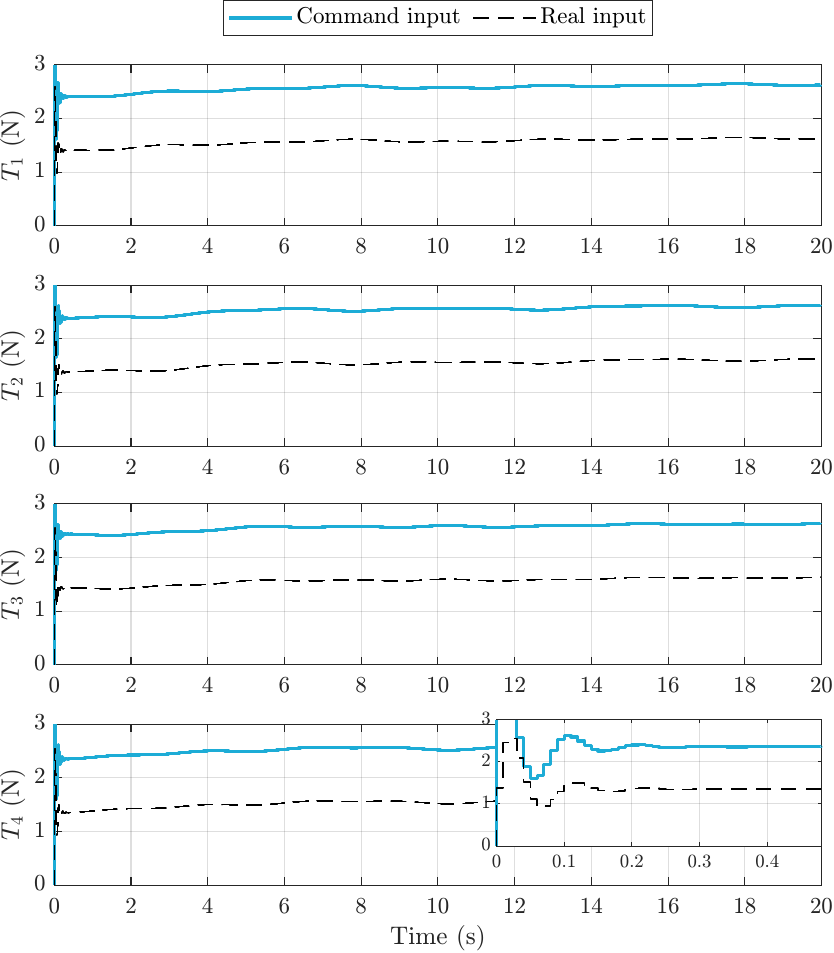}
  \end{center}
  \caption{Time history of control inputs in discrete FEL control of quadrotor.} 
  \label{fig:exc24:eff}
\end{figure}

The neural network prediction of the uncertainty is compared to the real one from Eq. (\ref{Exc24:Delta}) in Fig. \ref{fig:exc24:del}. The results show that the NN estimation is able to identify the uncertain dynamics accurately.
\begin{figure}[H]
  \begin{center}
    \includegraphics[width=0.9\textwidth]{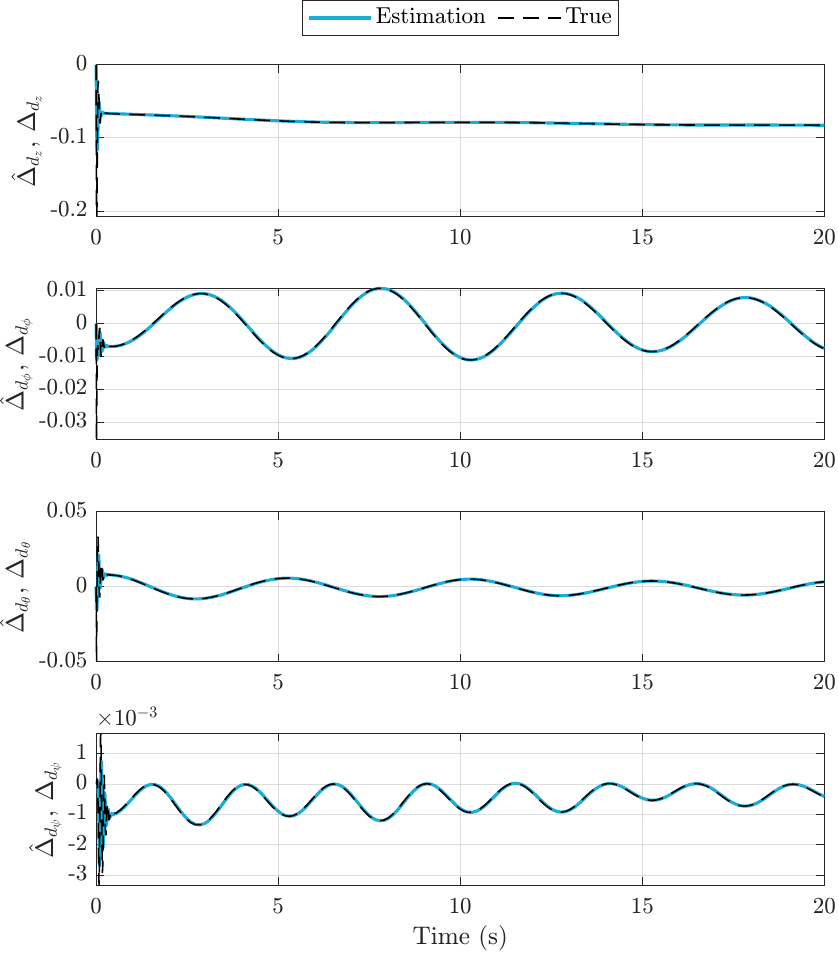}
  \end{center}
  \caption{Uncertainty estimation in discrete FEL control of a quadrotor.} 
  \label{fig:exc24:del}
\end{figure}

As a final point, the values of MAE and MAEE are presented in Table \ref{tab:ch2_exc04_MAEE}.

\begin{table}[H]
    \caption{Result of MAE and MAEE for the discrete FEL control of quadrotor.}
    \label{tab:ch2_exc04_MAEE}
    \centering
    \begin{tabular}{ccccc}
    \hline\noalign{\smallskip}
    MAE ($z$) & MAE ($\phi$) & MAE ($\theta$) & MAE ($\psi$) & MAEE \\
    \noalign{\smallskip}\hline\noalign{\smallskip}
    $3.69\times 10^{-4}$ & $2.52\times 10^{-4}$ & $2.84\times 10^{-4}$ & $3.48\times 10^{-5}$ & $5.54\times 10^{-4}$ \\
    \noalign{\smallskip}\hline\noalign{\smallskip}
    \end{tabular}
\end{table}

\item \textbf {Sensitivity analysis:}
We conducted a parameter study to analyze the impact of various parameters on the control performance, specifically the accuracy of uncertainty estimation measured by MAEE.
Fig. \ref{fig:exc24:param1} highlights the effect of the learning rate, sampling time and gain $c$ on the accuracy of the estimation. To this end, it is shown that as the learning rate increases, the gain must be decreased to tune the performance of the NN. Additionally, the sensitivity of the control scheme to different sampling times shows that for a fixed value of learning rate, there seems to be a minimum and maximum allowable region for $T_0$.
\begin{figure}[H]
  \begin{center}
    \includegraphics[width=0.9\textwidth]{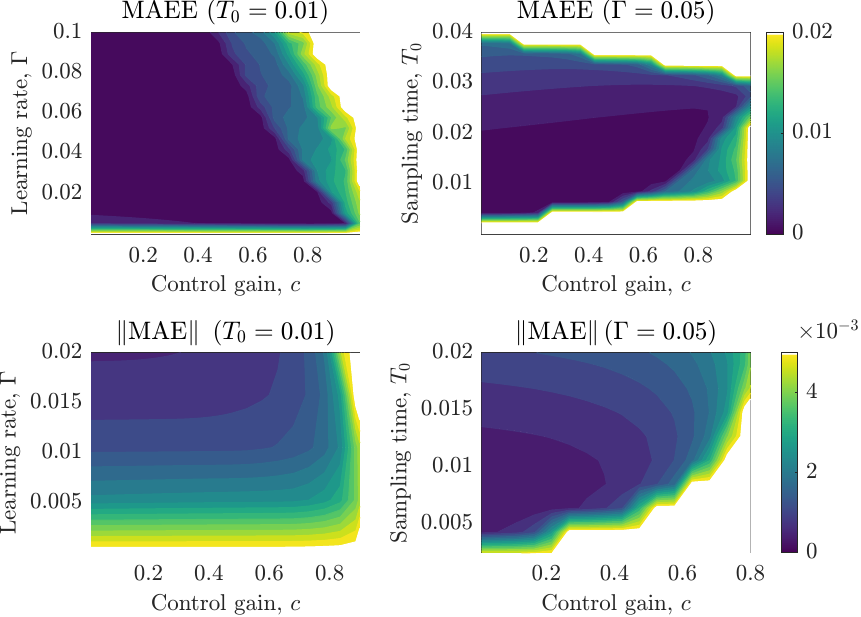}
  \end{center}
  \caption{Sensitivity analysis of MAEE for different values of $c$, $\Gamma$ and $T_0$ in discrete FEL control of quadrotor.} 
  \label{fig:exc24:param1}
\end{figure}
After examining Fig. \ref{fig:exc24:param1}, it might be concluded that there are limitations on the lower bounds of the learning rate and sampling time. However, comparing these parameters against each other in Fig. \ref{fig:exc24:param2} clarifies this ambiguity, demonstrating that when these parameters are varied in accordance with the other, MAEE tends to remain constant. Additionally, it shows that it is permissible to increase these parameters up to a certain value, beyond which the system dynamics prohibit further increments.
\begin{figure}[H]
  \begin{center}
    \includegraphics[width=0.48\textwidth]{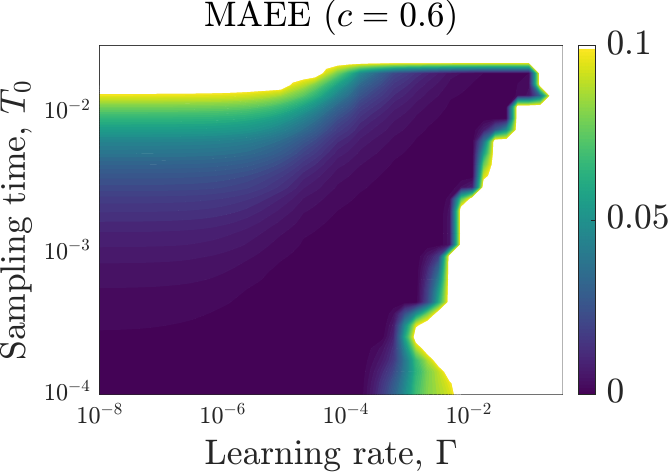}
  \end{center}
  \caption{Sensitivity analysis of MAEE for different values of $\Gamma$ and $T_0$ in discrete FEL control of quadrotor.} 
  \label{fig:exc24:param2}
\end{figure}
\end{enumerate}
\end{ex}

\subsection{Learning rules modification}
As previously mentioned, in the introduced (continuous/discrete) adaptive control scheme, the convergence of the NN weights to their ideal values is not trivial, and it requires a PE condition \citep{Kacprzyk.2010b}. More precisely, in the absence of persistent exciting input signals, the NN weight estimates might drift to very large values, which will result in a variation of high-gain control \citep{Zhang.1999,Ge.2002}.
Different approaches have been proposed in the literature to prevent parameter drift in such conditions. Some of the more common methods are briefly introduced in the following.
\begin{enumerate}
  \item \textbf{Dead-zone:} In this straightforward method, the previously mentioned updating rule is only used when the tracking error exceeds a predefined threshold \citep{VijayaKumar.2009}. Otherwise, the NN weights remain constant. Although such a method can successfully prevent parameter drift, as discussed in \citet{TzirkelHancock.1992,Fabri.1996}, the determination of an appropriate threshold requires the bounds of the control gain function and the NN estimation error ($\varepsilon_M$), which may not be generally known.
  \item \textbf{Projection:} The second simple method is to limit the NN weights to a predefined interval. It means that the time derivative of the parameters is set to zero when they reach the given bounds \citep{Boskovic.2004}. The main drawback of this method is the requirement of the lower and upper bounds of the NN parameters.
  \item \textbf{Sigma-modification:} The third method, which has been introduced by Ioannou and Kokotovic \citep{Ioannou.1983,Ioannou.2012} is a more useful approach. In this method, a modification term is incorporated in the updating rule of the NN parameters as $\dot{\hat{W}}=\Gamma \left(\mu e^T -\sigma W\right)$, where $\sigma$ is a positive constant \citep{Shin.2004}. Such an approach has been employed in many NN-based flight control systems such as \citep{Hovakimyan.2001,Ge.2002,Hovakimyan.2002,S.S.Ge.2009}.
  \item \textbf{$e$-modification:} Another popular approach has been introduced in \citet{Narendra.1987}, where the constant parameter $\sigma$ in the previous technique is replaced by a term proportional to $|e|$ \citep{Zhang.1999,Rysdyk.2005}. The boundedness of the NN parameters using $e$-modification has been presented in \citet{RolfRysdyk.1998}. Further, as a major advantage of the $e$-modification technique over the $\sigma$-modification, such a modification term is effectively attenuated by approaching the tracking error to zero, and (in the lack of the estimation error $\varepsilon$,) this method does not affect the convergence of the NN weights to their ideal values in the presence of persistent exciting training signals.
  \item \textbf{Alternate weights:} This approach has been first proposed in \citet{ChrisJ.B.Macnab.2006}. The $e$-modification method may not achieve acceptable performance in the presence of large oscillatory disturbances \citep{C.Nicol.2008}. The basic idea of this method is that different sets of NN weights are capable of uniformly approximating the same nonlinear function. An alternate set of weights with a smaller magnitude than $\hat{W}$ can be used to improve the training. By keeping the NN weights close to the smaller alternate weights, it is possible to provide a more efficient compromise between the approximation performance and keeping the NN weights bounded, while there is a need for two distinct sets of NN weights and their corresponding updating rules. This method has been employed in \citet{Coza.2011} to design a flight control system for a quadrotor air vehicle under the wind buffeting.
\end{enumerate}

\begin{ex}{Effect of learning modifications in FEL control}
In Example \ref{Example:GTM_FEL}, a basic FEL control for GTM was designed.
To investigate the impact of different types of learning modifications on tracking and estimation performance and preventing parameter drift, apply the previously introduced dead-zone, sigma-modification, and $e$-modification methods.

Consider the following update rules for each method:
\begin{align}
     \text{Dead-zone:} & \hspace{0.2cm} \dot{\hat{W}}=\begin{cases}
    \Gamma \mu e^T & \text{if \, $||e||>\frac{\varepsilon_M}{k}$}\\
    0 & \text{otherwise}
      \end{cases}, \\
      \sigma\text{-modification:} & \hspace{0.2cm} \dot{\hat{W}}=\Gamma \left(\mu e^T -\sigma \hat{W}\right), \\
      \text{$e$-modification:} & \hspace{0.2cm} \dot{\hat{W}}=\Gamma \left(\mu e^T -\sigma |e^T| \odot \hat{W}\right), \label{Eq:e-mod}
\end{align}
    %
where $\odot$ denotes the element-wise product. More precisely, it signifies multiplying each column of the weight matrix $\hat{W}$ by the corresponding element of $|e^T|$.

\begin{enumerate}[(a)]
    \item Plot the evolution of NN training weights over time for the three methods.
    \item Compare the tracking error and estimation error for closed-loop simulation of the three methods as well as the MAE and MAEE values.
    \item Evaluate the effect of $\sigma$ on the tracking performance (MAE) and estimation performance (MAEE) in the sigma-modification and $e$-modification methods.
    \item Evaluate the effect of NN estimation error ($\varepsilon_M$) and control gain ($k$) in the dead-zone method.
    \item In order to examine the effectiveness of modifications in preventing parameter drift compared to the basic FEL approach, take into account 70 neurons (to observe NN weight drift). Additionally, set $\sigma$ to 0.01 and $\varepsilon_M$ to 0.05. Subsequently, report the Mean Weight Norm (MWN) as well as MAE and MAEE for each method and compare it with the MWN of the basic FEL. MWN is defined as the average norm of the NN weights given by
        \begin{equation}\label{eq:MWN}
        \text{MWN} = \frac{1}{T}\mathop{\int }_{0}^{T}\left\| \hat{W} \right\|_\text{F} dt,
        \end{equation}
    where $\left\| \cdot \right\|_\text{F}$ is the Frobenius norm.
\end{enumerate}
\tcblower
\begin{enumerate}[(a)]
    \item Considering $\sigma = 0.001$, $\varepsilon_M = 0.01$ and $k=18$, Fig. \ref{fig:Ex_2.5_W} shows the update behavior of NN weights for the three methods. The changing trend of the weights is found to be nearly identical across the different learning modifications. 
    However, a notable distinction can be observed in the $e$-modification method, characterized by lower oscillations in its weight adaptation, which is due to the employment of an error-dependent \emph{damping} term in its update rule.
        \begin{figure}[H]
            \begin{center}
            \includegraphics[width=0.9\textwidth]{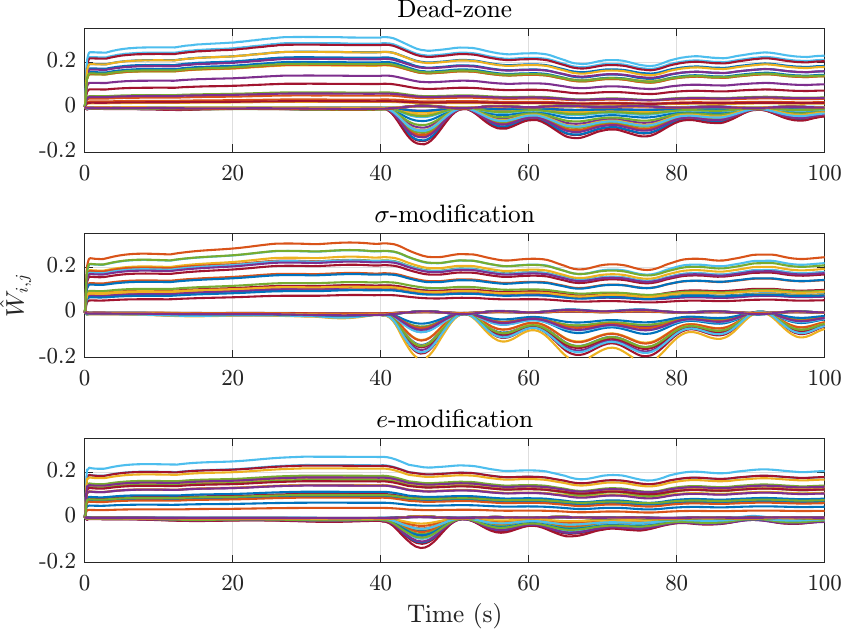}
            \end{center}
            \caption{Evolution of training weights for three learning modifications in FEL control of GTM.}
            \label{fig:Ex_2.5_W}
        \end{figure}
\item The time history of tracking errors is depicted in \ref{fig:EX_2.5_TE}. Comparing the results, it is evident that both the dead-zone and $\sigma$-modification methods exhibit larger errors when compared to the $e$-modification method.
        \begin{figure}[H]
            \begin{center}
            \includegraphics[width=0.9\textwidth]{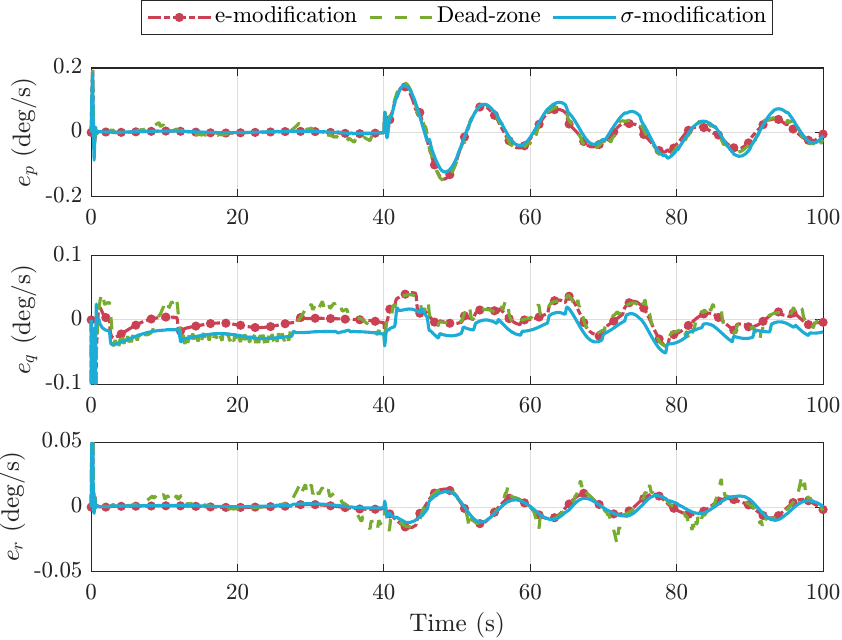}
            \end{center}
            \caption{History of tracking error for three learning modifications in FEL control of GTM.}
            \label{fig:EX_2.5_TE}
        \end{figure}
Additionally, the history of estimation error is depicted in \ref{fig:EX_2.5_EE}. Comparing the uncertainty estimation errors, it is evident that the dead-zone method exhibits a larger error when compared to the both $e$-modification and $\sigma$-modification methods, especially in the yaw rate channel.
        \begin{figure}[H]
            \begin{center}
            \includegraphics[width=0.9\textwidth]{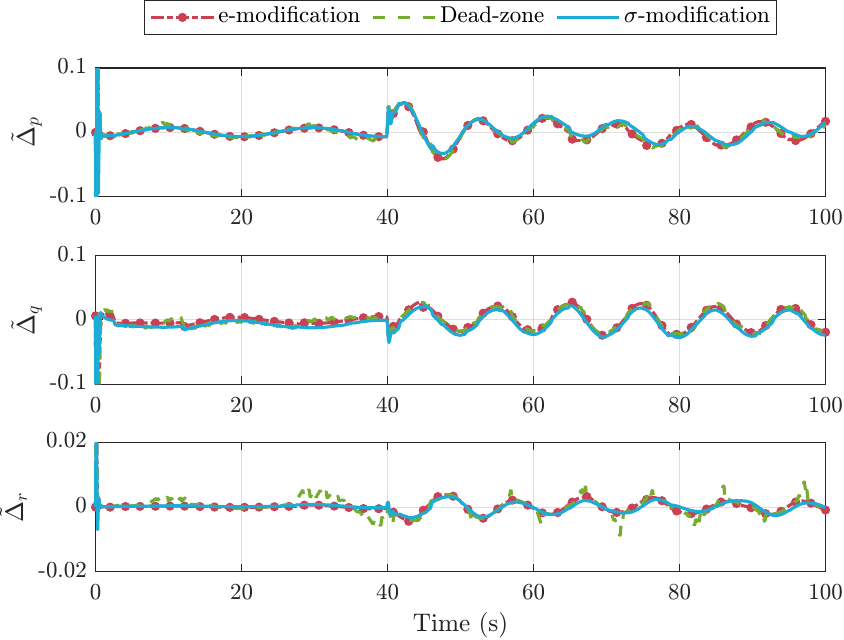}
            \end{center}
            \caption{History of estimation error for three learning modifications in FEL control of GTM.}
            \label{fig:EX_2.5_EE}
        \end{figure}
Furthermore, Table \ref{tab:EX_2.5_MAE} shows the MAE and MAEE values for this simulation. As can be seen, the MAE and MAEE values are higher when using learning modifications than when using the basic FEL method (Table \ref{tab:EX_2.2_MAE}).
This indicates that using modifications in the learning process reduces the quality of the estimation, which in turn affects tracking performance. It is noteworthy that the $e$-modification has lower MAE and MAEE values than the other modifications, indicating better performance in both estimation and tracking. This can be attributed to its ability to adjust the influence of $\hat{W}$ in the update rule based on the tracking error.
        \begin{table}[H]
            \caption{Result of MAE and MAEE for the FEL control of GTM utilizing learning modifications.}
            \label{tab:EX_2.5_MAE}
            \centering
            \begin{tabular}{c c c c c}
            \hline\noalign{\smallskip}
            Modification Method & MAE ($p$) & MAE ($q$) & MAE ($r$) & MAEE \\
            \noalign{\smallskip}\hline\noalign{\smallskip}
            dead-zone & $5.80\times 10^{-4}$ & $4.61\times 10^{-4}$  & $1.19\times 10^{-4}$ & $2.08\times 10^{-2}$\\
            $\sigma$-modification & $5.37\times 10^{-4}$ & $4.45\times 10^{-4}$  & $6.9\times 10^{-5}$ & $1.96\times 10^{-2}$\\
            $e$-modification & $4.62\times 10^{-4}$ & $3.31\times 10^{-4}$  & $6.7\times 10^{-5}$ & $1.92\times 10^{-2}$\\
            \noalign{\smallskip}\hline\noalign{\smallskip}
            \end{tabular}
        \end{table}
\item In order to assess the influence of the parameter $\sigma$ in the sigma-modification and $e$-modification methods on tracking and estimation performance, a sensitivity analysis is conducted. The results of the analysis are presented in Fig. \ref{fig:EX_2.5_SESA}. It can be observed that when the value of $\sigma$ increases, both the MAE and Mean MAEE increase in the sigma-modification method. This indicates that higher values of $\sigma$ in the sigma-modification method have a detrimental effect on the system's performance. On the other hand, due to the effect of $|e^T|$, increasing the value of $\sigma$ in the $e$-modification method leads to a reduction in both the MAE and MAEE. 
        \begin{figure}[H]
            \begin{center}
            \includegraphics[width=0.9\textwidth]{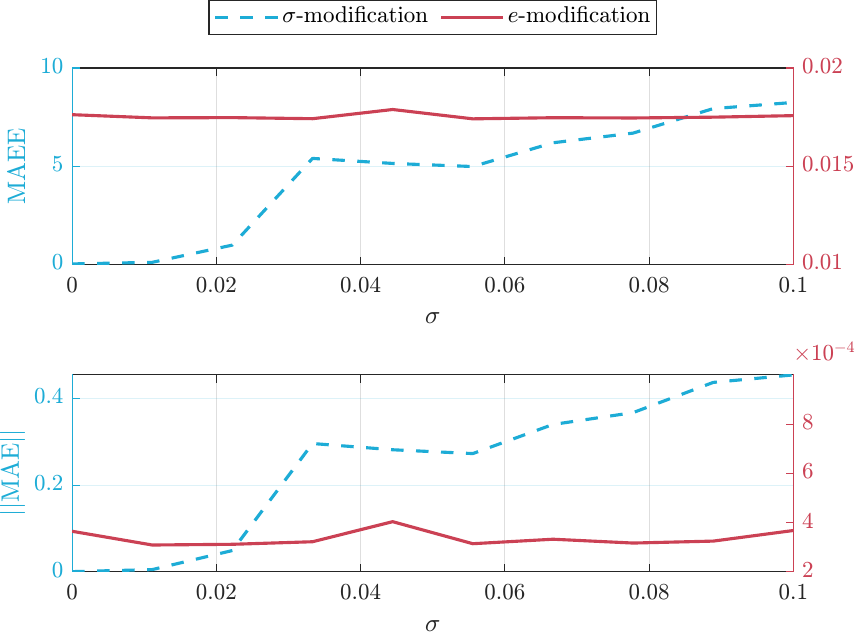}
            \end{center}
            \caption{Sensitivity analysis of $\sigma$ in modifications in FEL control of GTM (The right y-axis is utilized to represent the $e$-modification method and the left y-axis represents the $\sigma$-modification method).}
            \label{fig:EX_2.5_SESA}
        \end{figure}
        Furthermore, the comparison of MWN in both methods is presented in Fig. \ref{fig:EX_2.5_SESA2}. In the sigma-modification method, MWN tends to decrease as the value of sigma increases. This means that larger sigma values lead to a reduction in the MWN. On the other hand, in the $e$-modification method, the MWN remains relatively constant regardless of the sigma value used.
        \begin{figure}[H]
            \begin{center}
            \includegraphics[width=0.9\textwidth]{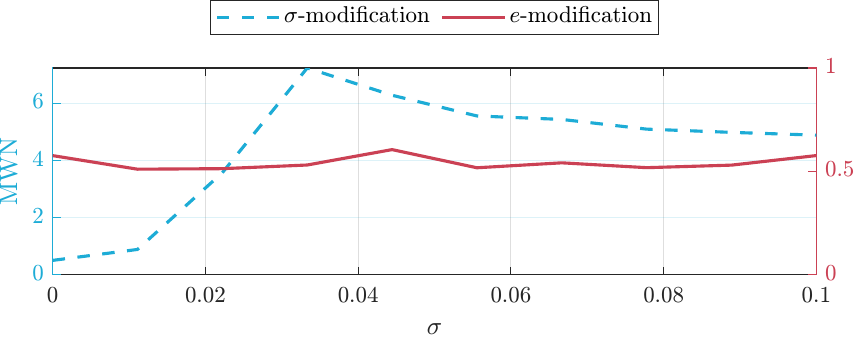}
            \end{center}
            \caption{Influence of variation of $\sigma$ on NN Weights in FEL control of GTM (The right y-axis is utilized to represent the $e$-modification method and the left y-axis represents the $\sigma$-modification method).}
            \label{fig:EX_2.5_SESA2}
        \end{figure}
\item
In the case of the dead-zone method, a parameter study was conducted to evaluate the effect of NN estimation error ($\varepsilon_M$) and control gain ($k$) on tracking and estimation performance. The results are shown in Figure \ref{fig:EX_2.5_DZSA}. They indicate that the performance of the dead-zone method depends on the value of $\varepsilon_M /k$. In cases with low control gains, especially with high $\varepsilon_M /k$ values, the MAE and mean MAEE tend to have higher values. This has a negative impact on the performance of the system when using the dead-zone method.
        \begin{figure}[H]
            \begin{center}
            \includegraphics[width=0.9\textwidth]{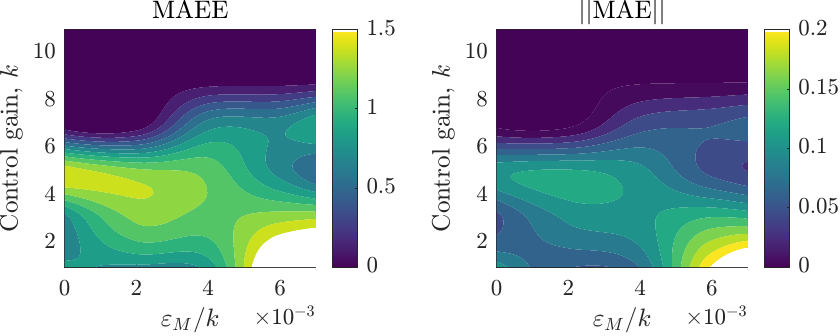}
            \end{center}
            \caption{Sensitivity analysis of $\varepsilon_M$ and control gain ($k$) in dead-zone modification in FEL control of GTM.}
            \label{fig:EX_2.5_DZSA}
        \end{figure}
    \item When utilizing 70 neurons in the FEL control of GTM, the NN weights exhibit drift over time and do not converge to constant values. The result of closed-loop simulations for various modifications on FEL is presented in Table \ref{tab:EX_2.5_PD}. It is evident that these modifications effectively prevent parameter drift and demonstrate improved performance in terms of estimation and tracking. Notably, among the modifications, the e-modification stands out with superior performance.
        \begin{table}[H]
            \caption{Result of MAE, MAEE and MWN for the FEL control of GTM using learning modifications in the case of parameter drift by utilizing 70 neurons.}
            \label{tab:EX_2.5_PD}
            \centering
            \begin{tabular}{c c c c c c}
            \hline\noalign{\smallskip}
            Method & MAE ($p$) & MAE ($q$) & MAE ($r$) & MAEE & MWN \\
            \noalign{\smallskip}\hline\noalign{\smallskip}
            basic FEL & $4.4\times 10^{-1}$ & $1.49\times 10^{-1}$  & $3.93\times 10^{-1}$ & $1.29\times 10^{1}$ & $1.98\times 10^{2}$\\
            dead-zone & $6.04\times 10^{-3}$ & $1.38\times 10^{-3}$  & $7.45\times 10^{-4}$ & $1.49\times 10^{-1}$ & $1.48\times 10^{0}$\\
            $\sigma$-modification & $2.06\times 10^{-2}$ & $5.86\times 10^{-2}$  & $3.27\times 10^{-3}$ & $1.63\times 10^{0}$ & $4.91\times 10^{0}$ \\
            $e$-modification & $6.96\times 10^{-4}$ & $8.43\times 10^{-4}$  & $6.19\times 10^{-4}$ & $4.47\times 10^{-2}$ & $3.82\times 10^{-1}$\\
            \noalign{\smallskip}\hline\noalign{\smallskip}
            \end{tabular}
        \end{table}
\end{enumerate}
\end{ex}

\subsection{Improved learning mechanisms}
Although the aforementioned approaches result in bounded NN parameters, satisfactorily, they do not ensure the convergence of the NN weights to their ideal values. Recently, a variety of modified learning approaches have been proposed in the literature, which causes the improvement in training the NN parameters. Some of the more attractive methods are as follows:
\begin{enumerate}
  \item \textbf{Composite learning:} Different composite learning approaches have been introduced in the literature where their fundamental idea is to include the estimation performance into the updating law \citep{Xu.2014b,Xu.2017,Xu.2021b}. This can lead to faster learning speed as well as higher precision \citep{Xu.2019}. More specifically, the state estimation can be constructed as $\dot{\hat{x}}=F(x)+B(x) u+\hat{W}^T \mu (x) - \beta \tilde{x}$, where $\tilde{x}=\hat{x}-x$, and $\beta$ is a positive constant. Thus, the updating rule (\ref{Learn_1}) can be modified as $\dot{\hat{W}}=\Gamma \mu (e^T -\Gamma_1 \tilde{x}^T)$ where $\Gamma_1$ is a positive definite matrix \citep{Xu.2019}. An improved learning method has been presented in \citet{Xu.2018}, where the basic updating rule (\ref{Learn_1}) is augmented by a novel prediction error signal constructed using online recorded data within a time interval $[t-\tau,t]$, which is equal to $\tilde{W}^T \int_{t-\tau}^{t} \mu (x) + \int_{t-\tau}^{t} \varepsilon$. As shown in \citet{Xu.2018}, the proposed approach, which has been applied to the longitudinal model of a hypersonic aircraft, can lead to better tracking with less chattering.
  \item \textbf{Concurrent learning:} A beneficial approach, which has been introduced in \citet{Chowdhary.2011,Chowdhary.2014}, utilizes a set of recorded data points concurrently with instantaneous data to improve the convergence of both parameter and tracking errors. The main benefit of the concurrent learning method is that PE or high adaptation gains are not required. More precisely, in the case of nonlinear systems with \emph{parametric} uncertainty (which can be modeled as $\Delta(x)={W^*}^T\mu(x)$), it has been proved that, if the training input signal is exciting in a sufficiently large finite time interval, both tracking error ($e$) and NN weight's estimation error ($\tilde{W}$) converge exponentially to zero. However, this approach requires precise estimation of the time derivative of the system states, which may be impractical in some cases.
  \item \textbf{Reinforced learning:} Another approach to improve the training performance is to reinforce the learning signal \citep{Lin.2007,Lin.2009}. This can be done by modifying the training rule (\ref{Learn_1}) using the output of another NN (commonly known as the critic network) as $\dot{\hat{W}}=\Gamma \mu \left(e + \|e\| \hat{W}_c^T \mu_c \right)^T$ where $\hat{W}_c$ represents the output weights of the critic network, which is tuned in such a way that guarantees the closed-loop stability \citep{Bu.2019}. The provided learning signal is more informative than the basic training rule (\ref{Learn_1}) thereby strengthening the control performance \citep{Luo.2015b}.
\end{enumerate}

\begin{ex}{Composite learning control of satellite}
Examine and compare two learning methods to enhance the performance of the FEL-based satellite attitude controller introduced in Example \ref{sat_fel}.
\begin{enumerate}[(a)]
\item Design two controllers based on the following approaches:
\begin{itemize}
    \item Method 1: Incorporate an additional term into the Lyapunov function based on the state estimation error.
    \item Method 2: Introduce an additional term based on the prediction of the neural network.
\end{itemize}
\item Analyze the closed-loop response of the designed control system to the reference input given in Eq. \eqref{sat_desired}.
		\item Plot the changes in training weights over time.
		\item Plot estimation of uncertainties and report MAE as well as MAEE for both methods.
		\item Examine how the design parameters of the first method (number of neurons, learning rate, and ratio of composite error signal to tracking error signal) and the second method (number of neurons, learning rate, and control gain) impact the Euclidean norm of MAE and MAEE.
	\end{enumerate}
	\tcblower
	\begin{enumerate}[(a)]
		\item \textbf{Control design and stability analysis:}
		\begin{itemize}
			\item \textit{First method:} To implement this method, we need to define new variables. At first, the estimation error $ e_D $ is defined as
			\begin{equation}
				e_D = \hat{x} - x,
			\end{equation}
			where $ \hat{x} $ denotes the estimation of $ x $. Then, the filtered estimation error would be defined as follows
			\begin{equation}
				s_D = \dot{e}_D + \lambda e_D.
			\end{equation}
Accordingly, the proposed second-order observer model is described by
			\begin{equation}
				\ddot{\hat{x}} = F + Bu + \hat{\Delta} - k_D s_D.
			\end{equation}
			For the first method, the control command is constructed as
			\begin{equation}\label{u_M1}
				u = B^{-1} \left( -F - \hat{\Delta} - ks - \lambda \dot{e} + \ddot{x}_d \right).
			\end{equation}
			Now, consider the following Lyapunov function
			\begin{equation}
				V = \dfrac{k_e}{2} s^T s + \dfrac{k_c}{2} s_D^T s_D + \dfrac{1}{2 \Gamma} \tr \left( \tilde{W}^T \tilde{W} \right).
			\end{equation}
			Taking the time-derivative of $ V $, we have
			\begin{equation}
				\begin{split}
					\dot{V} & = k_e s^T \dot{s} + k_c s_D^T \dot{s}_D + \dfrac{1}{\Gamma} \tr \left( \tilde{W} \dot{\hat{W}} \right) \\
					& = k_e s^T \left( \ddot{e} + \lambda \dot{e} \right) + k_c s^T_D \left( \ddot{e}_D + \lambda \dot{e}_D \right) + \dfrac{1}{\Gamma} \tr \left( \tilde{W}^T \dot{\hat{W}} \right) \\
					& = k_e s^T \left( F + Bu + \Delta - \ddot{x}_d + \lambda \dot{e} \right) \\
					& \quad + k_c s^T_D \left( F + B u + \hat{\Delta} - k_D s_D - F - B u - \Delta  \right) \\
					& \quad + \dfrac{1}{\Gamma} \tr \left( \tilde{W}^T \dot{\hat{W}} \right).
				\end{split}
			\end{equation}
			Using Eq. \eqref{u_M1}, we have
			\begin{equation}
				\begin{split}
					\dot{V} & = k_e s^T \left( -\tilde{\Delta} + \varepsilon - k s \right) + k_c s_D^T \left( \tilde{\Delta} - \varepsilon - k_D s_D \right) \\
					& \quad + \dfrac{1}{\Gamma} \tr \left( \tilde{W}^T \dot{\hat{W}} \right).
				\end{split}
			\end{equation}
			If the NN weights' update rule is defined as
			\begin{equation}
				\dot{\hat{W}} = \Gamma\ \mu \left( k_e s^T - k_c s_D^T \right) - \sigma \hat{W},
			\end{equation}
			then, we can simplify $ \dot{V} $ as follows
			\begin{equation}\label{Vdot1_1}
				\dot{V} = k_e s^T \varepsilon - k k_e s^T s - k_c s_D^T \varepsilon - k_c k_D s_D^T s_D - \dfrac{\sigma}{\Gamma} \tr \left( \tilde{W}^T \hat{W} \right).
			\end{equation}
			Regarding the last term of Eq. (\ref{Vdot1_1}) we get
			\begin{equation}\label{eq1_composite}
				-\tilde{W}^T \hat{W} = - \tilde{W}^T \left( \tilde{W} + W^* \right) = -\tilde{W}^T \tilde{W} - \tilde{W}^T W^*.
			\end{equation}
where again, for its last term we have
\begin{align}\label{eq2_composite}
\begin{split}
-\tilde{W}^T W^* &\leq \Vert \tilde{W} \Vert \Vert W^* \Vert \\
&= \sqrt{\tilde{W}^T \tilde{W}} \sqrt{{W^*}^T W^*} \\
&\leq \dfrac{\delta}{2} \tilde{W}^T \tilde{W} + \dfrac{1}{2\delta} {W^*}^T W^*,
\end{split}
\end{align}
where $ 0 < \delta < 2 $. Moreover, the first line follows from the Cauchy-Schwarz inequality, the second line takes the norm definitions, and the third line follows from Young's inequality. Combining Eqs. \eqref{eq1_composite} and \eqref{eq2_composite}, we have
			\begin{equation}
				- \tr \left( \tilde{W}^T \hat{W} \right) \leq - \left( 1 - \dfrac{\delta}{2} \right) \tr \left( \tilde{W}^T \tilde{W} \right) + \dfrac{1}{2 \delta} \tr \left( W^{*T} W^* \right).
			\end{equation}
			Since $ \dfrac{\sigma}{\Gamma} > 0 $, then we have
			\begin{equation}\label{eq3}
				- \dfrac{\sigma}{\Gamma} \tr \left( \tilde{W}^T \hat{W} \right) \leq - \dfrac{\sigma}{\Gamma} \left( 1 - \dfrac{\delta}{2} \right) \tr \left( \tilde{W}^T \tilde{W} \right) + \dfrac{\sigma}{\Gamma} \dfrac{1}{2 \delta} \tr \left( W^{*T} W^* \right).
			\end{equation}
			Substituting Eq. \eqref{eq3} into Eq. \eqref{Vdot1_1}, we have
			\begin{equation}
				\begin{split}
					\dot{V} & \leq - k k_e s^T s - k_D k_c s_D^T s_D - \dfrac{\sigma}{\Gamma} \left( 1 - \dfrac{\delta}{2} \right) \tr \left( \tilde{W}^T \tilde{W} \right) \\
					& \quad + k_e s^T \varepsilon  - k_c s_D^T \varepsilon + \dfrac{\sigma}{\Gamma} \dfrac{1}{2 \delta} \tr \left( W^{*T} W^* \right).
				\end{split}
			\end{equation}
			Using Cauchy-Schwarz and Young's inequalities
			\begin{subequations}
				\begin{equation}
					k_e s^T \varepsilon \leq \dfrac{k_e}{2} s^T s + \dfrac{\norm{\varepsilon}^2}{2} \leq \dfrac{k_e}{2} s^T s + \dfrac{\norm{\varepsilon_M}^2}{2},
				\end{equation}
				\begin{equation}
					-k_c s_D^T \varepsilon \leq \dfrac{k_c}{2} s_D^T s_D + \dfrac{\norm{\varepsilon}^2}{2} \leq \dfrac{k_e}{2} s_D^T s_D + \dfrac{\norm{\varepsilon_M}^2}{2},
				\end{equation}
			\end{subequations}
			where $ \norm{\varepsilon_M} $ represents the upper bound of $ \norm{\varepsilon}$. Then, we have
			\begin{equation}\label{m1vdot}
				\begin{split}
					\dot{V} & \leq \dfrac{k_e}{2} s^T s - k s^T s + \dfrac{k_e}{2} s_D^T s_D - k_D s_D^T s_D - \dfrac{2 C}{2 \Gamma} \tr \left( \tilde{W}^T \tilde{W} \right) + \norm{\varepsilon_M}^2 \\
					& \leq -\left( \dfrac{2k}{k_e} - 1 \right) \left( \dfrac{k_e}{2} s^T s \right) -\left( \dfrac{2k_D}{k_c} - 1 \right) \left( \dfrac{1}{2} s_D^T s_D \right) - 2C \left( \dfrac{1}{2 \Gamma} \tr \left( \tilde{W}^T \tilde{W} \right) \right) \\
					& \quad + \norm{\varepsilon_M}^2 \\
					& \leq -\bar{k} V + C',
				\end{split}
			\end{equation}
			where $ \bar{k} = \min \left\{ 2k/k_e - 1, 2k_D/k_c - 1, 2C \right\} $, $ C = \frac{\sigma}{\Gamma} \left( 1 - \frac{\delta}{2} \right) $, and $ C' = \norm{\varepsilon_M}^2 $. 
			\item \textit{Second method:} In this method, as stated before, the prediction error of neural network error is added to the Lyapunov function. Consider again the definition of filtered tracking error $ s $ as follows
			\begin{equation}
				s = \dot{e} + \lambda e.
			\end{equation}
			Taking the time derivative of $ s $, we have
			\begin{equation}
				\begin{split}
					\dot{s} & = \ddot{e} + \lambda \dot{e} \\
					& = \left( \ddot{x} - \ddot{x}_d \right) + \lambda \dot{e} \\
					& = F + B u + \Delta - \ddot{x}_d + \lambda \dot{e} \\
					& = F - F - \hat{\Delta} - \lambda \dot{e} - k s + \ddot{x}_d + \Delta - \ddot{x}_d + \lambda \dot{e} \\
					& = W^{*T} \mu + \varepsilon - \hat{W}^T \mu - ks.
				\end{split}
			\end{equation}
			Now, we can redefine $ C' $ based on the following relation
			\begin{equation}
				C' = \dot{s} + \chi,
			\end{equation}
			where $ \chi = \hat{W}^T \mu + k s $. In addition, $ P = \int_{t - \tau}^t \mu \left( \tau \right) \dd \tau $ and $C$ is defined as
			\begin{equation}
				C = \int_{t - \tau}^t C' \dd \tau = s \left( t \right) - s \left( t - \tau \right) + \int_{t - \tau}^t \chi \dd \tau.
			\end{equation}
			We can now rewrite $ C' $ using the following equation
			\begin{equation}
				C' = \dot{s} + \hat{W}^T \mu + ks = W^{*T} \mu + \varepsilon.
			\end{equation}
			Thus, $ C $ is defined as
			\begin{equation}
				C = \int_{t - \tau}^t C' \dd \tau = W^{*T} P + \int_{t - \tau}^t \varepsilon \dd \tau.
			\end{equation}
			We can define output estimation error or neural network as follows
			\begin{equation} \label{alpha}
				\alpha = C - \hat{W}^T P = -\tilde{W}^T P + \int_{t - \tau}^t \varepsilon \dd \tau.
			\end{equation}
			Similar to the previous method, the control command is designed as
			\begin{equation}\label{u_2nd}
				u = B^{-1} \left( -F - \hat{\Delta} - ks - \lambda \dot{e} + \ddot{x}_d \right).
			\end{equation}
			Now, consider the following Lyapunov function
			\begin{equation}
				V = \dfrac{1}{2} s^T s + \dfrac{1}{2} \tr \left( \tilde{W}^T \Gamma^{-1} \tilde{W} \right).
			\end{equation}
			Taking the time derivative of $ V $, we have
			\begin{equation}
				\dot{V} = s^T \dot{s} +  \tr \left( \tilde{W}^T \Gamma^{-1} \dot{\hat{W}} \right).
			\end{equation}
			If we define $ \dot{\hat{W}} = \Gamma \left( \mu s^T + k_W P \alpha^T \right) $, then we have
			\begin{equation}\label{V2dot1}
				\begin{split}
					\dot{V} & = s^T \left( \ddot{e} + \lambda \dot{e} \right) + \tr \left( \tilde{W}^T \left( \mu s^T + k_W P \alpha^T \right) \right) \\
					& = s^T \left( F + Bu + \Delta - \ddot{x}_d + \lambda \dot{e} \right) \\
					& + \tr \left( \tilde{W}^T \mu s^T \right) + k_W \tr \left( \tilde{W}^T P \alpha^T \right).
				\end{split}
			\end{equation}
			Substituting Eq. \eqref{u_2nd} into Eq. \eqref{V2dot1}, we have
			\begin{equation}
				\dot{V} = s^T \left( -\tilde{W}^T \mu + \varepsilon - k s \right) + \tr \left(  \tilde{W}^T \mu s^T \right) + k_W \tr \left( \tilde{W}^T P \alpha^T \right).
			\end{equation}
			Using trace properties, we have
			\begin{equation}
				\dot{V} = s^T \varepsilon - k s^T s + k_W \alpha^T \tilde{W}^T P.
			\end{equation}
			Substituting Eq. \eqref{alpha} into the previous equation, we have
			\begin{equation}
				\dot{V} = s^T \varepsilon - k s^T s - k_W P^T \tilde{W} \tilde{W}^T P + k_W \left( \int_{t - \tau}^t \varepsilon^T \dd \tau \right) \tilde{W}^T P.
			\end{equation}
			Defining two variables $ \eta $ and $ \gamma $ as
			\begin{equation}
				\begin{matrix}
					\eta = \tilde{W}^T P & , & \displaystyle \gamma = \int_{t - \tau}^t \varepsilon \dd \tau
				\end{matrix}.
			\end{equation}
			Using Cauchy-Schwarz and Young's inequalities, we have
			\begin{subequations}
				\begin{equation}
					s^T \varepsilon \leq \dfrac{s^T s}{2} + \dfrac{\norm{\varepsilon}^2}{2} \leq \dfrac{s^T s}{2} + \dfrac{\norm{\varepsilon_M}^2}{2},
				\end{equation}
				\begin{equation}
					\gamma^T \eta \leq  \dfrac{\eta^T \eta}{2} + \dfrac{\norm{\gamma}^2}{2} \leq \dfrac{\eta^T \eta}{2} + \dfrac{\norm{\gamma_M}^2}{2},
				\end{equation}
			\end{subequations}
			where $ \norm{\varepsilon_M} $ and $ \norm{\gamma_M} $ denote the upper bound of $ \norm{\varepsilon}$ and $ \norm{\gamma} $, respectively. Thus, we have
			\begin{equation}\label{m2vdot}
				\dot{V} \leq - \left( k - \dfrac{1}{2} \right) s^T s - \left( k_W - \dfrac{1}{2} \right) \eta^T \eta + \dfrac{\norm{\varepsilon_M}^2}{2} + \dfrac{\norm{\gamma_M}^2}{2},
			\end{equation}
			which means that if $ k $ and $ k_W $ are positive and constant term are bounded, the tracking error will be uniformly ultimately bounded. 
		\end{itemize}
		\item Considering $ k = 20 $, $ k_D = 30 $, $ k_e = 2 $, $ k_c = 20 $, $ \lambda = 10 $, $ \Gamma = 8 $, and $ \sigma = 0.0001 $ for the first method and $ k = 20 $, $ k_W = 300 $, $ \lambda = 10 $, and $ \Gamma = 8 $ for the second method, Fig. \ref{fig:sat_cl_states} shows the results for both methods.
		\begin{figure}[H]
			\centering
			\includegraphics[width = 0.9\textwidth]{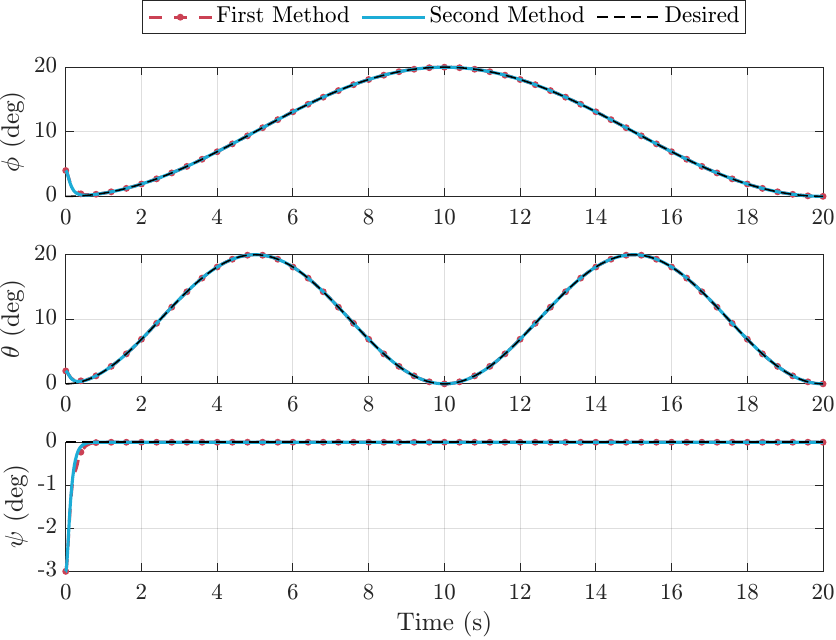}
			\caption{Time history of attitude in both composite learning methods of satellite.}
			\label{fig:sat_cl_states}
		\end{figure}
Both methods track the desired Euler angles with good precision. Figure \ref{fig:sat_cl_controls} shows the corresponding control inputs for both methods.
		\begin{figure}[H]
			\centering
			\includegraphics[width=0.9\textwidth]{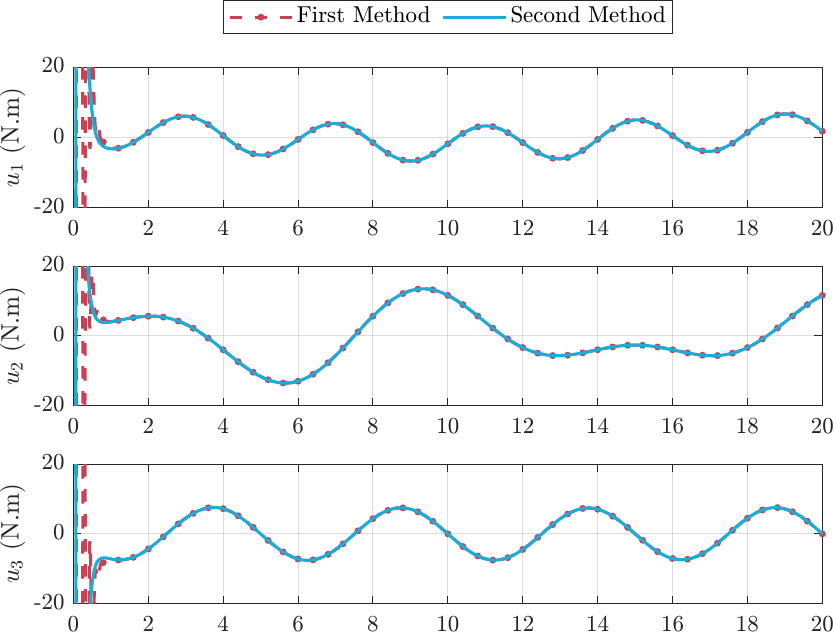}
			\caption{Time history of control inputs in first method of composite learning of satellite.}
			\label{fig:sat_cl_controls}
		\end{figure}
		It should be noted that the oscillation of $ u_2 $ in the first method results to high value of learning rate. In contrast, in the second method, there is no initial oscillation in $ u_2 $.
		\item Variation of neural networks' weights is shown in Fig. \ref{fig:sat_cl_Ws}. 
		\begin{figure}[H]
			\centering
			\includegraphics[width=0.9\textwidth]{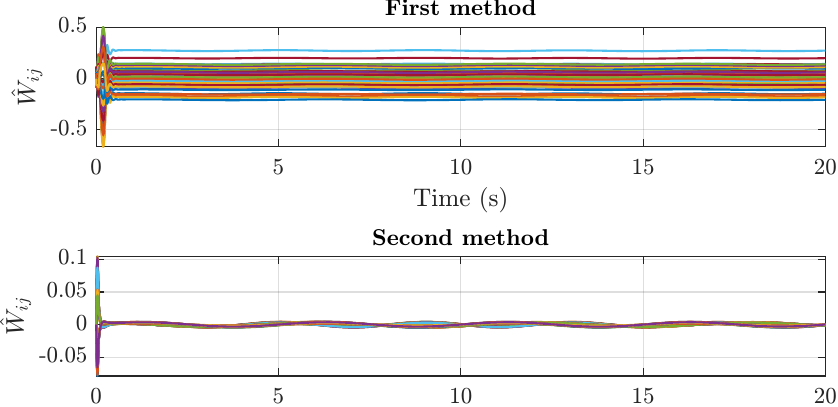}
			\caption{Evolution of training weights in both composite learning methods of satellite.}
			\label{fig:sat_cl_Ws}
		\end{figure}
		In contrast to the second method, in the first method, there is an oscillation at the beginning of the simulation.
		\item In Fig. \ref{fig:sat_cl_Est}, the performance of estimation of total uncertainties is shown. As this figure shows, the estimation of uncertainties are done with a good accuracy since the estimation error is negligible. However, it is worth mentioning that, in the second method, the fluctuation in initial times is significantly lower than that of the first method. This is an important advantage of this approach compared to the conventional composite learning technique. To be more precise, the increase in the observer error's weight in the updating rule of the NN in the basic composite learning technique results in an oscillatory estimation (mainly at initial times), while the new design can satisfactorily avoid such a shortcoming. 
		\begin{figure}[H]
			\centering
			\includegraphics[width=0.9\textwidth]{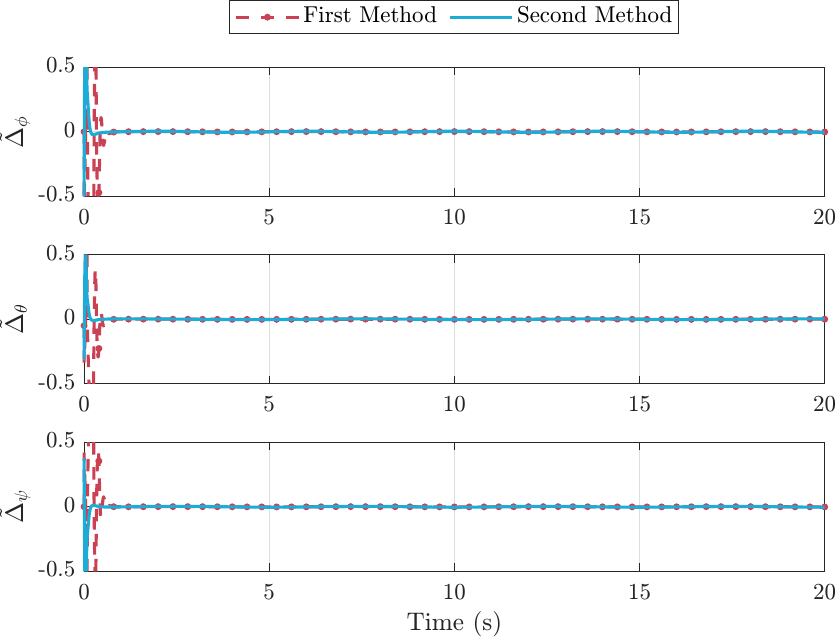}
			\caption{Uncertainty estimation in both composite learning methods of satellite.}
			\label{fig:sat_cl_Est}
		\end{figure}
The MAE and MAEE results in Table \ref{tab:composite_sat} show that the second method outperforms the first method in both tracking and estimation performance. Thus, the second method is able to track the desired Euler angles more accurately and estimate the uncertainties with a significantly lower estimation error.
		\begin{table}[H]
			\centering
			\caption{Results of MAE and MAEE for composite learning control of satellite.}
			\begin{tabular}{c c c c c}
    			\hline\noalign{\smallskip}
				Method & MAE ($ \phi $) & MAE ($ \theta $) & MAE ($ \psi $) & MAEE \\ 
				\noalign{\smallskip}\hline\noalign{\smallskip}
				First & $ 0.0005928 $ & $ 0.0002978 $ & $ 0.0004442 $ & $ 0.02875 $ \\
				Second & $ 0.0005215 $ & $ 0.0002631 $ & $ 0.0003907 $ & $ 0.008698 $ \\ 
				\noalign{\smallskip}\hline\noalign{\smallskip}
			\end{tabular}
			\label{tab:composite_sat}
		\end{table}
		\item For the first method, the effect of variation of number of neurons, learning rate, and ratio of composite error signal to tracking error signal, are investigated on the Euclidean norm of the MAE and the MAEE. The obtained results are shown in Fig. \ref{fig:sat_cl_sa_m1}. As is evident, increasing the number of neurons and the learning rate results in decreasing both the norm of MAE as well as MAEE. On the other hand, increasing the ratio of the composite error signal to tracking error signal decreases the MAEE, while it can weaken the tracking performance and result in an oscillatory response for larger value of the gain ratio due to reduction of the tracking error importance in the learning process. 
		\begin{figure}[H]
            \centering
            \includegraphics[width=0.9\textwidth]{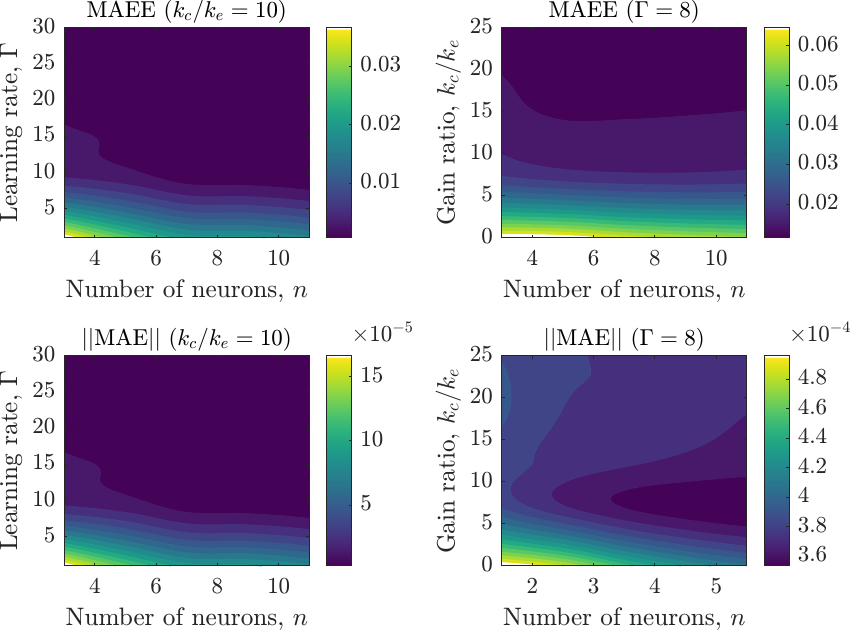}
            \caption{Sensitivity analysis of MAEE and norm of MAE for the number of neurons, learning rate, and the ratio of composite error signal to tracking error signal in the first method of composite learning of satellite.}
            \label{fig:sat_cl_sa_m1}
        \end{figure}
        For the second method, parameter study is done on variation of number of neurons, learning rate, and control gain in the second method. The results is shown in Fig. \ref{fig:sat_cl_sa_m2}. Similar to first method, in this method, increasing number of neurons and learning rate results in decreasing both norm of MAE as well as MAEE. Moreover, increasing control gain results increasing both norm of MAE as well as MAEE.
        \begin{figure}[H]
            \centering
            \includegraphics[width=0.9\textwidth]{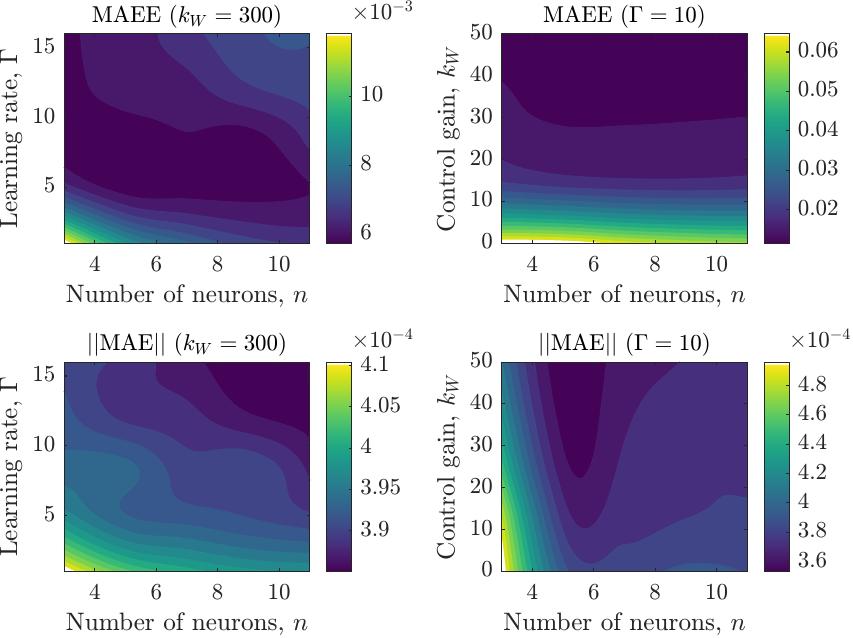}
            \caption{Sensitivity analysis of MAEE and norm of MAE for the number of neurons, learning rate, and the control gain in second method of composite learning of satellite.}
            \label{fig:sat_cl_sa_m2}
        \end{figure}
	\end{enumerate}
\end{ex}

The above-mentioned formulation corresponds to the \emph{indirect} FEL-based control where a NN attempts to identify model uncertainties, and then the control command is constructed using the estimated uncertainty \citep{HaojianXu.2003}.
FEL can also be satisfactorily employed in the framework of direct adaptive control systems. In this regard, it is possible to directly estimate the entire control command $u$ or the uncertain term $B^{-1}\Delta$ in (\ref{Ctrl_0}) by a NN. As a result, the updating rule of $\hat{W}$ will include the control gain matrix $B$. However, there is no fundamental difference between the direct and indirect approaches regarding the formulation of the updating rules and the stability analysis of the closed-loop system.
Direct methods may also be preferred in cases where the control gain function is entirely unknown (see Section \ref{Sec:DirectNeuralBackstepping}).

\begin{ex}[label=Example:quad_direct_FEL]{Direct FEL control of quadrotor UAV}
\begin{enumerate}[(a)]
\item Design a direct FEL-based controller for the quadrotor UAV represented in Example \ref{Example:quadFEL}.
\item Evaluate the closed-loop performance by using the following command trajectory
\begin{equation}
x_d=\left[\begin{matrix}z_d\\\phi_d\\\begin{matrix}\theta_d\\\psi_d\\\end{matrix}\\\end{matrix}\right]=\left[\begin{matrix}10-0.5t\\0.2\sin{\left(\frac{\pi}{10}t\right)}\\\begin{matrix}0.2\left(1-\cos{\left(\frac{\ \pi}{15}t\right)}\right)\\0\\\end{matrix}\\\end{matrix}\right].
\end{equation}
\item Plot the change of training weights over time.
\item Compare the outputs of NN with its expected true values.
\item Carry out a parameter study on MAE and MAEE.
\end{enumerate}
\tcblower
\begin{enumerate}[(a)]
\item \textbf{Controller design:}
Following the same procedure from Example \ref{Example:quadFEL}, The tracking error can be calculated as $e=x-x_d$. To generate the filtered error signal, we can define $s=\dot{e}+\lambda e$ with $\lambda >0$.
Therefore, the derivative of $s$ can be expressed as:
\begin{align}\label{direct_s_dot}
\begin{split}
  \dot{s} &= \ddot{e}+\lambda \dot{e} \\
      &= \bar{f} + \bar{B}u + \Delta(x,u) - \ddot{x}_d + \lambda \dot{e}.
\end{split}
\end{align}
Subsequently, using the previously introduced RBFNN structure to estimate $\bar{B}^{-1}\Delta(x,u)$ as $\hat{W}^T \mu (X)$, it is obtained that
\begin{equation}
\bar{B}^{-1}\Delta(x,u)={W^\ast}^T\mu(x)+\varepsilon.
\end{equation}
\\
As a result, we can construct the following control command
\begin{equation}\label{direct_u}
u=\bar{B}^{-1}(-\bar{f}(x)+{\ddot{x}}_d-\lambda_1\dot{e}-k_1s)-\hat{W}^T \mu (X),
\end{equation}
in which the NN output is directly included.
Now, we proceed to derive the update rule for NN parameters.

\textbf{Stability analysis:}
Consider the following Lyapunov function
\begin{equation}
V=\frac{1}{2}s^T s + \frac{1}{2} \text{tr} \left(\tilde{W}^T \Gamma^{-1} \tilde{W}\right),
\end{equation}
where $\tilde{W}$ and $\Gamma$ are as defined previously. Substituting the direct control command, $\dot{V}$ is expressed as
\begin{align}
\begin{split}
\dot{V} &= s^T \dot{s}+ \text{tr} \left(\tilde{W}^T \Gamma^{-1} \dot{\tilde{W}}\right) \\
&= - s^T \left(-\Delta+\bar{B}\hat{W}^T \mu (X)+ k_1 s\right)+ \text{tr} \left(\tilde{W}^T \Gamma^{-1} \dot{\tilde{W}}\right),
\end{split}
\end{align}
where $\bar{B}\hat{W}^T \mu (X)-\Delta =\bar{B} \left ( \tilde{W}^T \mu\left(X\right) -\varepsilon \right )$ and we assumed $\dot{\tilde{W}}\approx  \dot{\hat{W}}$. Considering direct FEL update rule $\dot{\hat{W}}=\Gamma \mu s^T\bar{B}$, results in
\begin{align}
\begin{split}
\dot{V} &= - s^T \left(\bar{B}{\tilde{W}}^T \mu - \bar{B}\varepsilon + k_1 s\right) + \text{tr} \left(\tilde{W}^T \mu s^T\bar{B}\right)\\
&=- s^T (k_1 s - \varepsilon_1).
\end{split}
\end{align}
This results in $\dot{V}<0$ for $|k_1 s|>|\varepsilon_1|$, guaranteeing bounded tracking error.
\\
\item\textbf{Simulation results:}
For the case with 15 neurons and $k_1=12$, $\Gamma=0.1$, $\lambda=8$, the time history of attitude and altitude is depicted in Fig. \ref{fig:exc27:stat}. The time constant of the first-order actuator is $\tau=0.01$.
\\
It is noteworthy to mention that the acceptable range of $\Gamma$ in direct method may differ from that in the indirect (basic FEL) method.
\begin{figure}[H]
  \begin{center}
    \includegraphics[width=0.9\textwidth]{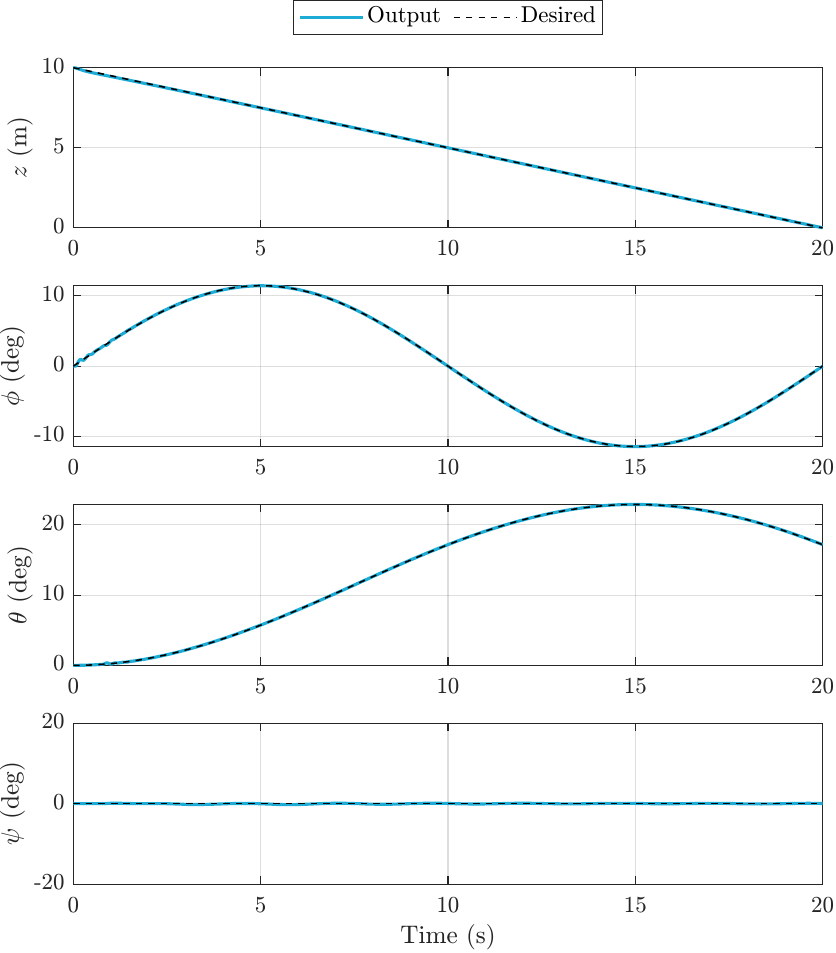}
  \end{center}
  \caption{Time history of attitude and altitude in direct FEL of quadrotor.} 
  \label{fig:exc27:stat}
\end{figure}
Generally, both attitude and altitude closely follow their desired trajectory. It should be mentioned that all states have zero initial conditions.
\\
Regarding the control effort, as depicted in Fig. \ref{fig:exc27:thrust}, owing to the presence of first-order filter, the applied control is weakened.
\begin{figure}[H]
  \begin{center}
    \includegraphics[width=0.9\textwidth]{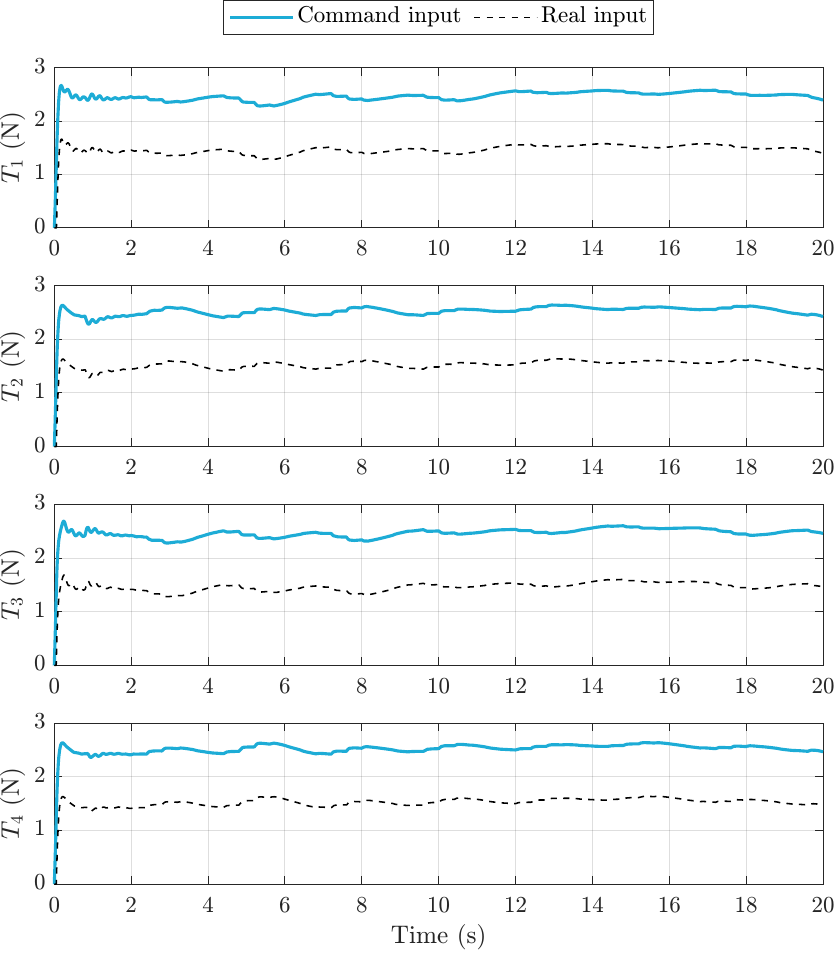}
  \end{center}
  \caption{Time history of control inputs in direct FEL control of quadrotor.} 
  \label{fig:exc27:thrust}
\end{figure}
The results of MAE and MAEE in direct FEL control of quadrotor are provided in Table \ref{Dir_FEL_quad_MAE}.
\begin{table}[H]
    \caption{Result of MAE in direct FEL control of quadrotor.}
    \label{tab:ch2_exc07_MAE}
    \centering
    \begin{tabular}{ccccc}
    \hline\noalign{\smallskip}
    MAE ($z$) & MAE ($\phi$) & MAE ($\theta$) & MAE ($\psi$) & MAEE\\
    \noalign{\smallskip}\hline\noalign{\smallskip}
    $1.62\times 10^{-4}$ & $1.11\times 10^{-3}$ & $5.91\times 10^{-3}$ & $2.89\times 10^{-4}$ & $9.74\times 10^{-2}$ \\
    \noalign{\smallskip}\hline\noalign{\smallskip}
    \end{tabular} \label{Dir_FEL_quad_MAE}
\end{table}

\item Additionally, Fig. \ref{fig:exc27:what} shows the evolution of training weights over time which seem to have converged to a constant value at the end of the simulation time.
\begin{figure}[H]
  \begin{center}
    \includegraphics[width=0.9\textwidth]{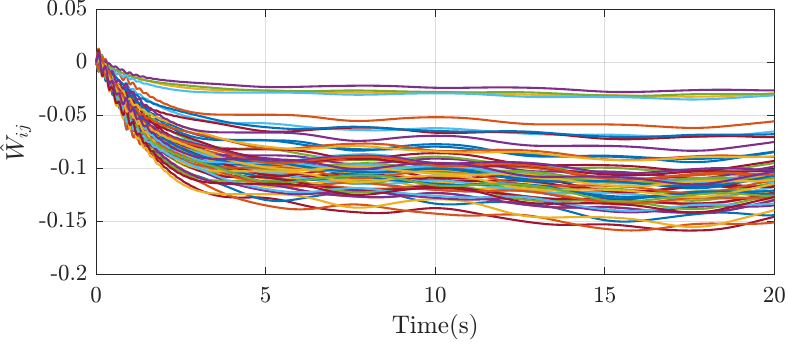}
  \end{center}
  \caption{Evolution of training weights in direct FEL control of quadrotor.} 
  \label{fig:exc27:what}
\end{figure}
\item By defining $\gamma =\bar{B}^{-1}\Delta(x,u)$ and $\hat{\gamma}=\hat{W}^T \mu (X)$ uncertainty estimation error is depicted in Fig. \ref{fig:exc27:del}, demonstrating an acceptable estimating performance.
\begin{figure}[H]
  \begin{center}
    \includegraphics[width=0.9\textwidth]{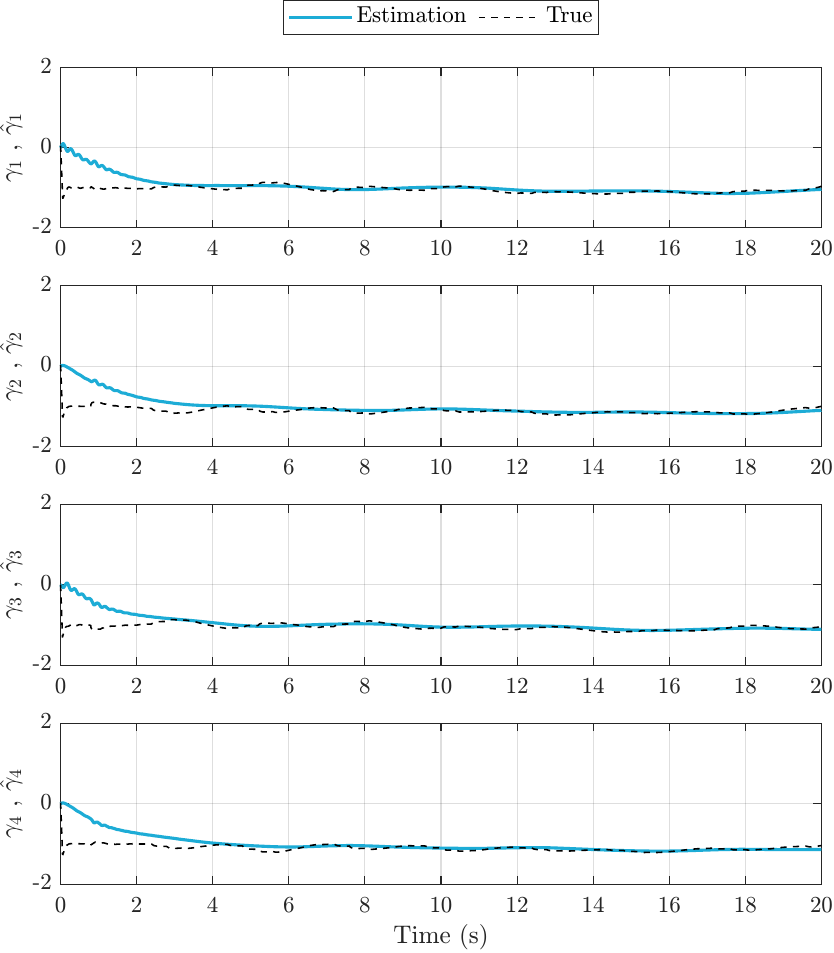}
  \end{center}
 \caption{Uncertainty estimation error in direct FEL control of quadrotor.} 
  \label{fig:exc27:del}
\end{figure}

\item\textbf{Sensitivity analysis:}
A parameter study is carried out to explore the impacts of various parameters on control performance, focusing on the precision of the tracking error (measured by MAE). As shown in Fig. \ref{fig:exc27:sens_anlys_MAE} increasing both $\Gamma$ and control gain leads to a decrease in MAE. Additionally increasing both $\lambda$ and number of neurons gives a better performance. 
\begin{figure}[H]
  \begin{center}
    \includegraphics[width=0.9\textwidth]{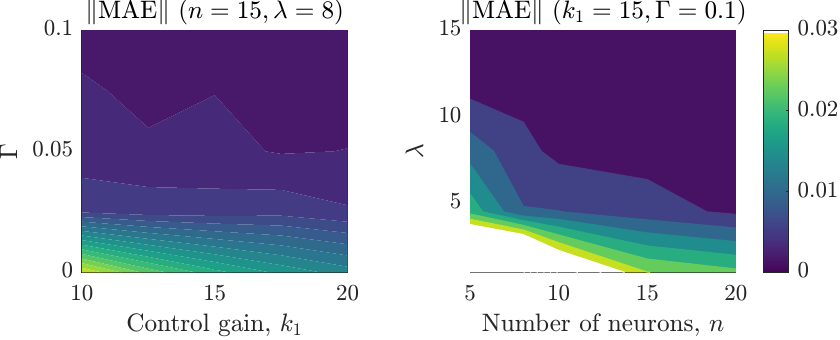}
 \end{center}
  \caption{Sensitivity analysis for $\Gamma$, number of neurons, $\lambda$ and the gain $k_1$ on the norm of MAE in direct FEL control of quadrotor.} 
  \label{fig:exc27:sens_anlys_MAE}
\end{figure}
Turning to MAEE, from Fig. \ref{fig:exc27:sens_anlys_MAEE} it can be observed that increasing $k_1$ results in an increase in MAEE, whereas increasing $\Gamma$ leads to a better performance. Additionally although MAE is unaffected by $\lambda$, increasing number of neurons gives a better performance.
\begin{figure}[H]
  \begin{center}
    \includegraphics[width=0.9\textwidth]{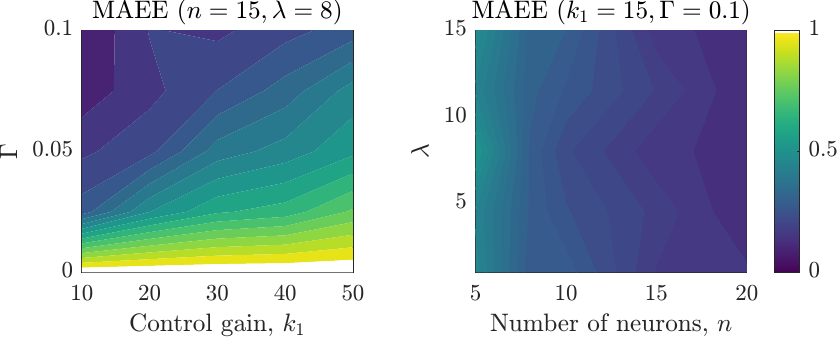}
 \end{center}
  \caption{Sensitivity analysis for $\Gamma$, number of neurons, $\lambda$ and the gain $k_1$ on MAEE in direct FEL control of quadrotor.} 
  \label{fig:exc27:sens_anlys_MAEE}
\end{figure}

\end{enumerate}
\end{ex}


\section{Pseudocontrol strategy}\label{Sec:PseudoControl}
In a somewhat similar manner to the introduced direct FEL-based neural control, in the case of \emph{non-affine} models, traditionally, the output of a baseline controller such as a Proportional-Integral-Derivative (PID) controller may be used to train a NN, which augments the output of the baseline controller to learn the inverse dynamics of the system, while there is considerable complexity in the closed-loop stability analysis \citep{Gomi.1993}. Accordingly, an auto-landing scheme has been proposed in \citet{Li.2004} for an aircraft under external disturbances using a FEL-based neural aided $H_{\infty}$ control. Similarly, a combination of a classical trajectory tracking control (using the loop shaping technique) with a FEL-based neural controller has been employed in \citet{Pashilkar.2006b} as a fault-tolerant auto-landing control method. Such a method has also been adopted in \citet{Li.2001} to control the attitude of a simplified model of a fighter aircraft using fully-tuned growing RBFNNs.


By incorporating a similar framework, type-2 Fuzzy Neural Networks (T2-FNN) have been employed in \citet{Khanesar.2015,Kayacan.2016} to augment a classical PD controller in the case of a set of Single-Input Single-Output (SISO) systems. FEL algorithm has been adopted, where the updating rule corresponding to the consequent part of the T2-FNN has been derived by minimizing $\int \left(\dot{e}+\lambda e \right)^2 dt$. Although in \citet{Khanesar.2015}, it has been assumed that the intended system has a second-order stabilizable dynamic model, the stability of the closed-loop system has been analyzed without making any assumption on the system characteristics (even the system's stabilizability!). Apparently, this is a consequence of estimating an \emph{explicit function of time} by a neural network as ${W^*}^T\mu(x)$, which is not generally feasible. This is a common issue in NN-based identification schemes (see Chapter 3).
The proposed control system has been applied to the trajectory tracking control of a quadrotor UAV. A self-organizing neuro-fuzzy-based control has been introduced in \citet{Ferdaus.2019} in which the consequent part of the fuzzy rules has been trained using a similar FEL scheme, where the designed controller has been applied to a hexacopter and a flapping-wing Micro Aerial Vehicle (MAV) to control the altitude and the attitude of the system.
It has been claimed that the controller's performance does not depend on \emph{any} features for the system. This is clearly an exaggerated statement since the most obvious feature required in a controlled system is the system's controllability. Again, it seems that the generality of the stability analysis is due to the aforementioned concern regarding the NN-based identification schemes.
Different from \citep{Ferdaus.2019}, in \citet{Ferdaus.2020}, the stability analysis has been provided for a $n$th order, affine, SISO model. A FEL scheme has been used to train the consequent parameters of a neuro-fuzzy control system, which augments a PID controller, while the updating rule is subject to the parameter drift issue.

\begin{figure}[!h]
  \begin{center}
    \includegraphics[width=\textwidth]{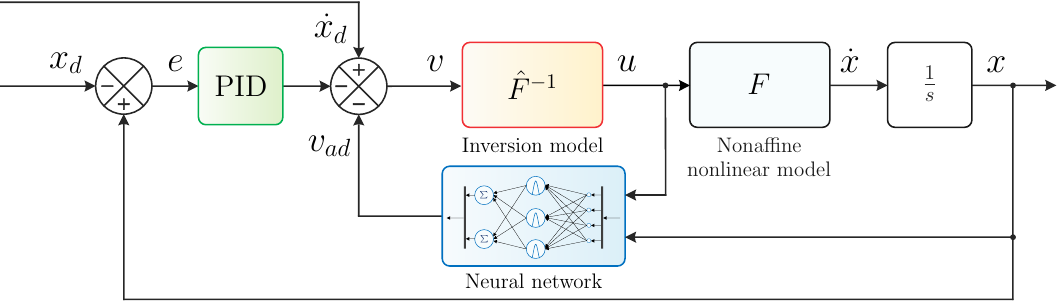}
  \end{center}
  \caption{Pseudocontrol scheme.} 
  \label{fig:pseudo}
\end{figure}

On the other hand, a simpler and popular approach to FEL-based direct adaptive control of non-affine systems, known as the \emph{pseudocontrol} strategy, has been widely employed in IFCSs. Generally speaking, in this approach, the control command is determined using a model inversion block where a neural network is utilized to cancel out the inversion error \citep{Kim.1997}, as shown in Fig. \ref{fig:pseudo}.
To be more precise, consider a generic non-affine nonlinear model of the system as follows
\begin{equation}\label{Eq:16_0}
  \dot{x}=F(x,u).
\end{equation}
As seen, unlike the previous subsection, here, there is no need for an affine model of the controlled system. Despite the possible complexities in the control of non-affine systems, the following design would be more effective in the case of nonconventional air vehicles with highly nonlinear dynamics, which could not be modeled as an affine model, satisfactorily. In particular, such a method could be an optimal choice in the case of a Hypersonic Flight Vehicle (HFV), which possesses a completely non-affine model \citep{Bu.2018}. Indeed, although, using some simplifications, HFVs are typically modeled as an approximate affine model (and the remaining nonlinear terms are treated as model uncertainty) to facilitate the control design, such an approach results in a conservative control system.
It should be noted that, in the case of a flight control problem, the dynamic model of the system is typically formulated as $\ddot{x}=F(\dot{x},x,u)$.
However, the following design can be applied to such a control problem, as well, using a simple change of variables and employing a composite error function consisting of both $e$ and $\dot{e}$.
Now, assuming the availability of an approximate inversion model, the control command can be computed as follows
\begin{equation}\label{Eq:16_01}
  u=\hat{F}^{-1}(x,\nu),
\end{equation}
where $\nu$ denotes the pseudocontrol input, which should be designed. Notice that, although there is no need for an accurate inversion model, the chosen inversion model should capture the control assignment structures. It means that, for example, the inversion model should include the fact that the elevator deflection affects the pitch rate.
In addition, it is assumed that $\hat{F}^{-1}(x,\nu)$ is a one-to-one function. This assumption can be realized if $\text{dim}(u)=\text{dim}(x)$ \citep{Chowdhary.2013}, which is reasonable in a typical flight control problem.
Accordingly, the pseudocontrol input $\nu$ can be designed as follows \citep{Lee.2005}
\begin{equation}\label{Eq:16_02}
  \nu = \dot{x}_d - k e - \nu_{ad},
\end{equation}
where $k$ is a positive constant which can be replaced with a PID controller, and $\nu_{ad}$ denotes an additional command to alleviate the inversion error. More precisely, defining $\Delta(x,u)= F(x,u) - \hat{F}(x,u)$, we have
\begin{equation}
  \dot{e}= - k e - \nu_{ad} + \Delta(x,u).
\end{equation}
Thus, if it is possible to have $\nu_{ad} = \Delta(x,u)$, the tracking error converges asymptotically to zero. However, $\Delta(x,u)$ is unknown. So, we estimate it using the feedback error learning scheme. In this regard, using a NN to identify $\Delta(x,u)$, we have $\Delta(x,u)={W^*}^T \mu (x,u) + \varepsilon$. Subsequently, $\nu_{ad}$ can be determined as $\nu_{ad} =\hat{W}^T \mu (x,u)$. Introducing a Lyapunov function $V$ as
\begin{equation}
  V=\frac{1}{2}e^T e + \frac{1}{2} \tr \left(\tilde{W}^T \Gamma^{-1} \tilde{W}\right),
\end{equation}
and using the updating rule $\dot{\tilde{W}}\approx \dot{\hat{W}}=\Gamma \mu (x,u) e^T$, the time derivative of $V$ is obtained as the following equation
\begin{align}
\begin{split}
\dot{V} & = - e^T \left(k e + \tilde{W}^T \mu (x,u) - \varepsilon\right)+ \tr \left(\tilde{W}^T \Gamma^{-1} \dot{\tilde{W}}\right) \\
 & = - e^T \left(k e - \varepsilon \right).
\end{split}
\end{align}
Accordingly, the introduced control strategy can satisfactorily ensure the bounded tracking error. Again, one of the above-mentioned modification techniques can be adopted in the proposed updating rule to prevent parameter drift.
Notice that the proposed control framework results in a control law as $u=\hat{F}^{-1}\left(x,\dot{x}_d - k e - \hat{W}^T \mu (x,u)\right)$, thereby requiring the contraction assumption. A modification to the introduced strategy has been given in \citet{Bu.2018} by taking advantage of the Mean Value theorem to relax this assumption, while the sign of $\partial F/\partial u$ should be known, and there are some concerns with the provided stability analysis.
Besides, authors in \citet{Bu.2018} have employed the pseudocontrol approach in the case of a SISO system in the normal feedback form with $\text{dim}(x)>\text{dim}(u)$, while there is a requirement for the availability of all the system states in the proposed control scheme. To this end, if we have $z_1=e=x_1-x_{1d}$, $z_2=\dot{z}_1$, and $\ddot{x}_1=f(x,u)$, one can define a filtered tracking error $s=\dot{e}+\lambda e$, which results in $\dot{s}=f(x,u)-\ddot{x}_{1d}+\lambda \dot{e}$. Thus, by replacing the real tracking error $e$ with the filtered tracking error $s$ in the introduced method, it is possible to design a similar pseudocontrol framework.

\begin{ex}{Attitude control of satellite using pseudocontrol}
Design a pseudocontrol-based controller for a non-affine satellite model. Use the model introduced in Example \ref{sat_fel} and consider the dead-zone nonlinearity (see Fig. \ref{fig:deadz}) in actuators' dynamic model.
	\begin{enumerate}[(a)]
		\item Design a neural pseudocontrol-based controller.
		\item Evaluate the performance of the designed control system in tracking the reference input given by Eq. \eqref{sat_desired}.
		\item Report MAE and MAEE.
		\item Plot the changes in training weights over time.
		\item Examine how design parameters, including the number of neurons, learning rate, and dead-zone's bound, impact Euclidean norm of MAE.
	\end{enumerate}
	\tcblower
	\begin{enumerate}[(a)]
		\item \textbf{Pseudocontrol design:} 
		The dead-zone phenomenon results in a non-affine dynamic model, while the pseudocontrol strategy can satisfactorily deal with such a problem. Since the dynamic equation is of second order, the pseudocontrol input is designed as
		\begin{equation}
			\nu = \ddot{x}_d - k s - \lambda \dot{e} - \nu_{ad},
		\end{equation}
		where $ \nu_{ad} $ is calculated using an NN as
		\begin{equation}
			\nu_{ad} = \hat{W}^T \mu.
		\end{equation}
		Considering the following update rule for the NN
		\begin{equation}
			\dot{\hat{W}} = \Gamma \mu s^T,
		\end{equation}
		where $ s = \dot{e} + \lambda e $.
		\item For the dead-zone limit of $L=20$ Nm, and using the values $k = 10$, $\lambda = 10$, and $\Gamma = 10$, Fig. \ref{fig:sat_pse_states} shows how well the satellite pseudocontrol system tracks the reference input for the Euler angles.
		\begin{figure}[H]
			\centering
			\includegraphics[width = 0.9\textwidth]{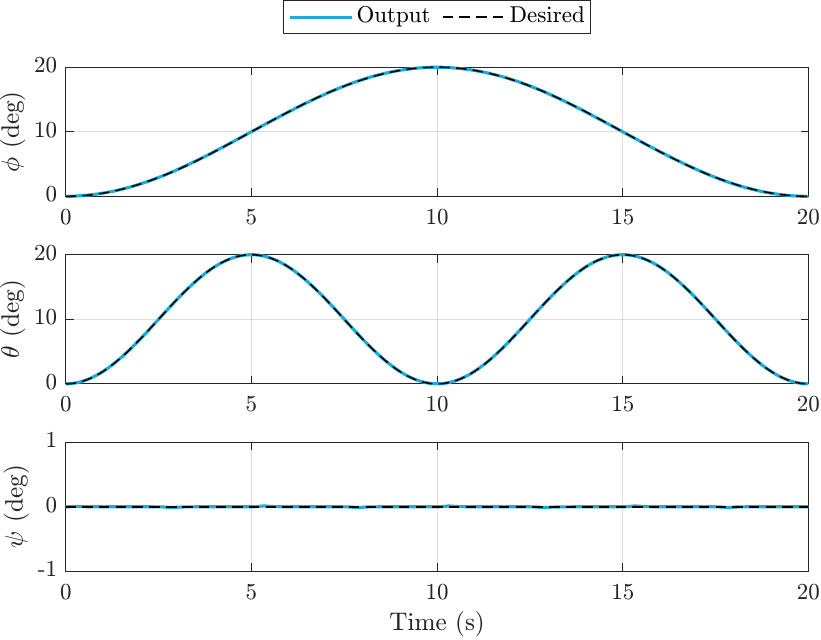}
			\caption{Time history of attitude in pseudocontrol of satellite.}
			\label{fig:sat_pse_states}
		\end{figure}
Using the pseudocontrol method, the satellite tracks the desired values of the Euler angles with high precision. Figure \ref{fig:sat_pse_controls} shows the corresponding control inputs. Note that the control input time history is affected by the dead-zone that has been applied.
		\begin{figure}[H]
			\centering
			\includegraphics[width=0.9\textwidth]{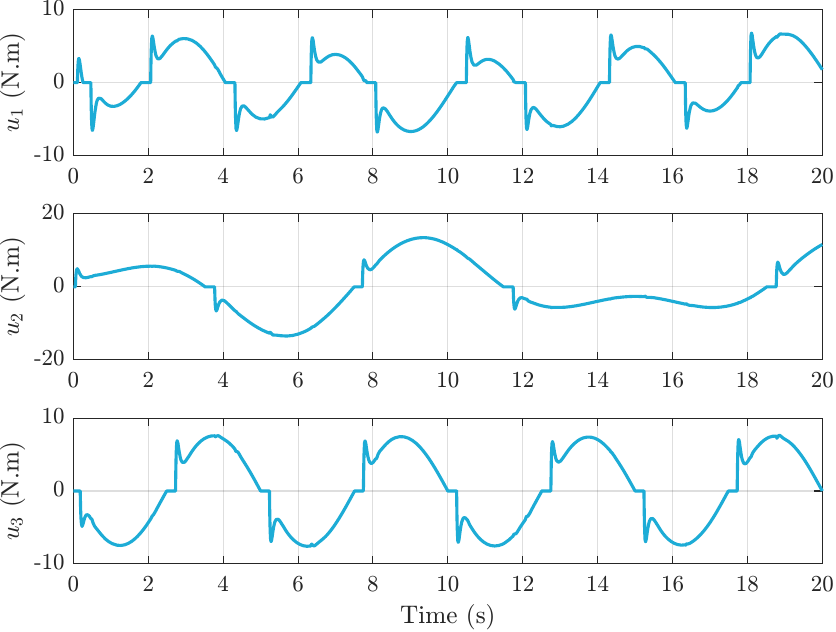}
			\caption{Time history of control inputs in pseudocontrol of satellite.}
			\label{fig:sat_pse_controls}
		\end{figure}
\item In Fig. \ref{fig:sat_pse_est} the performance of the estimation of total uncertainties is shown:
\begin{figure}[H]
    \centering
    \includegraphics[width=0.9\textwidth]{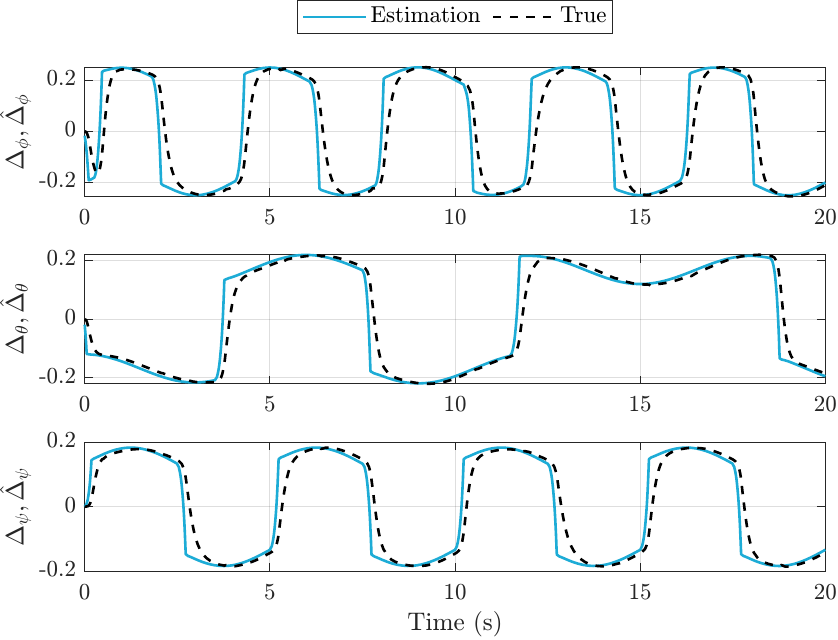}
    \caption{Uncertainty estimation in pseudocontrol of satellite.}
    \label{fig:sat_pse_est}
\end{figure}
By choosing the appropriate selection of design parameters, the estimation of uncertainties are done with high precision. In addition, results in Table \ref{tab:sat_pse_mae} show that the pseudocontrol system is able to track the desired values of the Euler angles with high precision. 
		\begin{table}[H]
			\centering
			\caption{Result of MAE in pseudocontrol of satellite.}
			\label{tab:sat_pse_mae}
			\begin{tabular}{c c c c}
				\hline\noalign{\smallskip}
				MAE ($\phi$) & MAE ($\theta$) & MAE ($\psi$) & MAEE \\ 
				\noalign{\smallskip}\hline\noalign{\smallskip}
				$ 1.514 \times 10^{-4} $ & $ 1.199 \times 10^{-4} $ & $ 1.093 \times 10^{-4} $ & $5.7 \times 10^{-2} $ \\
				\noalign{\smallskip}\hline\noalign{\smallskip}
			\end{tabular}
		\end{table}
\item Fig. \ref{fig:sat_pse_Ws} shows the variation of the neural network weights. Similar to the control inputs in Fig. \ref{fig:sat_pse_controls}, the effect of the dead-zone is noticeable in the network weights.
		\begin{figure}[H]
			\centering
			\includegraphics[width=0.9\textwidth]{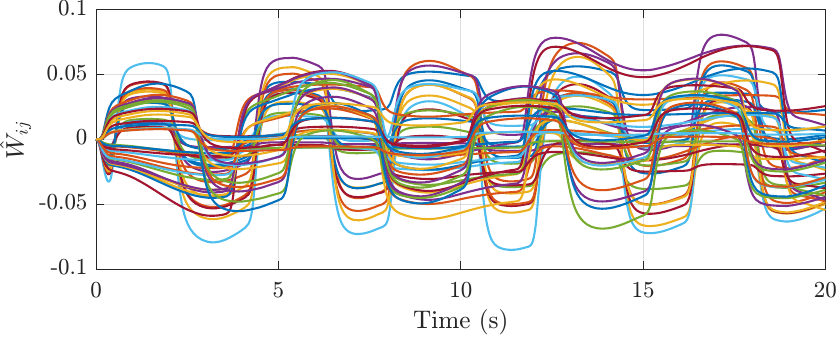}
			\caption{Evolution of training weights in pseudocontrol of satellite.}
			\label{fig:sat_pse_Ws}
		\end{figure}
\item For the sensitivity analysis, we first considered variations in $ \Gamma$ and $ n $ while $ L $ was constant. In the next step, we considered changes in $ L $ and $ n $ while maintaining $ \Gamma $ constant. The results are shown in Figure \ref{fig:sat_pse_sen}.
		\begin{figure}[H]
			\centering
			\includegraphics[width=0.9\textwidth]{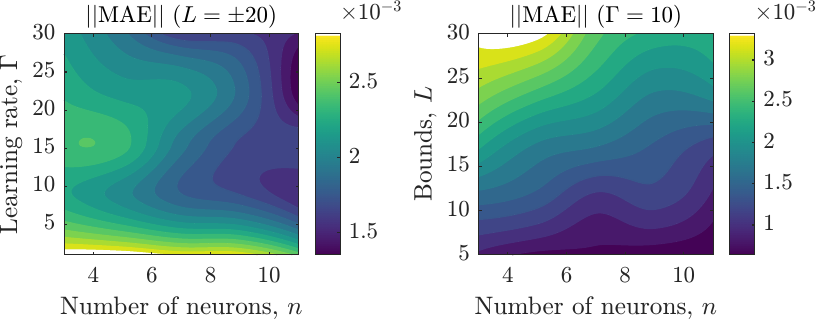}
			\caption{Sensitivity analysis for number of neurons, learning rate and dead-zone's bound on the norm of MAE in pseudocontrol of satellite.}
			\label{fig:sat_pse_sen}
		\end{figure}
As depicted in Fig. \ref{fig:sat_pse_sen}, increasing the number of neurons and learning rate (up to $\Gamma = 10$) both result in a reduction in MAE, while a large increase in the learning rate may cause undesirable oscillations in the system response. Additionally, as expected, lowering the dead-zone's bound contributes to a reduction in MAE.
	\end{enumerate}
\end{ex}

The pseudocontrol strategy has been employed in different flight control systems \citep{Rahideh.2012} such as the attitude control of a tailless fighter aircraft \citep{Calise.2000,Brinker.2000,Calise.2001,Brinker.2001}, the trajectory tracking control of a helicopter \citep{Lee.2005}, the attitude control of a tilt-rotor aircraft \citep{Rysdyk.2005}, etc.
A similar direct adaptive control has been utilized in \citet{NhanNguyen.2006,Nguyen.2008} to control the trajectory of a conventional fixed-wing aircraft under structural damages. A hybrid direct-indirect adaptive control has also been developed in them in which parallel FEL algorithms attempted to provide both the control augmentation signal and estimated uncertain dynamics.
in \citet{Chowdhary.2013}, an inner loop attitude control block based on the pseudocontrol strategy has been employed within a fault-tolerant guidance and control system for a conventional fixed-wing air vehicle. An acceleration (outer loop) guidance loop has been designed, which attempts to provide feasible acceleration command in the presence of structural damages and actuator faults. The performance of the proposed approach was verified in the presence of severe structural and actuator damages.

As an alternative to the above-mentioned approach to control non-affine systems, authors in \citet{KengPengTee.2008} attempted to directly estimate the desired control command rather than the inversion model error using a NN (in the case of a SISO system with stable zero dynamics). Under conservative assumptions on the value of $\partial F/\partial u$ (and its time derivative) and employing the implicit function theorem, one can assume that there is an ideal control command $u^*$ which ensures the closed-loop system stability, i.e. $F(x,u^*)=\dot{x}_d- k e$.
Subsequently, in lieu of utilizing an approximate inversion model, the Mean Value theorem has been adopted to provide an expression for $F(x,u)$ in terms of $F(x,u^*)$. Using such a formulation, a NN with a typical FEL scheme can be employed to estimate $u^*$. Although in this method, there is no need for an approximate inversion model of the system, different restrictive assumptions are required in the control design, which may not be satisfied in a practical flight control problem.
A somewhat similar approach has been utilized in \citet{DaoxiangGao.2014} in the framework of indirect adaptive control, where the singular perturbation theory has been adopted to move $u$ towards $u^*$ as $\epsilon \dot{u}= F(x,u^*)-F(x,u)$ with $\epsilon$ denotes a small positive constant. Such a method has been employed to control the longitudinal model of an HFV, where a set of NNs has been incorporated to estimate unknown dynamics. In this regard, there is no need for a strict feedback model and a backstepping design, while, again, restrictive assumptions should be made to ensure closed-loop stability.

\section{Neural backstepping control} \label{Sec:Backstepping}
\begin{figure}[!h]
  \begin{center}
    \includegraphics[width=\textwidth]{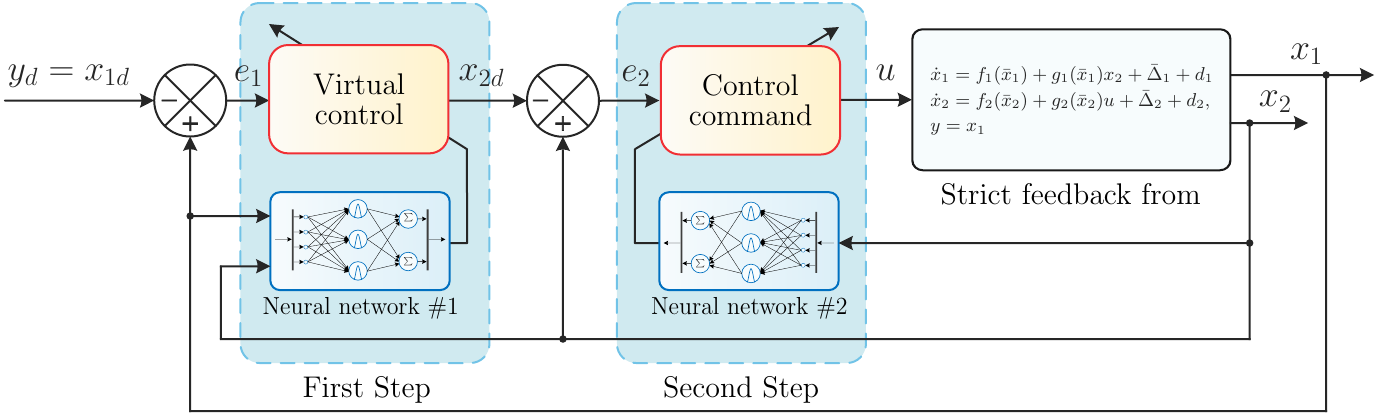}
  \end{center}
  \caption{Architecture of neural backstepping control with $n=2$.} 
  \label{fig:backstep}
\end{figure}

In the basic FEL scheme, it was assumed that $\text{dim}(x)=\text{dim}(u)$. Also, in both the above-mentioned control structures, the entire dynamic model of the system is assumed invertible. However, in many cases, the dimension of the system inputs is less than that of the system states. The backstepping control method can be effectively employed in such circumstances when the dynamic model can be formulated in a strict feedback form, as shown in Fig. \ref{fig:backstep}.
For simplicity, consider an uncertain nonlinear SISO system as follows:
\begin{gather}
  \dot{x}_i=f_i(\bar{x}_i)+g_i(\bar{x}_i)x_{i+1}+\bar{\Delta}_i+d_i, \quad 1\leq i \leq n-1, \label{Eq:16_1}\\
  \dot{x}_n=f_n(\bar{x}_n)+g_n(\bar{x}_n)u+\bar{\Delta}_n+d_n, \label{Eq:16_2}\\
  y = x_1 \label{Eq:16_3},
\end{gather}
where, $\bar{\Delta}_i$ and $d_i$ stand, respectively, for model uncertainties and external disturbances and $\bar{x}_i=[x_1,\ldots,x_i]^T$.
Without loss of generality, in the following, we assume that $n=2$. The introduced control method can be simply applied to higher-order systems. Defining $\Delta_i=\bar{\Delta}_i+d_i$ and the desired output as $y_d$, we have
\begin{equation}
  \dot{e}_1=\dot{y}-\dot{y}_d = f_1(\bar{x}_1)+g_1(\bar{x}_1)x_{2}+\Delta_1 - \dot{y}_d.
\end{equation}
Thus, a virtual control can be defined for $x_2$ as
\begin{equation}\label{Eq:17}
  x_{2d}=g_1^{-1}(\bar{x}_1)\left(\dot{y}_d - k_1 e_1 - f_1(\bar{x}_1) - \hat{\Delta}_1 \right),
\end{equation}
where $\hat{\Delta}_1$ represent the estimation of $\Delta_1$ and $k_1$ is a positive constant. Therefore, defining $e_2=x_2-x_{2d}$, we have
\begin{equation}
  \dot{e}_2 = \dot{x}_2-\dot{x}_{2d}= f_2(\bar{x}_2)+g_2(\bar{x}_2)u+ {\Delta}_2 - \dot{x}_{2d}.
\end{equation}
Finally, the control command can be defined as
\begin{equation} \label{Eq:18}
  u = g_2^{-1}(\bar{x}_2)\left(\dot{x}_{2d} - g_1(\bar{x}_1) e_1 - k_2 e_2 - f_2(\bar{x}_2) - \hat{\Delta}_2 \right),
\end{equation}
where $\hat{\Delta}_2$ denotes the estimation of $\Delta_2$ and $k_2$ is a positive constant.
Using feedforward NNs to estimate $\Delta_i$s, it is obtained that
\begin{equation}\label{Eq:18_1}
  \Delta_i={W^*}_i^T \mu_i (\bar{x}_i) + \varepsilon_i,
\end{equation}
such that $\|\varepsilon_i\|\leq \varepsilon_{M_i}$. To derive the updating rules of the NNs' parameters, one can define a Lyapunov function as follows
\begin{equation}
  V = \frac{1}{2}\left(e_1^2+e_2^2 + \tilde{W}_1^T \Gamma_1^{-1} \tilde{W}_1 + \tilde{W}_2^T \Gamma_2^{-1} \tilde{W}_2\right). \label{Eq:19_0}
\end{equation}
The time derivative of $V$ is obtained as
\begin{align}
\begin{split}
   \dot{V} = & e_1 \dot{e}_1 + e_2 \dot{e}_2 + \tilde{W}_1^T \Gamma_1^{-1} \dot{\tilde{W}}_1 + \tilde{W}_2^T \Gamma_2^{-1} \dot{\tilde{W}}_2 \\
     = & e_1 \left(g_1(\bar{x}_1)e_2 - \tilde{\Delta}_1 - k_1 e_1 \right) + e_2 \left(- g_1(\bar{x}_1) e_1 \right. \\
     & \left. - \tilde{\Delta}_2 - k_2 e_2 \right) + \tilde{W}_1^T \Gamma_1^{-1} \dot{\tilde{W}}_1 + \tilde{W}_2^T \Gamma_2^{-1} \dot{\tilde{W}}_2 \\
     = & - k_1 e_1^2 - k_2 e_2^2 + \tilde{W}_1^T \left( \Gamma_1^{-1} \dot{\tilde{W}}_1 - \mu_1 (x_1) e_1 \right) \\
     & + \tilde{W}_2^T \left( \Gamma_2^{-1} \dot{\tilde{W}}_2 - \mu_2 (\bar{x}_2) e_2 \right) + e_1 \varepsilon_1 + e_2 \varepsilon_2.
\end{split}
\end{align}
Thus, assuming $\dot{\tilde{W}}_i=\dot{\hat{W}}_i$, the updating rules of $W_i$s can be defined as follows
\begin{equation}
  \dot{\hat{W}}_i = \Gamma_i \left( \mu_i (\bar{x}_i) e_i - \sigma_i \hat{W}_i \right), \label{Eq:19}
\end{equation}
where the second term on the right-hand side of the equation corresponds to the $\sigma$-modification. Using the updating rules (\ref{Eq:19}), it is easy to show that $\dot{V}\leq -k V + C$ with $k$ and $C$ denote positive constant. As will be discussed in Section \ref{Sec:Stability}, this can ensure that all signals in the closed-loop system are uniformly ultimately bounded.

Such a control method can similarly be employed in cases where $x_i$s are some vectors rather than scalar functions. A neural backstepping controller for an uncertain Multiple-Input Multiple-Output (MIMO) dynamic model of a helicopter has been introduced in \citet{Ge.2010} to control the attitude of the vehicle considering actuator dynamics, while each step of the design process deals with the control of a four-dimensional state vector.
A neural backstepping controller has been adopted in \citet{Zheng.2020b} to control a planar Vertical Take-off and Landing (VTOL) air vehicle, where a gradient descent training algorithm has been replaced the updating rule (\ref{Eq:19}).
Although it has been claimed that such a training method results in better control performance, there is a need for the exact value of the uncertain term that is estimated by the NN, while in this paper, it has been computed by approximating the time derivatives of the system states and using the dynamic equations of the air vehicle.

It should be noted that although the proposed adaptive backstepping control leads to bounded tracking error in the presence of model uncertainties and external disturbances, it suffers from the \textit{explosion of terms}. More precisely, the control command (\ref{Eq:18}) includes the time derivative of $\dot{x}_{2d}$, which requires the time derivative of $g_1(x_1)$, $f_1(x_1)$, and $\hat{\Delta}_1$. This issue becomes more problematic by increasing the relative degree of the system.

\begin{ex}[label=Ex:HFV1]{Backstepping control of HFV}
Consider the following dynamic equations of attitude subsystem in the longitudinal mode of an HFV that is given in strict feedback form
\begin{align}\label{Eq:HFV_Dyn}
\begin{split}
\dot{x}_1 & = f_1 (x_1) + g_1 x_2 \\
\dot{x}_2 & = x_3 \\
\dot{x}_3 & = f_3 (x) + g_3 u \\
y &= x_1, \\
\end{split}
\end{align}
where $x = \begin{bmatrix}x_1 & x_2 & x_3 \end{bmatrix}^T = \begin{bmatrix}\gamma & \theta & q \end{bmatrix}^T$ with $\gamma$, $\theta$, and $q$ denoting the Flight Path Angle (FPA), pitch angle, and pitch rate, respectively. Also, $u = \delta_e$ denotes the system input. Furthermore, the derivation of equations and the value of the parameters can be found in Appendix \ref{sec:HFV}.

In this example, the control objective is to achieve the desired altitude. To accomplish this, a cascade control loop is employed. The outer loop is responsible for altitude control and generating the desired flight path angle. Meanwhile, the inner loop consists of the attitude controller, which utilizes the desired flight path angle to generate the elevator deflection command. Furthermore, consider that the vehicle velocity is constant during the simulation and use $\Phi = 0.25$ for velocity (propulsion) subsystem. Additionally, while the HFV velocity in this example is assumed to be constant, it can actually be controlled by implementing a simple PID controller. A schematic of proposed control architecture is illustrated in Fig. \ref{fig:HFV_Backstep_schem}.
\begin{figure}[H]
  \begin{center}
    \includegraphics[width=\textwidth]{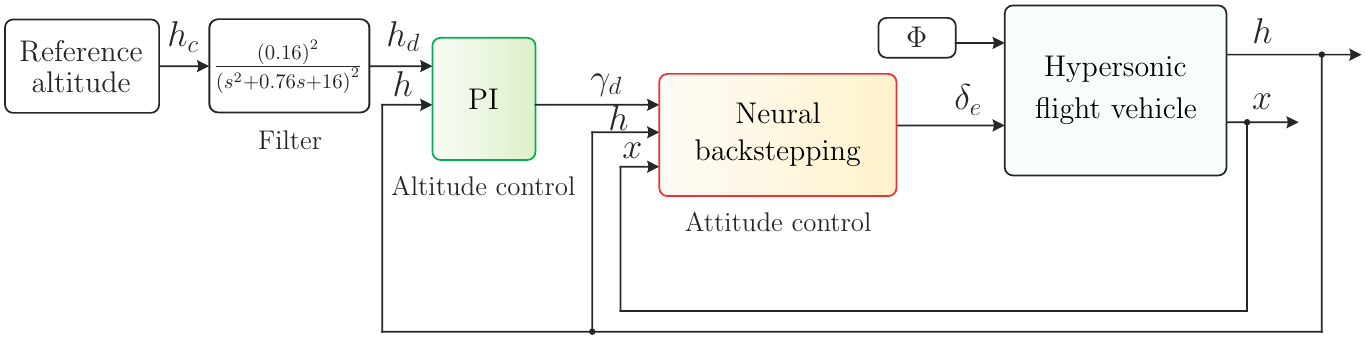}
  \end{center}
  \caption{Control architecture for longitudinal control of HFV using neural backstepping.} 
  \label{fig:HFV_Backstep_schem}
\end{figure}

\begin{enumerate}[(a)]
    \item Design a PI controller for the outer loop that converts the desired altitude $h_d$ into the desired flight path angle $\gamma_d$. During the design, make use of the equation $\dot{h} = V \sin(\gamma)$, where $V$ represents the speed of the vehicle. Additionally, take into account the small flight path angle approximation $(sin(\gamma) \approx \gamma)$.
    \item Design a neural backstepping control for the inner loop, assuming that $\gamma_d$ is available, $f_1$ and $f_3$ are unknown functions of system states and inputs, and $g_1$ and $g_3$ are known. Therefore, it is necessary to employ two neural networks in your design to estimate $f_1$ and $f_3$. These neural networks can have a similar RBF structure as the previous examples.
    \item Evaluate the performance of the closed-loop system's response to the altitude reference square command with period $100$ s and magnitude $1000$ ft. Additionally, in order to achieve a smoother response, utilize the following filter for the desired altitude
    \begin{equation}
        \frac{h_d}{h_c} = \frac{\left (0.16\right)^2}{\left (s^2+0.76s+16\right)^2}.
    \end{equation}
    Taking into account the $h_0 = 86000$ ft, $v_0 = 7850$ ft/s, $\theta_0 = 3.5$ deg and $\gamma_0 = q_0 = 0$ as initial conditions for the simulation.
    Notice that the controller should not use any information about $f_1$ and $f_3$. Also, consider $50$\% loss of effectiveness in the elevator from $t = 80$ s, while adding a bias $2$ deg to the input command from $t = 120$ s.
    Concerning the hidden nodes, consider $15$ hidden neurons and select the center and width of each neuron as a (constant) random number, where $x_{0_{i}} \in [-1, 1]$ and $\sigma_{i} \in [1, 2]$. Also, the input of the second NN can be defined as $X = \begin{bmatrix}x_1 & x_2 & x_3 & u \end{bmatrix}^T$. Notice that the input vector $X$ must be normalized to the range of $[-1, 1]$, as mentioned in Example \ref{Example:quadFEL}. Additionally, report the MAE for closed-loop simulation.
    \item Plot the change of training weights over time and check the convergence of them to constant values.
    \item Report the MAEE and compare the value of the estimated uncertain dynamics ($ \hat{\Delta}_i=\hat{W}_i^T \mu_i (X) $) with the real uncertain term
        \begin{equation}\label{Exc25:Delta}
        \Delta_1 = f_1, \quad \Delta_3 = f_3 + g_3 (u-u_c),
        \end{equation}
        where $u_c$ and $u$ denote the commanded elevator deflection and actual deflection of the elevator.
        \item Evaluate the effect of design parameters such as the number of hidden nodes of the NN and the learning rate ($\Gamma$) for the second NN, on the tracking performance (MAE) as well as estimation performance (MAEE). Furthermore, To simplify the comparison, utilize the Euclidean norm when analyzing the MAE values.
\end{enumerate}
\tcblower
\begin{enumerate}[(a)]
\item Considering the small FPA approximation in the given altitude equation is obtained as follows
\begin{equation} \label{gamma_approx_HFV}
    \gamma \approx \frac{\dot{h}}{V}.
\end{equation}
Accordingly, by defining the altitude tracking error as $e_h = h - h_d$, the PI control law is formulated to generate the desired FPA as follows
\begin{equation} \label{gamma_d_HFV}
    \gamma_d = \frac{\dot{h}_d-k_h e_h-k_i \int e_h dt}{V}.
\end{equation}
Combining Eqs. \ref{gamma_approx_HFV} and \ref{gamma_d_HFV}, assuming that $\gamma \to \gamma_d$ leads to
\begin{equation}  \label{pid_HFV}
-k_h e_h - k_i \int e_h dt + \dot{h}_d = \dot{h}.
\end{equation}
Using the definition of $e_h$, the dynamics of altitude tracking error is expressed as
\begin{equation}
\dot{e}_h + k_h e_h + k_i \int e_h dt = 0,
\end{equation}
or equivalently
\begin{equation}
\ddot{e}_h + k_h \dot{e}_h + k_i e_h = 0.
\end{equation}
If we choose positive values for the proportional gain ($k_h$) and integral gain ($k_i$), and we control the FPA to follow the desired FPA ($\gamma_d$), then Eq. (\ref{pid_HFV}) is valid. This means that the altitude tracking error will converge to zero exponentially since the characteristic equation is Hurwitz.

For optimal tracking performance, the control gains $k_h$ and $k_i$ in the simulation have been determined through trial and error. The values chosen are $k_h = 0.5$ and $k_i = 0.1$.
\item \textbf {Controller design:}
    By Defining the tracking errors as $e_1 = x_1 - x_{1d}$, $e_2 = x_2 - x_{2d}$ and $e_3 = x_3 - x_{3d}$,  the derivative of $e_1$ can be expressed as follows
    \begin{equation}
        \label{Eq:HFV1_e1dot_1}
         \dot{e}_1 = \dot{x}_1 - \dot{x}_{1d} =  f_1 + g_1 x_2 - \dot{x}_{1d} = f_1 + g_1(e_2+x_{2d}) - \dot{x}_{1d}.
     \end{equation}
    Thus, a virtual control can be defined for $x_2$ as
    \begin{equation}
    \label{Eq:HFV1_x2d}
      x_{2d}=g_1^{-1}\left(\dot{x}_{1d} - k_1 e_1 - \hat{\Delta}_1 \right),
    \end{equation}
    where $\hat{\Delta}_1$ represent the estimation of $\Delta_1 = f_1$ and $k_1$ is a positive constant. Therefore, by defining $\tilde{\Delta}_1 = \hat{\Delta}_1 - \Delta_1$ and utilizing Eqs. (\ref{Eq:HFV1_e1dot_1}) and (\ref{Eq:HFV1_x2d}) the $\dot{e}_1$ can be obtained as follows
    \begin{equation}
        \label{Eq:HFV1_e1dot_2}
         \dot{e}_1 =  g_1 e_2 - k_1 e_1 - \tilde{\Delta}_1.
     \end{equation}
    Additionally, the derivative of $e_2$ can be expressed as follows
    \begin{equation}
        \label{Eq:HFV1_e2dot_1}
         \dot{e}_2 = \dot{x}_2 - \dot{x}_{2d} =  x_3 - \dot{x}_{2d} = e_3+x_{3d} - \dot{x}_{2d}.
     \end{equation}
    Thus, a virtual control can be defined for $x_3$ as
    \begin{equation}
    \label{Eq:HFV1_x3d}
      x_{3d}=\left(\dot{x}_{2d} - k_2 e_2 - g_1 e_1 \right),
    \end{equation}
    Notice that $g_1 e_1$ is added to ensure stability when applying the Lyapunov method. Therefore, by utilizing Eqs. (\ref{Eq:HFV1_e2dot_1}) and (\ref{Eq:HFV1_x3d}) the $\dot{e}_2$ can be obtained as follows
    \begin{equation}
        \label{Eq:HFV1_e2dot_2}
         \dot{e}_2 =  e_3 - k_2 e_2 - g_1 e_1.
     \end{equation}
     Additionally, the derivative of $e_3$ can be expressed as follows
    \begin{equation}
        \label{Eq:HFV1_e3dot_1}
         \dot{e}_3 = \dot{x}_3 - \dot{x}_{3d} =  f_3 + g_3 u - \dot{x}_{3d}.
     \end{equation}
    Finally, the control command can be defined as
    \begin{equation}
    \label{Eq:HFV1_u}
      u = g_3^{-1}\left(\dot{x}_{3d} - e_2 - k_3 e_3 - \hat{\Delta}_3 \right),
    \end{equation}
    where $\hat{\Delta}_3$ denotes the estimation of $\Delta_3 = f_3 + g_3 (u-u_c)$ and $k_3$ is a positive constant. Notice that $e_2$ is added to ensure stability when applying the Lyapunov method.\\
    Therefore, by defining $\tilde{\Delta}_3 = \hat{\Delta}_3 -\Delta_3$ and utilizing Eqs. (\ref{Eq:HFV1_e3dot_1}) and (\ref{Eq:HFV1_u}) the $\dot{e}_3$ can be obtained as follows
    \begin{equation}
        \label{Eq:HFV1_e3dot_2}
         \dot{e}_3 =  -e_2 - k_3 e_3 - \tilde{\Delta}_3.
     \end{equation}
    Furthermore, using RBF NNs to estimate $\Delta_i$s as $ \hat{\Delta}_i=\hat{W}_i^T \mu_i (X)$, it is obtained that
    \begin{equation}\label{HFV1_NNDelta}
      \Delta_i={W^*}_i^T \mu_i (X) + \varepsilon_i.
    \end{equation}
    \textbf {Stability analysis:} In order to derive the updating rules of the NNs' parameters, one can define a Lyapunov function as follows
    \begin{equation}
      V = \frac{1}{2}\left(e_1^2+e_2^2 +e_3^2 + \tilde{W}_1^T \Gamma_1^{-1} \tilde{W}_1 + \tilde{W}_3^T \Gamma_3^{-1} \tilde{W}_3\right). \label{Eq:HFV1_Lyap}
    \end{equation}
    The time derivative of $V$ can be obtained according to Eqs. (\ref{Eq:HFV1_e1dot_2}), (\ref{Eq:HFV1_e2dot_2}) and (\ref{Eq:HFV1_e3dot_2}) as follows
    \begin{align}
    \begin{split}
       \dot{V} = & e_1 \dot{e}_1 + e_2 \dot{e}_2 + e_3 \dot{e}_3 + \tilde{W}_1^T \Gamma_1^{-1} \dot{\tilde{W}}_1 + \tilde{W}_3^T \Gamma_3^{-1} \dot{\tilde{W}}_3 \\
         = & e_1 \left(g_1 e_2 - k_1 e_1 - \tilde{\Delta}_1 \right) + e_2 \left(e_3 - k_2 e_2 - g_1 e_1 \right) + \\ 
         & e_3 \left(-e_2 - k_3 e_3 - \tilde{\Delta}_2 \right) + \tilde{W}_1^T \Gamma_1^{-1} \dot{\tilde{W}}_1 + \tilde{W}_3^T \Gamma_3^{-1} \dot{\tilde{W}}_3 \\
         = & - k_1 e_1^2 - k_2 e_2^2 - k_3 e_3^2 + \tilde{W}_1^T \left( \Gamma_1^{-1} \dot{\tilde{W}}_1 - \mu_1 e_1 \right) \\
         & + \tilde{W}_3^T \left( \Gamma_3^{-1} \dot{\tilde{W}}_3 - \mu_3 e_3 \right) + e_1 \varepsilon_1 + e_3 \varepsilon_3.
    \end{split}
    \end{align}
    Thus, assuming $\dot{\tilde{W}}_i=\dot{\hat{W}}_i$, the updating rules of $W_i$s can be defined as follows
    \begin{gather} 
     \label{Eq:HFV1_NNUR1}
         \dot{\hat{W}}_1 = \Gamma_1 \left( \mu_1 e_1 - \sigma_1 \hat{W}_1 \right), \\
     \label{Eq:HFV1_NNUR3}
         \dot{\hat{W}}_3 = \Gamma_3 \left( \mu_3 e_3 - \sigma_3 \hat{W}_3 \right),
    \end{gather}
    where the second term on the right-hand side of the equation corresponds to the $\sigma$-modification. Using the updating rules (\ref{Eq:HFV1_NNUR1}) and (\ref{Eq:HFV1_NNUR3}), it is easy to show that $\dot{V}\leq -k V + C$ with $k$ and $C$ denote positive constant. As will be discussed in Section \ref{Sec:Stability}, this can ensure that all signals in the closed-loop system are uniformly ultimately bounded, thereby resulting in a bounded tracking error.
    \item \textbf {Simulation results:}
            The controller and NNs parameters used in the simulation can be found in Table \ref{tab:Ex:HFV1_Param}.
        \begin{table}[H]
            \caption{Controller and Neural Network Parameters for Neural Backstepping Control of HFV}
            \label{tab:Ex:HFV1_Param}
            \centering
            \begin{tabular}{c c}
            \hline\noalign{\smallskip}
            Parameter & Value \\
            \noalign{\smallskip}\hline\noalign{\smallskip}
            $k_1$ & 25\\
            $k_2$ & 20\\
            $k_3$ & 20\\
            $\Gamma_1$ & 10\\
            $\Gamma_3$ & 10\\
            $\sigma_1$ & 0.0001 \\
            $\sigma_3$ & 0.0001 \\
            \noalign{\smallskip}\hline\noalign{\smallskip}
            \end{tabular}
        \end{table}
        The time history of altitude is depicted in Fig. \ref{fig:Ex_2.9_h}. It is observed that the altitude of HFV closely follows its desired trajectory in the presence of uncertainties and actuator fault. 
        \begin{figure}[H]
            \begin{center}
            \includegraphics[width=0.9\linewidth]{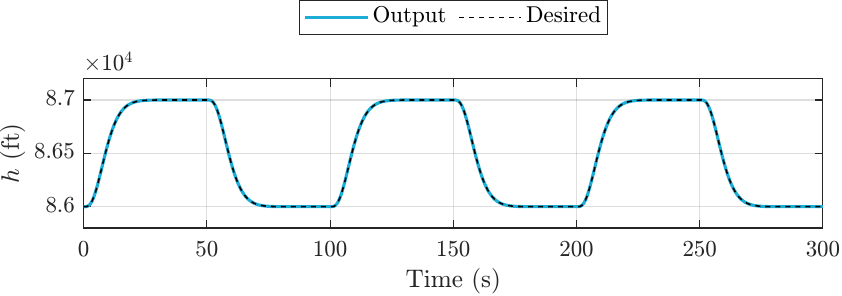}
            \end{center}
            \caption{Time history of altitude in neural backstepping control of HFV.} 
            \label{fig:Ex_2.9_h}
        \end{figure}
        Furthermore, the time history of HFV states is depicted in Fig.  \ref{fig:Ex_2.9_x}. The trajectory of the FPA demonstrates a consistently smooth pattern throughout the simulation. Similarly, the pitch angle is tracked smoothly, with only minor oscillations observed during the occurrence of faults at 80 and 120 seconds. Furthermore, the pitch rate suffers from high-frequency fluctuations during the occurrence of faults. These fluctuations can be attributed to abrupt changes in the aircraft's dynamics resulting from the faults. However, following the fluctuations, the pitch rate gradually stabilizes and follows a relatively smooth trajectory.
        \begin{figure}[H]
            \begin{center}
            \includegraphics[width=0.9\textwidth]{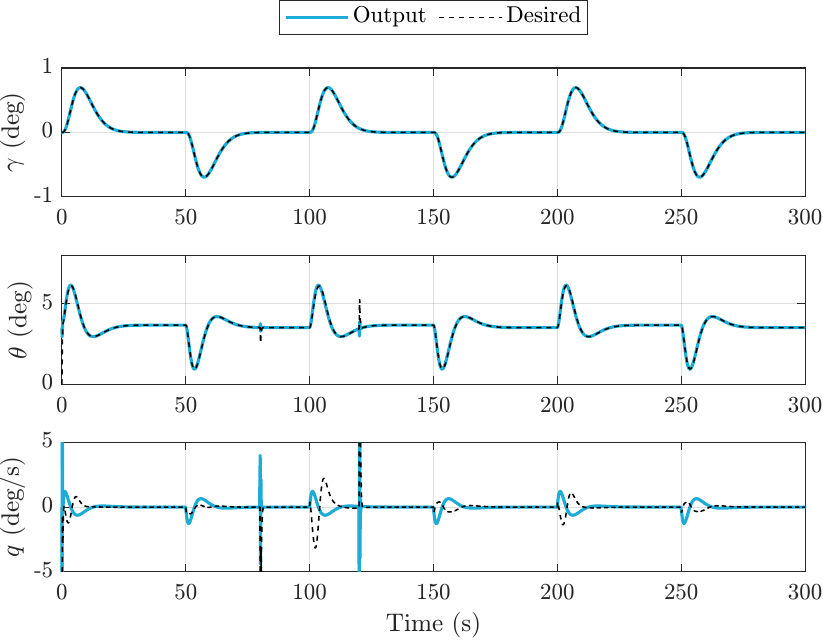}
            \end{center}
            \caption{Time history of states in neural backstepping control of HFV.} 
            \label{fig:Ex_2.9_x}
        \end{figure}
        Additionally, the corresponding control input (elevator deflection) is shown in Fig. \ref{fig:Ex_2.9_u}. The occurrence of the fault results in abrupt changes in the elevator command.
        \begin{figure}[H]
            \begin{center}
            \includegraphics[width=0.9\textwidth]{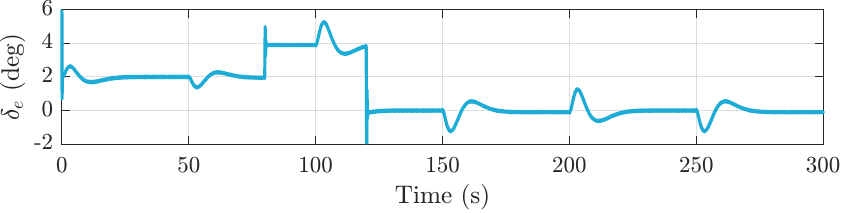}
            \end{center}
            \caption{Time history of control input in neural backstepping control of HFV.} 
            \label{fig:Ex_2.9_u}
        \end{figure}
        Furthermore, the MAE of tracking for this simulation is given in table \ref{tab:EX_2.9_MAE}.
        \begin{table}[H]
            \caption{Mean absolute error of angular rates tracking in Neural Backstepping control of HFV.}
            \label{tab:EX_2.9_MAE}
            \centering
            \begin{tabular}{ccc}
            \hline\noalign{\smallskip}
            MAE ($\gamma$) & MAE ($\theta$) & MAE ($q$)  \\
            \noalign{\smallskip}\hline\noalign{\smallskip}
            $1.1\times 10^{-6}$ & $2.87\times 10^{-4}$  & $5.52\times 10^{-3}$ \\
            \noalign{\smallskip}\hline\noalign{\smallskip}
            \end{tabular}
        \end{table}
    \item The evolution of training weights over time is shown in Fig. \ref{fig:Ex_2.9_W}. It is noticeable that the weights of the neural networks gradually settle at a fixed value after experiencing abrupt weight changes during altitude command switches. Furthermore, sudden weight changes can be observed when a fault occurs.
        \begin{figure}[H]
            \begin{center}
\includegraphics[width=0.9\textwidth]{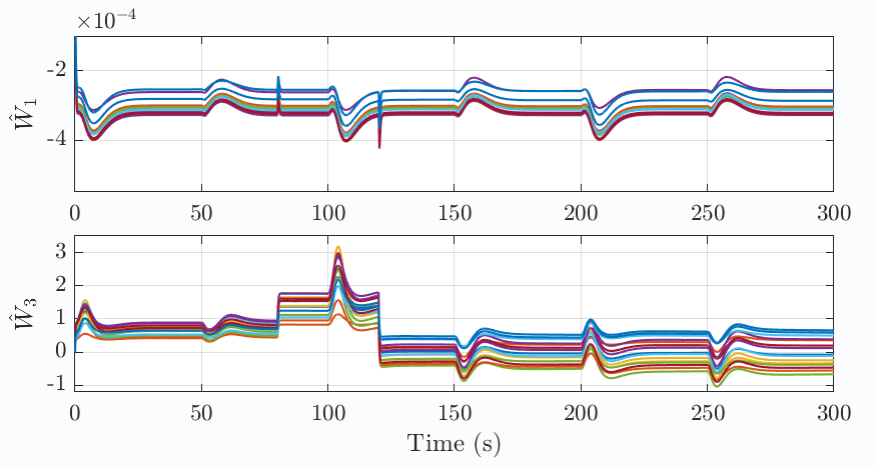}
            \end{center}
            \caption{Evolution of training weights in neural backstepping control of HFV.}
            \label{fig:Ex_2.9_W}
        \end{figure}
    \item Regarding uncertainty estimation, the MAEE obtained in this simulation is $2\times 10^{-5}$ and $1.177\times 10^{-1}$ for $\Delta_1$ and $\Delta_3$, respectively. Furthermore, the comparison of real uncertainty and the NN estimation is shown in Fig \ref{fig:Ex_2.9_Delta}. It can be observed that the neural networks effectively capture the uncertainty and fault effects, demonstrating acceptable estimation performance.
        \begin{figure}[H]
            \begin{center}
            \includegraphics[width=0.9\textwidth]{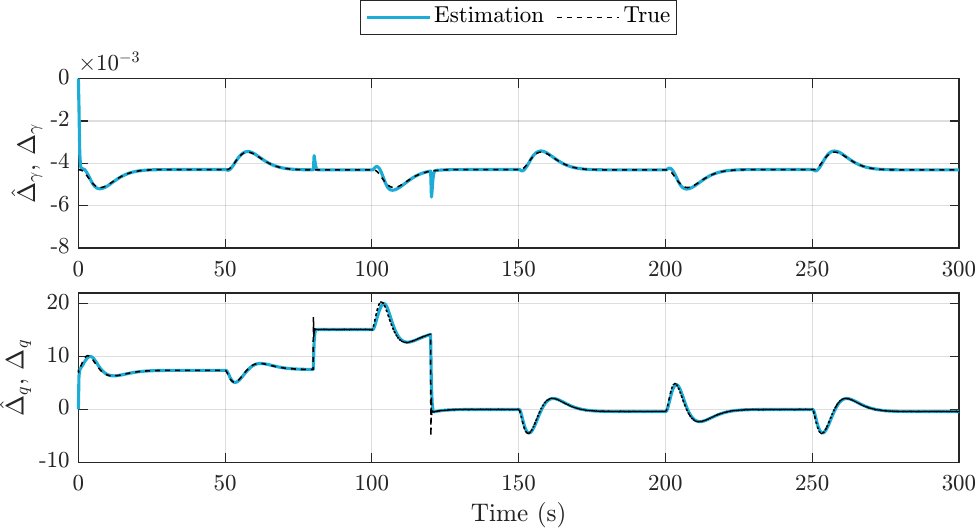}
            \end{center}
            \caption{Uncertainty estimation in neural backstepping control of HFV.}
            \label{fig:Ex_2.9_Delta}
        \end{figure}
    \item \textbf{Sensitivity analysis:} A parameter study is conducted to investigate the effects of various parameters on control performance, specifically focusing on the accuracy of uncertainty estimation (measured by MAEE) and the tracking error (measured by MAE). As shown in Fig. \ref{fig:Ex_2.9_SA}, lower learning rates and a lower number of neurons have a clear impact on increasing both the MAEE and MAE (Euclidean norm of MAE). Additionally, a higher number of neurons combined with a higher learning rate can lead to unintended effects on MAE and MAEE. Furthermore, it is worth noting that the high values of MAE and MAEE in cases with high learning rates and a large number of neurons are due to numerical issues and oscillations that arise when faults occur. However, this effect can be mitigated by using smaller sample times, which helps to alleviate the impact of numerical issues and oscillations.
        \begin{figure}[H]
            \begin{center}
            \includegraphics[width=0.9\textwidth]{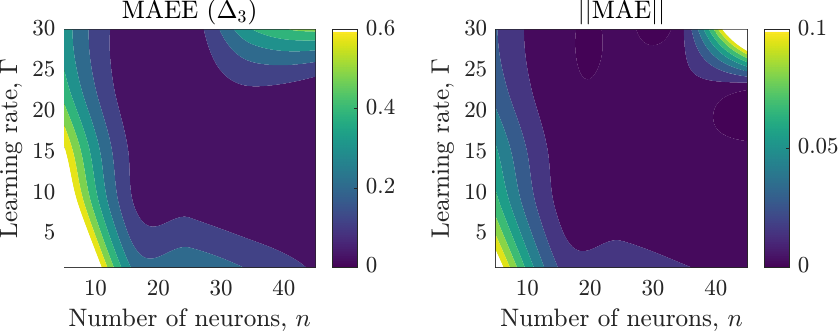}
            \end{center}
            \caption{Sensitivity analysis of MAEE and MAE for the number of neurons and learning rate of second NN in  Neural Backstepping control of HFV.}
            \label{fig:Ex_2.9_SA}
        \end{figure}
\end{enumerate}
\end{ex}

\subsection{Dynamic surface control}
To solve the above-mentioned issue, Dynamic Surface Control (DSC) has been introduced in \citet{Swaroop.2000} in which the virtual control is passed through a first-order filter. More precisely, if $x_{2c}$ is defined by (\ref{Eq:17}), then the desired value of $x_2$ is obtained as
\begin{equation}\label{Eq:19_1}
  \tau \dot{x}_{2d}+x_{2d}=x_{2c}, \quad x_{2d}(0) = x_{2c}(0),
\end{equation}
where $\tau$ represents the filter time constant. Subsequently, the filtering error is also incorporated in the Lyapunov function of the system to be compensated by the designed control commands.
Using such a technique, the problem of the explosion of terms in the traditional backstepping control can be effectively avoided, though at the cost of reducing the \emph{global} stability of the system obtained using the backstepping control to the \emph{semi-global} stability in the case of DSC \citep{Swaroop.2000}.

Several NN-based DSC methods have been introduced in the literature for different aerial vehicles \citep{Butt.2013,Zong.2014,Xu.2016d,Fu.2018}.
Such an approach has been proposed in \citet{Butt.2013b} to control the flight path angle and velocity of a flexible HFV, where the employment of the integral of the tracking error in the control law improves the tracking performance.
DSC has been employed in \citet{MouChen.2014} to control the attitude of a Near-Space Vehicle (NSV) in which recurrent wavelet NNs have been utilized at each step and trained using a composite learning method to compensate for external disturbances and model uncertainties.
Also, such a scheme has been adopted in \citet{Zhou.2015} to control the longitudinal dynamics of an air-breathing HFV considering model uncertainties and external disturbances compensated by fully tuned RBFNNs.
In addition, DSC has been applied to the longitudinal mode of an HFV in \citet{Xu.2015}. In comparison with conventional DSC, which results in a semi-globally uniformly ultimately bounded stability, global tracking has been achieved through aggregating the neural function approximation and a robust term (using a switching function), which brings the system states into the neural approximation domain from outside. The robust term has been designed to estimate the upper bound of uncertain terms in a similar way as will be discussed in Chapter 3. However, the determination of the \emph{active region} of NNs (which is required in designing the switching functions \citep{Xu.2021b,Xu.2021}) is not trivial.

\subsection{Command filtered backstepping} \label{Sec:ComFilBackstepping}
To simplify the stability analysis of DSC, a command filtered backstepping has been proposed in \citet{Farrell.2009} for a nonlinear system without uncertainty. The introduced method attempts to eliminate the filter effects using a set of compensating signals. This idea has been extended to nonlinear systems with parametric uncertainties in \citet{Dong.2012}. To clarify the main idea, consider again the aforementioned control problem. Assuming that the virtual control signals $x_{2c}$ and $x_{2d}$ are defined, respectively, by (\ref{Eq:17}) and (\ref{Eq:19_1}), and by defining the auxiliary variable $\xi_1$ as
\begin{equation}
  \dot{\xi}_1=-k_1 \xi_1 + g_1(\bar{x}_1)\left(x_{2d}-x_{2c}\right), \quad \xi_1(0)=0,
\end{equation}
a compensated tracking error can be defined as $\epsilon_1=y-y_{d}-\xi_1$. Thus, we have
\begin{align}
\begin{split}
  \dot{\epsilon}_1 & = \dot{y}-\dot{y}_{d}-\dot{\xi}_1 \\
                   & =  f_1(\bar{x}_1)+g_1(\bar{x}_1)x_{2}+\Delta_1 - \dot{y}_d + k_1 \xi_1 - g_1(\bar{x}_1)\left(x_{2d}-x_{2c}\right) \\
                   & = - k_1 \epsilon_1 + g_1(\bar{x}_1) e_2 - \tilde{\Delta}_1.
\end{split}
\end{align}
Accordingly, the control command can be defined as
\begin{equation}\label{Eq_u_CFB}
    u = g_2^{-1}(\bar{x}_2)\left(\dot{x}_{2d} - g_1(\bar{x}_1) \epsilon_1 - k_2 e_2 - f_2(\bar{x}_2) - \hat{\Delta}_2 \right),
\end{equation}
which leads to
\begin{align}
\begin{split}
  \dot{e}_2= \dot{x}_2-\dot{x}_{2d}= - g_1(\bar{x}_1)\epsilon_1 - k_2 e_2 - \tilde{\Delta}_2.
\end{split}
\end{align}
Thus, using the following updating rules
\begin{gather}
  \dot{\hat{W}}_1 = \Gamma_1 \left( \mu_1 (\bar{x}_1) \epsilon_1 - \sigma_1 \hat{W}_1 \right), \\
  \dot{\hat{W}}_2 = \Gamma_2 \left( \mu_2 (\bar{x}_2) e_2 - \sigma_2 \hat{W}_2 \right),
\end{gather}
and defining a Lyapunov function as
\begin{equation}
  V = \frac{1}{2}\left(\epsilon_1^2+e_2^2 + \tilde{W}_1^T \Gamma_1^{-1} \tilde{W}_1 + \tilde{W}_2^T \Gamma_2^{-1} \tilde{W}_2\right),
\end{equation}
it can be shown that $\dot{V}\leq -k V + C$, where $k$ and $C$ are positive constants. This results in bounded $\epsilon_1$ and $e_2$.
As discussed in \citet{Dong.2012,Xu.2016c}, assuming that $g_1(x_1)$ is bounded, it can be simply proved that $\xi_1$ is also bounded, thereby resulting in a bounded tracking error.
A command filtered backstepping control has been designed in \citet{Xu.2016c} for the longitudinal dynamics of an HFV considering input constraints and additive actuator faults. The control gain functions ($g_i$) have also been considered unknown, where it has been assumed that model uncertainties, as well as the control gain functions, can be written into a parametric form with partially unknown parameters. Considering the neural network-based representation, the above assumption means that the residual terms $\varepsilon_i$ in (\ref{Eq:18_1}) are equal to zero, while in the case of complex air vehicles with nonparametric uncertainties, such an assumption becomes infeasible.
Similarly, a command filtered backstepping control has been adopted in \citet{Sonneveldt.2007,Sonneveldt.2009} to control the trajectory of an F-16 fighter aircraft model with parametric uncertainties, where second-order filters have been used to impose both the magnitude and rate limits on the system states (see Chapter 3).
Another analogous formulation has been presented in \citet{Xu.2014b} in which the time derivative of ${\xi}_i$ consists of ${\xi}_{i+1}$, where $i$ represents the step of the backstepping control design process. Using this formulation, the control command $u$ can be written in terms of $e_1$ rather than $\epsilon_1$.

To this end, the control command of Eq. (\ref{Eq_u_CFB}) can be rewritten as
\begin{equation}\label{Eq_u_CFBDong}
    u = g_2^{-1}(\bar{x}_2)\left(\dot{x}_{2d} - g_1(\bar{x}_1) e_1 - k_2 e_2 - f_2(\bar{x}_2) - \hat{\Delta}_2 \right).
\end{equation}

Consider the following set of auxiliary variables
\begin{align} 
  \dot{\xi}_1&=-k_1 \xi_1 + g_1(\bar{x}_1)\left(x_{2d}-x_{2c}\right) + g_1(\bar{x}_1) \xi_2,& \xi_1(0)&=0, \\
  \dot{\xi}_2&=-k_2 \xi_2 - g_1(\bar{x}_1)\xi_1,& \xi_2(0)&=0.
\end{align}

Similar to before, another compensated tracking error can be expressed as $\epsilon_2=e_2-\xi_2$. Hence, using Eq. (\ref{Eq_u_CFBDong}) we get
\begin{align}
\begin{split}
  \dot{\epsilon}_2 & = \dot{x}_2-\dot{x}_{2d}-\dot{\xi}_2 \\
                   & =  f_2(\bar{x}_2)+g_2(\bar{x}_2)u+\Delta_2 - \dot{x}_{2d} + k_2 \xi_2 + g_1(\bar{x}_1)\xi_1 \\
                   & = - k_2 \epsilon_2 + g_1(\bar{x}_1) \epsilon_1 - \tilde{\Delta}_2.
\end{split}
\end{align}

Therefore, by employing the subsequent update rules
\begin{gather}
  \dot{\hat{W}}_1 = \Gamma_1 \left( \mu_1 (\bar{x}_1) \epsilon_1 - \sigma_1 \hat{W}_1 \right), \\
  \dot{\hat{W}}_2 = \Gamma_2 \left( \mu_2 (\bar{x}_2) \epsilon_2 - \sigma_2 \hat{W}_2 \right),
\end{gather}
and establishing a Lyapunov function in the form of
\begin{equation}
  V = \frac{1}{2}\left(\epsilon_1^2+\epsilon_2^2 + \tilde{W}_1^T \Gamma_1^{-1} \tilde{W}_1 + \tilde{W}_2^T \Gamma_2^{-1} \tilde{W}_2\right),
\end{equation}
it can be proved that $\dot{V}$ remains below the threshold of $-kV+C$, with both $k$ and $C$ being positive constants. As a consequence, $\epsilon_1$ and $\epsilon_2$ remain constrained within bounds.

\begin{ex}{Command filtered backstepping control of HFV}
\begin{enumerate}[(a)]
    \item Design a command-filtered backstepping control for attitude control of the longitudinal mode of HFV introduced in Example \ref{Ex:HFV1}. 
    \item In order to investigate the effect of the learning rate in the command-filtered method, compare the tracking performance as well as estimation performance for closed-loop simulation of the following scenarios:
    \begin{itemize}
        \item  basic neural backstepping,
        \item command-filtered backstepping utilizing $\Gamma_3 = 10$,
        \item command-filtered backstepping utilizing $\Gamma_3 = 20$.
    \end{itemize}
    Additionally, report and compare MAE and MAEE values for each scenario.
    \item To investigate the robustness of command-filtered backstepping compared to basic backstepping against measurement noise, compare the tracking performance and estimation performance for the closed-loop simulations of both methods, with the following settings:
    \begin{itemize}
        \item Add white noise to the altitude and attitude states with a noise power (height of Power Spectral Density (PSD)) of $10^{-4}$ and $10^{-12}$, respectively.
        \item Set the sample time of the white noise block to 0.1.
        \item Use $\Gamma_3 = 10$.
    \end{itemize}
\end{enumerate}
\tcblower
\begin{enumerate}[(a)]
\item \textbf{Controller design:}
We designed the controller using two different formulations. For brevity, we will only present the results of the primary method. The alternative method uses the actual tracking error in its virtual and actual control command. Expressing the control law in terms of the actual tracking error $e$, rather than the compensated tracking error $\epsilon$, is more intuitive for the designer - the control input vanishes when the errors converge to zero. However, the simulation results for the two methods are very similar, so we will not present them both.
\begin{itemize}
     \item \emph{Primary method}:\\
To design a command-filtered backstepping controller for the attitude subsystem of the longitudinal mode of the HFV described by Eq. (\ref{Eq:HFV_Dyn}), the desired values of $x_2$ and $x_3$ are obtained as follows
\begin{gather}
\label{Eq:HFV2_x2d}
  \tau \dot{x}_{2d}+x_{2d}=x_{2c}, \quad x_{2d}(0) = x_{2c}(0),\\
\label{Eq:HFV2_x3d}
  \tau \dot{x}_{3d}+x_{3d}=x_{3c}, \quad x_{3d}(0) = x_{3c}(0),
\end{gather}
where $\tau$ represents the filter time constant. The compensated tracking errors can be defined as $\epsilon_1=x_1-x_{1d}-\xi_1$ and $\epsilon_2=x_2-x_{2d}-\xi_2$, where the auxiliary variables $\xi_1$ and $\xi_2$ are defined as follows
\begin{align} 
  \dot{\xi}_1&=-k_1 \xi_1 + g_1\left(x_{2d}-x_{2c}\right) + g_1 \xi_2,& \xi_1(0)&=0, \\
  \dot{\xi}_2&=-k_2 \xi_2 + x_{3d}-x_{3c},& \xi_2(0)&=0.
\end{align}

Thus, by defining $x_{2c}$ and $x_{3c}$ as
\begin{align}
    \label{Eq:HFV2_x2c}
      x_{2c}&=g_1^{-1}\left(\dot{x}_{1d} - k_1 e_1 - \hat{\Delta}_1 \right),\\
    \label{Eq:HFV2_x3c}
      x_{3c}&=\dot{x}_{2d} - k_2 e_2 - g_1 \epsilon_1.
\end{align}

Then we have
\begin{align} \label{Eq:HFV2_epsilon1dot}
\begin{split}
  \dot{\epsilon}_1 & = \dot{x}_1-\dot{x}_{1d}-\dot{\xi}_1 \\
                   & =  f_1+g_1 x_{2} - \dot{x}_{1d} + k_1 \xi_1 - g_1\left(x_{2d}-x_{2c}\right) -g_1 \xi_2 \\
                   & =  g_1 \epsilon_2 - k_1 \epsilon_1 -\tilde{\Delta}_1.
\end{split}
\end{align}

Similarly, the derivative of $\epsilon_2$ is obtained as follows
\begin{align}\label{Eq:HFV2_epsilon2dot}
\begin{split}
  \dot{\epsilon}_2 & = \dot{x}_2-\dot{x}_{2d}-\dot{\xi}_2 \\
                   & =  x_{3} - \dot{x}_{2d} + k_2 \xi_2 - \left(x_{3d}-x_{3c}\right) \\
                   & =  e_3 - k_2 \epsilon_2 - g_1 \epsilon_1.
\end{split}
\end{align}

Accordingly, the control command can be defined as
\begin{equation}
     u = g_3^{-1}\left(\dot{x}_{3d} - \epsilon_2 - k_3 e_3 - \hat{\Delta}_3 \right),
\end{equation}
which leads to
\begin{align} \label{Eq:HFV2_e3dot}
\begin{split}
  \dot{e}_3= \dot{x}_3-\dot{x}_{3d}= -\epsilon_2 - k_3 e_3 - \tilde{\Delta}_3.
\end{split}
\end{align}

\textbf{Stability analysis}: In order to derive the updating rules of the NNs' parameters, one can define a Lyapunov function as follows:
    \begin{equation}
      V = \frac{1}{2}\left(\epsilon_1^2+\epsilon_2^2 +e_3^2 + \tilde{W}_1^T \Gamma_1^{-1} \tilde{W}_1 + \tilde{W}_3^T \Gamma_3^{-1} \tilde{W}_3\right). \label{Eq:HFV2_Lyap}
    \end{equation}
    The time derivative of $V$ can be obtained according to Eqs. (\ref{Eq:HFV2_epsilon1dot}), (\ref{Eq:HFV2_epsilon2dot}) and (\ref{Eq:HFV2_e3dot}) as follows
    \begin{align}
    \begin{split}
       \dot{V} & = \epsilon_1 \dot{\epsilon}_1 + \epsilon_2 \dot{\epsilon}_2 + e_3 \dot{e}_3 + \tilde{W}_1^T \Gamma_1^{-1} \dot{\tilde{W}}_1 + \tilde{W}_3^T \Gamma_3^{-1} \dot{\tilde{W}}_3 \\
         & = \epsilon_1 \left(g_1 \epsilon_2 - k_1 \epsilon_1 -\tilde{\Delta}_1 \right) + \epsilon_2 \left(e_3 - k_2 \epsilon_2 - g_1 \epsilon_1 \right) \\ 
         & \quad + e_3 \left(-\epsilon_2 - k_3 e_3 - \tilde{\Delta}_3 \right) + \tilde{W}_1^T \Gamma_1^{-1} \dot{\tilde{W}}_1 + \tilde{W}_3^T \Gamma_3^{-1} \dot{\tilde{W}}_3 \\
         & = - k_1 \epsilon_1^2 - k_2 \epsilon_2^2 - k_3 e_3^2 + \tilde{W}_1^T \left( \Gamma_1^{-1} \dot{\tilde{W}}_1 - \mu_1 \epsilon_1 \right) \\
         & \quad + \tilde{W}_3^T \left( \Gamma_3^{-1} \dot{\tilde{W}}_3 - \mu_3 e_3 \right) + \epsilon_1 \varepsilon_1 + e_3 \varepsilon_3.
    \end{split}
    \end{align}
    Thus, assuming $\dot{\tilde{W}}_i=\dot{\hat{W}}_i$, the updating rules of $W_i$s can be defined as follows
     \begin{gather} \label{Eq:HFV2_NNUR1}
      \dot{\hat{W}}_1 = \Gamma_1 \left( \mu_1 \epsilon_1 - \sigma_1 \hat{W}_1 \right), \\
      \label{Eq:HFV2_NNUR2}
      \dot{\hat{W}}_3 = \Gamma_3 \left( \mu_3 e_3 - \sigma_3 \hat{W}_3 \right),
    \end{gather}
    where the second term on the right-hand side of the equation corresponds to the $\sigma$-modification. Using the updating rules (\ref{Eq:HFV2_NNUR1}) and (\ref{Eq:HFV2_NNUR2}), it is easy to show that $\dot{V}\leq -k V + C$ with $k$ and $C$ denoting positive constants. This results in bounded $\epsilon_1$, $\epsilon_2$, and $e_3$. Assuming that $g_1$ is bounded, it can be simply proved that $\xi_1$ is also bounded, thereby resulting in a bounded tracking error.\\
    
     \item \emph{Alternative method (\citeauthor{Dong.2012} approach)}:\\
The compensated tracking errors can be defined as $\epsilon_i=x_i-x_{id}-\xi_i$, and $i=1,2,3$, where the auxiliary variables $\xi_i$ are defined as follows
\begin{align} 
  \dot{\xi}_1&=-k_1 \xi_1 + g_1\left(x_{2d}-x_{2c}\right) + g_1 \xi_2,& \xi_1(0)&=0, \\
  \dot{\xi}_2&=-k_2 \xi_2 + (x_{3d}-x_{3c}) -g_1\xi_1 + \xi_3,& \xi_2(0)&=0, \\
  \dot{\xi}_3&=-k_3 \xi_3 - \xi_2,& \xi_3(0)&=0.
\end{align}

Then we have
\begin{align} \label{Eq:HFV2_epsilon1dotDong}
\begin{split}
  \dot{\epsilon}_1 & = \dot{x}_1-\dot{x}_{1d}-\dot{\xi}_1 \\
                   & =  f_1+g_1 x_{2} - \dot{x}_{1d} + k_1 \xi_1 - g_1\left(x_{2d}-x_{2c}\right) -g_1 \xi_2 \\
                   & =  g_1 \epsilon_2 - k_1 \epsilon_1 -\tilde{\Delta}_1.
\end{split}
\end{align}

Similarly, the derivative of $\epsilon_2$ is obtained as follows
\begin{align}\label{Eq:HFV2_epsilon2dotDong}
\begin{split}
  \dot{\epsilon}_2 & = \dot{x}_2-\dot{x}_{2d}-\dot{\xi}_2 \\
                   & =  x_{3} - \dot{x}_{2d} + k_2 \xi_2 - \left(x_{3d}-x_{3c}\right) +g_1\xi_1 - \xi_3\\
                   & = e_3 - k_2 \epsilon_2 - g_1 (e_1-\xi_1) - \xi_3 \\
                   & =  \epsilon_3 - k_2 \epsilon_2 - g_1 \epsilon_1,
\end{split}
\end{align}
where, $x_{3d}$ was substituted from Eq. (\ref{Eq:HFV1_x3d}).
Accordingly, the control command can be defined as
\begin{equation}
     u = g_3^{-1}\left(\dot{x}_{3d} - e_2 - k_3 e_3 - \hat{\Delta}_3 \right),
\end{equation}
which is using $e_2$ instead of $\epsilon_2$.
Substituting it into Eq. (\ref{Eq:HFV1_e3dot_1}) gives
\begin{align}\label{Eq:HFV2_epsilon3dotDong}
\begin{split}
  \dot{\epsilon}_3 &= \dot{x}_3-\dot{x}_{3d} -\dot{\xi}_3 \\
  &= -\epsilon_2 - k_3 e_3 - \tilde{\Delta}_3 +k_3\xi_3 + \xi_2\\
  &= -\epsilon_2 - k_3 \epsilon_3 - \tilde{\Delta}_3.
\end{split}
\end{align}

\textbf{Stability analysis:} A Lyapunov function can be considered as follows:
\begin{equation}
  V = \frac{1}{2}\left(\epsilon_1^2+\epsilon_2^2 +\epsilon_3^2 + \tilde{W}_1^T \Gamma_1^{-1} \tilde{W}_1 + \tilde{W}_3^T \Gamma_3^{-1} \tilde{W}_3\right). \label{Eq:HFV2_Lyap_1}
\end{equation}
From Eqs. (\ref{Eq:HFV2_epsilon1dotDong}), (\ref{Eq:HFV2_epsilon2dotDong}) and (\ref{Eq:HFV2_epsilon3dotDong}), $\dot{V}$ can be obtained as
\begin{align}
\begin{split}
   \dot{V} & = \epsilon_1 \dot{\epsilon}_1 + \epsilon_2 \dot{\epsilon}_2 + \epsilon_3 \dot{\epsilon}_3 + \tilde{W}_1^T \Gamma_1^{-1} \dot{\tilde{W}}_1 + \tilde{W}_3^T \Gamma_3^{-1} \dot{\tilde{W}}_3 \\
	 & = \epsilon_1 \left(g_1 \epsilon_2 - k_1 \epsilon_1 -\tilde{\Delta}_1 \right) + \epsilon_2 \left(\epsilon_3 - k_2 \epsilon_2 - g_1 \epsilon_1 \right) \\ 
	 & \quad + \epsilon_3 \left(-\epsilon_2 - k_3 \epsilon_3 - \tilde{\Delta}_3 \right) + \tilde{W}_1^T \Gamma_1^{-1} \dot{\tilde{W}}_1 + \tilde{W}_3^T \Gamma_3^{-1} \dot{\tilde{W}}_3 \\
	 & = - k_1 \epsilon_1^2 - k_2 \epsilon_2^2 - k_3 \epsilon_3^2 + \tilde{W}_1^T \left( \Gamma_1^{-1} \dot{\tilde{W}}_1 - \mu_1 \epsilon_1 \right) \\
	 & \quad + \tilde{W}_3^T \left( \Gamma_3^{-1} \dot{\tilde{W}}_3 - \mu_3 \epsilon_3 \right) + \epsilon_1 \varepsilon_1 + e_3 \varepsilon_3.
\end{split}
\end{align}
Hence, we have
 \begin{gather} \label{Eq:HFV2_NNUR1Dong}
  \dot{\hat{W}}_1 = \Gamma_1 \left( \mu_1 \epsilon_1 - \sigma_1 \hat{W}_1 \right), \\
  \label{Eq:HFV2_NNUR2Dong}
  \dot{\hat{W}}_3 = \Gamma_3 \left( \mu_3 \epsilon_3 - \sigma_3 \hat{W}_3 \right),
\end{gather}
where in this formulation, both of the NNs are using the compensated tracking errors in their update rule. 
\end{itemize}
    \item
    \textbf{Learning rate comparison:} 
    The controller and NNs parameters used in the simulation can be found in Table \ref{tab:Ex:HFV1_Param}. Additionally, the time constant of the filter is set to $\tau = 0.01$.\\
    The time history of altitude is depicted in Fig. \ref{fig:Ex_2.10_h}. It is observed that the altitude of HFV closely follows the reference altitude in all scenarios. 
        \begin{figure}[H]
            \begin{center}
            \includegraphics[width=0.9\textwidth]{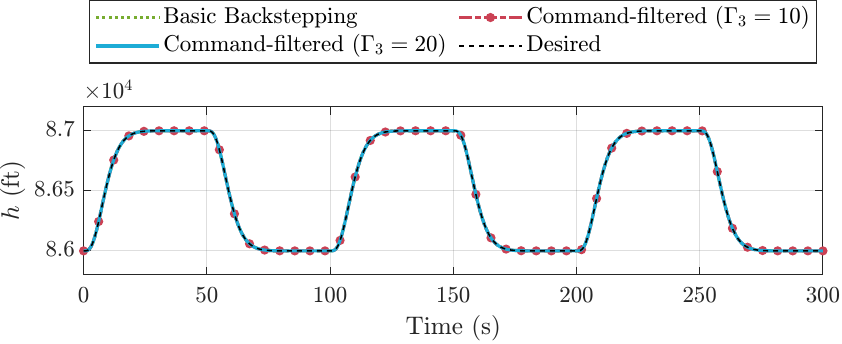}
            \end{center}
            \caption{Time history of altitude in command-filtered backstepping control of HFV.} 
            \label{fig:Ex_2.10_h}
        \end{figure}
    Furthermore, the time history of attitude dynamics states is depicted in Fig. \ref{fig:Ex_2.10_x}.
    It can be observed that when a fault occurs, the basic backstepping method attenuates the oscillations faster than the command-filtered method. 
    This is because the command filter introduces a first-order filter that causes a lag in the system's response. As a result, when significant changes like a fault occur, the command-filtered method requires more time to effectively mitigate and suppress the oscillations due to the inherent lag in its response. This lag is the price to be paid when using the command-filtered method to diminish the explosion of terms effect. Furthermore, when higher learning rates are utilized, the oscillations have a larger amplitude and take more time to attenuate.
        \begin{figure}[H]
            \begin{center}
            \includegraphics[width=0.9\textwidth]{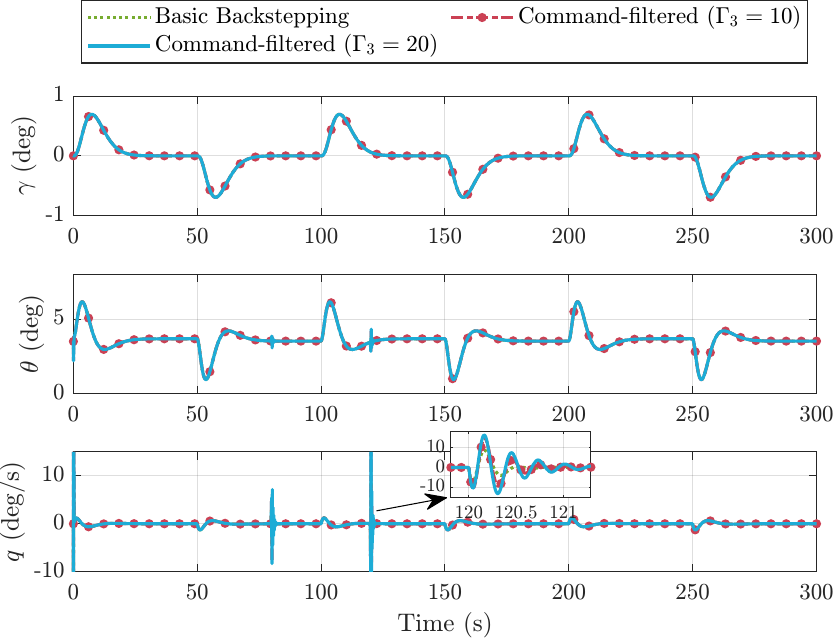}
            \end{center}
            \caption{Time history of states in command-filtered backstepping control of HFV.} 
            \label{fig:Ex_2.10_x}
        \end{figure}
        Additionally, the corresponding control input (elevator deflection) is shown in Fig. \ref{fig:Ex_2.10_u}.
        \begin{figure}[H]
            \begin{center}
            \includegraphics[width=0.9\textwidth]{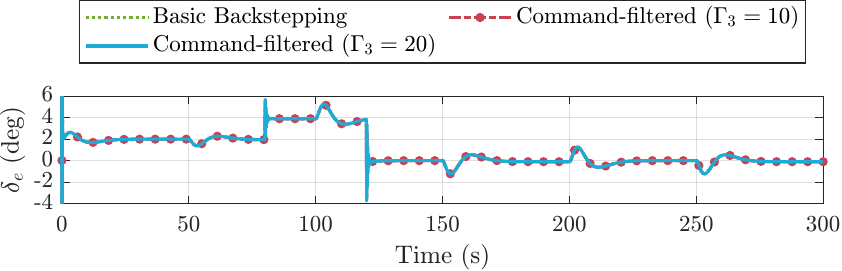}
            \end{center}
            \caption{Time history of control input in command-filtered backstepping control of HFV.} 
            \label{fig:Ex_2.10_u}
        \end{figure}
        Furthermore, the MAE and MAEE values for this simulation are given in Table \ref{tab:EX_2.10_MAE}. It can be inferred that the tracking and estimation errors in basic and command-filtered backstepping with $\Gamma_3 = 10$ are similar. However, when higher learning rates are employed, both the tracking and estimation errors decrease, with a notable reduction in the MAEE.
        \begin{table}[H]
            \caption{Comparison of MAE and MAEE in command-filtered backstepping control of HFV.}
            \label{tab:EX_2.10_MAE}
            \centering
            \resizebox{\textwidth}{!}{
                \begin{tabular}{c c c c c c}
                \hline\noalign{\smallskip}
                Control Method & MAE ($\gamma$) & MAE ($\theta$) & MAE ($q$) & MAEE ($\gamma$)  & MAEE ($q$) \\
                \noalign{\smallskip}\hline\noalign{\smallskip}
                Basic backstepping & $1.1\times 10^{-6}$ & $2.87\times 10^{-4}$  & $5.52\times 10^{-3}$ & $2.3\times 10^{-5}$ & $1.1766\times 10^{-1}$ \\
                Command-filtered ($\Gamma_3 = 10$) & $1\times 10^{-6}$ & $2.65\times 10^{-4}$  & $5.52\times 10^{-3}$ & $2.4\times 10^{-5}$ & $1.1044\times 10^{-1}$\\
                Command-filtered ($\Gamma_3 = 20$) & $1\times 10^{-6}$ & $1.45\times 10^{-4}$  & $3.26\times 10^{-3}$ & $1.6\times 10^{-5}$ & $6.913\times 10^{-2}$\\
                \noalign{\smallskip}\hline\noalign{\smallskip}
                \end{tabular}
            }
        \end{table}
        Furthermore, the time history of the uncertainty estimation error is shown in Fig. \ref{fig:Ex_2.10_Delta}.
        Looking at the first 2 seconds of the response, it can be observed that the command-filtered backstepping with $\Gamma_3$ = 20 has the best estimation performance, followed by basic backstepping, and then the command-filtering with $\Gamma_3$ = 10.
        Moreover, when a fault occurs, the analysis of the estimation performance yields similar results to those of the states. In other words, the basic backstepping scenario demonstrates the previously mentioned oscillation suppression performance compared to the other methods.
        \begin{figure}[H]
            \begin{center}
            \includegraphics[width=0.9\textwidth]{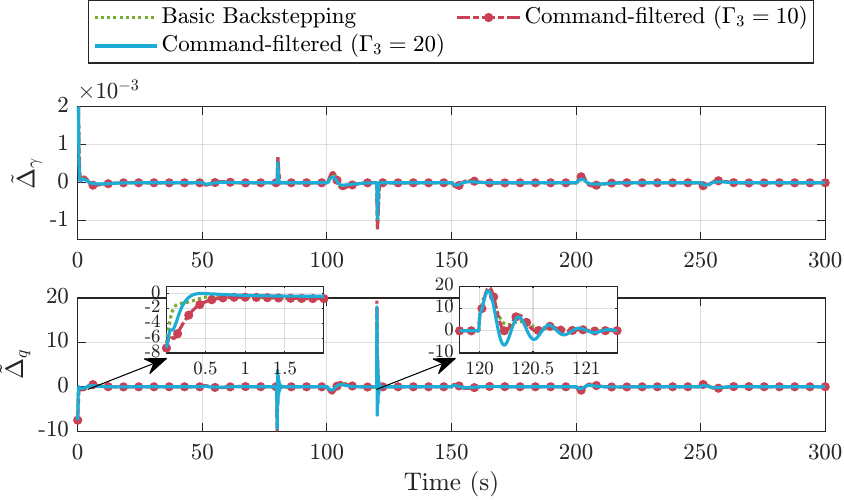}
            \end{center}
            \caption{Uncertainty estimation error in command-filtered backstepping control of HFV.}
            \label{fig:Ex_2.10_Delta}
        \end{figure}
        \item
        \textbf{Noise robustness assessment:} 
        The time history of altitude is depicted in Fig. \ref{fig:Ex_2.10_hn}. Both methods demonstrate commendable altitude tracking performance despite the presence of measurement noise.
        \begin{figure}[H]
            \begin{center}
            \includegraphics[width=0.9\textwidth]{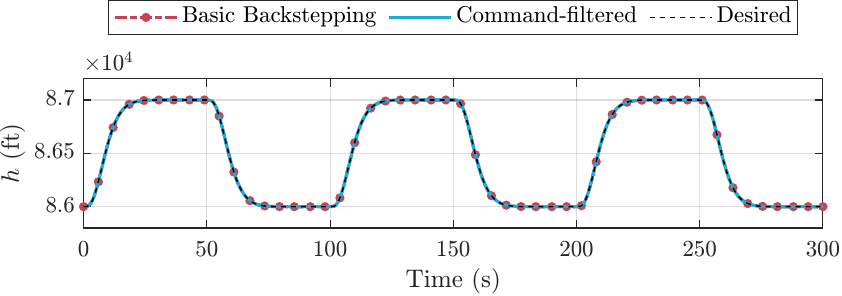}
            \end{center}
            \caption{Time history of altitude in command-filtered backstepping control of HFV in the presence of noise.} 
            \label{fig:Ex_2.10_hn}
        \end{figure}
     Furthermore, the time history of attitude dynamics states is depicted in Fig. \ref{fig:Ex_2.10_xn}. It can be observed that the command-filtered method exhibits superior robustness to noise due to its filtering capability when compared to basic backstepping. As a result of this capability, the command-filtered method becomes more well-suited for practical applications.
        \begin{figure}[H]
            \begin{center}
            \includegraphics[width=0.9\textwidth]{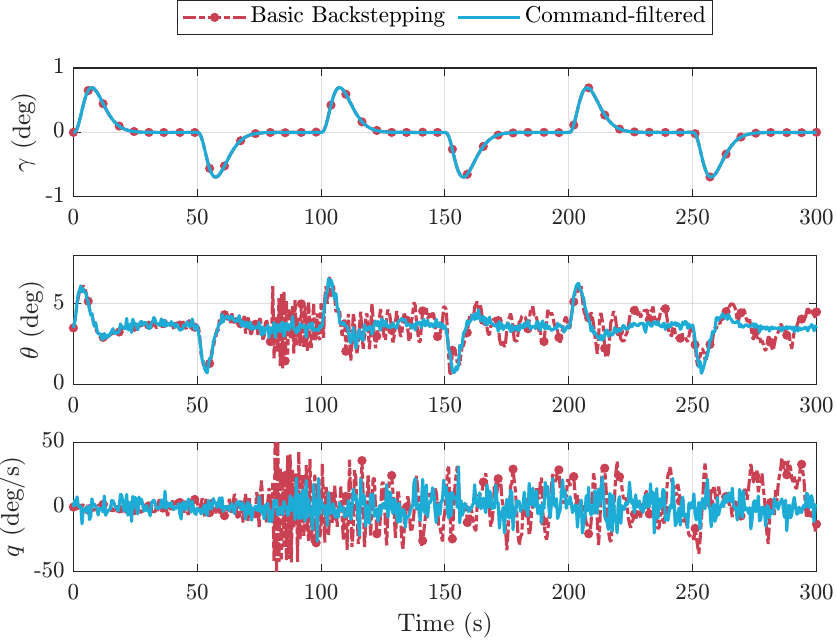}
            \end{center}
            \caption{Time history of states in Command-filtered Backstepping control of HFV in the presence of noise.} 
            \label{fig:Ex_2.10_xn}
        \end{figure}
        Additionally, Fig. \ref{fig:Ex_2.10_un} illustrates the corresponding control input (elevator deflection), revealing that the basic backstepping is more sensitive to noise than the command-filtered method. Moreover, the control command in basic backstepping demonstrates a more pronounced amplification of noise.
        \begin{figure}[H]
            \begin{center}
            \includegraphics[width=0.9\textwidth]{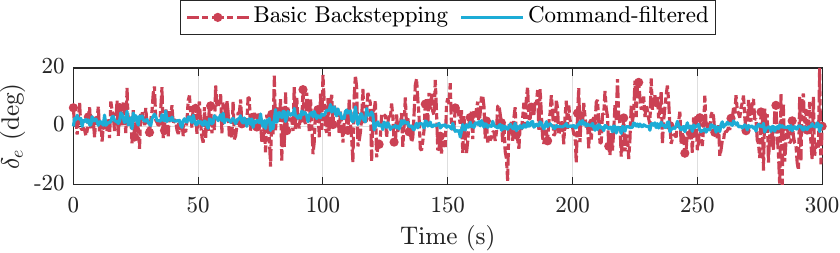}
            \end{center}
            \caption{Time history of control input in command-filtered backstepping control of HFV in the presence of noise.} 
            \label{fig:Ex_2.10_un}
        \end{figure}
         Furthermore, the MAE and MAEE values for this simulation are given in table \ref{tab:EX_2.10_MAEn}. It can be concluded that in the presence of noise, the command-filtered method exhibits significantly lower MAE and MAEE values compared to basic backstepping, as expected.
        \begin{table}[H]
            \caption{Comparison of MAE and MAEE in command-filtered backstepping control of HFV in the presence of noise.}
            \label{tab:EX_2.10_MAEn}
            \centering
            \begin{tabular}{c c c c c c}
            \hline\noalign{\smallskip}
            Control Method & MAE ($\gamma$) & MAE ($\theta$) & MAE ($q$) & MAEE ($\gamma$)  & MAEE ($q$) \\
            \noalign{\smallskip}\hline\noalign{\smallskip}
            Basic Backstepping & $2.9\times 10^{-5}$ & $1.285\times 10^{-2}$  & $1.0495\times 10^{0}$ & $2.43\times 10^{-4}$ & $1.3694\times 10^{1}$\\
            Command-filtered & $1.5\times 10^{-5}$ & $6.15\times 10^{-3}$ & $1.701\times 10^{-1}$ & $1.23\times 10^{-4}$ & $2.667\times 10^{0}$\\
            \noalign{\smallskip}\hline\noalign{\smallskip}
            \end{tabular}
        \end{table}
        Moreover, the time history of uncertainty estimation error is shown in Fig. \ref{fig:Ex_2.10_Deltan}, revealing the reduced sensitivity to noise exhibited by the command-filtered method, consequently leading to lower estimation error compared to basic backstepping.
        \begin{figure}[H]
            \begin{center}
            \includegraphics[width=0.9\textwidth]{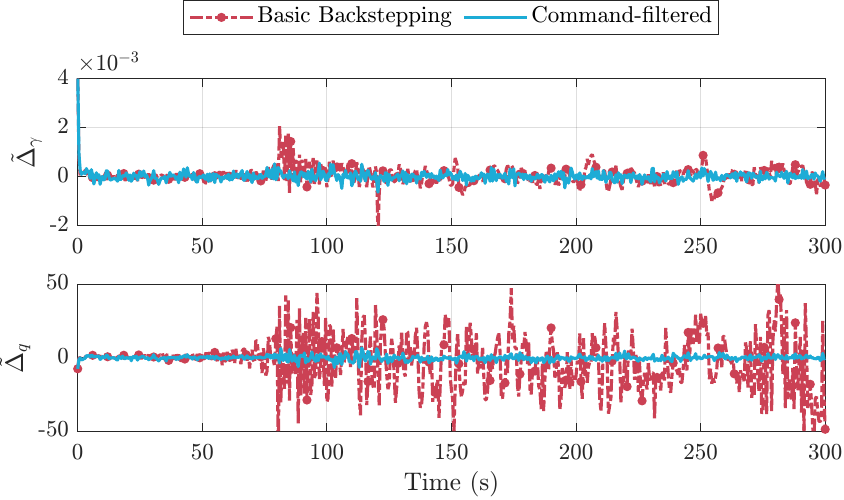}
            \end{center}
            \caption{Uncertainty estimation error in command-filtered backstepping control of HFV in the presence of noise.}
            \label{fig:Ex_2.10_Deltan}
        \end{figure}
    \end{enumerate}
In this example, a command filtered backstepping controller was designed to avoid the explosion of terms problem in traditional backstepping designs.
The simulation results showed that when actuator faults occurred, the basic backstepping controller was able to suppress oscillations faster than the command filtered approach. This is attributed to the lag introduced by the filtering.
Moreover, increasing the learning rate in the command filtered controller was shown to improve both tracking and estimation accuracy. 
Additionally, the command filtered method demonstrated reduced sensitivity to measurement noise compared to basic backstepping. This results from its inherent filtering capability that attenuates the noise propagation.

\end{ex}

\subsection{Backstepping augmented by the First-Order Sliding Mode Differentiators (FOSMD)}
Another improved approach to approximate the time derivative of the virtual control signal $x_{2d}$ is to employ a first-order sliding mode differentiator rather than employing a first-order filter. Using the FOSMD, the differentiation error tends to zero or a compact neighborhood of zero (depending on the signal's characteristics) after a finite-time transient process \citep{Levant.1998}.
Considering a known function $l(t)$, the FOSMD formulation is obtained as follows
\begin{gather}
  \dot{\varsigma}_0=-\varrho_0 \left|\varsigma_0-l(t)\right|^{0.5} \text{sign} \left(\varsigma_0-l(t)\right)+\varsigma_1, \\
  \dot{\varsigma}_1=-\varrho_1 \text{sign} \left(\varsigma_1-\dot{\varsigma_0}\right),
\end{gather}
where $\varsigma_0$ and $\varsigma_1$ represent the states of the differentiator, and $\varrho_0$ and $\varrho_1$ denote design parameters. Therefore, $\dot{\varsigma}_0-\dot{l}(t)$ remains bounded if $\dot{\varsigma}_0(0)-\dot{l}(0)$ and $\varsigma_0(0)-l(0)$ are bounded.

This approach has been adopted in the backstepping control design in \citet{Xu.2018,Xu.2019} to control the longitudinal mode of an HFV.
As shown in \citet{Xu.2018}, using the FOSMD, the stability analysis is more concise compared to the traditional backstepping, DSC, and command filtered design. Besides, a neural backstepping control approach using FOSMD has been proposed in \citet{Wu.2017} for the longitudinal dynamic model of a sweep-back wings morphing aircraft subject to input–output constraints.
It is notable that Higher-Order Sliding Mode Differentiators (HOSMD), which result in superior performance compared to FOSMDs \citep{Levant.2003}, can also be employed in the structure of the neural backstepping scheme \citep{Yu.2020}.

\begin{ex}{Backstepping control of HFV using FOSMD}
\begin{enumerate}[(a)]
\item Design a neural backstepping control utilizing sliding-mode differentiators for attitude control of the longitudinal mode of HFV introduced in Example \ref{Ex:HFV1}.
\item Compare the tracking error and estimation error for closed-loop simulation of both methods as well as MAE and MAEE values.
\end{enumerate}
\tcblower
\begin{enumerate}[(a)]
\item \textbf{Controller design:}
In order to design a neural backstepping using FOSMD for the attitude subsystem of the longitudinal mode of HFV described by Eq. (\ref{Eq:HFV_Dyn}), the estimated derivative of $x_{2d}$ is obtained as follows
\begin{gather}
  \label{HFV3_FOSMD_x2d}
  \dot{\varsigma}_{2_0}=-\varrho_{1} \left|\varsigma_{2_0}-x_{2d}\right|^{0.5} \text{sign} \left(\varsigma_{2_0}-x_{2d}\right)+\varsigma_{2_1}, \\
  \dot{\varsigma}_{2_1}=-\varrho_{2} \text{sign} \left(\varsigma_{2_1}-\dot{\varsigma}_{2_0}\right).
\end{gather}

The differentiator states $\varsigma_{2_0}$ and $\varsigma_{2_1}$ and design parameters $\varrho_{1}$ and $\varrho_{2}$ are used to derive $x_{2d}$ using FOSMD. As a result, $\dot{\varsigma}_{2_0}-\dot{x}_{2d}$ will remain bounded if the initial conditions $\dot{\varsigma}_{2_0}(0)-\dot{x}_{2d}(0)$ and $\varsigma_{2_0}(0)-x_{2d}(0)$ are bounded.
Similarly, the estimated derivative of $x_{3d}$ is expressed as
\begin{gather}
  \label{HFV3_FOSMD_x3d}
  \dot{\varsigma}_{3_0}=-\varrho_{3} \left|\varsigma_{3_0}-x_{3d}\right|^{0.5} \text{sign} \left(\varsigma_{3_0}-x_{3d}\right)+\varsigma_{3_1}, \\
  \dot{\varsigma}_{3_1}=-\varrho_{4} \text{sign} \left(\varsigma_{3_1}-\dot{\varsigma}_{3_0}\right),
\end{gather}
where $\varsigma_{3_0}$ and $\varsigma_{3_1}$ represent the states of the differentiator, and $\varrho_{3}$ and $\varrho_{4}$ denote design parameters for derivation of $x_{3d}$ using FOSMD. Therefore, $\dot{\varsigma}_{3_0}-\dot{x}_{3d}$ remains bounded if $\dot{\varsigma}_{3_0}(0)-\dot{x}_{3d}(0)$ and $\varsigma_{3_0}(0)-x_{3d}(0)$ are bounded.

It is worth noting that the aforementioned derivatives are estimations of the exact derivatives of $x_{2d}$ and $x_{3d}$.  The errors in these derivative estimations are bounded, and the bounded derivative estimation errors are defined as follows
\begin{gather} 
 \label{Eq:HFV3_DEE}
     \dot{x}_{2d} = \dot{\varsigma}_{2_0} + \varepsilon_{d2},\\
     \dot{x}_{3d} = \dot{\varsigma}_{3_0} + \varepsilon_{d3}.
\end{gather}
Subsequently, for backstepping control design, the $x_{2d}$ and $\dot{e}_1$ are the same as Eqs. (\ref{Eq:HFV1_x2d}) and (\ref{Eq:HFV1_e1dot_2}), respectively. However, due to the utilization of FOSMD for estimating $\dot{x}_{2d}$ and $\dot{x}_{3d}$, $x_{3d}$ is obtained as
\begin{equation}
\label{Eq:HFV3_x3d}
  x_{3d}= \dot{\varsigma}_{2_0} - k_2 e_2 - g_1 e_1,
\end{equation}
which is identical to Eq. (\ref{Eq:HFV1_x3d}), except that we replaced $\dot{x}_{2d}$ with $\dot{\varsigma}_{2_0}$.

Thus, the derivative of $e_2$ can be expressed as follows
\begin{align}     \label{Eq:HFV3_e2dot}
\begin{split}
     \dot{e}_2 & = x_3 - \dot{x}_{2d} = e_3+x_{3d} - \dot{x}_{2d} \\
     &= e_3 + \dot{\varsigma}_{2_0} - k_2 e_2 - g_1 e_1  -  \dot{x}_{2d} \\
     &= e_3 - k_2 e_2 - g_1 e_1 - \varepsilon_{d2}.
\end{split}
\end{align}
Additionally, the control command is defined as
\begin{equation}
\label{Eq:HFV3_u}
  u = g_3^{-1}\left(\dot{\varsigma}_{3_0} - e_2 - k_3 e_3 - \hat{\Delta}_3 \right).
\end{equation}
Therefore, $\dot{e}_3$ can be obtained as follows
\begin{align}     \label{Eq:HFV3_e3dot}
\begin{split}
     \dot{e}_3 &= \dot{x}_3 - \dot{x}_{3d} = f_3 + g_3 u - \dot{x}_{3d} \\
     &= -e_2 - k_3 e_3 - \tilde{\Delta}_3 - \varepsilon_{d3}.
\end{split}
\end{align}
Likewise, the NN estimation can be considered as Eq. (\ref{HFV1_NNDelta}).

\textbf {Stability analysis:}
In order to analyze the closed-loop stability and derive the updating rules of the NNs' parameters, one can define a Lyapunov function expressed as
\begin{equation}
  V = \frac{1}{2}\left(e_1^2+e_2^2 +e_3^2 + \tilde{W}_1^T \Gamma_1^{-1} \tilde{W}_1 + \tilde{W}_3^T \Gamma_3^{-1} \tilde{W}_3\right). \label{Eq:HFV3_Lyap}
\end{equation}

The time derivative of $V$ can be obtained according to Eqs. (\ref{Eq:HFV1_e1dot_2}), (\ref{Eq:HFV3_e2dot}) and (\ref{Eq:HFV3_e3dot}) as follows
\begin{align}
\begin{split}
   \dot{V} &= e_1 \dot{e}_1 + e_2 \dot{e}_2 + e_3 \dot{e}_3 + \tilde{W}_1^T \Gamma_1^{-1} \dot{\tilde{W}}_1 + \tilde{W}_3^T \Gamma_3^{-1} \dot{\tilde{W}}_3 \\
     &= e_1 \left(g_1 e_2 - k_1 e_1 - \tilde{\Delta}_1 \right) + e_2 \left(e_3 - k_2 e_2 - g_1 e_1 - \varepsilon_{d2} \right) \\ 
     &\quad+ e_3 \left(-e_2 - k_3 e_3 - \tilde{\Delta}_3 - \varepsilon_{d3} \right) + \tilde{W}_1^T \Gamma_1^{-1} \dot{\tilde{W}}_1 + \tilde{W}_3^T \Gamma_3^{-1} \dot{\tilde{W}}_3 \\
     & = - k_1 e_1^2 - k_2 e_2^2 - k_3 e_3^2 + \tilde{W}_1^T \left( \Gamma_1^{-1} \dot{\tilde{W}}_1 - \mu_1 e_1 \right) \\
     & \quad+ \tilde{W}_3^T \left( \Gamma_3^{-1} \dot{\tilde{W}}_3 - \mu_3 e_3 \right) + e_1 \varepsilon_1 - e_2 \varepsilon_{d2} + e_3 \left(\varepsilon_3 - \varepsilon_{d3} \right).
\end{split}
\end{align}
 Using the updating rules (\ref{Eq:HFV1_NNUR1}) and (\ref{Eq:HFV1_NNUR3}), it is easy to show that $\dot{V}\leq -k V + C$ with $k$ and $C$ denote positive constant. This results in a bounded tracking error.\\
 \item
\textbf{Simulation results:} The controller and NNs parameters used in the simulation are identical to those in Table \ref{tab:Ex:HFV1_Param}. Moreover, the FOSMD parameters used in the simulation are given in Table \ref{tab:Ex:HFV3_Param}.
        \begin{table}[H]
            \caption{Differentiator parameters for neural backstepping control of HFV utilizing FOSMD}
            \label{tab:Ex:HFV3_Param}
            \centering
            \begin{tabular}{c c}
            \hline\noalign{\smallskip}
            Parameter & Value \\
            \noalign{\smallskip}\hline\noalign{\smallskip}
            $\varrho_{1}$ & 4\\
            $\varrho_{2}$ & 0.001\\
            $\varrho_{3}$ & 25\\
            $\varrho_{4}$ & 0.01\\
            \noalign{\smallskip}\hline\noalign{\smallskip}
            \end{tabular}
        \end{table}
        The time history of altitude is depicted in Fig. \ref{fig:Ex_2.11_h}. It can be noted that in the backstepping method with FOSMD, the altitude of HFV closely follows the reference altitude.
        \begin{figure}[H]
            \begin{center}
            \includegraphics[width=0.9\textwidth]{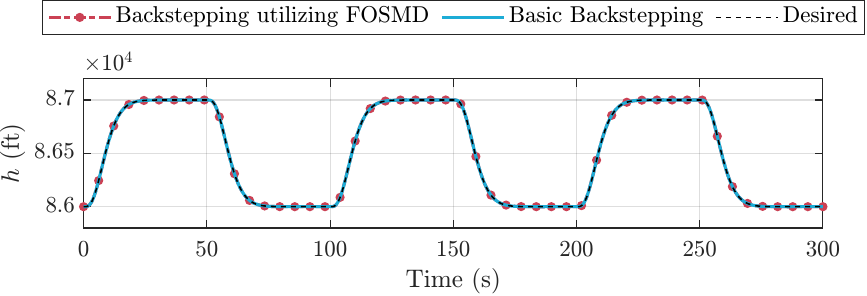}
            \end{center}
            \caption{Time history of altitude in neural backstepping control of HFV utilizing FOSMD.} 
            \label{fig:Ex_2.11_h}
        \end{figure}
         Furthermore, the time history of HFV states is depicted in Fig. \ref{fig:Ex_2.11_x}.
         Each simulation has a different desired state that we could not show in the plot, but it is clear that the states are closely following their desired states.
        However, when analyzing the time history of pitch rate in the backstepping control augmented by FOSMD, it is noticeable that the initial oscillations and oscillations that occur during fault instances have a larger magnitude compared to basic backstepping.
        \begin{figure}[H]
            \begin{center}
            \includegraphics[width=0.9\textwidth]{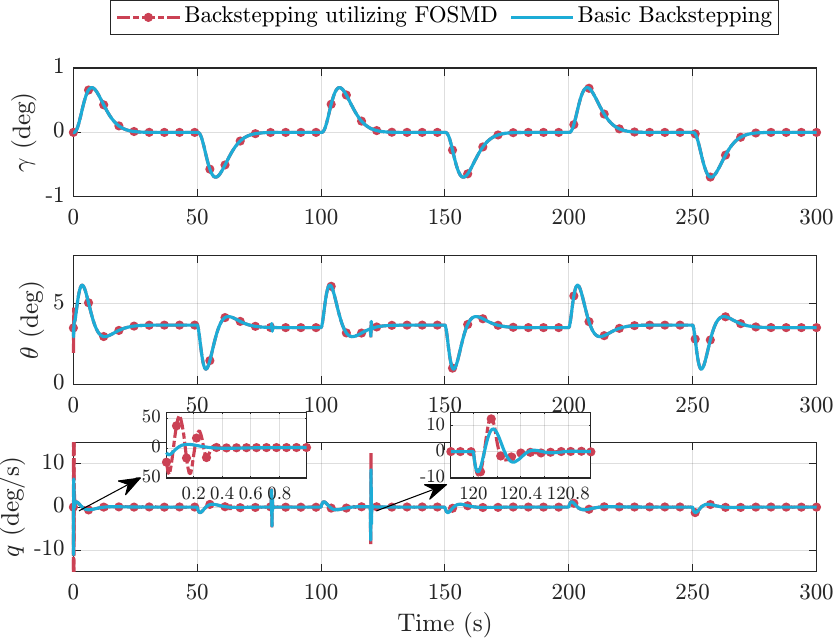}
            \end{center}
            \caption{Time history of states in neural backstepping control of HFV utilizing FOSMD.} 
            \label{fig:Ex_2.11_x}
        \end{figure}
        The time history of elevator deflection, which is the control input for the system, is shown in Fig. \ref{fig:Ex_2.11_u}. It is worth noting that the utilization of FOSMD in the backstepping method results in oscillations in the control input. This phenomenon, known as \emph{chattering} in the sliding mode control framework, is caused by utilizing the signum function in the FOSMD.
        \begin{figure}[H]
            \begin{center}
            \includegraphics[width=0.9\textwidth]{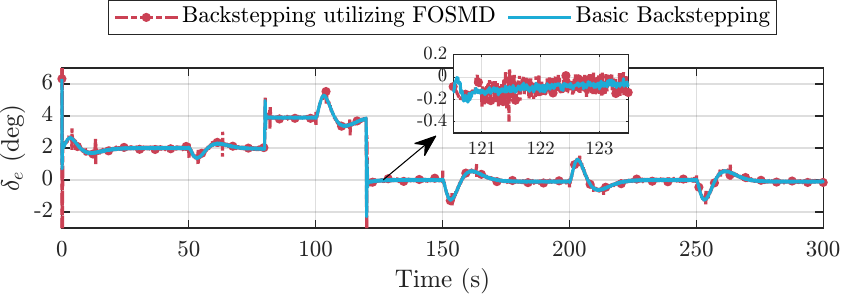}
            \end{center}
            \caption{Time history of control input in neural backstepping control of HFV utilizing FOSMD.} 
            \label{fig:Ex_2.11_u}
        \end{figure}
         Furthermore, the MAE and MAEE values for this simulation are given in table \ref{tab:EX_2.11_MAE}. It can be inferred that the tracking error and estimation error in the backstepping utilizing FOSMD have higher MAE and MAEE values compared to basic backstepping.
         \begin{table}[H]
            \caption{MAE and MAEE in backstepping control of HFV utilizing FOSMD.}
            \label{tab:EX_2.11_MAE}
            \centering
            \resizebox{\textwidth}{!}{
            \begin{tabular}{c c c c c c}
            \hline\noalign{\smallskip}
            Control Method & MAE ($\gamma$) & MAE ($\theta$) & MAE ($q$) & MAEE ($\gamma$)  & MAEE ($q$) \\
            \noalign{\smallskip}\hline\noalign{\smallskip}
            Basic Backstepping & $1\times 10^{-6}$ & $2.87\times 10^{-4}$  & $5.52\times 10^{-3}$ & $2.3\times 10^{-5}$ & $1.1766\times 10^{-1}$\\
            Backstepping with FOSMD & $1.2\times 10^{-6}$ & $6.26\times 10^{-3}$ & $5.89\times 10^{-3}$ & $4.6\times 10^{-5}$ & $1.7772\times 10^{-1}$\\
            \noalign{\smallskip}\hline\noalign{\smallskip}
            \end{tabular}}
        \end{table}
        Moreover, the time history of uncertainty estimation error is shown in Fig. \ref{fig:Ex_2.11_Delta}. During the initial 2 seconds of the simulation, the basic backstepping approach achieves a faster convergence to zero estimation error. Additionally, in the FOSMD method, there is an increased presence of oscillations in the estimation error around zero, which is attributed to the influence of the signum function used in FOSMD and the associated control command.
        \begin{figure}[H]
            \begin{center}
            \includegraphics[width=0.9\textwidth]{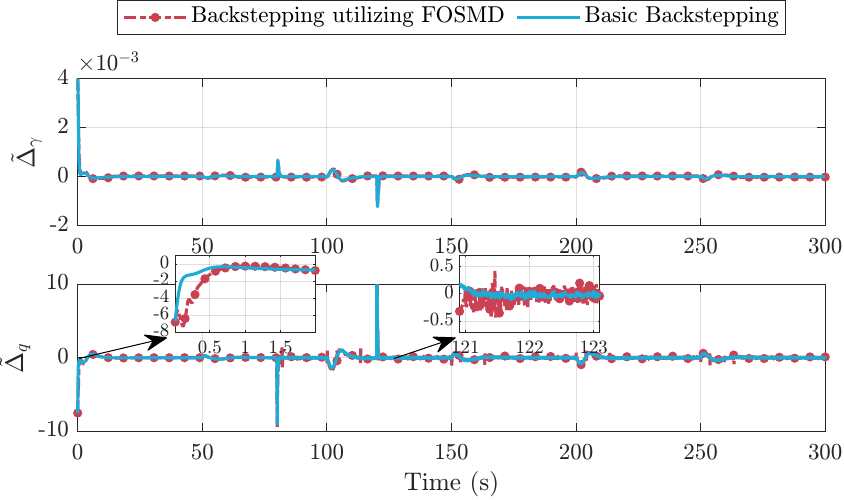}
            \end{center}
            \caption{Uncertainty estimation error in neural backstepping control of HFV utilizing FOSMD.}
            \label{fig:Ex_2.11_Delta}
        \end{figure}
       In general, augmenting FOSMD into backstepping control provides a solution to address the issue of explosion of terms. However, this approach comes at the cost of increased oscillations in the control command, resulting in higher MAE and MAEE values. These increased error metrics indicate a diminished performance in terms of both estimation accuracy and tracking capabilities.
       \end{enumerate}
\end{ex}

\subsection{Direct neural-backstepping control}\label{Sec:DirectNeuralBackstepping}
In addition to the above techniques, there are a variety of direct adaptive backstepping flight control systems in the literature, which can satisfactorily prevent the problem of the explosion of terms.
To be more precise, consider again the nonlinear model (\ref{Eq:16_1})-(\ref{Eq:16_3}) with $n=2$. Defining
\begin{gather}
  x_{2d}^{*}=g_1^{-1}(\bar{x}_1)\left(\dot{y}_d - k_1 e_1 - f_1(\bar{x}_1) - \Delta_1 \right), \\
  u^*= g_2^{-1}(\bar{x}_2)\left(\dot{x}_{2d} - g_1(\bar{x}_1) e_1 - k_2 e_2 - f_2(\bar{x}_2) - \Delta_2 \right), \label{Eq:20_0}
\end{gather}
and using two distinct neural networks to identify them as $x_{2d}^{*}={W^*}_1^T \mu_1(\bar{x}_1)+\varepsilon_1$ and $u^{*}={W^*}_2^T \mu_2(\bar{x}_2)+\varepsilon_2$, we have
\begin{gather}
 x_{2d}={\hat{W}}_1^T \mu_1(\bar{x}_1)=x_{2d}^{*}-\varepsilon_1+{\tilde{W}}_1^T \mu_1(\bar{x}_1), \\
 u={\hat{W}}_2^T \mu_2(\bar{x}_2)=u^*-\varepsilon_2+{\tilde{W}}_2^T \mu_2(\bar{x}_2). \label{Eq:20_01}
\end{gather}

Thus, considering a Lyapunov function candidate as (\ref{Eq:19_0}), we have
\begin{align}
\begin{split}
   \dot{V} &=  e_1 \dot{e}_1 + e_2 \dot{e}_2 + \tilde{W}_1^T \Gamma_1^{-1} \dot{\tilde{W}}_1 + \tilde{W}_2^T \Gamma_2^{-1} \dot{\tilde{W}}_2 \\
     &=  e_1 \left(g_1(\bar{x}_1)\left(e_2-\varepsilon_1+{\tilde{W}}_1^T \mu_1(\bar{x}_1)\right) - k_1 e_1 \right) \\
     &\quad + e_2 \left(- g_1(\bar{x}_1) e_1 - k_2 e_2 + g_2(\bar{x}_2)\left({\tilde{W}}_2^T \mu_2(\bar{x}_2)-\varepsilon_2\right)\right) \\
     & \quad + \tilde{W}_1^T \Gamma_1^{-1} \dot{\tilde{W}}_1 + \tilde{W}_2^T \Gamma_2^{-1} \dot{\tilde{W}}_2 \\
     &= - k_1 e_1 \left(e_1+\frac{g_1(\bar{x}_1)}{k_1}\varepsilon_1 \right) - k_2 e_2 \left(e_2+\frac{g_2(\bar{x}_2)}{k_2}\varepsilon_2 \right) \\ &\quad + \tilde{W}_1^T \left( \Gamma_1^{-1} \dot{\tilde{W}}_1 + \mu_1 (x_1) g_1(\bar{x}_1) e_1 \right) \\
        &\quad + \tilde{W}_2^T \left( \Gamma_2^{-1} \dot{\tilde{W}}_2 + \mu_2 (\bar{x}_2) g_2(\bar{x}_2) e_2 \right).
\end{split}
\end{align}

Accordingly, by introducing the following updating rules
\begin{gather}
  \dot{\hat{W}}_i = - \Gamma_i \left(\mu_i (\bar{x}_i) g_i(\bar{x}_i) e_i + \sigma_i \hat{W}_i\right), \quad i=1,2, \label{Eq:20_1}
\end{gather}
and assuming that $g_i$s are nonzero and \emph{bounded}, again, it can be concluded that $\dot{V}<-kV+C$, which ensures that all signals in the closed-loop system remain bounded (see Section \ref{Sec:Stability}).
As seen, despite the simpler formulation of the direct method compared to previously proposed indirect backstepping schemes, the boundedness of control gain functions ($g_i$s) is necessary to guarantee the closed-loop stability.


It is notable that, using the aforementioned direct neural backstepping scheme, the control \emph{singularity} problem in the control of dynamic systems with unknown $g_i$s (induced by approaching $\hat{g}_i$s to zero) is also avoided \citep{Bu.2015}.
A direct neural backstepping control has been designed in \citet{Bu.2015} to control the longitudinal mode of an air-breathing HFV with unknown $g_i$s, where it is necessary to have $\overline{g}_i \geq g_i>0$ ($\overline{g}_i$ denotes a positive constant).
To this end, concerning the appropriate Lyapunov function candidate, $1/g_i$ is multiplied by $1/2 e_i^2$ to eliminate the requirement for $g_i$ in updating rules, and extra terms in $\dot{V}$ raised by this reformulation have been compensated by defining an appropriate ideal control command ($u^*$).
Besides, a filtered tracking error including the integral of the error has been considered as the error function to remove the steady-state error, while the tracking error corresponding to the second step of the backstepping design was not considered in the first step.

\begin{ex}[label=Ex:HFVdir]{Direct backstepping control of HFV}
\begin{enumerate}[(a)]
\item Design a direct neural backstepping control for attitude control of the longitudinal mode of HFV introduced in Example \ref{Ex:HFV1}. To this end, three neural networks are required to generate the virtual and actual control commands. To prevent numerical problems, the neural networks can be employed to estimate only the unknown terms combined with known components.
\item Evaluate the performance of the closed-loop system.
\item Plot the change of training weights over time.
\item Compare the outputs of NNs with their true values.
\item Plot the change of MAE and MAEE for different control gains and learning rates.
\end{enumerate}
\tcblower
Consider the dynamic equations of attitude subsystem in the longitudinal mode of a HFV, given in Example \ref{Ex:HFV1}.
\begin{enumerate}[(a)]

\item \textbf{Controller design:}
According to Eq. (\ref{Eq:HFV1_e1dot_1}), the ideal value of $x_{2d}$ is defined as:
\begin{align} \label{Eq:Ch2Ex12:x2d_star}
\begin{split}
x_{2d}^{*}&=g_1^{-1}(\bar{x}_1)\left(\dot{x}_{1d} - k_1 e_1 -f_1(\bar{x}_1)\right) \\
&= g_1^{-1}(\bar{x}_1)\left(\dot{x}_{1d} - k_1 e_1 \right) + \Delta_1(\bar{x}_1),
\end{split}
\end{align}
where $k_1$ is a positive constant.
Similarly, for $x_{3d}$ we have
\begin{align} \label{Eq:HFV_direct_x3d_star}
\begin{split}
x_{3d}^*&={\dot{x}}_{2d}-k_2e_2-g_1({\bar{x}}_1)e_1 \\
&= -k_2e_2-g_1({\bar{x}}_1)e_1 + \Delta_2(\bar{x}_2),
\end{split}
\end{align}
where $k_2$ is a positive constant.
Finally, the true value of the control command can be defined as:
\begin{align} \label{Eq:HFV_direct_u_star}
\begin{split}
u^*&= g_3^{-1}(\bar{x}_3)\left(\dot{x}_{3d} -e_2 - k_3 e_3 -f_3(\bar{x}_3) \right) -\delta u\\
&= g_3^{-1}(\bar{x}_3)\left( -e_2 - k_3 e_3 \right) + \Delta_3 \left(\bar{x}_3,u^*\right),
\end{split}
\end{align}
where $k_3$ is a positive constant, $\delta u = u-u_c$ with $u$ and $u_c$ denoting the actual and command elevator input, and the existence of the solution requires a contraction mapping assumption. 
Besides, $\Delta_i$s denote the uncertain terms in the control variables that should be identified. These terms are given by:
\begin{align} \label{Eq:HFV_direct_Delta_i}
\begin{split}
\Delta_1&= -g_1^{-1}(\bar{x}_1)f_1(\bar{x}_1), \\
\Delta_2&= -g_2^{-1}(\bar{x}_2) \dot{x}_{2d}, \\
\Delta_3&= -g_3^{-1}(\bar{x}_3) \left( \dot{x}_{3d}-f_3(\bar{x}_3) \right) - \delta u.
\end{split}
\end{align}

To put it differently, it's worth mentioning that every NN has the responsibility of predicting a specific component of the command rather than the entire expression. This approach is advantageous because it results in a reliable controller that doesn't require significant adjustments, such as $\sigma$-modification.

Three neural networks are used to estimate the uncertain terms in both virtual and actual control commands, i.e., $\Delta_i$s, 
in a \emph{direct} manner. As a result, we have
\begin{gather}
x_{2d}=g_1^{-1}(\bar{x}_1)\left(\dot{x}_{1d} - k_1 e_1 \right)+ {\hat{W}}_1^T \mu_1(\bar{x}_1)=x_{2d}^{*}-\varepsilon_1+{\tilde{W}}_1^T \mu_1(\bar{x}_1), \label{Eq:HFV_direct_x_2}\\
x_{3d}=-k_2e_2-g_1({\bar{x}}_1)e_1 + {\hat{W}}_2^T \mu_2(\bar{x}_2)=x_{3d}^{*}-\varepsilon_2+{\tilde{W}}_2^T \mu_2(\bar{x}_2), \label{Eq:HFV_direct_x_3}\\
\begin{split}
    u_c& =g_3^{-1}(\bar{x}_3)\left( -e_2 - k_3 e_3 \right) +{\hat{W}}_3^T \mu_3(\bar{x}_3,u_c)\\
    &=u^*-\varepsilon_3+{\tilde{W}}_3^T \mu_3(\bar{x}_3,u_c) - \delta u.
\end{split} \label{Eq:HFV_direct_u}
\end{gather}

By utilizing Eqs. (\ref{Eq:Ch2Ex12:x2d_star}) and (\ref{Eq:HFV_direct_x_2}), knowing that $x_{2}=e_2 + x_{2d}$, the derivative of $e_1$ can be written as:
\begin{align}
\begin{split}
  \dot{e}_1 & = f_1(\bar{x}_1)+g_1(\bar{x}_1)x_{2} - \dot{\gamma}_d \\
                   & =  f_1(\bar{x}_1)+g_1(\bar{x}_1)\left( e_2 + x_{2d}^{*}-\varepsilon_1+{\tilde{W}}_1^T \mu_1(\bar{x}_1) \right) - \dot{\gamma}_d  \\
                   & =  -k_1 e_1 + g_1(\bar{x}_1)\left(e_2-\varepsilon_1+{\tilde{W}}_1^T \mu_1(\bar{x}_1)\right). \\
\end{split}
\end{align}

Using Eqs. (\ref{Eq:HFV_direct_x3d_star}) and (\ref{Eq:HFV_direct_x_3}), $\dot{e}_2$ can be obtained as follows:
\begin{align}
\begin{split}
  \dot{e}_2 & = x_{3} - \dot{x}_{2d} , \\
                  & =  e_3 + x_{3d}^{*}-\varepsilon_2+{\tilde{W}}_2^T \mu_2(\bar{x}_2)  - \dot{x}_{2d}, \\
                   & =  -k_2 e_2 + e_3-\varepsilon_2+{\tilde{W}}_2^T \mu_2(\bar{x}_2) - g_1(\bar{x}_1) e_1, \\
\end{split}
\end{align}
By combining Eqs. (\ref{Eq:HFV_direct_u_star}) and (\ref{Eq:HFV_direct_u}), $\dot{e}_3$ can be obtained as follows:
\begin{align}
\begin{split}
  \dot{e}_3 & = f_3(\bar{x}_3)+g_3(\bar{x}_3)u - \dot{x}_{3d} \\
                  & =  f_3(\bar{x}_3)+g_3(\bar{x}_3)u_c + g_3(\bar{x}_3)\delta u - \dot{x}_{3d} \\
                  & =  f_3(\bar{x}_3)+g_3(\bar{x}_3)\left( u^*-\varepsilon_3+{\tilde{W}}_3^T \mu_3(\bar{x}_3,u_c) - \delta u \right) + g_3(\bar{x}_3)\delta u - \dot{x}_{3d} \\
                   & =  -k_3 e_3 + g_3(\bar{x}_3)\left(-\varepsilon_3+{\tilde{W}}_3^T \mu_3(\bar{x}_3,u_c)\right) -  e_2. \\
\end{split}
\end{align}
\textbf{Stability analysis:} To obtain the updating rules for the parameters of the NNs, we establish a Lyapunov function in the following manner:
\begin{equation}
  V = \frac{1}{2}\sum_{i=1}^{3}\left(e_i^2 + \tilde{W}_i^T \Gamma_i^{-1} \tilde{W}_i \right).
\end{equation}
The time derivative of $V$ can be obtained as:
\begin{align}
\begin{split}
   \dot{V} &=  e_1 \dot{e}_1 + e_2 \dot{e}_2 +e_3 \dot{e}_3+ \tilde{W}_1^T \Gamma_1^{-1} \dot{\tilde{W}}_1 +  \tilde{W}_2^T \Gamma_2^{-1} \dot{\tilde{W}}_2 + \tilde{W}_3^T \Gamma_3^{-1} \dot{\tilde{W}}_3 \\
     &=  e_1 \left(g_1(\bar{x}_1)\left(e_2-\varepsilon_1+{\tilde{W}}_1^T \mu_1(\bar{x}_1)\right) - k_1 e_1 \right) \\
     &\quad + e_2 \left(e_3-k_2e_2-g_1({\bar{x}}_1)e_1-\varepsilon_2+{\tilde{W}}_2^T \mu_2(\bar{x}_2)\right) \\
     & \quad+ e_3 \left(g_3(\bar{x}_3)\left(-\varepsilon_3+{\tilde{W}}_3^T \mu_3(\bar{x}_3,u_c)\right) - k_3 e_3 -e_2 \right) \\
     & \quad+ \tilde{W}_1^T \Gamma_1^{-1} \dot{\tilde{W}}_1 + \tilde{W}_3^T \Gamma_3^{-1} \dot{\tilde{W}}_3,
\end{split}
\end{align}
which simplifies to
\begin{align}
\begin{split}
   \dot{V} &= - k_1 e_1 \left(e_1+\frac{g_1(\bar{x}_1)}{k_1}\varepsilon_1 \right) - k_2 e_2 \left(e_2+\frac{\varepsilon_2}{k_2} \right)- k_3 e_3 \left(e_3+\frac{g_3(\bar{x}_3)}{k_3}\varepsilon_3 \right) \\ & \quad+ \tilde{W}_1^T \left( \Gamma_1^{-1} \dot{\tilde{W}}_1 + \mu_1 (x_1) g_1(\bar{x}_1) e_1 \right) \\
        &\quad  + \tilde{W}_2^T \left( \Gamma_2^{-1} \dot{\tilde{W}}_2 + \mu_2 (x_1) e_2 \right) 
        \\ &
        \quad+ \tilde{W}_3^T \left( \Gamma_3^{-1} \dot{\tilde{W}}_3 + \mu_3 (\bar{x}_3,u_c) g_3(\bar{x}_3) e_3 \right).
\end{split}
\end{align}
Accordingly, by introducing the following updating rules
\begin{gather}
  \dot{\hat{W}}_1 = - \Gamma_1 \left(\mu_1 (\bar{x}_1) g_1(\bar{x}_1) e_1 + \sigma_1 \hat{W}_1\right),\\
  \dot{\hat{W}}_2 = - \Gamma_2 \left(\mu_2 (\bar{x}_2) e_2 + \sigma_2 \hat{W}_2\right), \\
  \dot{\hat{W}}_3 = - \Gamma_3 \left(\mu_3 (\bar{x}_3,u_c) g_3(\bar{x}_3) e_3 + \sigma_3 \hat{W}_3\right),
\end{gather}
and assuming nonzero and \emph{bounded} $g_1$ and $g_3$, it can be concluded that $\dot{V}<-kV+C$, which ensures that all signals in the closed-loop system are guaranteed to be bounded.
Finally, it should be noted that the desired flight path angle is obtained using Eq. (\ref{gamma_d_HFV}).

\item \textbf{Simulation results:}
To evaluate the validity of the proposed direct backstepping control, a numerical simulation is performed with 15 neurons and parameters listed in Table \ref{tab:Ex:Ch2Ex12_Param}.
\begin{table}[H]
    \caption{Parameters used in direct backstepping control of HFV.}
    \label{tab:Ex:Ch2Ex12_Param}
    \centering
    \begin{tabular}{cc}
    \hline\noalign{\smallskip}
    Parameter & Value \\
    \noalign{\smallskip}\hline\noalign{\smallskip}
    $[k_1, k_2, k_3]$ & $[20, 7, 20]$ \\ 
    $[\Gamma_1, \Gamma_2, \Gamma_3]$ & $[200, 1, 100]$ \\ 
    $[\sigma_1, \sigma_2, \sigma_3]$ & $[0, 0.01, 0]$ \\ 
    \noalign{\smallskip}\hline\noalign{\smallskip}
    \end{tabular}
\end{table}

The time history of altitude is plotted next to the command altitude in Fig. \ref{fig:ex2_12:alt}, showing an acceptable tracking performance.
\begin{figure}[H]
  \begin{center}
    \includegraphics[width=0.9\textwidth]{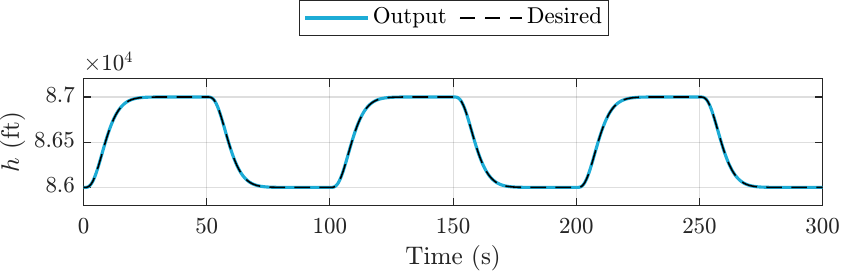}
  \end{center}
  \caption{Time history of altitude in direct neural backstepping control of HFV.}
  \label{fig:ex2_12:alt}
\end{figure}

Fig. \ref{fig:ex2_12:states} illustrates a comparison between the state trajectories and their desired values, i.e., virtual controls. Specifically, the trajectory of FPA exhibits a consistently smooth pattern. Additionally, the pitch angle $\theta$ is tracked smoothly, with only slight oscillations occurring during fault instances at 80 and 120 seconds. Additionally, pitch rate $q$ suffers from an initial high-frequency fluctuation, which also occurs during the times of the faults.
\begin{figure}[H]
  \begin{center}
    \includegraphics[width=0.9\textwidth]{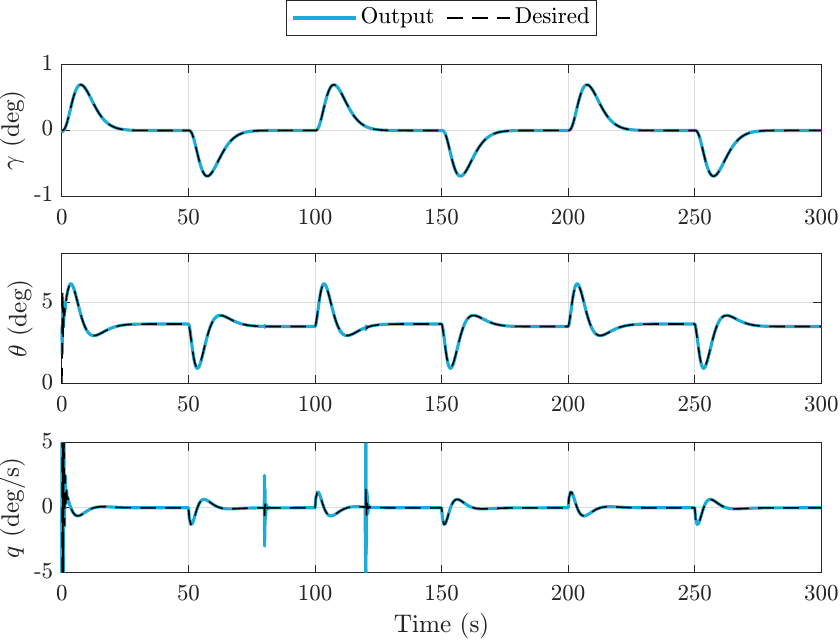}
  \end{center}
  \caption{Time history of states in direct neural backstepping control of HFV.} 
  \label{fig:ex2_12:states}
\end{figure}

Further, the control history of the elevator deflection is shown in Fig. \ref{fig:ex2_12:effort}, with a close view of the rapid oscillations of the control surface. These oscillations contribute to the training process of the neural networks, up to a point where they reach the proper value of the deflection.
\begin{figure}[H]
  \begin{center}
    \includegraphics[width=0.9\textwidth]{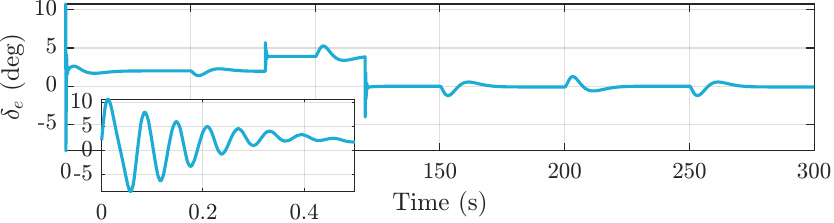}
  \end{center}
  \caption{Time history of control input in direct neural backstepping control of HFV.} 
  \label{fig:ex2_12:effort}
\end{figure}

\item Fig. \ref{fig:ex2_12:weights} shows the evolution of training weights over time for each of the neural networks. The first and third networks stabilize at constant weights after experiencing abrupt changes in weights at instances of altitude command switches.
In contrast, the second network uses $\sigma$-modification to prevent its weights from drifting over time. This results in a contraction-like pattern of weight evolution, which is different from the constant trend observed in the other networks, or when $\sigma_2=0$.
Furthermore, the third NN is particularly sensitive to faults and experiences more fluctuations during fault occurrences than the other NNs.
\begin{figure}[H]
  \begin{center}
    \includegraphics[width=0.9\textwidth]{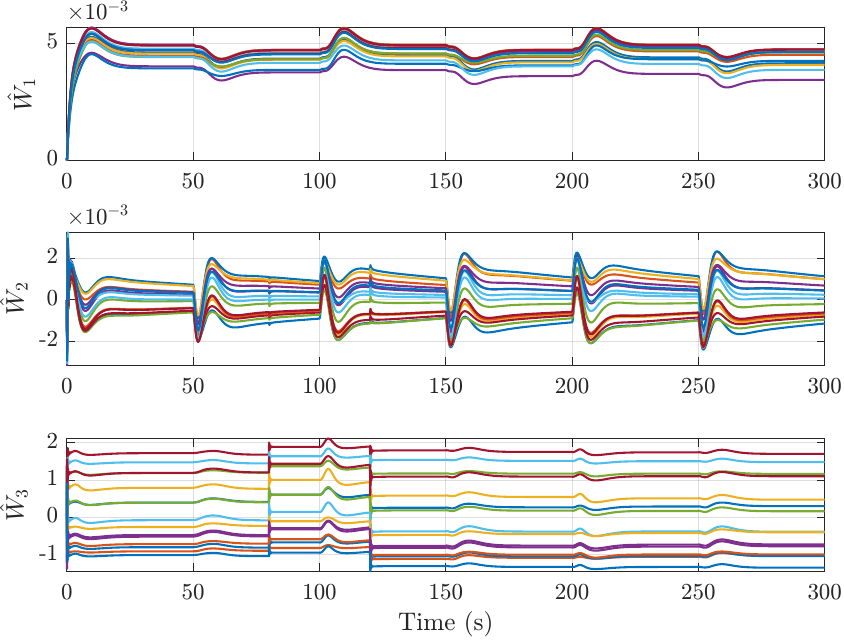}
  \end{center}
  \caption{Evolution of training weights in direct neural backstepping control of HFV.} 
  \label{fig:ex2_12:weights}
\end{figure}

\item Each NN is trained to estimate a part of the system's dynamics and/or the numerical derivative of a desired state, as expressed in Eq. (\ref{Eq:HFV_direct_Delta_i}). As shown in Fig. \ref{fig:ex2_12:delta}, the three networks are accurately estimating the expected outputs. Moreover, the third NN, which is directly being used in the control command, is showing a large initial estimation discrepancy. This discrepancy is responsible for the initial rapid oscillation observed in the control command, uncovering its underlying cause.
\begin{figure}[H]
  \begin{center}
    \includegraphics[width=0.9\textwidth]{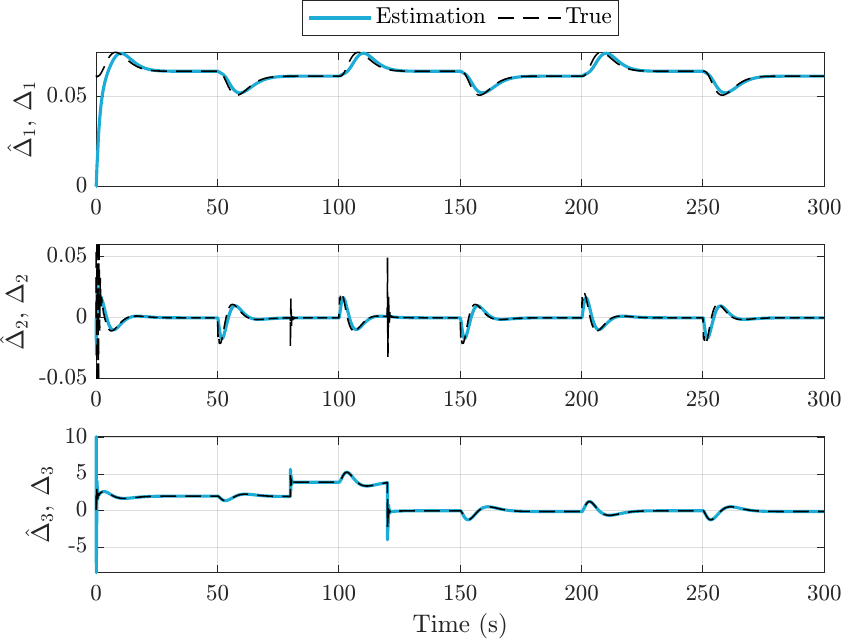}
  \end{center}
  \caption{Comparison of the uncertainty estimation of each NN with its true value in direct neural backstepping control of HFV.} 
  \label{fig:ex2_12:delta}
\end{figure}

\item \textbf{Sensitivity analysis:}
Two sets of parameter studies are performed to see how different values of the control gains $k_i$ and learning rates $\Gamma_i$ affect the accuracy of the control law. The accuracy is being measured using two metrics: the MAE of the tracking error and the MAEE of the uncertainty estimation.

The results of the parameter study are shown in Figs. \ref{fig:ex2_12:sens1} and \ref{fig:ex2_12:sens2}. The results show that the control law is sensitive to low values of $k_2$ and a value around 7 seems to be suitable.
Additionally, to improve the tracking and estimation accuracy, it is better to use higher values of $k_3$. Moreover, increasing $k_1$ up to around 10 can enhance performance, but beyond that point, the errors are unaffected by further increases.
\begin{figure}[H]
  \begin{center}
    \includegraphics[width=0.9\textwidth]{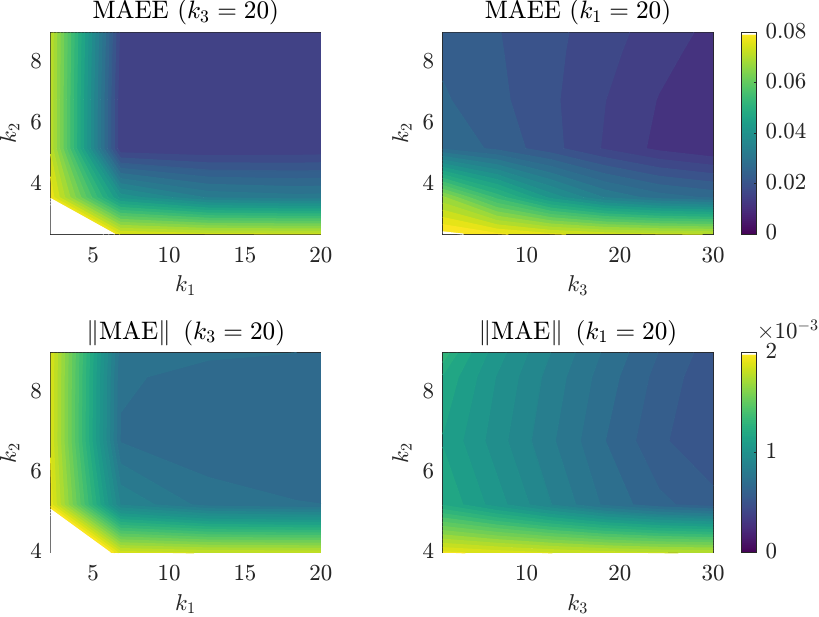}
  \end{center}
  \caption{Variation of MAEE and MAE for different values of control gain.} 
  \label{fig:ex2_12:sens1}
\end{figure}

From Fig. \ref{fig:ex2_12:sens2}, the following conclusions can be drawn:
\begin{itemize}
    \item For the first NN, the tracking error is unaffected by the learning rate. However, increasing the learning rate improves the estimation accuracy, as $\hat{\Delta}_1$ reaches its true value more quickly.
    \item For the second NN, high values of $\Gamma_2$ cause the proposed control to diverge. A value around 1 seems to be acceptable.
    \item For the third NN, it is important to balance increasing the learning rate to improve tracking performance with using lower values to improve estimation accuracy.
    \item The first and third NNs outperform the proposed control law. Using low learning rates for each of them would deteriorate the performance.
\end{itemize}

\begin{figure}[H]
  \begin{center}
    \includegraphics[width=0.9\textwidth]{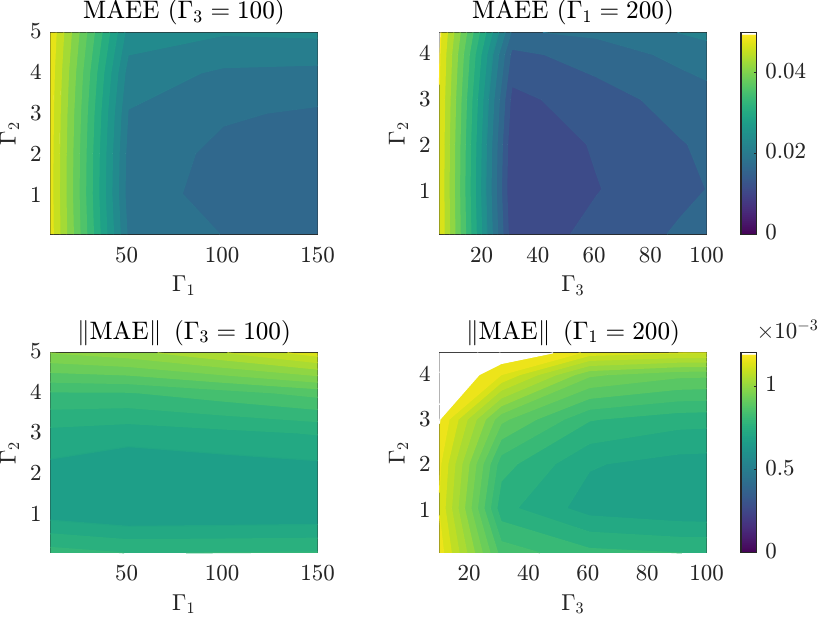}
  \end{center}
  \caption{Variation of MAEE and MAE for different values of learning rates for each NN.} 
  \label{fig:ex2_12:sens2}
\end{figure}
\end{enumerate}
\end{ex}

On the other hand, by introducing an output feedback form and utilizing High Gain Observers (HGOs) to estimate the time derivatives of the system output (Chapter 3), it is possible to derive the control command with no requirement for a backstepping scheme.  In this regard, a filtered tracking error is defined (as discussed in Section \ref{Sec:PseudoControl}) to provide a unified error dynamic model. Such a method has been adopted in \citet{Xu.2011,Xu.2015d} to control an HFV, where only one neural network is required in the altitude control block to determine the actual control command.
In a somewhat similar manner, a direct neuroadaptive control scheme has been integrated with the funnel control method in \citet{Bu.2018b} to control an air-breathing HFV considering non-affine dynamics. The altitude subsystem has been transformed into a simplified normal output feedback model, where only one NN is required to determine the control command.
Further, the non-affine dynamics of the vehicle have been handled by incorporating a low-pass filter in the last step of the design to define a new virtual control input in affine form.

\subsection{Discrete-time backstepping control}
In addition to the above-mentioned continuous-time backstepping control methods, several studies in the literature have addressed the design of a discrete-time neural backstepping controller. As discussed in Section \ref{Sec:FEL}, the general formulation of the NN updating rules in the discrete-time domain is similar to that of the continuous-time controller, while the stability analysis of the closed-loop system is quite different \citep{Xu.2011b}. A discrete-time direct neural backstepping control has been proposed in \citet{Xu.2012b} by incorporating HONNs to estimate uncertain terms in control commands. A similar control formulation has been given in \citet{Xu.2014} using Extreme Learning Machines (ELMs). To simplify the control structure, in \citet{Xu.2013,Xu.2013b}, dynamic equations corresponding to the altitude dynamics of an HFV have been aggregated into a prediction model as $x_1(k+n)= \bar{f}(x)+ \bar{g}(x) u$, where $\bar{f}$ and $\bar{g}$ represent uncertain nonlinear functions and $n$ denotes the system order. Subsequently, a single NN has been employed to tackle uncertain terms in the control command.
Such a method can be considered as the discrete-time equivalent of the above-mentioned approach to control output feedback models in the continuous-time domain.
Alternatively, an equivalent prediction model of an HFV has been defined in \citet{Xu.2015c} in which $x_i(k+n-i+1)$ is obtained as a function of $x_{i+1}(k+n-i)$. Using such a change in the formulation of system dynamics, all the information of the desired trajectory in future $n$ steps is involved in designing the controller, thereby improving the closed-loop performance. The designed controller in all these papers has been applied to the longitudinal mode of an HFV model.
An issue with the proposed discrete-time control schemes is the requirement to estimating the future virtual signals by NNs.
one can design $x_{2d}(k+3)$, $x_{3d}(k+2)$, and $x_{4d}(k+1)$ as a function of $x_{1d}(k+4)$ and subsequently, estimate them by distinct NNs. There are also direct and indirect approaches in discrete-time domain designs. In the following, we will describe these design methods.

\subsubsection{Indirect method}
To design an indirect neural-backstepping controller in the discrete-time domain, consider again a second order dynamic model which is discretized through Euler's expansion with sample time $T_{s}$ as follows:
\begin{gather}
  x_1\left(k+1\right)=x_{1}\left(k\right)+T_{s}\left[f_{1}\left(k\right)+g_{1}\left(k\right)x_{2}\left(k\right)+\Delta_{1}\left(k\right)\right], \label{Eq:HFV_discrete_indirect_x1k1}\\ 
  x_2\left(k+1\right)=x_{2}\left(k\right)+T_{s}\left[f_{2}\left(k\right)+g_{2}\left(k\right)u\left(k\right)+\Delta_{2}\left(k\right)\right]. \label{Eq:HFV_discrete_indirect_x2k1}
\end{gather}
Our goal is to determine the reference values in such a way that the error tends to zero. The idea that will be introduced is similar to the case where we have a geometric sequence in the form of $P(K+1)=cP(k)$ and we want this sequence to tend to zero. To achieve this demand, $|c|<1$ must be established. So by defining the tracking errors as $e_1(k)= x_1(k)-x_{1d}(k)$, $e_2(k)=x_2(k)-x_{2d}(k)$ the actual and virtual control commands can be expressed as follows:
\begin{gather}
u(k)=(T_{s}g_{2}(k))^{-1}\left[-T_{s}f_{2}(k)-T_{s}\hat{\Delta}_{2}(k)-x_{2}(k)+x_{2d}(k+1)+c_{2}e_{2}(k)\right], \label{Eq:HFV_discrete_indirect_uk}\\
x_{2d}(k)=(T_{s}g_{1}(k))^{-1}\left[-T_{s}f_{1}(k)-T_{s}\hat{\Delta}_{1}(k)-x_{1}(k)+x_{1d}(k+1)+c_{1}e_{1}(k)\right], \label{Eq:HFV_discrete_indirect_x2dk}
\end{gather}
where $|c_{1}|<1$, $|c_{2}|<1$ and the values of $\hat{\Delta}_{1}$ and $\hat{\Delta}_{2}$ are determined through two separate neural networks. In other words, it can be written:
\begin{gather}
\hat{\Delta}_{1}(k)=\hat{W}^T_{1}(k)\mu_1(k), \\
\hat{\Delta}_{2}(k)=\hat{W}^T_{2}(k)\mu_2(k).
\end{gather}
By looking carefully in Eq. (\ref{Eq:HFV_discrete_indirect_x2dk}), we find that all the required terms are available on the right side of the equation to construct $x_{2d}(k)$. However, to design $u(k)$, it is necessary to specify $x_{2d}(k+1)$.
\begin{align}\label{Eq:HFV_discrete_indirect_x2dk1}
\begin{split}
 x_{2d}(k+1)&=\left(T_{s}g_{1}\left(k+1\right)\right)^{-1} \left[-T_{s}f_{1}(k+1)-T_{s}\hat{\Delta}_{1}(k+1)-\hat{x}_{1}(k+1)\right.\\
 &\quad \left.+x_{1d}(k+2)+c_{1}e_{1}(k+1)\right].
\end{split}
\end{align}
To specify the above expression, we must first calculate $\hat{x}_{1}(k+1)$ based on the dynamic equation of the system presented in Eq. (\ref{Eq:HFV_discrete_indirect_x1k1}). $\hat{x}_{1}(k+1)$ means the estimated value of ${x}_{1}(k+1)$ because the system dynamics has an uncertainty term $\Delta_{1}$ and it should be estimated by NN. Subsequently, $f_{1}(k+1)$ and $\hat{\Delta}_{1}(k+1)$ can be determined (based on the fact that $\hat{x}_{1}(k+1)$ is as an input argument of these functions). On the other hand, the values of $x_{1d}(k+i),i=0,1,2,...$ are always specific according to the control objective. In addition, $e_{1}(k+1)$ can also be written as follows:
\begin{gather} \label{Eq:HFV_discrete_direct_e1k1_1}
  e_1(k+1)=\hat{x}_{1}(k+1)-x_{1d}(k+1).
\end{gather}
Therefore, all the required terms of the Eq. (\ref{Eq:HFV_discrete_indirect_x2dk1}) are available. consequently, the control command can be designed. The updating law for NN weights can also be expressed as follows:
\begin{gather}\label{Eq:HFV_discrete_inirect_updaterule1}
\hat{W}_{1}(k+1)=(1-\sigma_1)\hat{W}_{1}(k)+\lambda_1\mu_{1}(k)e_{1}(k+1), \\\label{Eq:HFV_discrete_inirect_updaterule2}
\hat{W}_{2}(k+1)=(1-\sigma_2)\hat{W}_{2}(k)+\lambda_2\mu_{2}(k)e_{2}(k+1),
\end{gather}
where $0<\sigma_i<1$ and $\lambda_{i}$ is a learning rate which is a positive constant.

\subsubsection{Direct method}
In designing the direct approach, we assume that we have no information about $f_{i}$ and $g_i$ but there exist known constants such that:
\begin{gather}
\underbar{$g_{i}$} \le g_{i} \le\bar{g_{i}}.
\end{gather}
For simplicity, we define $G_{i}(k)=T_{s}g_{i}$ and  $\bar{G}_{i}(k)=T_{s}\bar{g}_{i}$. The discrete direct neural backstepping design is illustrated in the following two steps:\\
\emph{Step $1$}: From Eqs. (\ref{Eq:HFV_discrete_indirect_x1k1}) and (\ref{Eq:HFV_discrete_direct_e1k1_1}), $e_{1}(k+1)$ can be written as follows:
\begin{gather}\label{Eq:HFV_discrete_indirect_e1k1n} 
e_{1}(k+1)=x_{1}(k)+T_{s}[f_{1}(k)+g_{1}(k)x_{2}(k)]-x_{1d}(k+1).
\end{gather}

The uncertainty $\text{UNC}_{1}$ is defined and can be estimated by an NN as:
\begin{align}\label{Eq:HFV_discrete_unc1}
\begin{split}
\text{UNC}_{1}(k)&=\frac{-x_{1}-T_{s}f_{1}(k)+x_{1d}(k+1)}{G_{1}(k)}+\frac{e_{1}(k)}{\bar{G}_{1}(k)} \\
&=W_{1}^{*T}(k)\mu_1(k)+\epsilon_{1}(k).
\end{split}
\end{align}
Therefore, since $x_{2}(k)$ can be considered the virtual input of Eq. (\ref{Eq:HFV_discrete_indirect_e1k1n}), it can be designed as follows:
\begin{align}\label{Eq:HFV_discrete_direct_x2d}
\begin{split}
x_{2d}(k)=\frac{(c_{1}-1)e_{1}(k)}{\bar{G}_{1}}+\hat{W}_{1}^{T}(k)\mu_1(k).
\end{split}
\end{align}
where $0 < c_{1} < 1$ is a design parameter. 
Combining Eqs. (\ref{Eq:HFV_discrete_indirect_e1k1n}) and (\ref{Eq:HFV_discrete_direct_x2d}), the following equation can be obtained:
\begin{align} \label{Eq:HFV_discrete_direct_e1k1n2}
\begin{split}
e_{1}(k+1) &= x_{1}(k)+T_{s}[f_{1}(k)+g_{1}(k)x_{2}(k)]-x_{1d}(k+1) \\
 &= G_{1}(k)e_{2}(k)+G_{1}(k)\left[x_{2d}(k)-\frac{-x_{1}(k)-T_{s}f_{1}(k)+x_{1d}(k+1)}{G_{1}(k)}\right] \\
 &= G_{1}(k)e_{2}(k)+G_{1}(k)\left[\frac{(c_{1}-1)e_{1}(k)}{\bar{G_{1}}}+\hat{W}_{1}^{T}(k)\mu_1(k)\right.\\
 & \quad -\left. \frac{-x_{1}(k)-T_{s}f_{1}(k)+x_{1d}(k+1)}{G_{1}(k)}\right] \\
 &= G_{1}(k)e_{2}(k)+\frac{G_{1}(k)}{\bar{G}_{1}}c_{1}e_{1}+G_{1}(k)(\tilde{W}_{1}^{T}(k)\mu_1(k)-\epsilon_{1}(k)),
\end{split}
\end{align} 
where $\tilde{W}(k)=\hat{W}_{1}(k)-W^{*}_{1}$. The adaption law can be designed as:
\begin{gather}\label{Eq:HFV_disceret_updaterule}
\hat{W}_{1}(k+1)=(1-\sigma_1)\hat{W}_{1}(k)-\lambda_1\mu_{1}(k)e_{1}(k+1),
\end{gather}
where $0<\sigma_1<1$ and $\lambda_{1}$ is a learning rate which is a positive constant.

\emph{Step $2$}: 
$e_{2}(k+1)$ can be written as follows:
\begin{gather}
e_{2}(k+1)=x_{2}(k)+T_{s}[f_{2}(k)+g_{2}(k)u(k)]-x_{2d}(k+1).
\end{gather}
The second uncertainty term is considered as:
\begin{align}\label{Eq:HFV_discrete_unc2}
\begin{split}
\text{UNC}_{2}(k)&=\frac{-x_{2}-T_{s}f_{2}(k)+x_{2d}(k+1)}{G_{2}(k)}+\frac{e_{2}(k)}{\bar{G}_{2}(k)} \\
&=W_{2}^{*T}(k)\mu_2(k)+\epsilon_{2}(k).
\end{split}
\end{align}
The controller command is designed given by:
\begin{align}\label{Eq:HFV_discrete_controlcommand}
\begin{split}
u(k)=\frac{(c_{2}-1)e_{2}(k)}{\bar{G}_{2}}+\hat{W}_{2}^{T}(k)\mu_2(k),
\end{split}
\end{align}
where $0 < c_{2} < 1$ is a design parameter. Based on Eq. (\ref{Eq:HFV_discrete_direct_e1k1n2}) the error dynamics for $e_{2}(k+1)$ is derived as:
\begin{align} \label{Eq:HFV_discrete_direct_e2k1n2}
\begin{split}
e_{2}(k+1)=\frac{G_{2}(k)}{\bar{G}_{2}}c_{2}e_{2}+G_{2}(k)(\tilde{W}_{2}^{T}(k)\mu_2(k)-\epsilon_{2}(k)).
\end{split}
\end{align} 
Finally the adaption law can be described same as Eq. (\ref{Eq:HFV_disceret_updaterule}):
\begin{gather}\label{Eq:HFV_disceret_updaterule2}
\hat{W}_{2}(k+1)=(1-\sigma_2)\hat{W}_{2}(k)-\lambda_2\mu_{2}(k)e_{2}(k+1),
\end{gather}
where $0<\sigma_2<1$ and $\lambda_{2}$ is a learning rate which is a positive constant.\\
In the end, it is worth mentioning that in Eqs.\ref{Eq:HFV_discrete_inirect_updaterule1}, \ref{Eq:HFV_discrete_inirect_updaterule2}, \ref{Eq:HFV_disceret_updaterule} and \ref{Eq:HFV_discrete_inirect_updaterule2} which express the rules of updating the weights of the NN, similar to the equations of the continuous cases, when the design approach is direct, the sign behind the learning rate is negative, and for the indirect approach, this sign is positive.

\begin{ex}{Discrete-time backstepping control of HFV: Indirect method}
Repeat Example \ref{Ex:HFV1} using discrete-time domain neural backstepping control of a HFV. To this end:
\begin{enumerate}[(a)]
\item Design a indirect neural backstepping discrete time controller.   
\item Investigate the closed-loop performance.
\item Plot the evolution of training weights over time.
\item Compare the estimated and actual uncertainties.
\item Check out the effect of sampling time and learning rate on the closed-loop performance.
\end{enumerate}

\tcblower
Consider the dynamic equations of attitude subsystem in the longitudinal mode of a HFV, given in Example \ref{Ex:HFV1}.
\begin{enumerate}[(a)]
\item\textbf{Controller design:}
First, we determine the error equations as follows:
\begin{align} \label{Eq:Ch2Ex13:discreteError}
\begin{split}
e_{1}(k)&=x_{1}(k)-x_{1d}(k) \\
e_{2}(k)&=x_{2}(k)-x_{2d}(k) \\
e_{3}(k)&=x_{3}(k)-x_{3d}(k).
\end{split}
\end{align}
As mentioned earlier, the values of $u(k)$, $x_{3d}(k)$ and $x_{2d}(k)$ should be determined in such a way that the error geometric sequence converges to zero. Therefore, we can write
\begin{align} \label{Eq:HFV_indirect_discrete_desired_value}
\begin{split}
u(k)&=(T_{s}g_{3}(k))^{-1}[-T_{s}\hat{f}_{3}(k)-x_{3}(k)+x_{3d}(k+1)+c_{3}e_{3}(k)] \\
x_{3d}(k)&=(T_{s}g_{2}(k))^{-1}[-T_{s}\hat{f}_{2}(k)-x_{2}(k)+x_{2d}(k+1)+c_{2}e_{2}(k)] \\
x_{2d}(k)&=(T_{s}g_{1}(k))^{-1}[-T_{s}\hat{f}_{1}(k)-x_{1}(k)+x_{1d}(k+1)+c_{1}e_{1}(k)].
\end{split}
\end{align}
where $T_s$ is a sample time and $k_i$ are constants between -1 and 1.
To design $x_{3d}(k)$, we need to calculate $x_{2d}(k+1)$ first. Therefore, we have
\begin{equation}\label{Eq:HFV_exp13_indirect_x2dk1}
x_{2d}(k+1)=(T_{s}g_{1}(k+1))^{-1}[-T_{s}\hat{f}_{1}(k+1)-\hat{x}_{1}(k+1)+x_{1d}(k+2)+c_{1}e_{1}(k+1)].
\end{equation}
All terms on the right side of Eq. (\ref{Eq:HFV_exp13_indirect_x2dk1}) except $x_{1d}(k+2)$ are unknown and must be calculated. $\hat{x}_{1}(k+1)$ can be obtained by using the dynamic equation of the system
\begin{align} \label{Eq:HFV_exp13_indirect_x1k1}
\begin{split}
\hat{x}_{1}(k+1)=x_{1}(k)+T_{s}[\hat{f}_{1}(k)+g_{1}(k)x_{2}(k)].
\end{split}
\end{align}
All terms on the right side of Eq. (\ref{Eq:HFV_exp13_indirect_x1k1}) are known, so $\hat{x}_{1}(k+1)$ is obtained. Now, since the input of the first NN ($\hat{f}_{1}$ estimator) is only based on the state variable $x_1$, by applying $\hat{x}_{1}(k+1)$ to this network, we can get the required output, which is $\hat{f}_{1}(k+1)$, and use it to calculate $e_{1}(k+1)$ error
\begin{align} \label{Eq:HFV_indirect_e1k1hat}
\begin{split}
e_{1}(k+1)=\hat{x}_{1}(k+1)-x_{1d}(k+1).
\end{split}
\end{align}
Therefore, we also calculated $e_{1}(k+1)$ and designed $x_{2d}(k+1)$. Now, we can also calculate $\hat{x}_{2}(k+1)$ and $e_{2}(k+1)$ as
\begin{align}
\begin{split}
\hat{x}_{2}(k+1)&=x_{2}(k)+T_{s}[\hat{f}_{2}(k)+g_{2}(k)x_{2}(k)] \\
e_{2}(k+1)&=\hat{x}_{2}(k+1)-x_{2d}(k+1).
\end{split}
\end{align}
In the next step, to calculate $u(k)$, we need to design $x_{3d}(k+1)$, which can be written as follows:
\begin{align} \label{Eq:HFV_indirect_x3dk1}
\begin{split}
x_{3d}(k+1)&=(T_{s}g_{2}(k+1))^{-1}[-T_{s}\hat{f}_{2}(k+1)-\hat{x}_{2}(k+1)\\
&+x_{2d}(k+2)+c_{2}e_{2}(k+1)].
\end{split}
\end{align}
According to the dynamics of HFV, we considered the value of $f_2$ to be known to be zero, so we do not use a network to estimate $\hat{f_2}$, and the term $-T_{s}\hat{f}_{2}(k+1)$ is zero. Therefore, we only need to specify $x_{2d}(k+2)$ as
\begin{equation}\label{Eq:HFV_indirect_exp13_x2dk2}
x_{2d}(k+2)=(T_{s}g_{1}(k+2))^{-1}[-T_{s}\hat{f}_{1}(k+2)-\hat{x}_{1}(k+2)+x_{1d}(k+3)+c_{1}e_{1}(k+2)].
\end{equation}
First we need to calculate $\hat{x}_{1}(k+2)$. $\hat{x}_{1}(k+1)$ was determined earlier. Therefore, based on the dynamics of the system, we can write
\begin{align} \label{Eq:HFV_indirect_exp13_x1k2}
\begin{split}
\hat{x}_{1}(k+2)=\hat{x}_{1}(k+1)+T_{s}[\hat{f}_{1}(k+1)+g_{1}(k+1)\hat{x}_{2}(k+1)].
\end{split}
\end{align}
All terms on the right side of Eq. \eqref{Eq:HFV_indirect_exp13_x1k2} are known, so $\hat{x}_{1}(k+2)$ is obtained. On the other hand, $\hat{x}_{1}(k+2)$ is used again as the input of the first NN to determine $\hat{f_{1}}(k+2)$. Also, $e_{1}(k+2)$ can be written as follows:
\begin{align}
\begin{split}
e_{1}(k+2)=\hat{x}_{1}(k+2)-x_{1d}(k+2).
\end{split}
\end{align}
Subsequently, $x_{2d}(k+2)$ and finally $x_{3d}(k+1)$ can be obtained. Now, to calculate the control command $u(k)$, it is only necessary to estimate the value of $\hat{f_{3}}(k)$ by the second neural network and calculate the error $e_{3}(k)=x_{3}(k)-x_{3d}(k)$.
\begin{align} \label{Eq:HFV_indirect_exp13_uk}
\begin{split}
u(k)=(T_{s}g_{3}(k))^{-1}[-T_{s}\hat{f}_{3}(k)-x_{3}(k)+x_{3d}(k+1)+c_{3}e_{3}(k)]. \\
\end{split}
\end{align}
Now, through the obtained control command and the system dynamics, $\hat{x}_{3}(k+1)$ and $e_{3}(k+1)$ can be determined as
\begin{align}
\begin{split}
\hat{x}_{3}(k+1)& =x_{3}(k)+T_{s}[\hat{f}_{3}(k)+g_{3}(k)u(k)] \\
e_{3}(k+1)& =\hat{x}_{3}(k+1)-x_{3d}(k+1).
\end{split}
\end{align}
In the final step, we must mention how to update the NN weights. Based on Eq. \eqref{Eq:HFV_disceret_updaterule}, we can write
\begin{align}
\begin{split}
\hat{W}_{1}(k+1) & =(1-\sigma_1)\hat{W}_{1}(k)+\lambda_1\mu_{1}(k)e_{1}(k+1) \\
\hat{W}_{3}(k+1) & = (1-\sigma_3)\hat{W}_{3}(k) +\lambda_3\mu_{3}(k)e_{3}(k+1)\\
\hat{f}_{1}(k) & = \mu_1(k)\hat{W}_{1}(k) \\
\hat{f}_{3}(k) & =\mu_3(k)\hat{W}_{3}(k),
\end{split}
\end{align}
where $0<\sigma_{i}<1$ and $\lambda_{i}>0$.

\item\textbf{Simulation results:}
In order to investigate the validity of the proposed discrete indirect backstepping control scheme, a numerical simulation is performed with 15 neurons and parameters listed in Table \ref{tab:Ex:Ch2Ex13_Param_1}.
\begin{table}[H]
    \caption{Parameters used in discrete indirect neural backstepping control of HFV.}
    \label{tab:Ex:Ch2Ex13_Param_1}
    \centering
    \begin{tabular}{cc}
    \hline\noalign{\smallskip}
    Parameter & Value \\
    \noalign{\smallskip}\hline\noalign{\smallskip}
    $[c_1, c_2, c_3]$ & $[0.8, 0.5, 0.5]$ \\ 
    $[\lambda_1, \lambda_3]$ & $[0.4, 0.7]$ \\ 
    $[\sigma_1, \sigma_3 ]$ & $[0.1, 0.1]$ \\ 
    \noalign{\smallskip}\hline\noalign{\smallskip}
    \end{tabular}
\end{table}

The time history of altitude and its square wave setpoint are depicted in Fig. \ref{fig:HFV_EX2.13a_h}, showing a promising and acceptable tracking performance.
\begin{figure}[H]
  \begin{center}
    \includegraphics[width=0.9\textwidth]{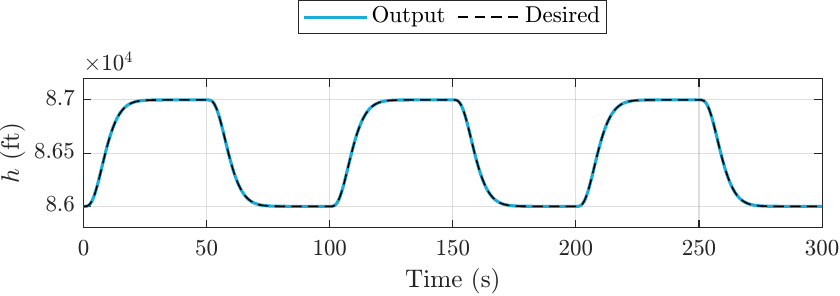}
  \end{center}
  \caption{Time history of altitude in discrete indirect neural backstepping control of HFV.}
  \label{fig:HFV_EX2.13a_h}
\end{figure}

Fig. \ref{fig:HFV_EX2.13a_x} compares the system state trajectories to their desired values, i.e., virtual controls. The FPA and pitch angle ($\theta$) of the system follow the desired values well, and the occurrence of a actuator fault does not significantly affect them. However, the pitch rate performance is slightly different. At the moment of actuator failure, pitch rate of the HFV is greatly affected, but it recovers after a relatively short period of time.
\begin{figure}[H]
  \begin{center}
    \includegraphics[width=0.9\textwidth]{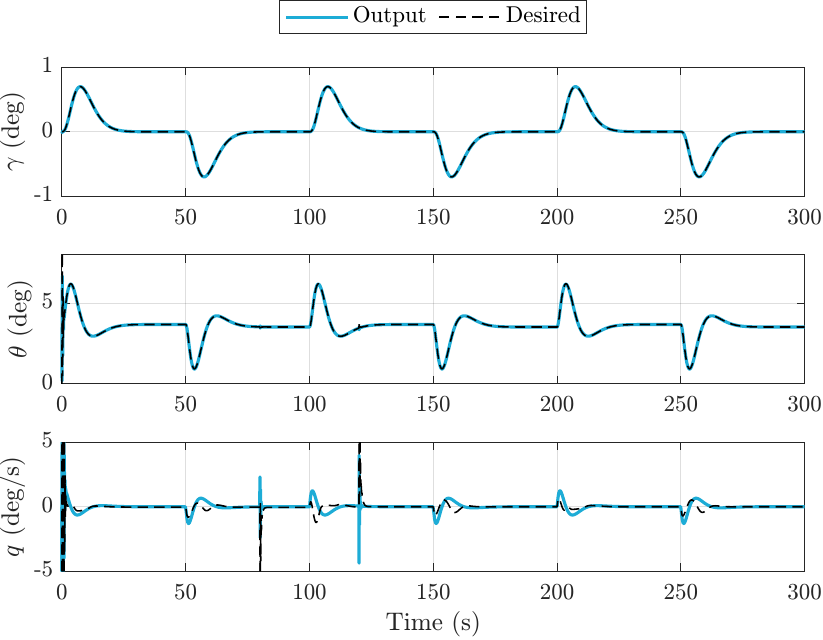}
  \end{center}
  \caption{Time history of states in discrete indirect neural backstepping control of HFV.} 
  \label{fig:HFV_EX2.13a_x}
\end{figure}
Furthermore, the elevator deflection as a control input is shown in Fig. \ref{fig:HFV_EX2.13a_de}. As it can be seen, the control command is within the permissible range for elevator deflection, and there are no extreme momentary changes in it, except in the initial moments of the failure, which are not too much.
\begin{figure}[H]
  \begin{center}
    \includegraphics[width=0.9\textwidth]{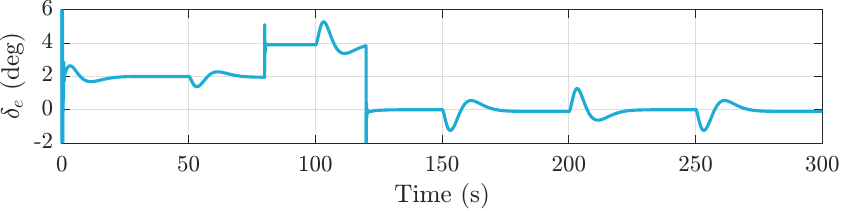}
  \end{center}
  \caption{Time history of control input in discrete indirect neural backstepping control of HFV.}
  \label{fig:HFV_EX2.13a_de}
\end{figure}

\item Fig. \ref{fig:HFV_EX2.13a_W} shows the evolution of the NN training weights over time. The weights of the first network stabilize at nearly constant weights after experiencing abrupt changes at the instances of altitude command switches.
In contrast, the weights of the second network are initially large, and the control command is also affected by this. However, as time goes on, the weights of the second network converge, albeit slowly. In addition, the weights of the second network are affected by the faults and converge to different values compared to their previous state.
\begin{figure}[H]
  \begin{center}
    \includegraphics[width=0.9\textwidth]{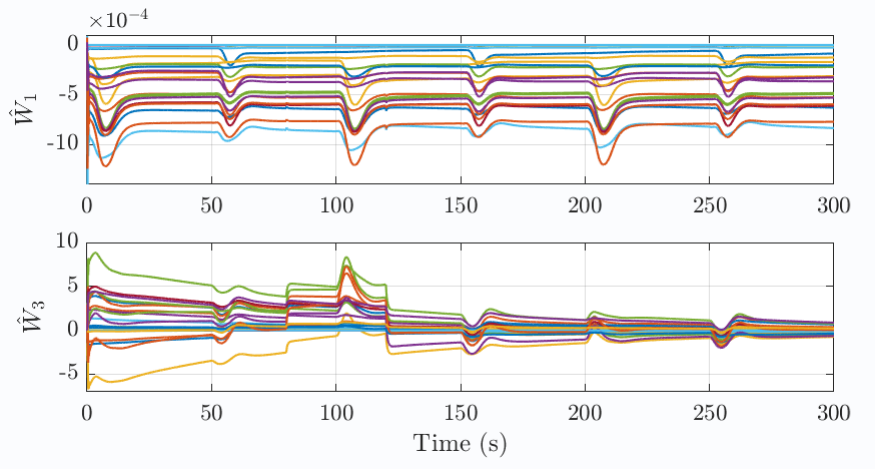}
  \end{center}
  \caption{Evolution of training weights in discrete indirect neural backstepping control of HFV.}
  \label{fig:HFV_EX2.13a_W}
\end{figure}

\item As shown in Fig. \ref{fig:HFV_EX2.13a_Delta}, two neural networks have been able to accurately estimate the unmodeled dynamic behavior of the system. In addition, it can be seen that the estimation accuracy did not drop significantly despite the occurrence of faults in the system and the network was able to overcome this challenge as well.
\begin{figure}[H]
  \begin{center}
    \includegraphics[width=0.9\textwidth]{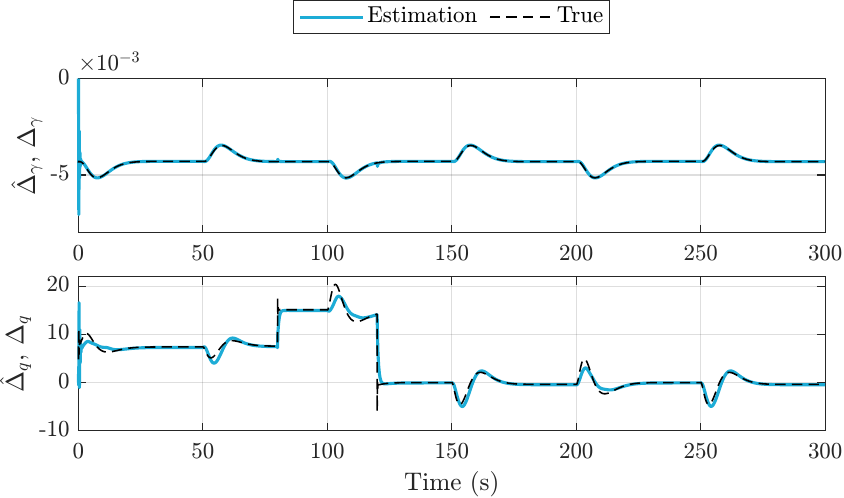}
  \end{center}
  \caption{Comparison of the uncertainty estimation of two NN with its real value in discrete indirect neural backstepping control of HFV.} 
  \label{fig:HFV_EX2.13a_Delta}
\end{figure}

\item \textbf{Sensitivity analysis:}
In this section, we will investigate the effect of sampling time and learning rate in the performance of the control structure. The results of the parameter study are shown in Figs. \ref{fig:HFV_EX2.13_SA} and \ref{fig:HFV_EX2.13_SA2}. The results show that the accuracy of the first uncertainty estimation (MAEE($\Delta_1$)) is improved by increasing the sampling time $T_s$ and learning rate $\lambda_1$. The effect of increasing the learning rate seems reasonable because with the increase of this parameter, the contribution of the NN becomes more prominent, and this can be the fact for improvement of uncertainty estimation. On the other hand, increasing the sampling time and its positive effect on improving the estimation quality seems a bit strange. First of all, it should be mentioned that if the sampling time exceeds a certain value, the estimation will diverge and the system will become unstable. In fact, the range shown in the sensitivity analysis figure is the permissible threshold for determining the sampling time. In justifying the positive effect of increasing the sampling time, it can be said that since in some equations in the design process the sampling time appears in the denominator of the fraction, this can cause the related term to become too large and as a result, the response oscillates. Therefore, the appropriate increase in sampling time can solve this problem. On the other hand, it seems that the intensity of uncertainty term and the influence of the first NN in estimating other states of the system ($x_{1}(k+1),x_{1}(k+2),x_{2}(k+1),x_{3}(k+1)$) could have led to the occurrence of such a phenomenon. These two parameters ($T_S$ and $\lambda_1$) have a similar effect and interpretation for the MAE criterion.

\begin{figure}[H]
  \begin{center}
    \includegraphics[width=0.9\textwidth]{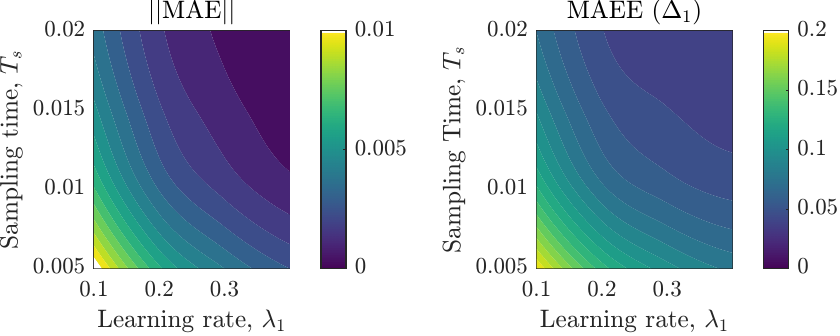}
  \end{center}
  \caption{Variation of MAE and MAEE  for different values of sampling time and learning rate $\lambda_1$ in discrete indirect neural backstepping control of HFV.} 
  \label{fig:HFV_EX2.13_SA}
\end{figure}
The situation in Figure \ref{fig:HFV_EX2.13_SA2} is slightly different. In this figure, we can see that by increasing the learning rate $\lambda_3$, the estimation accuracy, MAEE($\Delta_3$), and control performance ($\parallel\text{MAE}\parallel$) have improved, but when the learning rate exceeds 0.7, the control performance gradually decreases. From the aspect of sampling time, we also see that the effect of this parameter is almost neutral for a learning rate less than 0.5, and after that, the studied criteria improve to some extent with the increase of the sample time.
\begin{figure}[H]
  \begin{center}
    \includegraphics[width=0.9\textwidth]{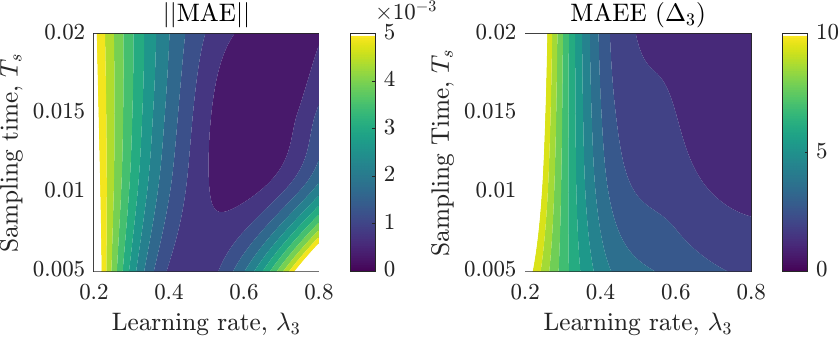}
  \end{center}
  \caption{Variation of MAE and MAEE  for different values of sampling time and learning rate $\lambda_3$ in discrete indirect neural backstepping control of HFV.} 
  \label{fig:HFV_EX2.13_SA2}
\end{figure}
\end{enumerate}
\end{ex}
\begin{ex}{Discrete-time backstepping control of HFV: direct method}
Repeat Example \ref{Ex:HFV1} using discrete-time domain neural backstepping control of a HFV. To this end:
\begin{enumerate}[(a)]
\item Design a direct neural backstepping discrete time controller.   
\item Investigate the closed-loop performance.
\item Plot the evolution of training weights over time.
\item Check out the effect of sampling time, learning rate and control gain on the closed-loop performance.
\end{enumerate}

\tcblower
Consider the dynamic equations of attitude subsystem in the longitudinal mode of a HFV, given in Example \ref{Ex:HFV1}.

\begin{enumerate}[(a)]
\item \textbf{Controller design:}
The discrete direct neural backstepping control design is illustrated in the following three steps.\\
\emph{Step $1$}: According to the error definition, it can be written from Eq. (\ref{Eq:HFV_indirect_e1k1hat})
\begin{align} \label{Eq:Ch2Ex14:discreteError1}
\begin{split}
e_{1}(k+1)=x_{1}(k)+T_{s}[\hat{f}_{1}(k)+g_{1}(k)x_{2}(k)]-x_{1d}(k+1).
\end{split}
\end{align}
Based on Eq. (\ref{Eq:HFV_discrete_unc1}), the uncertainty term can be defined as:
\begin{align}
\begin{split}
\text{UNC}_{1}(k)&=\frac{-x_{1}-T_{s}f_{1}(k)+x_{1d}(k+1)}{G_{1}(k)}+\frac{e_{1}(k)}{\bar{G_{1}}} \\
&=W_{1}^{*T}(k)\mu_1(k)+\epsilon_{1}(k).
\end{split}
\end{align}
Since $x_{2}(k)$ is considered as the virtual control in Eq. (\ref{Eq:Ch2Ex14:discreteError1}), the value of $x_{2d}(k)$ is defined as follows:
\begin{align}
\begin{split}
x_{2d}(k)=\frac{(c_{1}-1)e_{1}(k)}{\bar{G_{1}}}+\hat{W}_{1}^{T}(k)\mu_1(k),
\end{split}
\end{align}
where $0 \le c_{1} \le 1$ is a design parameter. Now, from Eq. (\ref{Eq:HFV_discrete_direct_e1k1n2}), the dynamics of $e_{1}(k+1)$ can be expressed as follows:
\begin{align}
\begin{split}
e_{1}(k+1)= G_{1}(k)e_{2}(k)+\frac{G_{1}(k)}{\bar{G}_{1}}c_{1}e_{1}+G_{1}(k)(\tilde{W}_{1}^{T}(k)\mu_1(k)-\epsilon_{1}(k)).
\end{split}
\end{align}
The NN updating rule can be written as:
\begin{gather}
\hat{W}_{1}(k+1)=(1-\sigma_1)\hat{W}_{1}(k)-\lambda_1\mu_{1}(k)e_{1}(k+1),
\end{gather}
where $0<\sigma_1<1$ and $\lambda_{1}$ is a positive learning parameter.

\emph{Step $2$}: 
$e_{2}(k+1)$ can be expressed as follows:
\begin{gather}
e_{2}(k+1)=x_{2}(k)+T_{s}[f_{2}(k)+g_{2}(k)x_{3}(k)]-x_{2d}(k+1)
\end{gather}
The second uncertainty term is given by
\begin{align}\label{Eq:HFV_discrete_unc2_1}
\begin{split}
\text{UNC}_{2}(k)&=\frac{-x_{2}-T_{s}f_{2}(k)+x_{2d}(k+1)}{G_{2}(k)}+\frac{e_{2}(k)}{\bar{G_{2}}} \\
&=W_{2}^{*T}(k)\mu_2(k)+\epsilon_{2}(k),
\end{split}
\end{align}
Accordingly, the second virtual control command, $x_{3d}(k)$, is designed as:
\begin{align}\label{Eq:HFV_discrete_controlcommand_1}
\begin{split}
x_{3d}(k)=\frac{(c_{2}-1)e_{2}(k)}{\bar{G_{2}}}+\hat{W}_{2}^{T}(k)\mu_2(k)
\end{split}
\end{align}
where $0 \le c_{2} \le 1$ is a design parameter. Using Eq. \eqref{Eq:HFV_discrete_direct_e1k1n2}, we have
\begin{align} \label{Eq:HFV_discrete_direct_e2k1}
\begin{split}
e_{2}(k+1)=G_{2}(k)e_{3}(k)+\frac{G_{2}(k)}{\bar{G}_{2}}c_{2}e_{2}+G_{2}(k)(\tilde{W}_{2}^{T}(k)\mu_2(k)-\epsilon_{2}(k)).
\end{split}
\end{align} 
Again, the adaption law can be described same as Eq. (\ref{Eq:HFV_disceret_updaterule})
\begin{gather}
\hat{W}_{2}(k+1)=(1-\sigma_2)\hat{W}_{2}(k)-\lambda_2\mu_{2}(k)e_{2}(k+1),
\end{gather}
where $0<\sigma_2<1$ and $\lambda_{2}$ is a positive learning parameter.

\emph{Step $3$}: 
$e_{3}(k+1)$ can be expressed as follows
\begin{gather}
e_{3}(k+1)=x_{3}(k)+T_{s}[f_{3}(k)+g_{3}(k)u(k)]-x_{3d}(k+1).
\end{gather}
The third uncertainty term can be expressed as follows:
\begin{align}\label{Eq:HFV_discrete_unc3}
\begin{split}
\text{UNC}_{3}(k)&=\frac{-x_{3}-T_{s}f_{3}(k)+x_{3d}(k+1)}{G_{3}(k)}+\frac{e_{3}(k)}{\bar{G_{3}}} \\
&=W_{3}^{*T}(k)\mu_3(k)+\epsilon_{3}(k).
\end{split}
\end{align}
The control command, $u(k)$, is designed as:
\begin{align}
\begin{split}
u(k)=\frac{(c_{3}-1)e_{3}(k)}{\bar{G_{3}}}+\hat{W}_{3}^{T}(k)\mu_3(k),
\end{split}
\end{align}
where $0 \le c_{3} \le 1$ is a designing parameter. The error dynamic for $e_{3}(k+1)$ is derived as:
\begin{align}
\begin{split}
e_{3}(k+1)=\frac{G_{3}(k)}{\bar{G}_{3}}c_{3}e_{3}+G_{3}(k)(\tilde{W}_{3}^{T}(k)\mu_3(k)-\epsilon_{3}(k)).
\end{split}
\end{align} 
Finally, the updating law can be described same as Eq. (\ref{Eq:HFV_disceret_updaterule})
\begin{gather}
\hat{W}_{3}(k+1)=(1-\sigma_3)\hat{W}_{3}(k)-\lambda_3\mu_{3}(k)e_{3}(k+1),
\end{gather}
where $0<\sigma_3<1$ and $\lambda_{3}$ is a positive learning parameter.
\item \textbf{Simulation results:}
In order to check the performance of the proposed discrete direct backstepping control scheme, a numerical simulation has been done with 20 neurons and the following parameters.
\begin{table}[H]
    \caption{Parameters used in discrete direct backstepping control of HFV.}
    \label{tab:Ex:Ch2Ex13_Param}
    \centering
    \begin{tabular}{cc}
    \hline\noalign{\smallskip}
    Parameter & Value \\
    \noalign{\smallskip}\hline\noalign{\smallskip}
    $[c_1, c_2, c_3]$ & $[0.9, 0.7, 0.5]$ \\ 
    $[\lambda_1, \lambda_2, \lambda_3]$ & $[2,2,4]$ \\ 
    $[\sigma_1, \sigma_2, \sigma_3 ]$ & $[0.1,0.1,0.1]$ \\ 
    \noalign{\smallskip}\hline\noalign{\smallskip}
    \end{tabular}
\end{table}

The investigation of the main control objective, i.e., HFV altitude control and appropriate tracking of the desired value, is shown in Fig. \ref{fig:HFV_EX2.14a_h}. As can be seen in this figure, the designed controller is able to achieve the main control objective. Also, the occurrence of faults in the elevator does not cause its quality to decline.
\begin{figure}[H]
  \begin{center}
    \includegraphics[width=0.9\textwidth]{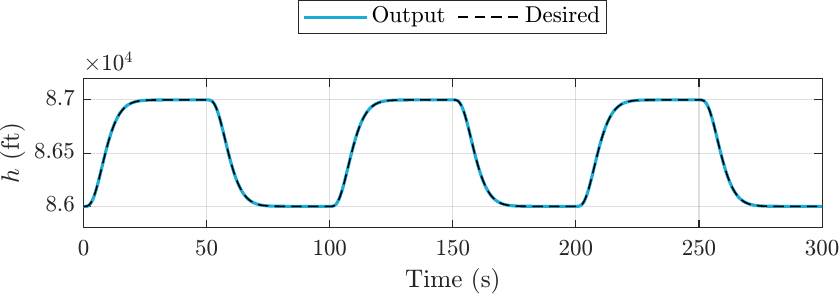}
  \end{center}
  \caption{\small Time history of altitude in discrete indirect neural backstepping control of HFV.}
  \label{fig:HFV_EX2.14a_h}
\end{figure}

Fig. \ref{fig:HFV_EX2.14a_x} shows a comparison between the system state trajectories and their desired values, i.e., virtual control inputs. FPA and pitch angle($\theta$)   follow the desired values perfectly, and imposing a fault has no effect on them. However, pitch rate behavior is almost different. At the beginning of the simulation, the reference value $x_{3d}$ and $x_{3}$ follow each other with a distance. This behavior continues even after the first fault occurs. Finally, after the occurrence of the second fault, the performance becomes satisfactory. The reason that can be considered for this phenomenon is that at the beginning of the simulation, the input richness of the second NN is not very suitable for the correct estimation of uncertainty. But after the system is disrupted for a few moments due to the fault, the richness of the NN input data increases and its performance in estimating the uncertainty and consequently constructing the $x_{3d}$ increases.
\begin{figure}[H]
  \begin{center}
    \includegraphics[width=0.9\textwidth]{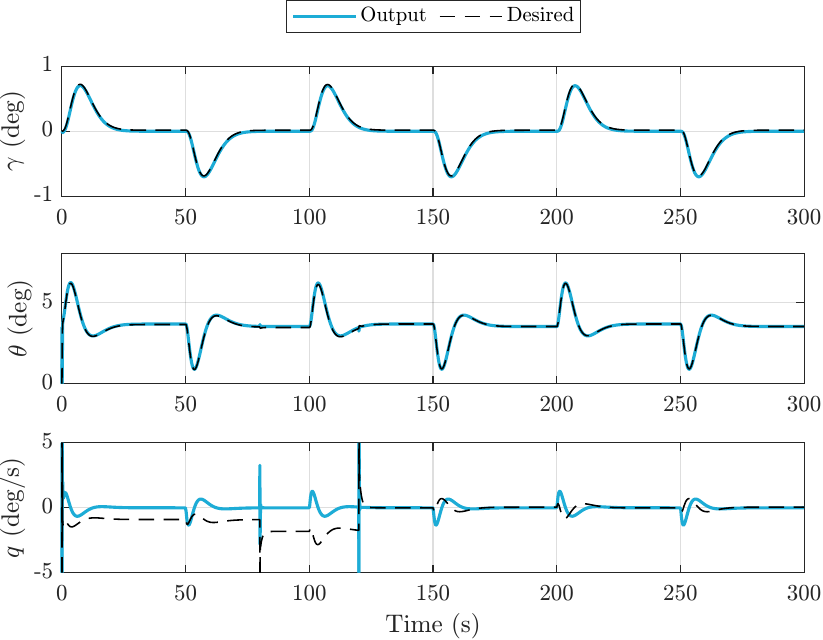}
  \end{center}
  \caption{\small Time history of states in discrete indirect neural backstepping control of HFV.} 
  \label{fig:HFV_EX2.14a_x}
\end{figure}
In the next figure elevator deflection as a control input is shown. As can be seen, the angle of elevator deflection is on two positive degrees when the system has no faults and is tracking a constant height. When the fault occurs, the level of the elevator deflection has also changed according to the size of the fault, which actually indicates the correct reaction of the controller in the face of the new condition. In addition, the transient changes of the control command at the moments of altitude change are also evident. These changes of the control command are also consistent with the physics of the problem. For example, when it is necessary to increase the altitude, the deviation of the elevator increases, and as a result plane rises. After it reaches the desired height, it tries to stabilize the altitude by reducing and then adjusting the angle of the elevator deflection.
\begin{figure}[H]
  \begin{center}
    \includegraphics[width=0.9\textwidth]{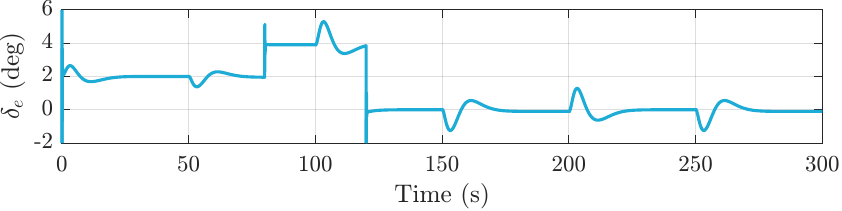}
  \end{center}
  \caption{\small Time history of control inputs in discrete indirect neural backstepping control of HFV.} 
  \label{fig:HFV_EX2.14a_de}
\end{figure}

\item Fig. \ref{fig:HFV_EX2.14a_W} shows the evolution of NN training weights over time. The weights of the first NN are not affected by the imposed faults and only have transient changes during the altitude switch and then converge to the same previous value. In case of altitude switch, the weights of the second and third NN have the same behavior as the first one. However, the weights of these two networks change due to the fault in the system and converge to new levels.
\begin{figure}[H]
  \begin{center}
    \includegraphics[width=0.9\textwidth]{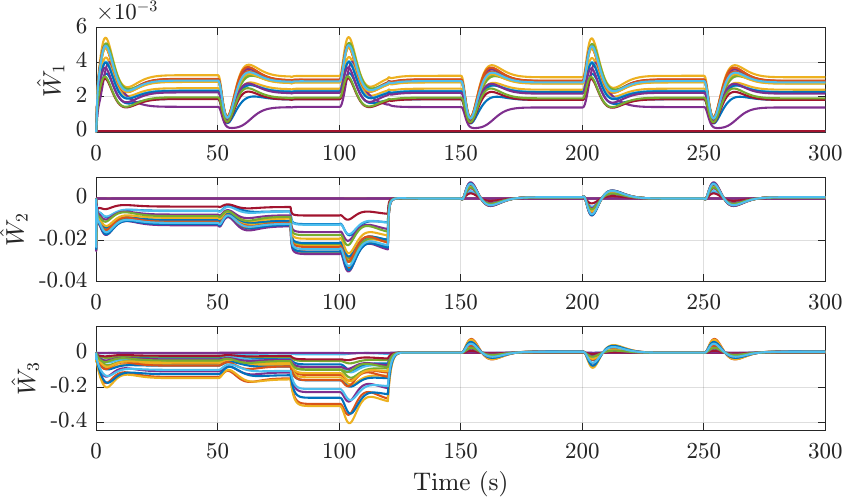}
  \end{center}
  \caption{\small Evolution of training weights in discrete indirect neural backstepping control of HFV.} 
  \label{fig:HFV_EX2.14a_W}
\end{figure}

\item \textbf{Sensitivity analysis:}
In this section, we will investigate the effect of sampling time $T_s$, learning rate $\lambda_3$ and control gain $c_3$ in the performance of the control structure. For this purpose, we use the MAE to evaluate the close-loop system performance. The results of the parameter study are shown in Figs.\ref{fig:HFV_EX2.14_SA} and \ref{fig:HFV_EX2.14_SA2}. The reason for choosing parameters $\lambda_3$ and $c_3$ is their direct effect in generating the control command. The results of the sensitivity analysis in Fig.\ref{fig:HFV_EX2.14_SA} show that performance of the control system is slightly affected by the sampling time, and in general, the smaller the sampling time, the better performance. Also, this figure shows that a suitable range can be found for the learning rate $\lambda_3$ so that the tracking error is minimized.
\begin{figure}[H]
  \begin{center}
    \includegraphics[width=0.5\textwidth]{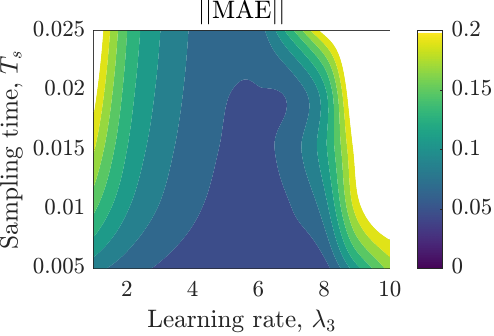}
  \end{center}
  \caption{Variation of MAE for different values of sampling time and learning rate $\lambda_3$ in discrete direct neural backstepping control of HFV.} 
  \label{fig:HFV_EX2.14_SA}
\end{figure}
Fig.\ref{fig:HFV_EX2.14_SA2} presents the effect of control gain $c_3$ and sampling time on the performance of the control system. The control gain is proportional to the error damping. Pay attention, the lower this coefficient is, the higher the intensity of damping. Because in the ideal case, $e_{3}(k+1)=c_{3}e_{3}(k)$. Therefore, as it is clear from the figure below, the lower this gain, the better performance of the control system. On the other hand, in this sensitivity analysis, we see a more obvious effect of sampling time, so that reducing sampling time has a more significant effect on reducing the MAE criterion.

\begin{figure}[H]
  \begin{center}
    \includegraphics[width=0.5\textwidth]{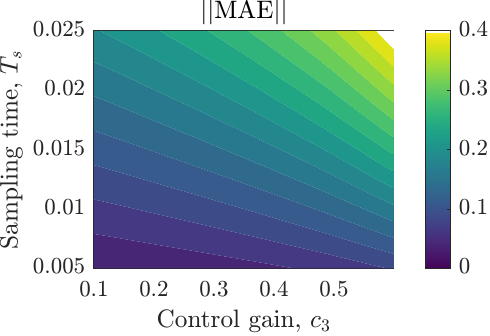}
  \end{center}
  \caption{Variation of MAE for different values of sampling time and control gain $c_3$ in discrete direct neural backstepping control of HFV.} 
  \label{fig:HFV_EX2.14_SA2}
\end{figure}
\end{enumerate}
\end{ex}

\subsubsection{Digress to stability analysis for discrete time controller}
In general, the stability analysis of discrete-time method is more complicated than the continuous one. To have an overview of this issue, in this section, the process of proving the stability of the controller designed with a direct approach for an HFV will be presented. References like \citet{xu2011adaptive,xu2013neural} are suggested for further study. The following proof is based on \citet{xu2011adaptive}.

Suppose we rewrite the HFV dynamics as follows:
\begin{gather}
\begin{split}
x_{0}(k+1)&= x_{0}(k)+T_{s}[f_{0}(k)+g_{0}(k)x_{1}(k)],\\
x_{1}(k+1)&= x_{1}(k)+T_{s}[f_{1}(k)+g_{1}(k)x_{2}(k)+\Delta_{1}], \\
x_{2}(k+1)&= x_{2}(k)+T_{s}[f_{2}(k)+g_{2}(k)x_{3}(k)+\Delta_{2}], \\
x_{3}(k+1)&= x_{3}(k)+T_{s}[f_{3}(k)+g_{3}(k)u(k)+\Delta_{3}].
\end{split}
\end{gather}
In this description, the dynamics of the  altitude is also added to the set of equations, but this dynamic has no uncertainty term. Same as before, the other states of the system i.e. $x_{1}$, $x_{2}$ and $x_{3}$ are flight path angle, pitch angle and pitch rate.
The error dynamic and its one step ahead prediction can be defined as follows:
\begin{gather}
\begin{split}
e_{i}(k)&= x_{i}(k)-x_{id}(k),i=0,1,2,3,\\
e_{i}(k+1)&= x_{i}(k)+T_{s}[f_{i}(k)+g_{i}(k)x_{i+1}(k)]-x_{id}(k+1),\; i=0,1,2, \\
e_{3}(k+1)&= x_{3}(k)+T_{s}[f_{3}(k)+g_{3}(k)u(k)]-x_{3d}(k+1).
\end{split}
\end{gather}

Take $x_{1}$ as a virtual control input and determine its desired value as:
\begin{gather}
\begin{split}
x_{1d}(k)=[T_{s}g_{0}(k)]^{-1}[-x_{0}(k)+c_{0}e_{0}(k)+x_{0d}(k+1)-T_{s}f_{0}(k)],
\end{split}
\end{gather}
where $0<c_{0}<1$. Consequently the following equation can be obtained.
\begin{gather}
\begin{split}
e_{0}(k+1)=c_{0}e_{0}(k)+T_{s}g_{0}(k)e_{1}(k).
\end{split}
\end{gather}
$U_{1}$ is defined as the uncertainty term and approximated by NN as:
\begin{gather}
\begin{split}
U_{1}(k)&= \frac{\bar{g}_1}{g_{1}(k)}\left[-x_{1}(k)-T_{s}f_{1}(k)-T_{s}\Delta_{1}(k)+x_{1d}(k+1)\right]\\
& \quad -[-x_{1}(k)-T_{s}f_{1}(k)]
=  W_{1}^{*T}\mu_{1}(k)+\epsilon_{1}(k).
\end{split}
\end{gather}
Now take $x_{2d}$ as a virtual control input and determine its value as
\begin{gather}
\begin{split}
x_{2d}(k)=[T_{s}\bar{g}_{1}(k)]^{-1}[-x_{1}(k)+c_{1}e_{1}(k)-T_{s}f_{1}(k)+\hat{W}_{1}^{T}(k)\mu_{1}(k)],
\end{split}
\end{gather}
where $0<c_{1}<1$. Consequently, the following equation can be obtained.
\begin{gather}
\begin{split}
e_{1}(k+1)=\frac{g_{1}(k)}{\bar{g_{1}}}c_{1}e_{1}(k)+T_{s}g_{1}(k)e_{2}(k)+\frac{g_{1}(k)}{\bar{g_{1}}}[\tilde{W}_{1}^{T}\mu_{1}(k)-\epsilon_{1}(k)],
\end{split}
\end{gather}
where $\tilde{W}_{1}(k)=\hat{W}_{1}(k)-W_{1}^{*}(k)$. The update law can be described as:
\begin{gather}
\begin{split}
\hat{W}_{1}(k+1)=\hat{W}_{1}(k)-\lambda_{1}e_{1}(k+1)\mu_{1}(k)-\sigma_{1}\hat{W}_{1}(k).
\end{split}
\end{gather}

In the next step, we define $U_{2}$ as the uncertainty term and approximate it using an NN as follows:
\begin{gather}
\begin{split}
U_{2}(k)=x_{2d}(k+1)=W_{2}^{*T}\mu_{2}(k)+\epsilon_{2}(k).
\end{split}
\end{gather}

Similarly, we take $x_{3d}$ as a virtual control input and define
\begin{gather}
\begin{split}
x_{3d}(k)=[T_{s}\bar{g}_{2}(k)]^{-1}\left[-x_{2}(k)+c_{2}e_{2}(k)+\hat{W}_{2}^{T}(k)\mu_{2}(k)\right],
\end{split}
\end{gather}
where $0<c_{2}<1$. Actually, the construction of $U_{2}(k)$ and $x_{3d}(k)$ is the same as $U_{1}(k)$ and $x_{2d}(k)$ but here, according to the dynamics of HFV, $f_{2}=0$, $\Delta_{2}=0$ and $g_{2}=1$ are placed.

Consequently, the following equation can be obtained.
\begin{gather}
\begin{split}
e_{2}(k+1)=c_{2}e_{2}(k)+T_{s}g_{2}(k)e_{3}(k)+\left[\tilde{W}_{2}^{T}\mu_{2}(k)-\epsilon_{2}(k)\right],
\end{split}
\end{gather}
where $\tilde{W}_{2}(k)=\hat{W}_{2}(k)-W_{2}^{*}(k)$. 

The update law can be described as: 
\begin{gather}
\begin{split}
\hat{W}_{2}(k+1)=\hat{W}_{2}(k)-\lambda_{2}e_{2}(k+1)\mu_{2}(k)-\sigma_{2}\hat{W}_{2}(k).
\end{split}
\end{gather}

In the final step $U_{3}$ is defined as uncertainty term and approximated by NN as:
\begin{gather}
\begin{split}
U_{3}(k)=& \frac{\bar{g}_3}{g_{3}(k)}[-x_{3}(k)-T_{s}f_{3}(k)-T_{s}\Delta_{3}(k)+x_{3d}(k+1)]\\
& -[-x_{3}(k)-T_{s}f_{3}(k)]
=  W_{3}^{*T}\mu_{3}(k)+\epsilon_{3}(k).
\end{split}
\end{gather}

The actual control command is defined as:
\begin{gather}
\begin{split}
u(k)=\left[T_{s}\bar{g}_{3}(k)]^{-1}[-x_{3}(k)+c_{3}e_{3}(k)-T_{s}f_{3}(k)+\hat{W}_{3}^{T}(k)\mu_{3}(k)\right],
\end{split}
\end{gather}
where $0<c_{3}<1$. Consequently the following equation can be obtained.
\begin{gather}
\begin{split}
e_{3}(k+1)=\frac{g_{3}(k)}{\bar{g_{3}}}c_{3}e_{3}(k)+\frac{g_{3}(k)}{\bar{g_{3}}}[\tilde{W}_{3}^{T}\mu_{3}(k)-\epsilon_{3}(k)],
\end{split}
\end{gather}
where $\tilde{W}_{3}(k)=\hat{W}_{3}(k)-W_{3}^{*}(k)$. The update law can be described as:
\begin{gather}
\begin{split}
\hat{W}_{3}(k+1)=\hat{W}_{3}(k)-\lambda_{3}e_{3}(k+1)\mu_{3}(k)-\sigma_{3}\hat{W}_{3}(k).
\end{split}
\end{gather}

The following Lyapunov function is considered
\begin{equation} \label{discrete_stability_1} 
L(k) = L_0(k) + L_1(k) + L_2(k) + L_3(k),   
\end{equation} 
where
\begin{gather} \label{discrete_stability_2} 
\begin{split}
L_0(k) = {\left(\frac{e_0\left(k\right)}{{\overline{g}}_0}\right)}^2,\quad \quad \quad \; L_1(k)&= L_{10}(k) + L_{11}(k),\\
L_2(k)= L_{20}(k) + L_{21}(k),\quad  L_3(k)&= L_{30}(k) + L_{31}(k),  
\end{split}
\end{gather}
with
\begin{gather} \label{discrete_stability_3} 
\begin{split}
L_{10}(k)= {e_1}^2(k),\quad L_{11}(k)&=\frac{\widetilde{W}_{1}^{T}
\left(k\right){\widetilde{W}}_{1}\left(k\right)}{{\lambda }_1} ,\\
L_{20}(k)= {e_2}^2(k),\quad
L_{21}(k)&= \frac{\widetilde{W}_{2}^{T}
\left(k\right){\widetilde{W}}_{2}\left(k\right)}{{\lambda }_2},\\
L_{30}(k) = {e_3}^2(k),\quad
L_{31}(k)&= \frac{\widetilde{W}_{3}^{T}
\left(k\right){\widetilde{W}}_{3}\left(k\right)}{{\lambda }_3} . 
\end{split}
\end{gather}

Define the first difference of $L(k)$ as:
\begin{gather} \label{discrete_stability_4} 
\begin{split}
\Delta  L(k)&= L(k + 1)\ - L(k) \\
&= \Delta L_0(k) + \Delta L_1(k) + \Delta L_2(k) + \Delta L_3(k). 
\end{split}
\end{gather}

From Cauchy–Schwarz and Young's inequalities, we have $2ab\le \rho a^2+ {{\frac{1}{\rho}}}b^2$, where $\rho$ is constant and positive.

Regarding the first term in \eqref{discrete_stability_1} we have
\begin{gather} \label{discrete_stability_5} 
\begin{split}
\Delta L_0(k)&=\ {\left(e_0\left(k\ +\ 1\right)\slash {\overline{g}}_0\right)}^2-{\left(e_0\left(k\right)\slash {\overline{g}}_0\right)}^2 \\
&= {\left(1\slash {\overline{g}}_0\right)}^2\left[{\left(c_0e_0+T_s{\mathrm{g}}_0\left(k\right)e_1\left(k\right)\right)}^2-e_0{\left(k\right)}^2\right] \\
& \le -{\tau }_{11}L_0(k)+{\tau }_{12}L_{10}(k),      
\end{split}
\end{gather}
where
\begin{gather} \label{discrete_stability_6} 
\begin{split}
{\tau }_{11}=\frac{1-{c_0}^2-{\ \rho }_0{c_0}^2}{{{\overline{g}}_0}^2} ,\quad   {\tau }_{12}={T_s}^2\left(1+\frac{1}{{\ \rho }_0}\right). 
\end{split}
\end{gather}

Moreover, for the second term in the Lyapunov function, the following items can be calculated.
\noindent
\begin{gather} \label{discrete_stability_7} 
\begin{split}
\Delta L_{1}\left(k\right)&= e_{1}^{2}\left(k+1\right)-e_{1}^{2}\left(k\right)\ 
+\frac{\widetilde{W}_{1}^{T}
\left(k+1\right){\widetilde{W}}_1\left(k+1\right)}{{\lambda }_{1}}-\frac{\widetilde{W}_{1}^{T}
\left(k\right){\widetilde{W}}_1\left(k\right)}{{\lambda }_1}\\ 
& \le {\tau }_{21}L_{10}(k)-{\tau }_{22}L_{11}(k)+{\tau }_{23}+{\tau }_{24}L_{20}(k),
\end{split}
\end{gather}
\noindent where
\begin{gather} \label{discrete_stability_8} 
\begin{split}
{\tau }_{21}&=\frac{{\sigma }_1}{{\lambda }_1} ,\quad \quad \quad  {\tau }_{21}=1-c_1{\rho }^{-1}_{11}+\frac{T_s{\overline{g}}_{1\ }\rho_{13}}{{\lambda }_1} , \\ 
{\tau }_{23}&=\frac{{\lambda }_1\rho_{10}{\epsilon }^2_{1}+{\sigma }_1{\left\|{W}^*_1\right\|\ }^2}{{\lambda }_1} ,\quad    {\tau }_{24}=\frac{T_s{\overline{g}}_{1\ }\rho_{13}}{{\lambda }_1} .
\end{split}
\end{gather}
Similarly, for the third term, the following conclusion can be acquired:
\begin{gather} \label{discrete_stability_9} 
\begin{split}
 \Delta L_2\left(k\right)&= {e_2}^2\left(k+1\right)-{e_2}^2\left(k\right)
+\frac{\widetilde{W}_{2}^{T}
\left(k+1\right){\widetilde{W}}_{2}\left(k+1\right)}{{\lambda }_2}-\frac{\widetilde{W}_{2}^{T}
\left(k\right){\widetilde{W}}_{2}\left(k\right)}{{\lambda }_2}\\ 
& \le {\tau }_{31}L_{20}(k)-{\tau }_{32}L_{21}(k)+{\tau }_{33}+{\tau }_{34}L_{30}(k),
\end{split}
\end{gather}
\noindent where
\begin{gather} \label{discrete_stability_10} 
\begin{split}
{\tau }_{31}&=\frac{{\sigma }_2}{{\lambda }_2},\quad \quad \quad   {\tau }_{32}=1-c_2{\rho }^{-1}_{21}+\frac{T_s\rho_{13}}{{\lambda }_2} ,\\
{\tau }_{33}&=\frac{{\lambda }_2\rho_{20}{\epsilon }^2_{2}+{\sigma }_2{\left\|{W}^*_2\right\|\ }^2}{{\lambda }_2},\quad    {\tau }_{34}=\frac{T_s\rho_{23}}{{\lambda }_2}. \\
\end{split}
\end{gather}
and for the fourth term we get
\begin{gather} \label{discrete_stability_11_0} 
\begin{split}
\Delta L_3\left(k\right)&= {e_3}^2\left(k+1\right)-{e_3}^2\left(k\right)
+\frac{\widetilde{W}_{3}^{T}
\left(k+1\right){\widetilde{W}}_{3}\left(k+1\right)}{{\lambda }_3}-\frac{\widetilde{W}_{3}^{T}
\left(k\right){\widetilde{W}}_{3}\left(k\right)}{{\lambda }_3} \\
& \le {\tau }_{41}L_{30}(k)-{\tau }_{42}L_{31}(k)+{\tau }_{43},   
\end{split}
\end{gather}
where
\begin{gather} \label{discrete_stability_11} 
\begin{split}
{\tau }_{41}&=\frac{{\sigma}_3}{{\lambda}_3},\quad {\tau }_{42}=1-c_3{\rho }^{-1}_{31}, \\
{\tau}_{43}&=\frac{{\lambda}_3\rho_{30}{\epsilon}^2_{3}+{\sigma}_3 {\left\| {W}_{3}^{*} \right\|}^2}{{\lambda}_3} . 
\end{split}
\end{gather}
Combining Eqs. \eqref{discrete_stability_5} through \eqref{discrete_stability_11}, we have the following inequality:
 \begin{gather} \label{discrete_stability_12} 
\begin{split}
\Delta L\left(k\right)\le & -{\tau }_{11}L_0\left(k\right)+{\tau }_{12}L_{10}\left(k\right)-{\tau }_{21}L_{10}\left(k\right)\\
&-{\tau }_{22}L_{11}(k)+{\tau }_{23}+{\tau }_{24}L_{20}(k)\\
&-{\tau }_{31}L_{20}(k)-{\tau }_{32}L_{21}(k)+{\tau }_{33}+{\tau }_{34}L_{30}(k)\\ 
& -{\tau }_{41}L_{30}(k)-{\tau }_{42}L_{31}(k)+{\tau }_{43}\\ 
\le &-{\tau }_{11}L_0\left(k\right)-\left({\tau }_{21}-{\tau }_{12}\right)L_{10}\left(k\right)-{\tau }_{22}L_{11}\left(k\right)\\
& -\left({\tau }_{31}-{\tau }_{24}\right)L_{20}\left(k\right)-{\tau }_{32}L_{21}\left(k\right)-\left({\tau }_{41}-{\tau}_{34}\right)L_{30}\left(k\right)\\
& -{\tau }_{42}L_{31}\left(k\right)+{\tau }_{23}+{\tau }_{33}+{\tau }_{43}\\
\le & -{\mathrm{K}}_0L_0\left(k\right)-{\mathrm{K}}_1L_1\left(k\right)-{\mathrm{K}}_2L_2\left(k\right)-{\mathrm{K}}_3L_3\left(k\right)+{\mathrm{K}}_4,
\end{split}
\end{gather}
\noindent 
\noindent 
\noindent where the proper positive parameters $\rho _{ij}$ can be designed to have the following results:
\noindent 
\begin{gather} \label{discrete_stability_13} 
\begin{split}
K_{0}&={\tau}_{11}>0, \\
K_{1}&= \min \left\{{\tau }_{21}-{\tau }_{12},{\tau }_{22}\right\}>0,\\
K_{2}&= \min \left\{{\tau }_{31}-{\tau }_{24},{\tau }_{32}\right\}>0,\\
K_{3}&= \min \left\{{\tau }_{41}-{\tau }_{34},{\tau }_{42}\right\}>0,\\
K_{4}&={\tau }_{23}+{\tau }_{33}+{\tau }_{43}>0.
\end{split}
\end{gather}
\noindent 
Take $r_\mathrm{A}=\min \left\{ \mathrm{K}_0, \mathrm{K}_1, \mathrm{K}_2, \mathrm{K}_3 \right\}$, then $\Delta L(k)\le -r_\mathrm{A}L(k)+ \mathrm{K}_4$. According to \citet{ioannou1996robust},  we know $L(k)$ is bounded so $e_i\left(k\right)$, $i=0, 1, 2, 3$ and ${W}_j\left(k\right)$, $j= 1, 2, 3$  are semi-globally uniformly ultimately bounded.

\section{How to analyze the closed-loop stability?} \label{Sec:Stability}
As mentioned before, most of the current feedback error learning-neural control schemes in the literature can only guarantee the Uniformly Ultimately Bounded (UUB) stability of the closed-loop system.
To be more precise, it is not typically possible to prove the negative definiteness of the time derivative of the Lyapunov function, but it can be proved that
\begin{equation}\label{Lyap_3}
  \dot{V}\leq -k V + C,
\end{equation}
where $k$ and $C$ denote positive constants. As a result, it is obtained that \citep{Wang.2014b,Ge.2010}
\begin{equation}\label{Lyap_4}
  V (t) \leq \left(V(0)-\frac{C}{k}\right)e^{-k t}+\frac{C}{k}.
\end{equation}
In this regard, many studies in the literature attempted to propose a flight control system, which satisfies the local \citep{Lee.2001,Shin.2004}, semi-global \citep{Xu.2011,Xian.2015,Chen.2016,Ge.2010,Chen.2015c}, or global \citep{Xu.2015} UUB tracking.
There are fewer works that have addressed more stringent stability criteria including asymptotic, exponential, or finite-time stability.

\begin{figure}[ht]
\centering
\includegraphics[width=0.8\textwidth]{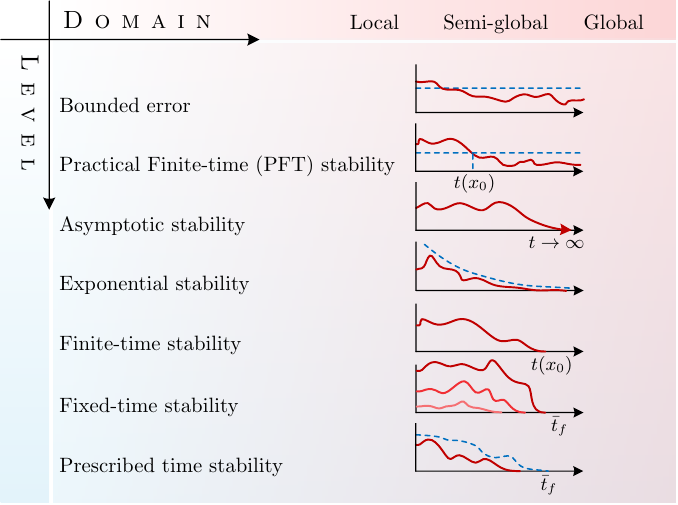}
\caption{Levels of stability in different domains}
\label{fig:stability}
\end{figure}

Fig. \ref{fig:stability} compares the rigor of different stability definitions, which can be guaranteed to different extents, i.e., locally, semi-globally, or globally.
\subsection{Asymptotic stability}\label{Sec:AsymptoticStability}
A variety of flight control systems are given in the literature, which can prove the convergence of the tracking error to zero as time tends to infinity.
As a straightforward approach, if we can assume that the estimation error $\varepsilon$ in Eq. (\ref{Lyap_3}) is zero (or negligible), it is obtained that $\dot{V}=-k_1 e^T e$, which guarantees the asymptotic convergence of the tracking error to zero. Such an assumption is reasonable in the case of dynamic systems with \emph{parametric} uncertainties. More precisely, in such a circumstance, it is possible to estimate uncertain terms as $\Delta(x)={W^*}^T \mu(x)$, where $\mu(x)$ and $W^*$ represent the vector of appropriate basis functions and the matrix of unknown weights, respectively.
According to this, a constrained adaptive backstepping control scheme has been proposed in \citet{Sonneveldt.2007} for a strict feedback system with parametric uncertainties. A set of modified tracking errors ($\bar{z}_i$) have been defined (see Chapter 3), and it has been proved that the time derivative of the Lyapunov function is obtained as $\dot{V}=-\sum c_i \bar{z}_i^2$ with $c_i$ denotes positive constants. This leads to convergence of the modified tracking errors to zero as time tends to infinity \citep{Krstic.1992}, while the actual tracking error may increase if the control inputs are saturated. Finally, the proposed control scheme was employed in \citet{Sonneveldt.2007} to control the attitude of a simplified model of an F-16 aircraft with multi-axis thrust vectoring, considering actuator faults and symmetric structural damages (only in the simulation phase).
A similar approach has been presented in \citet{Sonneveldt.2009} to provide a trajectory tracking control for an F-16 fighter aircraft under parametric uncertainties and system constraints. An adaptive backstepping controller has been introduced in \citet{Shi.2014} for the attitude control of an NSV in the presence of model uncertainties and multiplicative actuator faults. For this purpose, first, an adaptive neural state observer has been proposed (Chapter 3), and subsequently, the estimated states have been utilized in a backstepping control scheme. The asymptotic stability of the system has been proved assuming that the NN estimation error is negligible.
Obviously, such control approaches can not ensure the asymptotic stability of the system in the presence of \emph{nonparametric} uncertainties and external disturbances, which are usually present in practical flight control problems.

Several studies have been reported in the literature in which NNs are combined with \emph{discontinuous} feedback control methods such as variable structure or Sliding Mode Controllers (SMC) to guarantee the asymptotic stability of the closed-loop system. The fundamental idea of the combination of NN function approximation and robust terms (such as in SMC), which can result in the asymptotic stability of the closed-loop system, is given in Chapter 3.
A robust output feedback control with neural network function approximation has been designed in \citet{Xian.2015} for the attitude and altitude control of a quadrotor UAV in the presence of model uncertainties and external disturbances, where the attitude dynamics are constructed in terms of the unit quaternion. Although the asymptotic stability of the closed-loop system has been proved, the proposed control command leads to high-amplitude and oscillatory thrust forces, and the chattering phenomenon due to the employment of the signum function in the control command.
Indeed, such discontinuous controllers suffer from well-known limitations including a requirement for an infinite control bandwidth and chattering. Unfortunately, ad hoc fixes for these effects result in a loss of asymptotic stability \citep{Patre.2008}.
An adaptive SMC has been proposed in \citet{Razmi.2019}, where the parameters of the sliding surface were trained by NNs through error back-propagation learning. The hyperbolic tangent function has replaced the signum function to eliminate the chattering phenomenon, while the asymptotic stability of the system has been achieved by neglecting model uncertainties in the control design process.

\begin{ex}[label=Ex:SMCquad]{Inclusion of SMC into FEL-based control of quadrotor UAV}
Improve the designed control system in Example \ref{Example:quadFEL} by
\begin{enumerate}[(a)]
\item incorporating SMC into the design.
\item Ensure the asymptotic stability of the closed-loop system.
\item Evaluate the closed-loop system's ability to track the reference command
\begin{equation}\label{FEL_trajectory}
{{x}_{d}}\left( t \right)=\left[ \begin{matrix}
   {{z}_{d}}\left( t \right)  \\
   {{\phi }_{d}}\left( t \right)  \\
   {{\theta }_{d}}\left( t \right)  \\
   {{\psi }_{d}}\left( t \right)  \\
\end{matrix} \right]=\left[ \begin{matrix}
   2-1.5u_s(t-60) \,\text{m} \\
   40-25u_s(t-20) \,\text{deg} \\
   27+15u_s(t-40) \,\text{deg} \\
   5u_s(t-30) \,\text{deg}  \\
\end{matrix} \right],
\end{equation}
where $u_s(t)$ is the unit step function.

\item Plot the change of MAEE for different number of hidden nodes and different learning rates. Analyze the effect of SMC on the control performance compared with the basic FEL.
\end{enumerate}
\tcblower
Consider the following linearized dynamics for the quadrotor UAV
\begin{equation}
\ddot{x} =\bar{f}+\bar{B}u+\Delta\left(x,u\right),
\end{equation}
where $x$ is the state vector, $u$ is the control input, 
$\bar{f}$ and $\bar{B}$ represent the assumed linear dynamics and $\Delta\left(x,u\right)$ describes the whole uncertainties.
\begin{enumerate}[(a)]
\item \textbf{Sliding mode controller design:}
To design a SMC, first we define a sliding surface $s\left(x\right)=0$, representing the desired dynamics. A suitable choice is
\begin{equation}
s\left(x\right) = \dot{e} + \lambda^T e,
\end{equation}
where $\lambda>0$ is a constant parameter. Similar to Example \ref{Example:quadFEL}, $\dot{s}$ can be written as
\begin{align}\label{Ex214:sd}
\begin{split}
 \dot{s} &= \ddot{e}+\lambda^T \dot{e} \\
       &= \bar{f} + \bar{B}u + \Delta\left(x,u\right) - \ddot{x}_d + \lambda^T \dot{e}.
\end{split}
\end{align}

Similar to before, we approximate $\Delta$ with an NN as $\hat{\Delta}=\hat{W}^T \mu\left(X\right)$. 
Expressing the dynamics of the sliding surface by defining a constant-proportional ``reaching law'' gives
\begin{equation}\label{Ex214:reachlaw}
\dot{s} = -k_1s-Q \text{sign}\left(s\right).
\end{equation}

We now proceed to find an estimation for $Q$ with an adaptive law, denoted by $\hat{Q}$. As before, we also use the notation $\tilde{Q} = \hat{Q} - Q$ to represent the estimation error.

Now, combining Eqs. (\ref{Ex214:sd}) and (\ref{Ex214:reachlaw}), the control law is designed to maintain the sliding condition, along with an extra term to force the state trajectory to reach the sliding surface, expressed as
\begin{equation}\label{Eq:Ex214:u}
u=\bar{B}^{-1}\left(-\bar{f}-\hat{\Delta}+\ddot{x}_d-\lambda^T \dot{e}-k_1 s - \hat{Q} \text{sign}\left(s\right) \right),
\end{equation}
where, in the SMC literature, the former is commonly referred to as the equivalent control $u_{eq}$, and the latter is called the discontinuous control $u_{dis}$.


\item \textbf {Stability analysis:}
Consider the following Lyapunov function candidate
\begin{equation}
V=\frac{1}{2}s^T s + \frac{1}{2} \text{tr} \left(\tilde{W}^T \Gamma^{-1} \tilde{W}\right) + \frac{1}{2}\tilde{Q}^T\Lambda^{-1}\tilde{Q},
\end{equation}
where $\Lambda$ is a positive definite matrix.
The time derivative of $V$ is given by
\begin{align}
\begin{split}
\dot{V} &= s^T \dot{s}+ \text{tr} \left(\tilde{W}^T \Gamma^{-1} \dot{\tilde{W}}\right) + \tilde{Q}^T\Lambda^{-1}\dot{\tilde{Q}}\\
&= - s^T \left( \tilde{\Delta} + k_1s + \hat{Q} \text{sign}\left(s\right) \right)+ \text{tr} \left(\tilde{W}^T \Gamma^{-1} \dot{\hat{W}}\right) + \tilde{Q}^T\Lambda^{-1}\dot{\hat{Q}}.\\
\end{split}
\end{align}

It is notable that since the control law is unaware of $\Delta\left(x,u\right)$ or $Q$, we were unable to substitute $\dot{s}$ from Eq. (\ref{Ex214:reachlaw}). Using the update rule $\dot{\hat{W}} =\Gamma \mu s^T$, knowing that $\tilde{\Delta}=\tilde{W}^T \mu (x,u) - \varepsilon$, we have
\begin{equation}
\dot{V}  = - s^T \left( k_1s - \varepsilon + \hat{Q} \text{sign}\left(s\right) \right) + \tilde{Q}^T\Lambda^{-1}\dot{\hat{Q}}.
\end{equation}

Assuming the upper bound of NN approximation to be less than the unknown term in the reaching law, i.e., $\left\| \varepsilon \right\| \leq \varepsilon_M \leq Q$, we get
\begin{align}
\begin{split}
\dot{V}  &= - s^T \left( k_1s - \varepsilon + \hat{Q} \text{sign}\left(s\right) \right) + \tilde{Q}^T\Lambda^{-1}\dot{\hat{Q}}\\
&\leq -k_1s^Ts + \left|s\right|^T Q - \left(s \, \text{sign}\left(s\right)\right)^T\hat{Q}+\tilde{Q}^T\Lambda^{-1}\dot{\hat{Q}}.
\end{split}
\end{align}

From the property $\left|s\right|=s \, \text{sign}\left(s\right)$, we get
\begin{align}
\begin{split}
\dot{V} &\leq -k_1s^Ts  - \left|s\right|^T \tilde{Q}+\tilde{Q}^T\Lambda^{-1}\dot{\hat{Q}}.
\end{split}
\end{align}

Consider the following adaptive update rule for the estimate $\hat{Q}$, given by
\begin{equation}\label{Eq:Ex214:Qdot}
\dot{\hat{Q}}=\Lambda \left|s\right|^T.
\end{equation}

Thus, we have
\begin{equation}
\dot{V}  \leq -k_1s^Ts \leq 0.
\end{equation}

Therefore, the origin of the sliding surface is asymptotically stable. By properly selecting the sliding surface, the tracking error can be driven to zero, achieving asymptotic tracking of the desired state trajectory.

It is important to note that an alternative function, such as the hyperbolic tangent (tanh), can be employed instead of the signum function to reduce the chattering phenomenon, which is a high-frequency oscillation in the control command that can damage the actuators. We do not get into the stability analysis procedure for the tanh function; however, the comparison of its results will be presented here. To this end, we need to switch $\text{sign}\left(s\right)$ with $\tanh\left(s/\alpha\right)$ in the expressions presented in Eqs. (\ref{Eq:Ex214:u}) and (\ref{Eq:Ex214:Qdot}), selecting a suitable value of $\alpha=0.01$.

\item \textbf {Simulation results:}
To increase the difficulty of uncertainty estimation for the NN, we add parametric uncertainty to the model by using a length of $l=0.375 \text{m}$ instead of 0.25, and half of the true value of $c$ in the controller for the matrix $B_2$ described in Eq. (\ref{eq:exc21:FBlinear}).
Unlike Example \ref{Example:quadFEL} in which the gains were selected as scalars, here we consider them as vectors and for the sake of sensitivity analysis, we added a constant multiplier for each gain. To this end, we have $k_1=k_0 \left[ 0.5, 1, 0.5, 0.5\right]^T$, $\lambda=\lambda_0\left[ 0.2, 1.0, 0.5, 0.25\right]^T$, and $\Lambda=\Lambda_0\left[ 1, 0.1, 0.1, 0.01\right]^T$.

The state trajectory for the attitude and altitude of a case with 15 neurons, $k_0=40$, $\lambda_0=13$, $\Gamma=50$, and $\Lambda_0=2$, is shown in Fig. \ref{fig:exc210:stat}.
\begin{figure}[H]
  \begin{center}
    \includegraphics[width=0.9\textwidth]{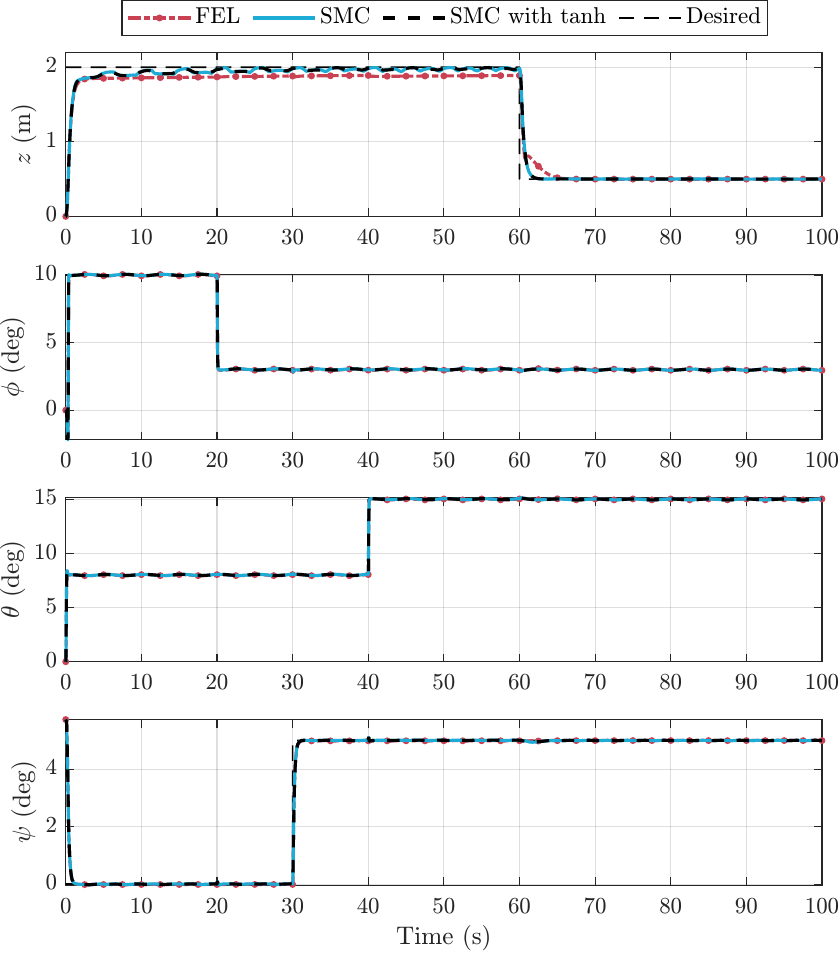}
  \end{center}
  \caption{Comparison of state trajectory between SMC and basic FEL.} 
  \label{fig:exc210:stat}
\end{figure}
Overall, what stands out from the figure is that the addition of SMC is aiding FEL-based controller to improve the tracking behavior. Specifically, although the altitude tracking in basic FEL method has a significant steady-state error, SMC appears to compensate for it. Additionally, it appeared that both signum function and tanh are performing similarly and their trajectories are approximately similar to each other.

Moving on to the control effort, SMC introduces high-frequency oscillations in its final control signal, as illustrated in Fig. \ref{fig:exc210:eff}. Comparing Figs. \ref{fig:exc210:stat} and \ref{fig:exc210:eff} shows that while sign and tanh act similar in tracking performance, the chattering phenomenon is completely removed by utilizing tanh in the control command and the update rule.
\begin{figure}[H]
  \begin{center}
    \includegraphics[width=0.9\textwidth]{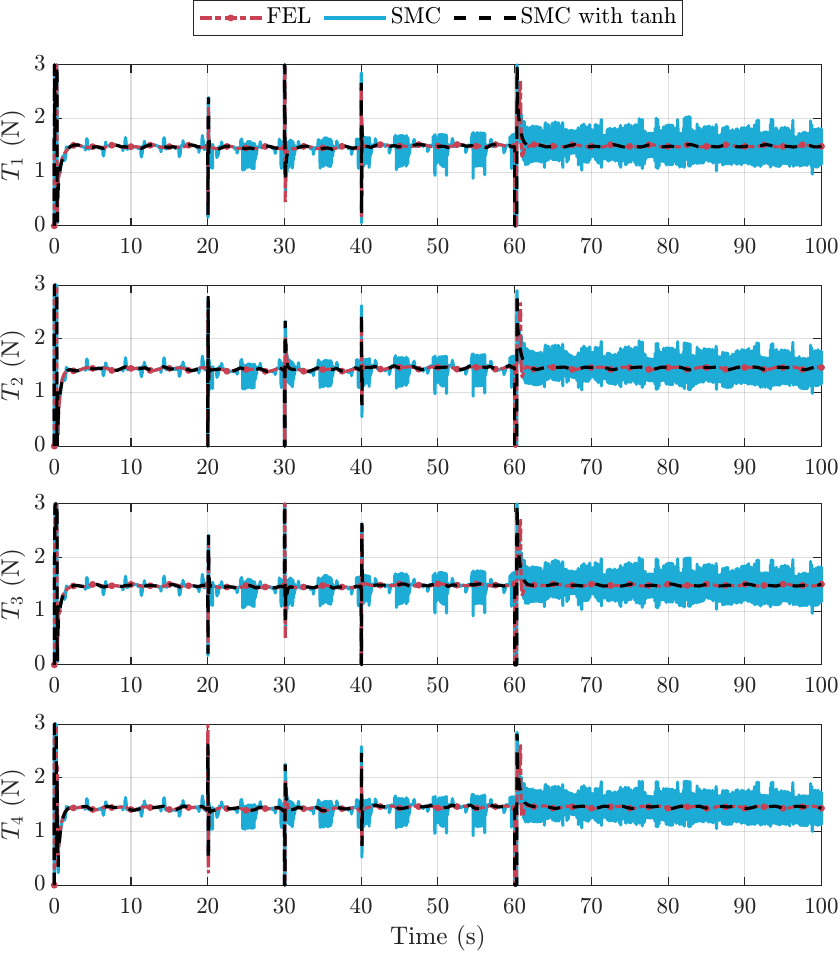}
  \end{center}
  \caption{Comparison of control inputs between SMC and basic FEL.} 
  \label{fig:exc210:eff}
\end{figure}

The comparison of the estimation error $\tilde{\Delta}_i$ for the two methods is shown in Fig. \ref{fig:exc210:del}. Both methods use the same NNs, so the learning of uncertainty is generally similar. However, SMC exhibits high-frequency behavior in the estimation error due to the noise in the commanded control. It should be noted that the estimation error is a function of the input control and the same high-frequency response in the control effort is also seen here.

Fig. \ref{fig:exc210:del} shows that the NN estimation is unable to fully learn the underlying system dynamics. To this end, we can conclude that the excitation is not persistent enough and the learning algorithm would need a richer input. It should be noted that the input consisted of pure step signals, with their derivatives being zero which seems to being unable to excite the whole dynamics of the quadrotor.
\begin{figure}[H]
  \begin{center}
    \includegraphics[width=0.9\textwidth]{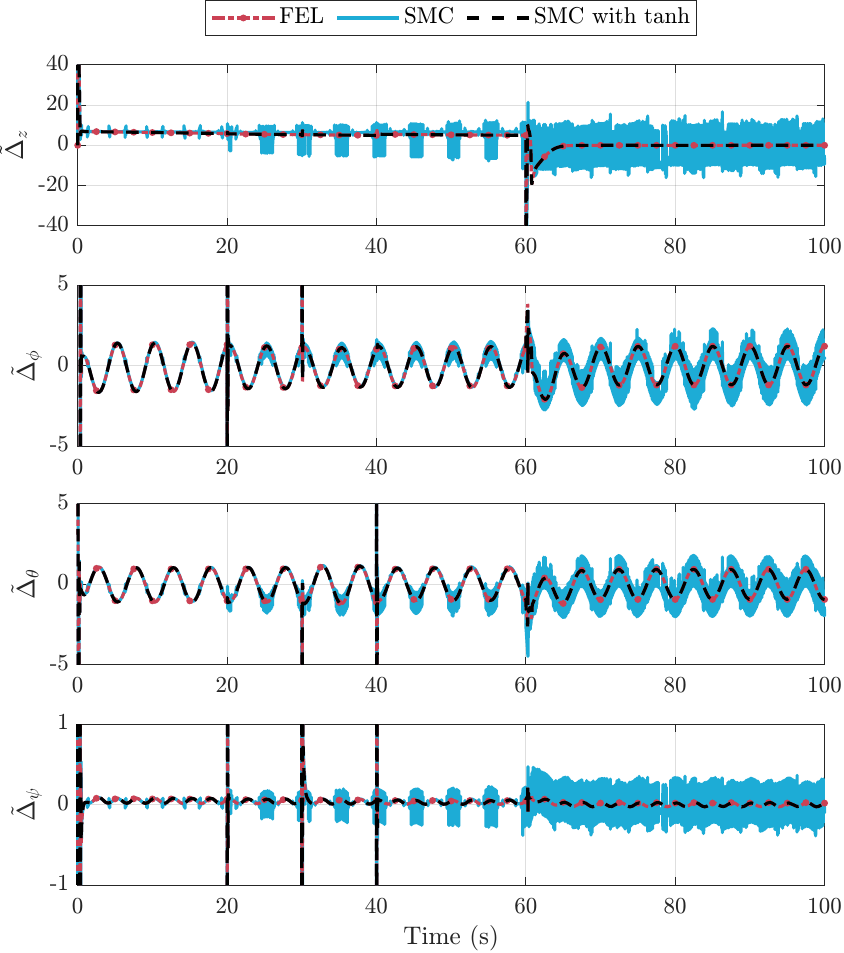}
  \end{center}
  \caption{Uncertainty estimation error for SMC compared to basic FEL.} 
  \label{fig:exc210:del}
\end{figure}

Additionally, Fig. \ref{fig:exc210:weigh} shows the evolution of coefficient $Q$ in the reaching law of SMC.
The coefficients does not seem to converge to a constant value due to the previously mentioned issue of lacking PE, as well as the fact that we did not use any modification terms (such as $\sigma$-modification) in the adaptive rule of Eq. (\ref{Eq:Ex214:Qdot}).
\begin{figure}[H]
  \begin{center}
    \includegraphics[width=0.9\textwidth]{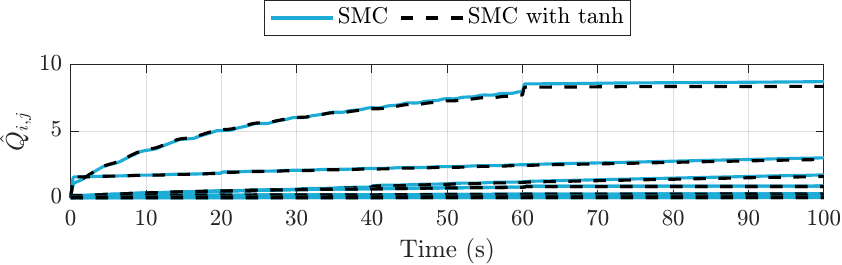}
  \end{center}
  \caption{Evolution of the adaptive parameter $Q$ in the SMC control law.} 
  \label{fig:exc210:weigh}
\end{figure}

To sum up, the mean tracking error and estimation performance is compared in Table \ref{tab:exc21_MAE_MAEE}. Examining the results reveals that while SMC performs better in terms of tracking the desired altitude, there is also a negative impact on the overall NN estimation, and the yaw angle tracking is slightly impaired.

\begin{table}[H]
    \caption{Comparison of tracking and estimation performance among SMC, SMC with tanh, and basic FEL.}
    \label{tab:exc21_MAE_MAEE}
    \centering
    \begin{tabular}{cccccc}
    \hline\noalign{\smallskip}
    & MAE ($z$) & MAE ($\phi$) & MAE ($\theta$) & MAE ($\psi$) & MAEE \\
    \noalign{\smallskip}\hline\noalign{\smallskip}
    FEL & $1.11\times 10^{-1}$ & $1.28\times 10^{-3}$ & $7.22\times 10^{-4}$ & $6.96\times 10^{-4}$ & $5.71$ \\
    tanh & $4.96\times 10^{-2}$ & $1.28\times 10^{-3}$ & $7.83\times 10^{-4}$ & $6.79\times 10^{-4}$ & $6.47$ \\
    SMC & $5.05\times 10^{-2}$ & $1.27\times 10^{-3}$ & $7.74\times 10^{-4}$ & $6.75\times 10^{-4}$ & $9.16$ \\
    \noalign{\smallskip}\hline\noalign{\smallskip}
    \end{tabular}
\end{table}

\item \textbf {Sensitivity analysis:}
In order to investigate the effect of employing different parameters on the control performance, we perform a parameter study on tracking and estimation error.

From Fig. \ref{fig:exc21:param_1}, the following results can be drawn:
\begin{itemize}
    \item The variation of learning rate does not have a considerable impact on MAE or MAEE, since the input signal is not PE.
    \item The increase of $\lambda$ coefficient in the sliding surface expression while improving the tracking error, reduces the estimation accuracy.
    \item Utilizing higher gains of $k_1$ can initially reduce the estimation error, but beyond a certain point, it will actually increase the error. This is because higher gains result in greater oscillations in the chattering, which is equivalent to a larger estimation error (MAEE). Additionally, higher gains tend to be advantageous for the tracking error; however, it may cause unwanted initial overshoot.
    \item Increasing the value of the coefficient $\Lambda$ in the adaptive rule of the reaching law leads to chattering and degrade the estimation accuracy. However, it can still improve the tracking error up to a certain point, regardless of further increases in the coefficient.
\end{itemize}
\begin{figure}[H]
  \begin{center}
    \includegraphics[width=0.9\textwidth]{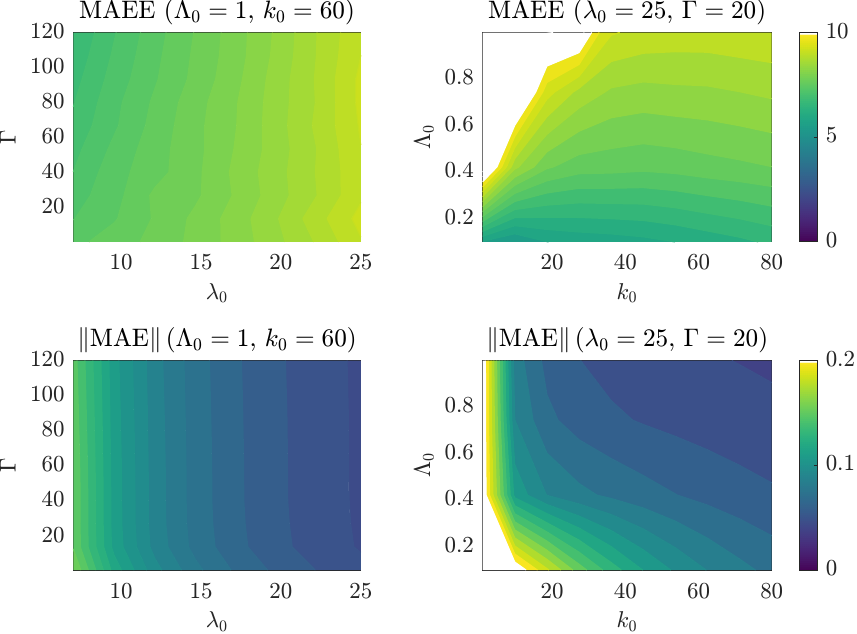}
  \end{center}
  \caption{Sensitivity analysis for learning rate, coefficient $\lambda$ in the sliding surface, coefficient $\Lambda$ in the reaching law and the gain $k_1$ for the SMC controller.} 
  \label{fig:exc21:param_1}
\end{figure}

\end{enumerate}
\end{ex}

To overcome the above-mentioned issues, a trajectory tracking control system has been introduced in \citet{JonghoShin.2012} for a rotorcraft UAV using Robust Integral of the Signum of the Error (RISE) feedback \citep{Patre.2008}, where a NN has been adopted to compensate for uncertain dynamics. The RISE control scheme is a differentiable control method that can compensate for additive disturbances and parametric uncertainties. By combining it with an NN-based FEL method, there is no need for linearity in the parameters to ensure the asymptotic stability of the system.
The proposed approach was employed in \citet{JonghoShin.2012} in a multi-loop control structure, where the desired attitude is determined in the outer loop and the attitude tracking control has been addressed in the inner loop. The semi-global asymptotic stability of the inner loop in tracking the desired attitude has been proved assuming that the first four time derivatives of the reference trajectory are bounded.
Another alternative has been given in \citet{Wang.2014b,Li.2020}, where the signum function of $e$ has been substituted by $e/\left(\sqrt{e^2+\omega^2}\right)$ with $\omega(t)$ denotes a vanishing positive function satisfying $\int_{0}^{\infty}\omega^2(t) dt<\infty$. Accordingly, asymptotic tracking can be achieved, while the updating rules are subject to possible parameter drift.
A class of NN-based optimal control methods with guaranteed asymptotic stability has also been introduced in the literature, which will be addressed in Chapter 5.

\subsection{Exponential stability}
As stated in \citet{Mo.2018}, in the case of a flight control problem, few studies have claimed to achieve exponential convergence \citep{Zhang.2018c}. This is even less so regarding uncertain dynamic systems. As mentioned earlier, an improved learning method has been introduced in \citet{Chowdhary.2014}, which can ensure the exponential parameter and tracking error convergence for a specific class of single-input nonlinear systems with parametric uncertainties assuming that a precise estimation of the time derivative of system states is available.

\subsection{Finite-time stability}
Moreover, a variety of indirect NN-based flight control systems has been developed in the literature employing SMC-based methods, which can guarantee the finite-time or \emph{practical finite-time} stability of the closed-loop system. The practical finite-time stability means that the tracking error converges to a small neighborhood of the origin in finite time \citep{Zhu.2011}.
To clarify the principal idea of the mentioned control methods, consider again the first control problem given in Section \ref{Sec:FEL}. Let's define a sliding manifold as
\begin{equation}\label{Eq:20_10}
  s=e+\eta \int_{0}^{t} \text{sig}^r(e) dt,
\end{equation}
where $\text{sig}^r(e)=[\text{sign}(e_1)|e_1|^r,\cdots,\text{sign}(e_n)|e_n|^r]^T$, $\eta>0$, and $0<r<1$. Now, if the control command is computed as
\begin{align}
\begin{split}
     u= B^{-1}  \big(\dot{x}_d-F(x)-\eta \text{sig}^r(e)-\hat{\Delta}(x)
                       -k_1 s -k_2 \text{sig}^r(s) \big), \label{Eq:20_11}
\end{split}
\end{align}
where the same NN function approximation as Eq. (\ref{NN_1}) with an updating rule as $\dot{\hat{W}}=\Gamma \left(\mu s^T-\sigma \hat{W}\right)$ is incorporated, using a Lyapunov function as
\begin{equation}\label{Eq:20_12}
  V=\frac{1}{2}s^T s + \frac{1}{2} \tr \left(\tilde{W}^T \Gamma^{-1} \tilde{W}\right),
\end{equation}
it is easy to prove the satisfaction of Eq. (\ref{Lyap_3}), thereby guaranteeing the boundedness of all signals in the closed-loop system. As a consequence, we have
\begin{equation}\label{Eq:20_13}
  \dot{s}\leq -k_1 s - k_2 \text{sig}^r(s) + \rho_M I,
\end{equation}
where $I$ and $\rho_M$ denote the identity matrix and a positive constant satisfying
\begin{equation}\label{Eq:20_14}
  \left\|\varepsilon-\tilde{W}^T\mu(x)\right\|_\infty \leq \rho_M.
\end{equation}
Thus, by appropriate choice of $k_1$ and $k_2$, one can simply show the convergence of $s$ to a compact neighborhood of the origin in finite time \citep{Xu.2018b}.
Such a method is extensively discussed within the framework of adaptive Terminal SMC (TSMC).
Further, by incorporating a robust term into the control command (in a similar manner as discussed in Chapter 3) to compensate for $\rho_M$, and replacing the first term in the Lyapunov function (\ref{Eq:20_12}) by $\|s\|$, it is possible to guarantee the finite-time stability of the closed-loop system \citep{Yu.2018}.
An FTC has been introduced in \citet{Yu.2018} to control the longitudinal mode of a conventional aircraft using a similar SMC, which ensures the finite-time convergence of $s$ to zero under appropriate control gains. Self-constructing fuzzy neural networks have been utilized to estimate the bound of uncertain dynamics caused by actuator faults and model uncertainties, while the minimum estimation error $\varepsilon$ under optimal network weights ($W^*$) has been neglected.
A TSMC augmented by neural approximation and disturbance observers (as given in Chapter 3) has been presented in \citet{Xu.2018b} for the trajectory tracking control of a quadrotor aerial robot under model uncertainties, input dead-zone, and external disturbances, where the practical finite-time stability of the system has been ensured. A sliding surface, which is equal to the time derivative of (\ref{Eq:20_10}) due to the second-order dynamics of the controlled system, has been utilized in the proposed design.
A formation flight control problem for a group of helicopter UAVs has been dealt with in \citet{Wang.2018b} using an analogous control scheme. TSMC combined with neural approximation and the above-mentioned robust term has been adopted in both the position and attitude control loops, which can ensure, respectively, the finite-time and practical finite-time stability of the position and attitude tracking error, while the control loops have been decoupled based on the multiple time-scale assumption.
The same problem has been addressed in \citet{Wang.2019c} using TSMC in the inner control loop and an adaptive NN-based control scheme in the outer loop, though, again, the control loops have been analyzed separately. The inter-vehicle collision avoidance has also been solved by incorporating an exponential potential function into the design process. The proposed control structure can guarantee the practical finite-time stability of the closed-loop system, while it requires only the \emph{relative} position of UAVs to their adjacent.

\begin{ex}{Finite-time attitude control of satellite}
Design a finite-time control system for a satellite that can achieve accurate tracking of the desired Euler angles in a finite amount of time, in the presence of uncertainties using the model in Example \ref{sat_fel}.
	\begin{enumerate}[(a)]
		\item Design the finite-time controller and analyze its closed-loop stability.
		\item Quantify the tracking error of the designed control system with respect to the reference input in Eq. \eqref{sat_desired}
		\item Plot the changes in training weights over time.
		\item Plot estimation of uncertainties and report MAE as well as MAEE for both methods.
		\item Examine the results obtained from finite-time control and asymptotic control.
	\end{enumerate}
	\tcblower
	The goal of this example is designing a finite-time attitude control of a satellite in the presence of moment of inertia uncertainties and external disturbances.
	\begin{enumerate}[(a)]
		\item Since the attitude equations are second-order, the filtered tracking error is defined as
		\begin{equation}
			s = \dot{e} + \eta \mathrm{sig}^r \left( e \right).
		\end{equation}
		For this approach, a finite-time control is designed according to the following equation
		\begin{equation}\label{u_ft}
			u = B^{-1} \left( -F - \hat{\Delta} - \eta \dv{}{t} \mathrm{sig}^r \left( e \right) - k_1 s - k_2 \mathrm{sig}^r \left( s \right) + \ddot{x}_d \right).
		\end{equation}
		\textbf{Stability analysis:} Now, a Lyapunov function candidate is defined as
		\begin{equation}
			V = \dfrac{1}{2} s^T s + \dfrac{1}{2 \Gamma} \tr \left( \tilde{W}^T \tilde{W} \right).
		\end{equation}
		Taking time-derivative of $ V $, we have
		\begin{equation}
			\begin{split}
				\dot{V} & = s^T \dot{s} + \dfrac{1}{\Gamma} \tr \left( \tilde{W}^T \dot{\hat{W}} \right) \\
				& = s^T \left( \ddot{e} + \eta \dv{}{t} \mathrm{sig}^r \left( e \right) \right) + \dfrac{1}{\Gamma} \tr \left( \tilde{W}^T \dot{\hat{W}} \right) \\
				& = s^T \left( \ddot{x} - \ddot{x}_d + \eta \dv{}{t} \mathrm{sig}^r \left( e \right) \right) + \dfrac{1}{\Gamma} \tr \left( \tilde{W}^T \dot{\hat{W}} \right) \\
				& = s^T \left( F + Bu + \Delta - \ddot{x}_d + \eta \dv{}{t} \mathrm{sig}^r \left( e \right) \right) + \dfrac{1}{\Gamma} \tr \left( \tilde{W}^T \dot{\hat{W}} \right),
			\end{split}
		\end{equation}
		where $ \dv{}{t} \mathrm{sig}^r \left( e \right) $ is defined as
		\begin{equation}
			\dv{}{t} \mathrm{sig}^r \left( e \right) = r \abs{e}^{r - 1} \dot{e}.
		\end{equation}
		If NN's update rule is considered as
		\begin{equation}\label{nn_update_ft}
			\dot{\hat{W}} = \Gamma \mu s^T - \sigma \hat{W},
		\end{equation}
		and substituting Eq. \eqref{u_ft}, we can show that
		\begin{equation}
			\dot{V} \leq -k V + C.
		\end{equation}
		
		\item \textbf{Simulation results:}
		Fig. \ref{fig:SAT_FT_State} shows the results of finite-time tracking of Euler angles using $ k_1 = 3 $, $ k_2 = 3 $, $ \eta = 3 $, $ \Gamma = 3 $, $ r = 0.9 $, and $ \sigma = 0.0001 $.
		It should be mentioned that there must be a compromise between the selection of control coefficients ($ k_1$, $ k_2 $, and $ \lambda $) and the learning rate ($ \Gamma $) of NN. Increasing $ \Gamma $ leads to higher oscillations during the initial times. Furthermore, increasing the control gains results in an increase in the control input magnitude at the initial moments.
		\begin{figure}[H]
			\centering
			\includegraphics[width=0.9\textwidth]{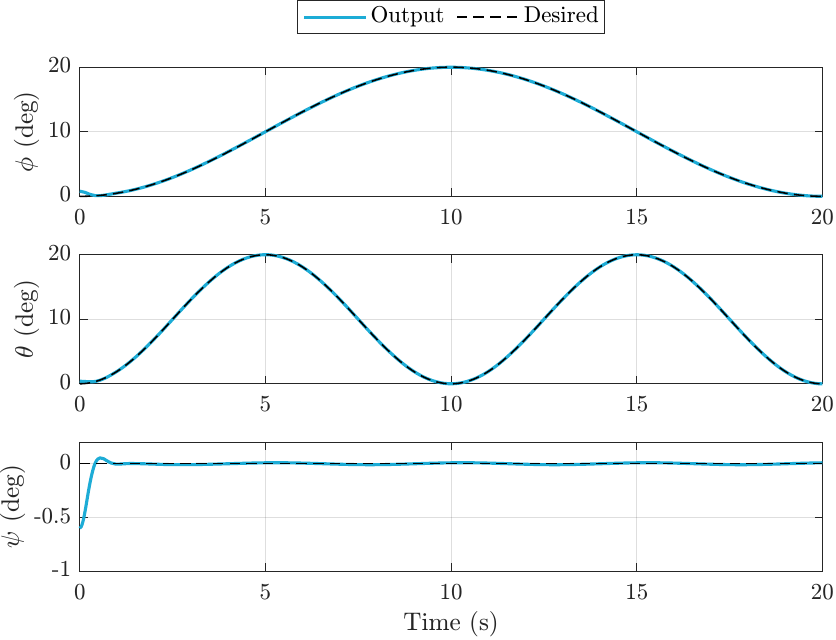}
			\caption{Time history of attitude in finite-time control of satellite.}
			\label{fig:SAT_FT_State}
		\end{figure}
		In addition, Fig. \ref{fig:SAT_FT_Control} shows the required control torques. As is clear, for finite-time control of the satellite, control inputs are within a limited and reasonable range. 		
		\begin{figure}[H]
			\centering
			\includegraphics[width=0.9\textwidth]{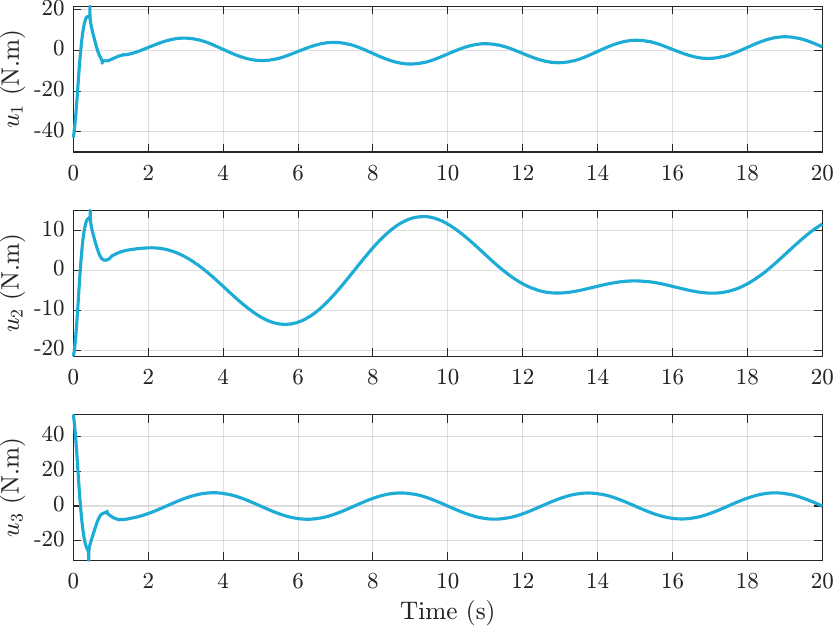}
			\caption{Time history of control inputs in finite-time control of satellite.}
			\label{fig:SAT_FT_Control}
		\end{figure}
		\item Variation of neural networks' weights is shown in Fig. \ref{fig:SAT_FT_W}.
		As previously mentioned, the impact of a high learning rate is also evident during the initial moments.
		\begin{figure}[H]
			\centering
			\includegraphics[width=0.9\textwidth]{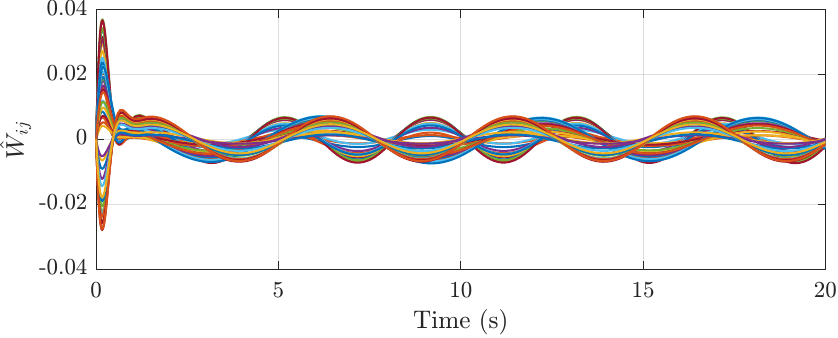}
			\caption{Evolution of training weights in finite-time control of satellite.}
			\label{fig:SAT_FT_W}
		\end{figure}

		\item Fig. \ref{fig:SAT_FT_Est} shows the performance of uncertainty estimation in finite-time. As this figure shows, the initial deviation is due to the high value of the learning rate.
		\begin{figure}[H]
			\centering
			\includegraphics[width=0.9\textwidth]{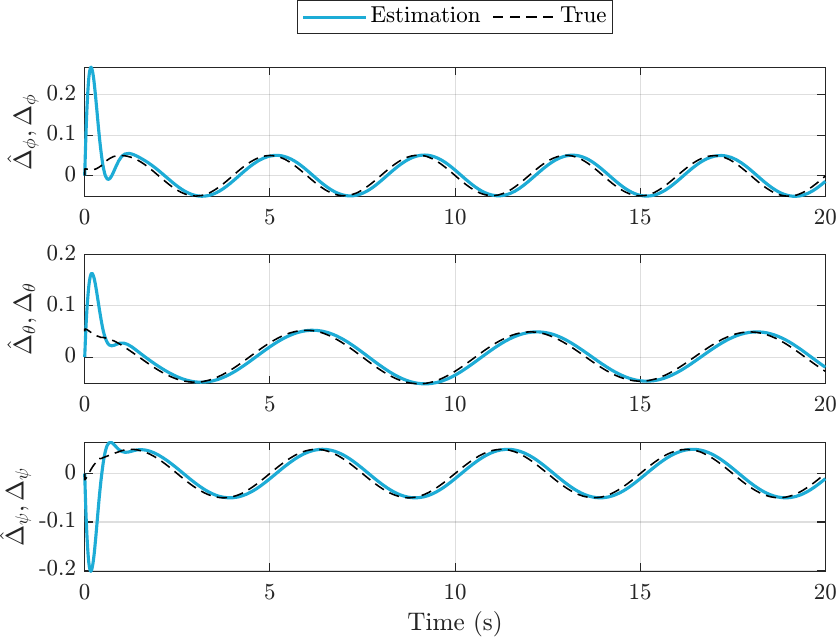}
			\caption{Uncertainty estimation in finite-time control of satellite.}
			\label{fig:SAT_FT_Est}
		\end{figure}
		Furthermore, MAE as well as MAEE results for both approaches are provided in Table \ref{tab:sat_ft}, showing that the proposed  finite-time control approach can achieve accurate tracking of the desired Euler angles.
		\begin{table}[H]
			\centering
			\caption{Results of MAE and MAEE in finite-time control of satellite.}
			\begin{tabular}{c c c c}
				\hline\noalign{\smallskip}
			    MAE ($ \phi $) & MAE ($ \theta $) & MAE ($ \psi $) & MAEE \\
			    \noalign{\smallskip}\hline\noalign{\smallskip}
				$ 2.837 \times 10^{-4} $ & $ 1.618 \times 10^{-4} $ & $ 1.146 \times 10^{-4} $ & $ 1.85 \times 10^{-2} $ \\ \noalign{\smallskip}\hline\noalign{\smallskip}
			\end{tabular}
			\label{tab:sat_ft}
		\end{table}
		\item 
		One way to examine the results obtained from finite-time control is to look at the upper bound on the reaching time.
		Using the introduced filtered tracking error, an upper bound of reaching time is finite. In contrast, in the FEL method, the reaching time is equal to infinity.
		To illustrate the difference between finite-time stability and asymptotic stability, Fig. \ref{fig:SAT_FT_SA_States} shows the time history of states for different values of $ r $:
		\begin{figure}[H]
			\centering
			\includegraphics[width=0.9\textwidth]{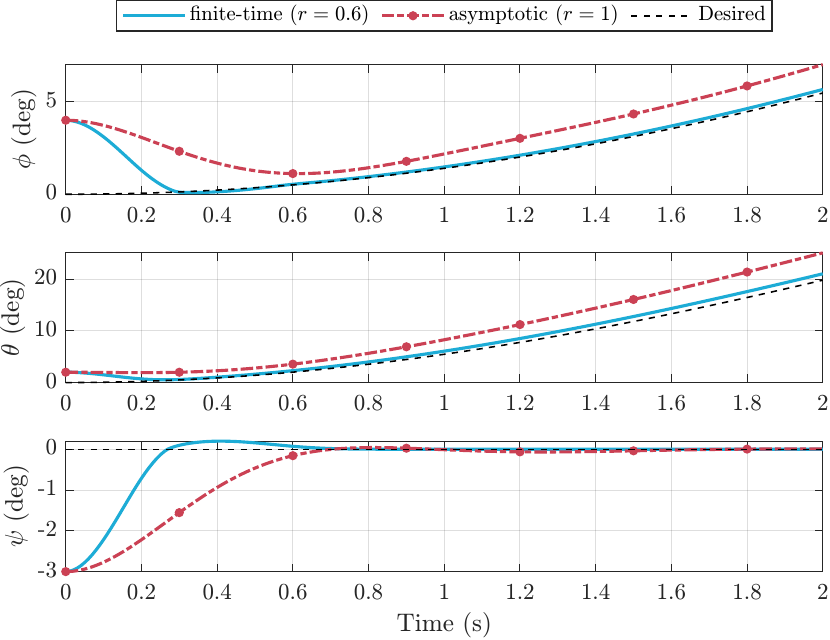}
			\caption{Finite-time convergence and asymptotic convergence.}
			\label{fig:SAT_FT_SA_States}
		\end{figure}
		As is clear from the figure, using the finite-time method, the states reach their desired values in a shorter amount of time.
	\end{enumerate}
\end{ex}

In addition to the aforementioned indirect adaptive control schemes, Direct T2-FNNs have been employed in \citet{Kayacan.2012,Kayacan.2016b,Camci.2018} as an augmentation for a PD controller in the case of SISO nonlinear systems, where the network parameters are updated by an SMC-based algorithm (with a sliding surface as $s=\dot{e}+\eta e$) using the output of the PD controller as the learning signal. The practical finite-time stability of the closed-loop system has been proved using a simple Lyapunov function as $V=1/2s^2$ assuming that a PD controller can stabilize the system \citep{Kayacan.2016b}. Six T2-FNNs have been used in \citet{Camci.2018} to control the trajectory of a 6-DOF quadrotor air vehicle, where each T2-FNN corresponds to a distinct system state. Due to the complexity of computing the gradient of the cost function with respect to the antecedent parameters of the FNN, particle swarm optimization has been adopted as a gradient-free approach to train them, while the consequent parameters have been updated using the mentioned SMC-based algorithm.
Notice that, despite the superior convergence properties of the above-mentioned approaches, as discussed earlier, they are subject to considerable  limitations of discontinuous control systems.

\section{Concluding remarks}
\subsection{Basic FEL neural control}
\begin{itemize}
    \item \emph{Feedback error learning}: FEL can be applied in both indirect and direct frameworks using either continuous-time or discrete-time formulations.
    \item \emph{Direct FEL control}: In the direct method, increasing learning rate beyond a certain threshold, results in deterioration in tracking performance.
    \item \emph{NN estimation power}: The basic controller in the neural FEL scheme can utilize the linearized dynamic model, and as shown in Examples \ref{Example:quadFEL} and \ref{Example:GTM_FEL}, NN is able to account for nonlinear terms as well as model uncertainties (under the contraction mapping assumption).
    \item \emph{Learning rate}: The increment of the learning rate improves the tracking performance and estimation accuracy at the expense of larger amplitude oscillations at initial times or after sudden changes in the model/environment.
    \item \emph{Persistent Excitation}: The convergence of NN weights to their ideal values requires the persistent excitation condition, which is not easy to guarantee in real applications. The lack of persistent exciting input signals may lead to the undesirable parameter drift phenomenon. In such a circumstance,  increasing the learning rate does not improve the performance. Additionally, the weights do not seem to converge to constant values. In particular, in Example \ref{Ex:SMCquad}, FEL was unable to identify the whole uncertainties and the robust term was utilized to improve the control performance.
    \item \emph{Training rule modifications}: Different modifications, including dead-zone, projection, sigma-modification, and $e$-modification, were employed in the NNs update rule in the FEL method to prevent parameter drift, while they could not ensure the convergence of the parameters to the ideal values and also led to reduced estimation quality and tracking performance. 
    \item \emph{$e$-modification}: The $e$-modification method excels among the modifications, demonstrating superior performance in both estimation and tracking.
    \item \emph{Improved learning algorithms}: Some improved learning approaches such as composite learning, concurrent learning and reinforced learning were introduced to enhance the learning process, thereby resulting in a better estimation of uncertain terms in the closed-loop control.  
    \item \emph{Composite learning}: Typically, the estimation error of a state observer is utilized as an aid in the learning rule, while it was shown that employment of the NN prediction error could satisfactorily eliminate the initial oscillations from the estimation.
\end{itemize}
\subsection{Pseudocontrol}
\begin{itemize}
    \item \emph{Approximate model inversion}: Pseudocontrol enables adaptive control of nonlinear non-affine systems using approximate model inversion in which a NN compensates for the model inversion error using the FEL approach.
    \item \emph{Contraction mapping} assumption is needed to ensure the existence and uniqueness of the control command in such a control framework.
    \item \emph{Non-affine model}: In the implementation of the pseudocontrol method for a satellite, a dead-zone was applied to the control surface, transitioning the satellite's model from an affine to a non-affine one. 
\end{itemize}
\subsection{Neural backstepping control}
\begin{itemize}
    \item \emph{Backstepping control} recursively stabilizes nonlinear systems with a strict feedback form.
    \item \emph{Explosion of complexity}: The key challenge in the basic backstepping method is the explosion of terms from repeated differentiation of virtual control commands.
    \item \emph{Modified Backstepping schemes}: Dynamic surface control, command-filtered backstepping, and employing sliding mode differentiators can resolve this issue, while resulting in semi-global stability.
    \item \emph{Command-filtering} results in the introduction of a lag into the system's response, leading to more oscillations after sudden changes in the model (like the fault occurrence). 
    \item \emph{Siding model differentiation}: FOSMD (or higher-order ones) can result in an accurate differentiation (in specific cases) with no lag time, while due to the utilization of the signum function, it suffers from the undesirable chattering phenomenon affecting estimation accuracy and tracking performance.
    \item \emph{Direct backstepping control}: In the direct method, NN can have the responsibility to estimate either the entire command or particular uncertain components. The latter will result in a more efficient design.
    \item \emph{Adaptive backstepping}: Neural backstepping provides a systematic way to handle nonparametric uncertainties in complex high-order systems.
\end{itemize}
\subsection{Stability analysis}
\begin{itemize}
    \item \emph{Error Boundedness}: Mostly, adaptive neural control systems can only ensure ultimate boundedness, not asymptotic or more stringent stability criteria. Stronger stability like asymptotic, exponential, and finite-time stability require extra assumptions or robust terms.
    \item \emph{Discontinuous control} methods (like SMC) combined with neural approximation can improve the tracking performance at the expense of infinite control bandwidth and chattering. The substitution of the signum function with hyperbolic tangent function reduces this undesirable phenomenon.
    \item \emph{Finite-time stability}: When employing a finite-time approach to design NN-based controller, the convergence time is restricted and, notably, dependent on the Lyapunov function's value at the initial instant. On the other hand, the stability analysis and the the control input structure become more intricate in such a condition to empower the system to attain finite-time convergence.
\end{itemize}


%
%
%

\appendix
\chapter{Case Studies: Derivation of Governing Equations}
\label{introA} 

\section{Quadrotor UAV}
\label{sec:A1}
A quadrotor is a type of unmanned aerial vehicle (UAV) that uses four rotors to generate thrust and control its motion. The dynamic model of a quadrotor can be derived using Newton-Euler equations, taking into account forces and moments acting on the vehicle body.

The position of the quadrotor is defined by the inertial coordinates $(x_I, y_I, z_I)$ in 3D space.  The Euler angles $(\phi, \theta, \psi)$ specify the orientation, representing roll, pitch, and yaw, respectively. The quadrotor motion is controlled by varying the thrusts $T_i$ of each rotor $i=1,\dotsc,4$.

The translational accelerations are given by \citep{Emami.2019c}
\begin{align}
\ddot{x}_I &=  \frac{F}{m} \left ( \cos\psi \sin\theta \cos\phi + \sin\psi \sin\phi \right ) + \bar{d}_1, \\
\ddot{y}_I &=  \frac{F}{m} \left ( \sin\psi \sin\theta \cos\phi - \cos\psi \sin\phi \right ) + \bar{d}_2, \\
\ddot{z}_I &=  \frac{F}{m} \left ( \cos\theta \cos\phi \right ) - g + \bar{d}_3.
\end{align}

Additionally, the rotational accelerations are expressed as
\begin{align}
\ddot{\phi} &= \frac{1}{I_x} \left (u_\phi + \left (I_y - I_z\right )\dot{\theta}\dot{\psi}\right ) + \bar{d}_4, \\
\ddot{\theta} &= \frac{1}{I_y} \left (u_\theta + \left (I_z - I_x\right )\dot{\phi}\dot{\psi}\right ) + \bar{d}_5, \\
\ddot{\psi} &= \frac{1}{I_z} \left (u_\psi + \left (I_x - I_y\right )\dot{\theta}\dot{\phi}\right ) + \bar{d}_6.
\end{align}

Furthermore, the control inputs are related to rotor thrusts as \begin{equation}\label{eq:exc21:FBnonlinear}
\begin{bmatrix} F \\ u_\phi \\ u_\theta \\ u_\psi \end{bmatrix}
= \begin{bmatrix}
1 & 1 & 1 & 1 \\
0 & l & 0 & -l \\
-l & 0 & l & 0 \\
c & -c & c & -c
\end{bmatrix}
\begin{bmatrix}
T_1 \\
T_2 \\
T_3 \\
T_4
\end{bmatrix},
\end{equation}
which gives
\begin{equation}\label{eq:Apdx:quad:fB}
\begin{split}
f(x) &=\begin{bmatrix}
-g\\
\dot{\theta} \dot{\psi} \frac{{\left(I_{y} -I_{z} \right)}}{I_{x} }\\
\dot{\phi} \dot{\psi} \frac{{\left(I_{z} -I_{x} \right)}}{I_{y} }\\
\dot{\theta} \dot{\phi} \frac{{\left(I_{x} -I_{y} \right)}}{I_{z} }
\end{bmatrix} \\
B(x) &=\underbrace{\begin{bmatrix}
\frac{\mathrm{cos}\phi \mathrm{cos}\theta }{m} & 0 & 0 & 0\\
0 & \frac{1}{I_{x} } & 0 & 0\\
0 & 0 & \frac{1}{I_{y} } & 0\\
0 & 0 & 0 & \frac{1}{I_{z} }
\end{bmatrix}}_{B_1}\underbrace{\begin{bmatrix}
1 & 1 & 1 & 1 \\
0 & l & 0 & -l \\
-l & 0 & l & 0 \\
c & -c & c & -c
\end{bmatrix}}_{B_2},
\end{split}
\end{equation}
where $F$ is total thrust (not to be confused with the vector $f(x)$ denoting internal system dynamics), $m$ is vehicle mass, $I_x$, $I_y$, $I_z$ are moments of inertia, $g$ is gravity acceleration, $\bar{d}_i$s represent unmodeled dynamics and disturbances, $l$ is arm length, $c = k_D/k_T$, in which $k_D$ and $k_T$ are drag and thrust coefficients obtained from experimental data of rotor, and $T_i$s are rotor thrusts, constructing the input vector $u$.

\textbf {Wind disturbance and ground effect:}
The aerodynamic disturbances acting on a quadrotor primarily consist of wind and ground effect \citep{Manconi.2022}.
The relative wind velocity $V_r$ in body frame is computed from the wind velocity $W$ in inertial frame and quadrotor velocity $V_I$ as:
\begin{align}
\begin{split}
V_r=R^T(W-V_I),
\end{split}
\end{align}
where $W=\left[W_x,W_y,W_z\right]^T$ is the wind velocity in inertial coordinates, $V_I=\left[\dot{x},\dot{y},\dot{z}\right]^T$ is the velocity in inertial coordinates, and $R$ is the rotation matrix from body to inertial frame given by
\begin{equation}\label{eq:rotationmat}
R = \begin{bmatrix}
\cos\psi\cos\theta & \cos\psi\sin\theta\sin\phi - \sin\psi\cos\phi & \cos\psi\sin\theta\cos\phi + \sin\psi\sin\phi\\
\sin\psi\cos\theta & \sin\psi\sin\theta\sin\phi + \cos\psi\cos\phi & \sin\psi\sin\theta\cos\phi - \cos\psi\sin\phi\\
-\sin\theta & \cos\theta\sin\phi & \cos\theta\cos\phi
\end{bmatrix}.
\end{equation}

Therefore, $V_r$ is defined in the body
coordinate. It is assumed that the relative velocity is the same across all areas of each rotor, which are considered as disks.

Subsequently, the relative velocity for each rotor is
\begin{align}
\begin{split}
V_1&=-l\Omega e_1+V_r, \\
V_2&=-l\Omega e_2+V_r, \\
V_3&=l\Omega e_1+V_r, \\
V_4&=l\Omega e_2+V_r. \\
\end{split}
\end{align}

In this context, the unit vectors $e_1=\left[1,\ 0,\ 0\right]^T$ and $e_2=\left[0,\ 1,\ 0\right]^T$ serve to extract specific components of the skew-symmetric matrix $\Omega$, which is expressed as:
\begin{align}
\begin{split}
\Omega=\text{skew}(\omega)=\left[\begin{matrix}0&-r&q\\r&0&-p\\-q&p&0\\\end{matrix}\right],
\end{split}
\end{align}
where $p$, $q$, and $r$ represent the rotational velocities in body axes, which make the vector $\omega$.

The following notation briefly represents the relative velocities for each rotor
\begin{align}
\begin{split}
V_{i,j}=V_i^T e_j.
\end{split}
\end{align}

\begin{figure}[!h]
  \begin{center}
    \includegraphics[width=0.7\textwidth]{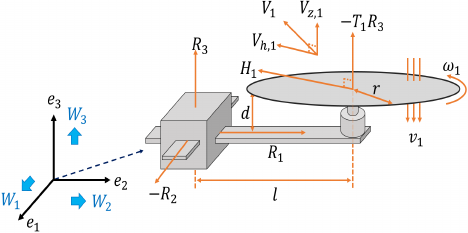}
  \end{center}
  \caption{Forces and velocities of a propeller \cite{Jeon.2020}.} 
  \label{fig:quadwind}
\end{figure}

As shown in Fig. \ref{fig:quadwind}, the relative velocity $V_i$ of each propeller moving through the air can be decomposed into two perpendicular components given by
\begin{equation}
V_{z,i} = \begin{bmatrix} 0\\ 0\\ V_{i,3} \end{bmatrix},\;
V_{h,i} = \begin{bmatrix} V_{i,1}\\ V_{i,2}\\ 0 \end{bmatrix}.
\end{equation}

The horizontal disturbance force $H$ acting on each rotor is given by
\begin{equation}\label{eq:quad_H_i}
H_i=nT_iV_{h,i},
\end{equation}
where $n$ is a positive constant. 

The ground effect refers to the increased efficiency of the rotors when a quadrotor flies close to the ground. This phenomenon occurs as the rotor blades function within their own downwash, which is confined near the ground surface.
This is modeled by defining an increased thrust in ground effect $T_{IGE}$ compared to the thrust out of ground effect $T_{OGE}$:
\begin{equation}
T_{IGE} = \frac{T_{OGE}}{1-\frac{r_b^2}{{16z}^2}},
\end{equation}
where $r_b$ is the rotor radius and $h$ is the height above ground. The ground effect ratio is defined as:
\begin{equation}
K_g = \frac{T_{IGE}}{T_{OGE}}.
\end{equation}

The horizontal disturbance force on each rotor $H_i$ in Eq. \eqref{eq:quad_H_i} is used to compute the net disturbance force as
\begin{align}
f_D = \sum_{i=1}^4 H_i
= nF\left[\begin{matrix}V_r^Te_1\\V_r^Te_2\\0\\\end{matrix}\right]+nlr\left[\begin{matrix}T_2-T_4\\T_3-T_1\\0\\\end{matrix}\right]+\left[\begin{matrix}0\\0\\K_{g}F\\\end{matrix}\right].
\end{align}

The disturbance torque on each rotor is
\begin{align}
\tau_{D,1} &= \begin{bmatrix} l, & 0, & d \end{bmatrix}^T \times H_1,\\
\tau_{D,2} &= \begin{bmatrix} 0, & l, & d \end{bmatrix}^T \times H_2,\\
\tau_{D,3} &= \begin{bmatrix} -l, & 0, & d \end{bmatrix}^T \times H_3,\\
\tau_{D,4} &= \begin{bmatrix} 0, & -l, & d \end{bmatrix}^T \times H_4,
\end{align}
where $d$ is the vertical distance of the rotor from the origin of the body frame..

Thus, the net disturbance torques are given by
\begin{align}
    \tau_D & = K_{g}cnF\left[\begin{matrix}{-V}_r^Te_2\\V_r^Te_1\\0\\\end{matrix}\right]+K_{g}cnlr\left[\begin{matrix}T_1-T_3\\T_2-T_4\\0\\\end{matrix}\right]\\
& \quad + nl\left[\begin{matrix}0\\0\\T_1V_{1,2}-T_2V_{2,1}-T_3V_{3,2}+T_4V_{4,1}\\\end{matrix}\right].
\end{align}

Finally, $\bar{d}_1, \cdots, \bar{d}_6$ can be expressed as 
\begin{align}
\begin{split}
\left[\begin{matrix}\bar{d}_1,\ \bar{d}_2,\ \bar{d}_3\end{matrix}\right]^T & = \frac{1}{m}Rf_D,
\\
\left[\begin{matrix}\bar{d}_4,\ \bar{d}_5,\ \bar{d}_6\end{matrix}\right]^T & = \tau_D.
\end{split}
\end{align}

The numerical values of quadrotor parameters are given in Table \ref{tab:quad}. This model provides the basis for developing control strategies, such as trajectory tracking algorithms, for quadrotor UAVs.
\begin{table}[H]
\caption{Numerical values for parameters of quadrotor  \citep{Emami.2019c, Manconi.2022}.}
\centering
\begin{tabular}{ll}
\hline
Parameter & Value \\
\hline
$m$ & 0.58 kg \\
$l$ & 0.25 m \\
$\left (I_x, I_y, I_z\right )$ & (0.01, 0.01, 0.02) kg$\cdot$m$^2$ \\
$k_D$ & 2.82$\times$10$^{-7}$ kg$\cdot$m$^2$ \\
$k_T$ & 1.5$\times$10$^{-5}$ kg$\cdot$m \\
$r_b$ & 0.2 m\\
$n$ & 0.0837\\
$d$ & 0.0188 m\\
\hline
\end{tabular}
\label{tab:quad}
\end{table}

\begin{figure}[!h]
  \begin{center}
    \includegraphics[width=0.45\textwidth]{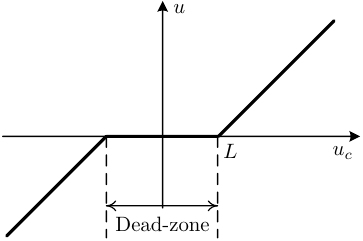}
  \end{center}
  \caption{Dead-zone nonlinearity.} 
  \label{fig:deadz}
\end{figure}

\textbf {Actuator dynamics:}
As shown in Fig. \ref{fig:deadz}, the control inputs have dead-zone nonlinearity given by
\begin{align}
u = G(u_c) &= \begin{cases}
k\left (u_c-L\right ), & u_c \geq L\\
0, & -L < u_c < L\\
k\left (u_c+L\right ), & u_c \leq -L
\end{cases},
\end{align}
where $u_c$ is the control command, $k$ is the slope of the dead-zone, and $L$ is breakpoint.
Additionally, a first order transfer function is employed with time constant $\tau$ to take account of lag in the actuator dynamics.

\section{Fixed-wing Aircraft (GTM) Model}
\label{app:GTM}
\subsection{Nonlinear Dynamic Model}
\label{app:GTM1}
\begin{figure}[H]
\centering
\includegraphics[scale=.4]{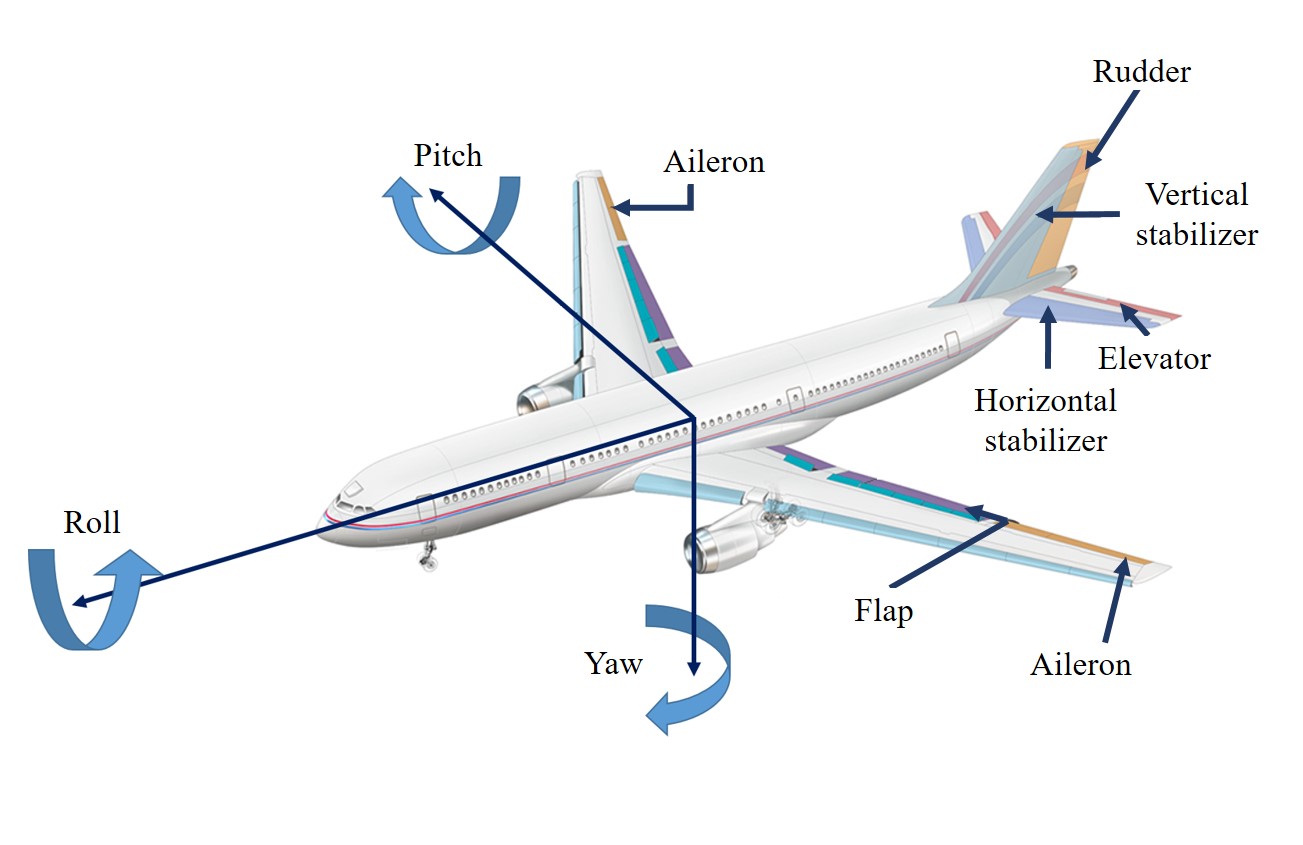}
\caption{Schematic view of a fixed-wing aircraft as well as the body-fixed frame and control surfaces}
\label{fig:FWAF}   
\end{figure}

The 6-DOF nonlinear dynamic model of a conventional fixed-wing aircraft can be derived using the Newton-Euler laws and kinematic equations. The schematic view of a conventional aircraft as well as the body-fixed frame is shown in Fig. \ref{fig:FWAF}.
The translational dynamic equations of a fixed-wing aircraft can be derived by applying Newton's second law and by expressing it in the body-fixed coordinate system, the translational dynamics model of the aircraft can be formulated as follows
\begin{align}
\begin{split}\label{Eq:TDE}
  \dot{u} = \frac{F_x}{m} - qw + rv \\
  \dot{v} = \frac{F_y}{m} - ru + pw \\
  \dot{w} = \frac{F_z}{m} - pv + qu, 
\end{split}
\end{align}
where $u$, $v$, $w$, $p$, $q$, and $r$ represent the body components of the translational and angular velocities with respect to inertial frame and expressed in the body-fixed coordinate system. Moreover, $F_x$, $F_y$, and $F_z$ are the body components of total forces acting on the aircraft in the body-fixed coordinate system.

The rotational dynamic equations of a fixed-wing aircraft can be derived by applying the Euler's law and by expressing it in the body-fixed coordinate system, the rotational dynamics model of the aircraft can be formulated as follows
\begin{align}
\begin{split}\label{Eq:RDE}
  \dot{p} &= \frac{1}{I_{xx}I_{zz}-I_{xz}^2}\left[(I_{yy}I_{zz}-I_{zz}^2-I_{xz}^2)qr\right.\\ 
  & \quad\left.-I_{xz}(I_{zz}+I_{xx}-I_{yy})pq+I_{zz}M_x-I_{xz}M_z\right] \\
    \dot{q} &= \frac{1}{I_{yy}}\left[(I_{zz}-I_{xx})pr+I_{xz}(p^2-r^2)+M_y\right] \\
    \dot{r} &= \frac{1}{I_{xx}I_{zz}-I_{xz}^2}\left[(-I_{xx}I_{yy}+I_{xx}^2+I_{xz}^2)pq\right.\\ 
  & \quad\left.+I_{xz}(I_{zz}+I_{xx}-I_{yy})rq+I_{xx}M_z-I_{xz}M_x\right], \\
\end{split}
\end{align}
where $M_x$, $M_y$, and $M_z$ are the body components of total moments acting on the aircraft in the body-fixed coordinate system and $I_{xx}$, $I_{yy}$, and $I_{zz}$ are the diagonal elements of the moment of inertia, respectively. Furthermore, $I_{xz}$ represents the off-diagonal element in the moment of inertia tensor that describes the product of inertia between the x-axis and z-axis and is commonly referred to as the "xz product of inertia", which is nonzero for conventional fixed-wing aircrafts. Nevertheless, in most fixed-wing aircraft, the value of $I_{xz}$ is significantly lower compared to the principal moments of inertia. Consequently, by disregarding it in Eq. (\ref{Eq:RDE}), a decoupled and more straightforward nominal rotational dynamics model can be obtained as follows
\begin{align}
\begin{split}\label{Eq:RDE2}
  \dot{p} &= \frac{1}{I_{xx}}\left[(I_{yy}-I_{zz})qr+M_x\right] \\
    \dot{q} &= \frac{1}{I_{yy}}\left[(I_{zz}-I_{xx})pr+M_y\right] \\
    \dot{r} &= \frac{1}{I_{zz}}\left[(I_{xx}-I_{yy})pq+M_z\right]. \\
\end{split}
\end{align}

The Euler rates can be obtained from the angular velocity components using the rotational kinematics equations as follows
\begin{equation}\label{Eq:RKE}
    \begin{bmatrix} \dot{\phi} \\  \dot{\theta}\\ \dot{\psi} \end{bmatrix}
    = \begin{bmatrix}
    1 \, & \sin\phi \, \tan\theta  \, &  \cos\phi \, \tan\theta\\
    0 \, & \cos\phi \, & -\sin\phi \\
    0 \, & \sin\phi\, \sec\theta \, & \cos\phi\, \sec\theta \\
    \end{bmatrix} 
    \begin{bmatrix} p \\  q \\ r \end{bmatrix},
\end{equation}
where $\phi$, $\theta$, and $\psi$ are Euler angles representing the roll, pitch, and yaw angles, respectively. Euler angles represent the aircraft's attitude and can be easily obtained through the integration of Euler rates. Additionally, the components of the aircraft's angular velocity in the body-fixed coordinate system ($p$, $q$, $r$) can be obtained from Eq. (\ref{Eq:RDE}).

In addition, to obtain the aircraft velocity with respect to inertial frame and expressed in the inertial reference coordinate system, the transformation matrix from the body-fixed coordinate system to the inertial reference coordinate system should be used. Consequently, the translational kinematics equations can be formulated as follows
\begin{equation}\label{Eq:TKE}
    \begin{bmatrix} \dot{x} \\  \dot{y}\\ \dot{z} \end{bmatrix}
    = R \begin{bmatrix} u \\ v \\ w \end{bmatrix},
\end{equation}
where $R$ is the rotation matrix as given in Eq. \eqref{eq:rotationmat}.
Consequently, the aircraft's position ($x$,$y$,$z$) can be easily obtained by integrating the right-hand side of Eq. (\ref{Eq:TKE}). Additionally, the components of the aircraft's translational velocity in the body-fixed coordinate system ($u$, $v$, $w$) can be obtained from Eq. (\ref{Eq:TDE}). 

Concerning the model of forces and moments on the right-hand side of Eqs. (\ref{Eq:TDE}) and (\ref{Eq:RDE}) it can be said that each component of the forces ($F_x$, $F_y$, $F_z$) and moments ($M_x$, $M_y$, $M_z$) acting on the aircraft is composed of aerodynamic, propulsive, and gravitational terms. The aerodynamic forces and moments are complex nonlinear functions of the flight condition and the deflection of control surfaces, including the elevator, rudder, and aileron. Various studies utilize a linear aerodynamic model based on stability and control derivatives, which serve as indicators revealing how the forces and moments acting on an aircraft alter with changes in flight conditions and deflection of control surfaces. In the case of conventional aircraft, these parameters are themselves nonlinear functions of the flight condition (Mach number, Reynolds number, etc.). Moreover, the propulsive forces and moments are nonlinear functions of the flight condition and the throttle setting.

To gain a deeper understanding of the modeling of aerospace vehicle dynamics, readers can also refer to \citet{zipfel.2014} for further details about reference frames, deriving equations from Newton's and Euler's laws, and the modeling of forces and moments.

A straightforward nominal affine model of rotational dynamics can be derived using Eq. (\ref{Eq:RDE2}) and a linear aerodynamic model as follows
\begin{equation} \label{Eq:ARDE}
    \underbrace{\begin{bmatrix} \dot{p} \\  \dot{q}\\ \dot{r}  \end{bmatrix} }_{\dot{x}}
    = \underbrace{\begin{bmatrix} 
    \frac{I_{yy}-I_{zz}}{I_{xx}}qr \\[0.8ex]
    \frac{I_{zz}-I_{xx}}{I_{yy}}pr \\[0.8ex]
    \frac{I_{xx}-I_{yy}}{I_{zz}}pq 
    \end{bmatrix}  }_{F}\\
    +
    \underbrace{\begin{bmatrix}
    0  & \frac{\bar{q}SbC_{l_{\delta_a}}}{I_{xx}} &  \frac{\bar{q}SbC_{l_{\delta_r}}}{I_{xx}} \\[0.8ex]
    \frac{\bar{q}S\bar{c}C_{m_{\delta_e}}}{I_{yy}} & 0 & 0 \\
    0 & \frac{\bar{q}SbC_{n_{\delta_a}}}{I_{zz}} & \frac{\bar{q}SbC_{n_{\delta_r}}}{I_{zz}} \\[0.8ex]
    \end{bmatrix} 
    }_{B}
    \underbrace{\begin{bmatrix} \delta_e \\  \delta_a \\ \delta_r \end{bmatrix}  }_{u} + \, \Delta,
\end{equation}
where $\bar{q}$, $S$, $b$, $\bar{c}$, $\delta_e$, $\delta_a$ and $\delta_r$ represent the dynamic pressure, reference area, wing span, mean aerodynamic chord, the elevator deflection, the aileron deflection, and the rudder deflection, respectively. Moreover, the stability derivatives in Eq. (\ref{Eq:ARDE}) are typically functions of the angle of attack ($\alpha$), sideslip angle  ($\beta$), and Reynolds number. The Reynolds number itself is a function of altitude and Mach number. However, in order to make it easier to implement the nominal model in the controller, one can make the assumption that stability derivatives remain constant.
The nominal rotational dynamics model Eq. (\ref{Eq:ARDE}) disregards damping derivatives such as $C_{l_p}$ and similar derivative, as well as the possible influence of propulsive moments.
Furthermore, these effects, alongside other unmodeled dynamics, are considered as uncertainties and are denoted as $\Delta$ in Eq. (\ref{Eq:ARDE}).
\subsection{GTM Simulation}
\label{app:GTM2}
\begin{figure}[H]
\centering
\includegraphics[scale=.2]{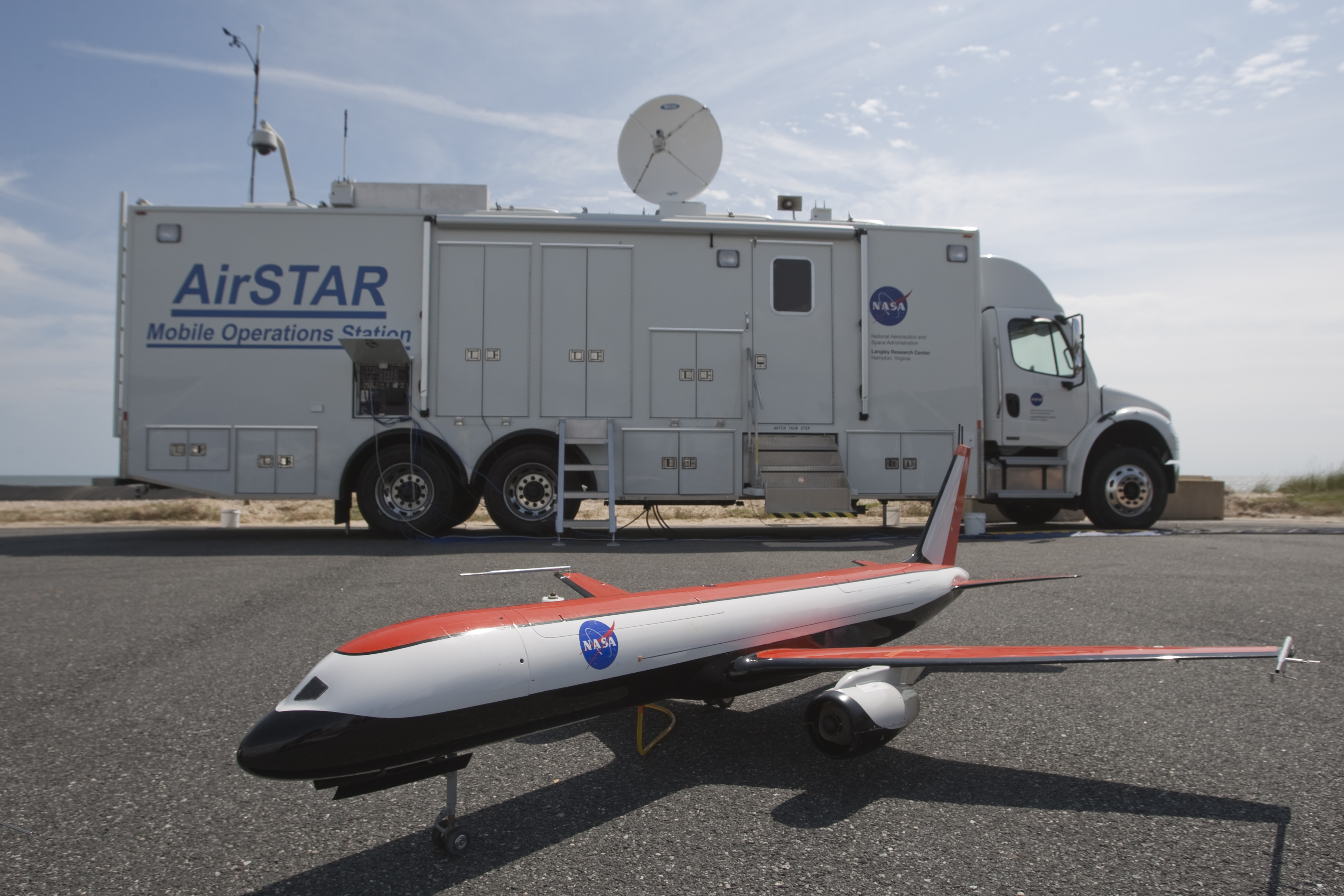}
\caption{General view of the GTM aircraft (the 5.5\% scale model) \citep{hovakimyan.2011}}
\label{fig:GTM_img}   
\end{figure}

The detailed nonlinear model of this aircraft at a 5.5-percent scale is developed by NASA Langley Research Center for nonlinear simulation experiments. The general view of the 5.5\% scale model of GTM aircraft is presented in Fig. \ref{fig:GTM_img}.
The GTM simulation has been implemented in Simulink and is available for download from \citet{GTMGit}. Furthermore, the implementation is described in \citet{Hueschen.2011}. This simulation includes an aerodynamic database that contains data for the aerodynamic body axes coefficients $C_X$, $C_Y$, $C_Z$, $C_l$, $C_m$, and $C_n$ with values defined as a function of angle of attack, sideslip angle, and Reynolds number. This simulation enables interfacing with a flight control system, utilizing sensor data, and commanding the surface actuators and throttle position of the engine model. The model also includes trim and linearization routines, which are used to generate a linearized model of aircraft dynamics around a trim condition. The GTM aircraft can be trimmed at an altitude from zero up to about 10000 ft above sea level and the EAS of about 50 knots up to 110 knots.

Linearized model of GTM  is represented in state-space form, including longitudinal and lateral-directional modes at the trim condition of 8000 ft. altitude, an EAS of 75 knots, flight path angle of $\gamma = 0$, angle of attack of $\alpha = 5.56 \, $deg throttle of $\delta_t = 42.99 \%$ and elevator deflection of $\delta_e = 0.1 \, $deg considering
$ x_{Lon} = \begin{bmatrix}u & w & q & \theta \end{bmatrix}^T$,
$ u_{Lon} = \begin{bmatrix}\delta_e & \delta_{s_L} & \delta_{s_R} & \delta_{th_L} & \delta_{th_R} \end{bmatrix}^T$,
$ x_{Lat} = \begin{bmatrix}v & p & r & \phi \end{bmatrix}^T$ and
$ u_{Lat} = \begin{bmatrix}\delta_a & \delta_r & \delta_{s_L} & \delta_{s_R} \end{bmatrix}^T$ 
is obtained as follows
\begin{equation}
    \begin{aligned}
    A_{Lon}
    &= \begin{bmatrix}
    -0.03017 \, & 0.1047  \, &  -13.07 \, & -32.07 \\
    -0.2407 \, & -2.135  \, &  136.4 \, & -2.211 \\
    0.01834 \, & -0.2764  \, &  -2.93 \, & 0.1193 \\
    0 \, & 0  \, &  1 \, & 0
    \end{bmatrix} \\
    B_{Lon} 
    &= \begin{bmatrix}
    -0.008233 \, & -0.0283  \, &  -0.0283 \, &  0.05902 \, &   0.05906 \\
    -0.5262 \, & 0.1943  \, &  0.1943 \, &  -0.00204 \, & -0.002011 \\
   -0.6891 \, & 0.01286  \, & 0.01397 \, &   0.007931 \, & 0.00791 \\
    0 \, & 0  \, &  0  \, &  0 \, &  0
    \end{bmatrix} \\
    C_{Lon} 
    &= \begin{bmatrix}
    1 \, & 0  \, &  0 \, & 0 \\
    0 \, & 1  \, &  0 \, & 0 \\
    0 \, & 0  \, &  1 \, & 0 \\
    0 \, & 0  \, &  0 \, & 1 \\
    \end{bmatrix},  \, \, \,
    D_{Lon} 
     = \begin{bmatrix}
    0 \, & 0  \, &  0 \, & 0 \, & 0 \\
    0 \, & 0  \, &  0 \, & 0 \, & 0 \\
    0 \, & 0  \, &  0 \, & 0 \, & 0 \\
    0 \, & 0  \, &  0 \, & 0 \, & 0 \\
    \end{bmatrix},
    \end{aligned}
\end{equation}

\begin{equation}
    \begin{aligned}
    A_{Lat}
    &= \begin{bmatrix}
    -0.446 \, & 13.98  \, &  -141.2\, & 31.95 \\
    -0.5779 \, & -5.159  \, &   1.952  \, & -1.662 \\
     0.1823  \, & -0.4267  \, &  -1.191 \, & -0.08073 \\
    0 \, & 1  \, &   0.09732 \, & 0
    \end{bmatrix} \\
    B_{Lat} 
    &= \begin{bmatrix}
     -0.01717 \, &  0.3723  \, &  0.01016 \, &  -0.01016 \\
    -0.7974 \, & 0.2039  \, &  -0.4289 \, &  0.4359\\
    -0.04728 \, & -0.3986 \, &  -0.049 \, &  0.04956\\
    0 \, & 0  \, &  0  \, &  0
    \end{bmatrix} \\
    C_{Lat} 
    &= \begin{bmatrix}
    1 \, & 0  \, &  0 \, & 0 \\
    0 \, & 1  \, &  0 \, & 0 \\
    0 \, & 0  \, &  1 \, & 0 \\
    0 \, & 0  \, &  0 \, & 1 \\
    \end{bmatrix},  \, \, \,
    D_{Lat} 
     = \begin{bmatrix}
    0 \, & 0  \, &  0 \, & 0  \\
    0 \, & 0  \, &  0 \, & 0  \\
    0 \, & 0  \, &  0 \, & 0  \\
    0 \, & 0  \, &  0 \, & 0 \\
    \end{bmatrix}. \\
    \end{aligned}
\end{equation}
Additionally, the longitudinal and lateral-directional modes of the GTM model under the given trim condition are summarized in the Table \ref{tab:GTMMode}
\begin{table}[H]
\caption{longitudinal and lateral-directional modes of the GTM}
\centering
\begin{tabular}{c c c c}
\hline\noalign{\smallskip}
\multirow{2}{*}{Mode} & \multirow{2}{*}{Poles} & Natural Frequency & Damping Ratio\\
 & & $\omega_{n}$ (rad/s) &  $\zeta$ \\
\hline\noalign{\smallskip}
Phugoid & $-0.0211 \pm 0.2758 i$ & 0.2766 & 0.0761\\
Short Period & $-2.5264 \pm 6.1310 i$ & 6.6311 & 0.3810 \\
Dutch Roll & $-0.7101 \pm 5.9641 i$ & 6.0062 & 0.1182\\
Roll & $-5.0598$ & - & -\\
Spiral & $-0.3162$ & - & -\\
\hline
\end{tabular}
\label{tab:GTMMode}
\end{table}

\section{Satellite}\label{app:sat}

\begin{figure}[b]
    \centering
    \includegraphics[width = 0.5\textwidth]{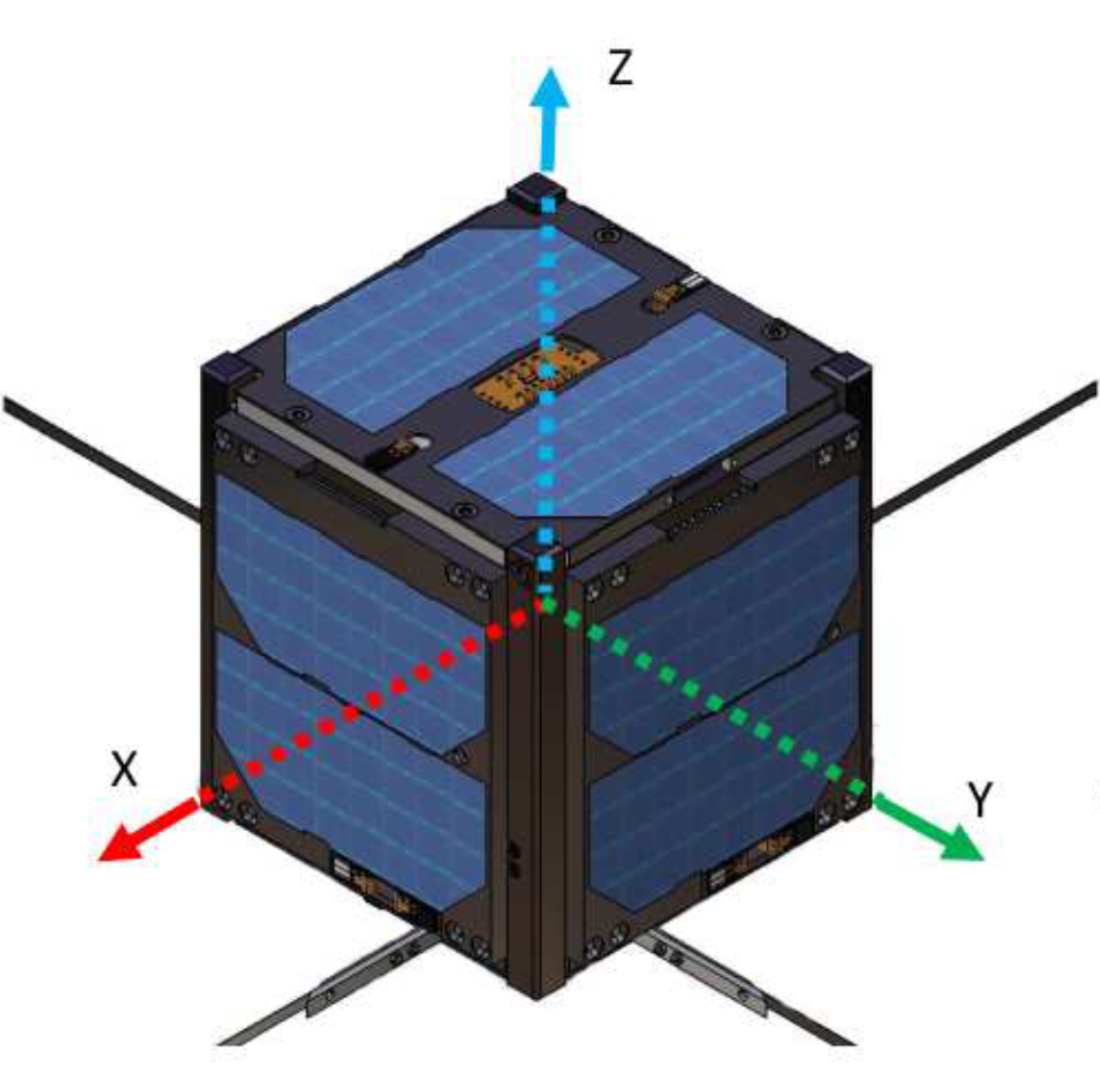}
    \caption{A schematic of a cubesat with body axes \citep{Hoffmann.2023}.}
    \label{fig:sat_cubesat}
\end{figure}

A schematic view of a cubesat is shown in Fig. \ref{fig:sat_cubesat}. To express kinematics equation of a satellite, Euler angles are considered. In addition using Euler's law, kinetics equation of a satellite could be expressed. To find the attitude equation of a satellite, using small angles approximation, we have
\begin{equation} \label{assump}
\dot{x} = \omega,
\end{equation}
where $ \omega = \mqty[p & q & r]^T $. Since the uncertainty is considered in the satellite's moment of inertia, we have
\begin{equation}
J = J_0 + \Delta J,
\end{equation}
where $ J_0 $ is the nominal value of the moment of inertia and $ \Delta J $ is the uncertainty term. Using Euler's equation
\begin{equation}
\left( J_0 + \Delta J \right) \dot{\omega} + \Omega \left( J_0 + \Delta J \right) \omega = u + d,
\end{equation}
where $ \Omega $, $ u $, and $ d $ are skew-symmetric matrix of $ \omega $, control torques, and external disturbances, respectively. Using Eq. \ref{assump}, the above equation is rewritten as follows
\begin{equation}
\left( J_0 + \Delta J \right) \ddot{x} + \text{skew}(\dot{x}) \left( J_0 + \Delta J \right) \dot{x} = u + d,
\end{equation}
where
\begin{equation}
\text{skew}(\dot{x}) = \left[ \dot{x} \times \right] =\begin{bmatrix} 0 & -\dot{x}_3 & \dot{x}_2 \\ \dot{x}_3 & 0 & -\dot{x}_1 \\ -\dot{x}_2 & \dot{x}_1 & 0 \end{bmatrix}.
\end{equation}
Rewriting the previous equation, we have
\begin{equation}
\ddot{x} = J_0^{-1} u - J_0^{-1} \text{skew}(\dot{x}) J_0 \dot{x} - J_0^{-1} \Delta J \ddot{x} - J_0^{-1} \text{skew}(\dot{x}) \Delta J \dot{x} + J_0^{-1} d.
\end{equation}
Defining the following variables
\begin{subequations}
\begin{equation}
B = J_0^{-1},
\end{equation}
\begin{equation}
F = -J_0^{-1} \text{skew}(\dot{x}) J_0 \dot{x},
\end{equation}
\begin{equation}
\Delta = -J_0^{-1} \Delta J \ddot{x} - J_0^{-1} \text{skew}(\dot{x}) \Delta J \dot{x} + J_0^{-1} d,
\end{equation}
\end{subequations}
Now we can write the attitude equation of satellite as follows
\begin{equation} \label{system}
\ddot{x} (t) = F (\dot{x}) + B u + \Delta (x, \dot{x}).
\end{equation}
For all satellite-related examples, the values in Table \ref{SattPar1} values have been considered for simulation
\begin{table}[H]\label{SattPar1}
    \centering
    \caption{Model and simulation parameters of satellite example.}
    \begin{tabular}{l c}
    \hline\noalign{\smallskip}
        {Parameter} & {Value} \\ 
        \noalign{\smallskip}\hline\noalign{\smallskip}
         $J_0$ & ${\mqty[\dmat[0]{100, 120, 150}]}$ \\[0.5cm]
         $ \Delta J $ & $ {\mqty[\dmat[0]{4, 5, 6}]} $ \\[0.5cm]
         $ d $ & $ {\mqty[0.05 \sin \left( \frac{\pi t}{2} \right) & 0.05 \cos \left( \frac{\pi t}{3} \right) & 0.05 \sin \left( \frac{2 \pi t}{5} \right) ]^T} $ \\ 
         \noalign{\smallskip}\hline\noalign{\smallskip}
    \end{tabular}
    \label{tab:sat_params}
\end{table}
\section{HFV}
\label{sec:HFV}
Hypersonic flight vehicles (HFVs) fly at extremely high speeds, posing significant modeling and control challenges. To enable effective controller design, the complex nonlinear dynamics are often simplified into a control-oriented model (COM). This section provides an overview of the COM approach for modeling the longitudinal dynamics of an air-breathing HFV.

\subsection{Model Derivation}
The COM of the hypersonic longitudinal dynamics is given by \citet{Parker.2007}. This model represents the longitudinal dynamics with five states $X = [V, h, \alpha, \gamma, q]^T$ and two control inputs $U = [\delta_e, \Phi]^T$, where $V$ is velocity, $h$ is altitude, $\alpha$ is the angle of attack, $\gamma$ is flight path angle (FPA), $q$ is pitch rate, $\delta_e$ is elevator deflection and $\Phi$ is fuel equivalence ratio.


The equations of motion are derived by considering the forces and moments acting on the vehicle \citep{Xu.2018}

\begin{align}
\dot{V} &= \frac{T\cos\alpha - D}{m} - g\sin\gamma, \\
\dot{h} &=V \sin\gamma, \\
\dot{\gamma} &= \frac{L + T\sin\alpha}{mV} - \frac{g}{V}\cos\gamma,  \label{Eq:A4:hfv_com_3} \\
\dot{\alpha} &= q - \dot{\gamma}, \\
\dot{q} &= \frac{M_{yy}}{I_{yy}},
\end{align}

\noindent where the aerodynamic forces and moments are modeled as \citep{Xu.2015}

\begin{align}
T &= T_\Phi(\alpha)\Phi + T_0(\alpha) \\
 &\approx [\beta_1 \Phi + \beta_2 ]\alpha^3 + [\beta_3 \Phi + \beta_4 ]\alpha^2 + [\beta_5 \Phi + \beta_6 ]\alpha + [\beta_7 \Phi + \beta_8 ], \\
D &\approx \bar{q}S(C_{D_0} + C_{D_\alpha}\alpha + C_{D_{\alpha^2}}\alpha^2), \\
L &\approx \bar{q}SC_{L_0} + \bar{q}SC_{L_\alpha}\alpha, \\
M_{yy} &\approx z_TT + \bar{q}S\bar{c}(C_{M_0} + C_{M_\alpha}\alpha + C_{M_{\alpha^2}}\alpha^2) + \bar{q}S\bar{c}C_{M_{\delta_e}}\delta_e, \\
\bar{q} &= \frac{1}{2}\rho V^2, \\
\rho &= \rho_0 \exp \left [ -\frac{h-h_0}{h_s} \right].
\end{align}

\noindent The thrust $T$ is a function of $\alpha$ and $\Phi$, fitted using curve parameters $\beta_i$, which are functions of altitude $h$ and dynamic pressure $\bar{q}$. The aerodynamic coefficients $C_{D_i}$, $C_{L_i}$, $C_{M_i}$ are considered known. This approximates the complex aerodynamics with simple polynomial models.

\subsection{Dynamics Transformation}
Since $\gamma$ is small during the cruise, the approximation $\sin\gamma \approx \gamma$ is applied \citep{Xu.2013}. Also, the thrust term $T\sin\alpha$ in Eq. (\ref{Eq:A4:hfv_com_3}) is neglected compared to lift $L$.

Applying these approximations results in the transformed dynamics used for controller design.
To this end, the dynamics are transformed into a cascaded form given by \citep{Xu.2015}
\begin{align}
\dot{V} &= f_v + g_v u_v, \\
\dot{x}_1 &= f_1(x_1) + g_1(x_1) x_2, \\
\dot{x}_2 &= f_2(x_1,x_2) + g_2(x_1,x_2) x_3, \\
\dot{x}_3 &= f_3(x_1,x_2,x_3) + g_3(x_1,x_2,x_3) \delta_e,
\end{align}
where $x_1 = \gamma$, $x_2 = \theta_p$, $x_3 = q$, and $u_v = \Phi$.

This transforms the original 5th order system into a SISO velocity subsystem and a 3rd order attitude subsystem in strict feedback form.

\subsubsection{Velocity Subsystem}
The velocity dynamics are typically much slower than the attitude dynamics. This enables a \emph{time-scale decomposition}, where the fast attitude subsystem is designed assuming the slow velocity state is constant.
The velocity subsystem is given by
\begin{align}
\dot{V} &= f_v + g_v u_v, \\
u_v &= \Phi,
\end{align}
where 
\begin{align}
f_v &= \frac{T_0\cos\alpha - D}{m} - g\sin\gamma, \\
g_v &= \frac{T_\Phi\cos\alpha}{m}.
\end{align}

This represents the engine thrust acting on the velocity state.

\subsubsection{Attitude Subsystem}
The attitude subsystem is given by
\begin{align}
\dot{x}_1 &= f_1(x_1) + g_1(x_1) x_2, \\
\dot{x}_2 &= f_2(x_1,x_2) + g_2(x_1,x_2) x_3, \\
\dot{x}_3 &= f_3(x_1,x_2,x_3) + g_3(x_1,x_2,x_3) \delta_e,
\end{align}
where
\begin{align}
f_1 &= \frac{L_0 - L_\alpha \gamma}{mV} - \frac{g}{V}\cos x_1, \\
g_1 &= \frac{L_\alpha}{mV}, \\
f_2 &= 0, \\
g_2 &= 1, \\
f_3 &= \frac{M_T + M_0(\alpha)}{I_{yy}}, \\
g_3 &= \frac{M_{\delta_e}}{I_{yy}},
\end{align}
where the dynamics capture the effect of flight path, pitch angles, and elevator on the angle rates. This represents the aerodynamic forces and moments acting on the attitude states.
The transformed model facilitates backstepping controller design for the velocity and attitude subsystems.

\subsection{Numerical values}
Table \ref{table:params} provides values for reference area, air density, altitudes, moments of inertia, mass, and gravity used in the model.
\begin{table}[h!]
\centering
\caption{Miscellaneous Coefficient Values \citep{Parker.2007}.}
\begin{tabular}{lcc}
\hline
Coefficient & Value & Units \\ \hline
S & 1.7000 $\times$ 10$^{1}$ & ft$^{2}$ \\
$\rho_{0}$ & 6.7429 $\times$ 10$^{-5}$ & slugs $\cdot$ ft$^{-3}$ \\
$h_{0}$ & 8.5000 $\times$ 10$^{4}$ & ft \\
$h_{s}$ & 2.1358 $\times$ 10$^{4}$ & ft \\
$I_{yy}$ & 5.0000 $\times$ 10$^{5}$ & lb $\cdot$ ft \\
m & 3.0000 $\times$ 10$^{2}$ & lb $\cdot$ ft$^{-1}$ \\
g & 3.2000 $\times$ 10$^{1}$ & ft $\cdot$ s$^{-2}$ \\
\hline
\end{tabular}
\label{table:params}
\end{table}

Table \ref{table:lift} gives the lift slope, elevator effectiveness, and zero angle of attack lift coefficients that appear in the lift force equation.
\begin{table}[h!]
\centering
\caption{Lift Coefficient Values \citep{Parker.2007}.}
\begin{tabular}{lcc}
\hline
Coefficient & Value & Units \\ \hline
$C_{L_\alpha}$ & 4.6773 $\times$ 10$^{0}$ & rad$^{-1}$ \\
$C_{L_{\delta_e}}$ & 7.6224 $\times$ 10$^{-1}$ & rad$^{-1}$ \\
$C_{L_0}$ & -1.8714 $\times$ 10$^{-2}$ & - \\
\hline
\end{tabular}
\label{table:lift}
\end{table}

Table \ref{table:drag} provides the quadratic drag, drag due to angle of attack, elevator effectiveness, and base drag coefficients used in the drag force model.
\begin{table}[ht!]
\centering
\caption{Drag Coefficient Values \citep{Parker.2007}.}
\begin{tabular}{lcc}
\hline
Coefficient & Value & Units \\ \hline
$C_{D_{\alpha^2}}$ & 5.8224 $\times$ 10$^{0}$ & rad$^{-2}$ \\
$C_{D_\alpha}$ & -4.5315 $\times$ 10$^{-2}$ & rad$^{-1}$ \\
$C_{D_{\delta_e^2}}$ & 8.1993 $\times$ 10$^{-1}$ & rad$^{-2}$ \\
$C_{D_{\delta_e}}$ & 2.7699 $\times$ 10$^{-4}$ & rad$^{-1}$ \\
$C_{D_0}$ & 1.0131 $\times$ 10$^{-2}$ & - \\
\hline
\end{tabular}
\label{table:drag}
\end{table}

Table \ref{table:moment} lists values for the engine moment arm, mean chord, and quadratic, linear, and constant pitch moment coefficients used in the pitching moment equation.
\begin{table}[ht!]
\centering
\caption{Moment Coefficient Values \citep{Parker.2007}.}
\begin{tabular}{lcc}
\hline
Coefficient & Value & Units \\ \hline
$z_T$ & 8.3600 $\times$ 10$^{0}$ & ft \\
$\bar{c}$ & 1.7000 $\times$ 10$^{1}$ & ft \\
$C_{M_{\alpha^2}}$ & 6.2926 $\times$ 10$^{0}$ & rad$^{-2}$ \\
$C_{M_\alpha}$ & 2.1335 $\times$ 10$^{0}$ & rad$^{-1}$ \\
$C_{M_0}$ & 1.8979 $\times$ 10$^{-1}$ & - \\
$C_{M_{\delta_e}}$ & -1.2897 $\times$ 10$^{0}$ & rad$^{-1}$ \\
\hline
\end{tabular}
\label{table:moment}
\end{table}

Table \ref{table:thrust} gives the polynomial curve fit coefficients used to model thrust as a function of angle of attack and fuel equivalence ratio.
\begin{table}[ht!]
\centering
\caption{Thrust curve fit coefficients \citep{Parker.2007}.}
\begin{tabular}{lcc}
\hline
Coefficient & Value & Units \\ \hline
$\beta_1$ & -3.7693 $\times$ 10$^{5}$ & lb$\cdot$ft$^{-1}$$\cdot$rad$^{-3}$ \\
$\beta_2$ & -3.7225 $\times$ 10$^{4}$ & lb$\cdot$ft$^{-1}$$\cdot$rad$^{-3}$ \\
$\beta_3$ & 2.6814 $\times$ 10$^{4}$ & lb$\cdot$ft$^{-1}$$\cdot$rad$^{-2}$ \\
$\beta_4$ & -1.7277 $\times$ 10$^{4}$ & lb$\cdot$ft$^{-1}$$\cdot$rad$^{-2}$ \\
$\beta_5$ & 3.5542 $\times$ 10$^{4}$ & lb$\cdot$ft$^{-1}$$\cdot$rad$^{-1}$ \\
$\beta_6$ & -2.4216 $\times$ 10$^{3}$ & lb$\cdot$ft$^{-1}$$\cdot$rad$^{-1}$ \\
$\beta_7$ & 6.3785 $\times$ 10$^{3}$ & lb$\cdot$ft$^{-1}$ \\
$\beta_8$ & -1.0090 $\times$ 10$^{2}$ & lb$\cdot$ft$^{-1}$ \\
\hline
\end{tabular}
\label{table:thrust}
\end{table}





%

\backmatter
\bibliographystyle{unsrt}
\bibliography{Ref}

\end{document}